\documentclass[aps,amsfonts,reprint,tightenlines,amssymb,superscriptaddress,twocolumn]{revtex4-2}                                       
\usepackage[T1]{fontenc}       
\usepackage{amsfonts} 
\usepackage{amsmath,amsbsy,amssymb,graphicx,enumerate,latexsym,bm,ulem,multirow,listings,here,qcircuit}                      
\usepackage{times}   
\usepackage{color}

\usepackage{qcircuit} 

\usepackage[pagebackref=false]{hyperref} 
\hypersetup{
    unicode=false,          
    pdftoolbar=true,        
    pdfmenubar=true,        
    pdffitwindow=false,     
    pdfstartview={FitH},    
    pdfkeywords={keyword1} {key2} {key3}, 
    pdfnewwindow=true,      
    colorlinks=true,       
    linkcolor=red,          
    citecolor=blue,        
    filecolor=magenta,      
    urlcolor=cyan,           
}

\begin{document}                 
\title{Quantum Error Mitigation via Quantum-Noise-Effect Circuit Groups } 
\author{Yusuke~Hama$^\ast$}  
\affiliation{Quemix Inc., 2-11-2 Nihombashi, Chuo-ku, Tokyo 103-0027, Japan}  
\author{Hirofumi~Nishi}  
\affiliation{
Laboratory for Materials and Structures, Institute of Innovative Research, Tokyo Institute of Technology, Yokohama 226-8503, Japan}
\affiliation{Quemix Inc., 2-11-2 Nihombashi, Chuo-ku, Tokyo 103-0027, Japan} 
\date{\today}
\maketitle
\section{Abstract}\label{Abs}  
 Near-term quantum computers have been built as intermediate-scale quantum devices and are fragile against quantum noise effects, namely, NISQ devices. Traditional quantum-error-correcting codes are not implemented on such devices and to perform quantum computation in good accuracy with these machines we need to develop alternative approaches for mitigating quantum computational errors. In this work, we propose quantum error mitigation (QEM) scheme for quantum computational errors which occur due to couplings with environments during gate operations, i.e., decoherence.
To establish our QEM scheme, first we estimate the quantum noise effects on single-qubit states and represent them as groups of quantum circuits, namely, quantum-noise-effect circuit groups.  
Then our QEM scheme is conducted by subtracting expectation values generated by the quantum-noise-effect circuit groups from that obtained by the quantum circuits for the quantum algorithms under consideration. 
As a result, the quantum noise effects are reduced, and we obtain approximately the ideal expectation values via the quantum-noise-effect circuit groups and the numbers of elementary quantum circuits composing them scale polynomial with respect to the products of the depths of quantum algorithms and the numbers of register bits. To numerically demonstrate the validity of our QEM scheme, we run noisy quantum simulations of qubits under amplitude damping effects for four types of quantum algorithms. 
Furthermore, we implement our QEM scheme on  IBM Q Experience processors and examine its efficacy.
Consequently, the validity of our scheme is verified via both the quantum simulations and the quantum computations on the real quantum devices.
Our QEM scheme is solely composed of quantum-computational operations (quantum gates and measurements), and thus, it can be conducted by any type of quantum device. In addition, it can be applied to error mitigation for many other types of quantum noise effects as well as noisy quantum computing of long-depth quantum algorithms.      
\section{Introduction}\label{Intro}   
 The research and development of quantum computers are currently an important and active field of quantum information science and technology \cite{QCFeynman,QCDeutsch,QCLloyd,DiVincenzoQC,QCQINandC,QSRMP2014,linke2017experimental,SCQRPP2017,
 SCQNISQ20191,SCQARCMP2020,ZhugroupSQC2020,trappedionNISQ2019,AVSQSiontrapQC2020}. 
On the one side, quantum computer devices have been engineered with state-of-the-art technologies using various kinds of elements including superconducting circuits  \cite{NakamuraTsaigroupSC,SCQRMP2001,circuitqedreview1,linke2017experimental,SCQRPP2017,SCQNISQ20191,SCQARCMP2020,ZhugroupSQC2020,TsaigroupSCCQC2021} and trapped ions  \cite{CiracZollergate,TionSMgate,trappedionsreview1A,trappedionsreview1B,trappedionsreview2,linke2017experimental,trappedionNISQ2019,AVSQSiontrapQC2020}.
On the other side, toward the application to, for example,  material science, quantum chemistry, optimization problems, and quantum machine learning,
many new kinds of quantum algorithms have been recently developed such as Variational Quantum Eigensolver (VQE) \cite{VQEnc2014,VQE0,VQEnature2017,QCchemistryRMP2020,hybridQCalgorithmJPSJ2021}, 
Quantum Approximate Optimization Algorithm (QAOA) \cite{QAOA2014,crooks2018performance,wang2018quantum,shaydulin2019evaluating,zhou2020quantum, hybridQCalgorithmJPSJ2021,TIsing3}, and quantum circuit learning  \cite{hybridQCalgorithmJPSJ2021,SchuldandPetruccionegroupQML,MitaraietalQML,QSTQML2019}.
These algorithms have characteristics such that they are constructed by the hybridization between quantum and classical computational procedures. Recently, in the task of sampling random quantum circuits, quantum supremacy has been demonstrated using the superconducting circuit device \cite{Qsup}. All these facts are implying important milestones for the advancement of the research and development of the quantum computers
and the broadening of quantum-computing applications to many fields of science and engineering.           

While the above successful results of the research and development of quantum computers have been reported,     
near-term quantum computers based on circuit models have been built as intermediate-scale quantum devices yet and are fragile against quantum noise effects: they are called noisy intermediate-scale quantum (NISQ) devices \cite{PreskillNISQ2018,SCQARCMP2020,trappedionNISQ2019}. 
Quantum noise effects (decoherence) are major obstacles for performing quantum computation and historically many great efforts have been made on reducing such effects \cite{EkertgroupQCdissipation,resch2021benchmarking}. 
One of the traditional and representative schemes for this is the quantum-error-correction (QEC) coding \cite{ShorPRAQEC1995,QCQINandC,NemotogroupQEC,lidar2013quantum,QECRoffe,SCQRPP2017,SCQARCMP2020,ZhugroupSQC2020,TsaigroupSCCQC2021}. 
Another important one is the dynamical decoupling which plays fundamental role in extending coherence times of qubits \cite{viola1998dynamical,viola1999dynamical,khodjasteh2005fault,lidar2013quantum,masuyama2018extending,SCQNISQ20191,trappedionNISQ2019,TsaigroupSCCQC2021}.
The QEC codes are, however, not implemented on NISQ devices and to obtain quantum computational results in good accuracy with NISQ devices we need to search for alternative approaches for mitigating quantum noise effects. 
This research field is called quantum error mitigation (QEM), and these days, it is one of the important themes of the research and development of quantum computation 
\cite{EMPRL2017,EMNature2019,EMPRX2017,EMPRX2018,EMarxiv2018,PhysRevA.98.062339,mcardle2019error,jattana2020general,xiong2020sampling,zlokapa2020deep,EMPRA2021,CandSQEMPRAp2021,QCchemistryRMP2020,hybridQCalgorithmJPSJ2021,OttenGrayQEM1,OttenGrayQEM2,QSEQEM, CliffordQEM,LearningBasedQEM,VirtualDistillationQEM,koczor2021exponential,PRXQuantum.2.010316,piveteau2021error,lostaglio2021error,suzuki2022quantum,piveteau2022quasiprobability, pascuzzi2022computationally, takagi2021optimal, larose2022mitiq, koczor2021dominant,cai2022quantum}.The difficulty of the treatment of quantum noises (e.g., amplitude damping, phase damping (dephasing), depolarizing channel) is that we cannot directly construct their inverse processes by quantum gates  
due to their non-unitarity.
On the other hand, it is possible to formulate quantum noise effects as quantum circuits by using ancilla bits and measurements on them \cite{QCQINandC,NorigroupNEQPRA2011,KaisgroupOpenQ2020,openQHubbard,openQsimnpj2020,drivendisspativePRB2020,koppenhofer2020quantum,de2021quantum,hama2020quantum}.
By utilizing the quantum circuits representing the quantum noise effects under consideration, we expect that we can establish QEM schemes for reducing such effects.
If this is established, we become able to mitigate the quantum noise effects by the gate operations and measurements, i.e., QEM conducted by all-quantum-computational operations.
In other words, we become able to programmably run quantum algorithms with mitigating the quantum noise effects solely by the quantum computational operations and realize high-accurate quantum computation.   
\begin{figure*}[!htb]  
\centering
\includegraphics[width=0.9 \textwidth]{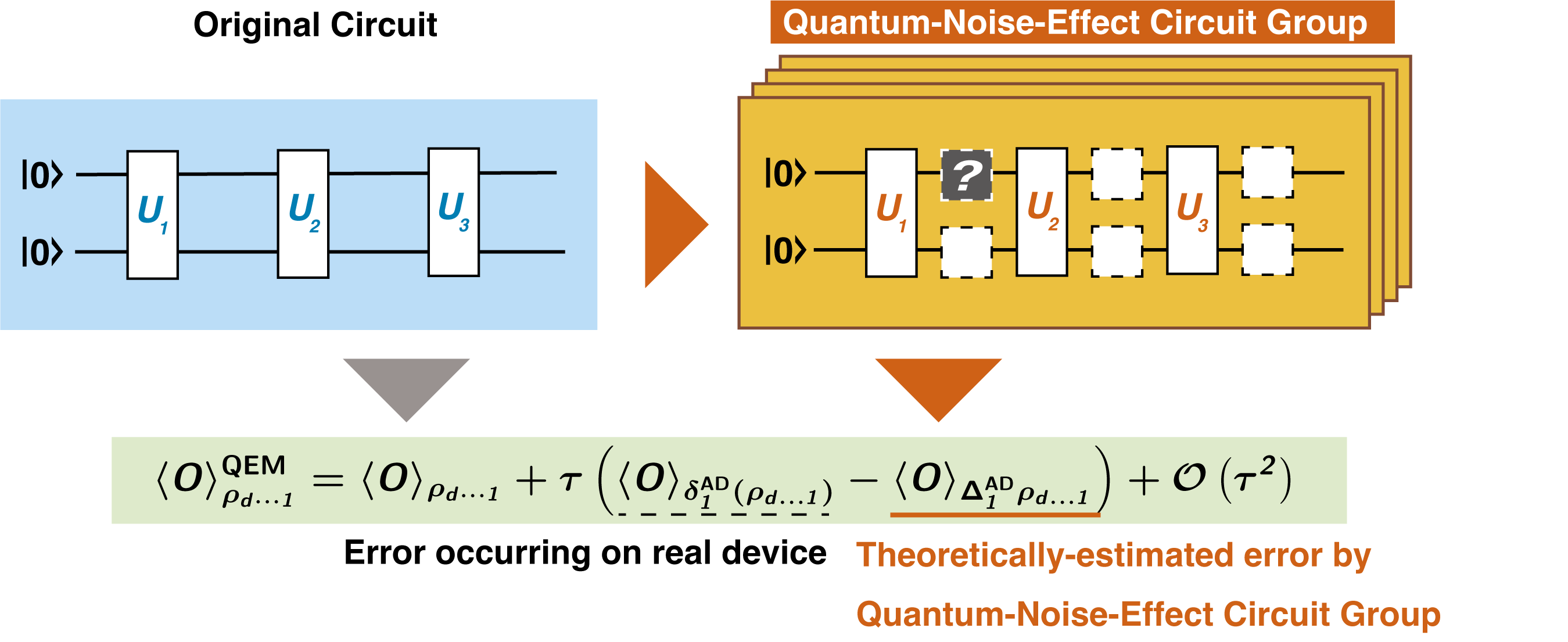}
\caption{ Schematic of our proposed QEM method. The original circuit represented by the blue rectangle (left side) describes the quantum circuit for the quantum algorithm to be ran and is composed of the unitary operations $U_k$
with $k=1,\ldots,d$ and $d$ denotes the depth of the quantum algorithm. It yields the expectation value $ \langle \hat{O}  \rangle_{ \rho^\text{real}_{d\cdots1}} $.
On the other hand, 
the quantum-noise-effect circuit group, which is represented by the orange rectangles (right side), is constructed from the original circuit by inserting an additional operation between $U_k$ and $U_{k+1}$ (gray box).
 It yields the theoretically-estimated quantum computational error $\langle \hat{O} \rangle_{(\Delta_1^{\text{AD}} \rho_{d\ldots 1})^{\text{real}}}$.
 By using these two expectation values, we obtain the equation for our QEM scheme in the green rectangle. Here we have taken  $d=3.$}
\label{QEMconcept}
\end{figure*}

In this work, we propose our QEM schemes for quantum computational errors which occur owing to couplings with environments (decoherence) during gate operations:
errors of state preparation (initialization) and measurement, imperfections of quantum gates, and cross talks among qubits are not taken into account. 
In particular, we make detailed analysis on quantum computational errors generated by amplitude damping (AD) of single-qubit states.
We show the schematic representation of our QEM scheme in Fig. \ref{QEMconcept} and it consists of two elements, the quantum circuit for the quantum algorithm under consideration (original circuit) represented by the blue rectangle and the ensemble of quantum circuits which yields the theoretical value of the quantum computational error due to the quantum noise effect, namely, quantum-noise-effect circuit group and is represented by the orange rectangles. 
By utilizing the quantum-noise-effect circuit groups, 
we formulate our QEM scheme as a perturbation theory with respect to a strength(s) of quantum noise(s) and perform it by subtracting the expectation values given by the quantum-noise-effect circuit groups from those generated by the quantum circuits for the quantum algorithms under consideration as expressed by the formula in the green rectangle; see also the right-hand side of the first line in Eq. \eqref{QEMformula1}.
As a result, the quantum noise effects are  mitigated and we approximately obtain the ideal expectation values.
Then, we discuss the numbers of elementary quantum circuits which compose the quantum-noise-effect circuit groups and show that they scale polynomial (linear) with respect to the products of the numbers of register bits 
 and the depths of the quantum algorithms (circuit depths or the numbers of unitary gates composing the quantum algorithm).  
 Finally, we numerically demonstrate the validity of our QEM scheme by running noisy quantum simulators of qubits under the AD effects for four types of quantum algorithms in the linear-order perturbation regime.
 The detailed explanation on how to extend our QEM scheme to other kinds of quantum noise effects including phase damping, generalized amplitude damping (thermalization), and depolarizing channel, 
 and extension of our QEM scheme to higher-order quantum noise effects are given in Supplementary Material.
  
The structure of this paper is given as follows. 
It begins by Sec. \ref{QEMS} with our modeling of the quantum computation under the influence of the quantum noise effect. 
After then we explain the formalisms of our QEM scheme.   
In Sec. \ref{nss}, which presents our main results, we demonstrate numerically our QEM schemes for the noisy quantum simulations for four types of quantum algorithms.
These simulations are done by both our original numerical code and Qiskit \cite{Qiskit}.
In Sec. \ref{QEMimplementation}, we discuss our quantum computation results for our QEM scheme run on the IBM Q Experience processors \cite{IBMQExp}. 
In Sec. \ref{sec:comparison_with_other_methods}, we make comparisons between our scheme and other QEM methods. 
Sec. \ref{conclusion} is devoted to the conclusion of this paper.

\section{QEM Schemes}\label{QEMS}   
\subsection{Modeling and Formulation}\label{MandF} 
\begin{figure}[h] 
\includegraphics[width=0.4 \textwidth]{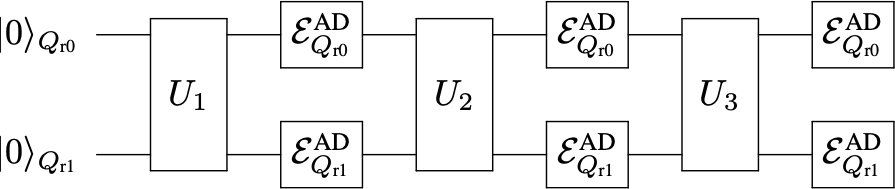}
\caption{Illustration of quantum computation under AD effects represented as a quantum circuit. Here we show it for $d=3$ and $N_q =2.$ 
The symbol $\mathcal{E}^{\text{AD}} _{ Q_{ \text{r}j}}$ expresses the occurrence of AD effect on the register bit $Q_{ \text{r}j} $ (the $j$-th register bit).   
 }  
\label{QCunderAD1} 
\end{figure}

Let us explain our modeling of quantum computation under the influence of quantum noise effects \cite{QCchemistryRMP2020,EMPRL2017,hybridQCalgorithmJPSJ2021,CandSQEMPRAp2021,OttenGrayQEM1,Qiskit}.
 In the following, we focus on the amplitude damping (AD) effect: generalized-amplitude-damping (GAD) effect at zero temperature. 
 As discussed later, it is straightforward to generalize the argument for the AD effect to other quantum noise effects such as phase damping (PD) and stochastic Pauli noises.
We schematically represent such a circumstance as a quantum circuit and show it in Fig. \ref{QCunderAD1}. 
There are $N_q$ register bits and the quantum algorithm to be run is represented by the unitary transformation $U^{ \text{QC}}$. 
It is comprised of $d$ unitary transformations described by $U^{ \text{QC}}= \prod_{k=1}^d  U_{ k} =U_{ d} \cdot U_{ d-1} \cdots U_{ 2} \cdot U_{ 1}$.
The unitary transformation $U_{ k} $ $(k=1,2,\ldots, d)$ is composed of single- and two-qubit gates.
We assume that the duration time (gate operation time) of the unitary transformation $U_ k$ is $\Delta t$ for any $k$.
During the time interval $\Delta t$, the register bits are influenced by the AD effects due to couplings with environments, e.g., electromagnetic field in the vacuum, phonons in solids, etc.  
The quantum master equation describing the AD process in the interaction picture is given by \cite{QCQINandC,carmichaeltxb,Agarwalltxb,opendynamicstext}
\begin{align}
\frac { \partial \rho (t)}{\partial t } & =  \gamma \mathcal{L} _{ \text{AD}} [\rho (t)] \notag\\ 
& =  \gamma \sum_{j=0}^{N_q-1} \left[  \tilde{\sigma}^-_j    \rho (t)   \tilde{\sigma}^+_j 
 - \frac{1}{2} \big{\{}   \tilde{\sigma}^+_j  \tilde{\sigma}^-_j ,  \rho (t)   \big{\}} \right] , 
 \label{ADQME}
\end{align}
where $ \rho (t)$ is the density matrix of the $N_q$ register bits at a time $t$ and $\gamma$ is the decay rate.    
The symbol $\mathcal{L} _{ \text{AD}} $ denotes the Lindblad superoperator of the AD process and the operators $\tilde{\sigma}^\pm_j = \frac{ X_j \mp i Y_j}{2}$ are the ladder (raising and lowering) operators acting on the register bit $Q_{ \text{r}j }$. 
$X_j $ and $ Y_j$ are $X$ and $Y$ gates acting on $Q_{ \text{r}j }$, respectively.
$\big{\{} A, B \big{\}}$ is the anti-commutator between the operators $A$ and $B$.
In our model, we assume that the $N_q$ register bits experience homogeneously the AD effect of single-qubit state given by the decay rate $\gamma$. 
At the initial time $t=0$, all the register bits are in $| 0 \rangle$ state (ground state), namely, $\rho (0) =  | 0 \rangle \langle 0|^{\otimes N_q }$. 
Let us write the total amount of quantum computational time (running time of the quantum algorithm under consideration) by $T(= d \cdot \Delta t)$ while we introduce the dimensionless time $ \tau = \gamma \Delta t$.
By assuming $ \tau   \ll 1$, in the following let us evaluate the density matrix at the time $T$, $\rho (T) $,  by using the quantum master equation \eqref{ADQME} and express it as a perturbation series with respect to $\tau$ 
given by 
\begin{align}
 \rho (T) & = \sum_{p=0} ^ \infty \frac{\tau^p}{p!} \cdot\Delta^{\text{AD}}_p \rho_{d\cdots1}  \notag\\
 & = \rho_{d\cdots1} + \tau\cdot\Delta^{\text{AD}}_1 \rho_{d\cdots1}  +  \mathcal{O}(\tau^2). 
 \label{ADDM}
\end{align}
Here  $ \rho_{d\cdots1} = U^{ \text{QC}} \cdot  \rho (0) \cdot \big{(} U^{ \text{QC}} \big{)}^\dagger$ describes the noise-free (ideal) quantum state of the register bits.
In other words, it is the ideal output quantum state generated by the quantum algorithm given by $U^{ \text{QC}}$. 
The quantity $\Delta^{\text{AD}}_p \rho_{d\cdots1}$ $(p \geq1)$ is the theoretically-evaluated $p$-th-order AD effect.
Let us focus on the first-order AD effect $\Delta^{\text{AD}}_1 \rho_{d\cdots1}$ which has the form
\begin{align}
\Delta^{\text{AD}}_1 \rho_{d\cdots1} & = \sum_{k=1}^{d} \Delta^{\text{AD}}_{1,k} \rho_{d\cdots1}, \notag\\
 \Delta^{\text{AD}}_{1,k} \rho_{d\cdots1} & =  \left(\prod_{l=k+1}^d U_l  \right) \cdot \tilde{\rho}^{\text{AD}}_{k\cdots1}      \cdot \left(\prod_{l=k+1}^d U_l \right)^\dagger , 
 \label{ADdeviationDM1}
\end{align}  
where
\begin{align}
\tilde{\rho}^{\text{AD}}_{k\cdots1} & = \mathcal{L} _{ \text{AD}} \big{[} \rho_{k\cdots1} \big{]}
= \sum_{j=0}^{N_q-1} \left[  \tilde{\sigma}^-_j    \rho_{k\cdots1}   \tilde{\sigma}^+_j 
 - \frac{1}{2} \big{\{} P^1_{j}  ,  \rho_{k\cdots1}   \big{\}} \right], \notag\\
  \rho_{k\cdots1} & = \left(\prod_{l=1}^k U_l  \right) \cdot  \rho (0)    \cdot \left(\prod_{l=1}^k U_l \right)^\dagger,
   \label{ADdeviationDM2}
\end{align}
with  $ \prod_{l=k+1}^d U_l  = U_d \cdot U_{d-1}  \cdots U_{k+2} \cdot U_{k+1}$ and  $ \prod_{l=1}^k U_l  = U_k \cdot U_{k-1}  \cdots U_2 \cdot U_1.$ 
In the above equation we have used $ \prod_{l=d+1}^d U_l= \boldsymbol{1}$, where $\boldsymbol{1}$ denotes the identity operator.
The operator $ P^1_{j} =  \tilde{\sigma}^+_j  \tilde{\sigma}^-_j = \frac{\boldsymbol{1}_j - Z_j}{2}$ describes the projection onto the quantum state $|1  \rangle_j$ with $\boldsymbol{1}_j$  and $Z_j$
denoting the identity operator and the $Z$ gate acting on $Q_{ \text{r}j }$, respectively:
On the other hand, the projection operator of the quantum state $|0  \rangle_j$ is given by $ P^0_{j} =  \tilde{\sigma}^-_j  \tilde{\sigma}^+_j = \frac{ \boldsymbol{1}_j + Z_j}{2}$.
\subsection{QEM Scheme }\label{QEM} 
\begin{figure}[h] 
\includegraphics[width=0.4 \textwidth]{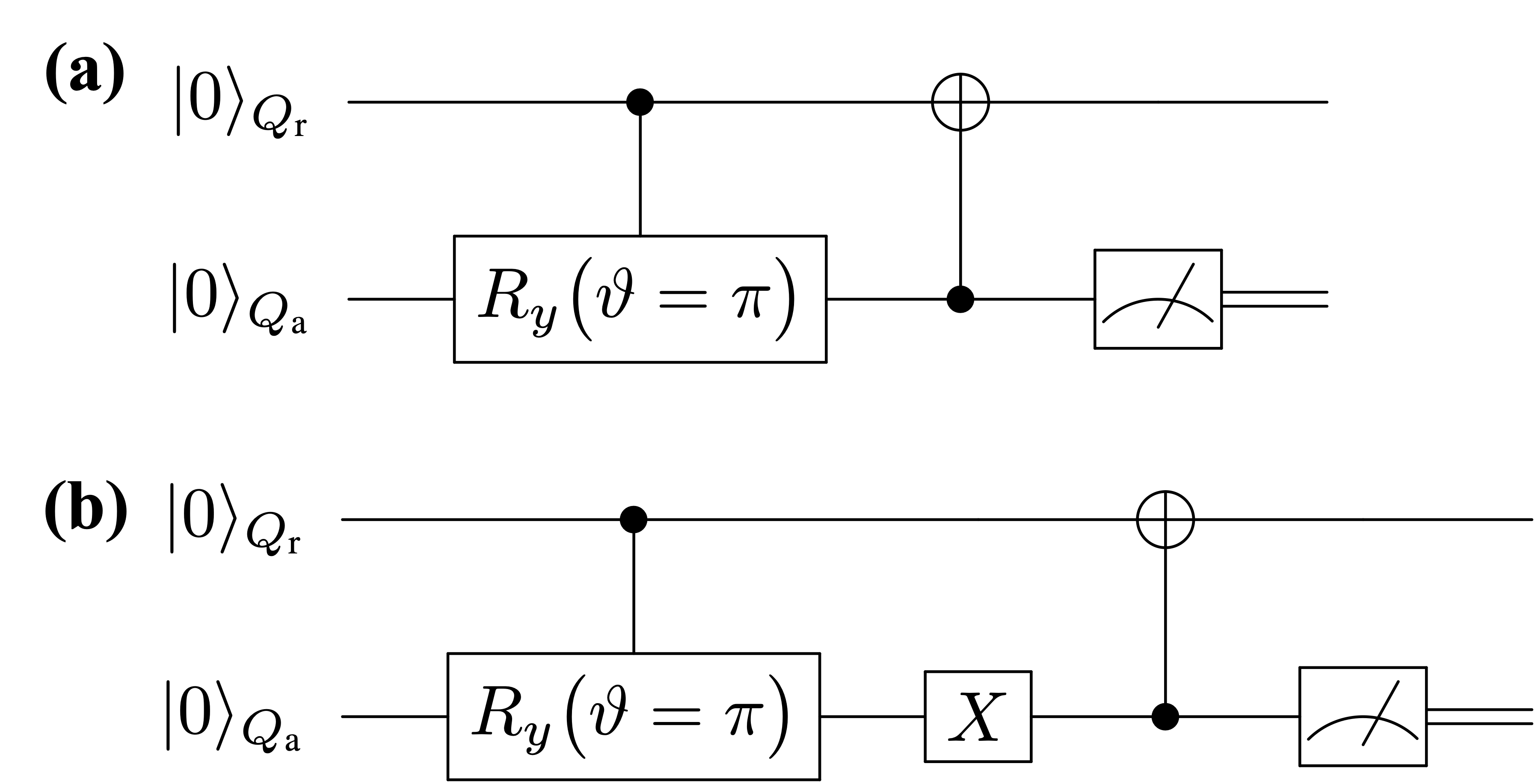}
\caption{
(a) Schematic of AD-effect circuit A. When we set $\vartheta =\pi $ and post-select the measurement result of the quantum state of $Q_\text{a}$ to be $|1 \rangle _{Q_\text{a}}$, 
we realize the operation of $ \tilde{\sigma}^-$ on $Q_{\text{r}0}$.
(b) Schematic of AD-effect circuit B.  By setting $\vartheta =\pi $ and post-selecting the measurement result of the quantum state of $Q_\text{a}$ to be $|0 \rangle _{Q_\text{a}}$,
we the operation of $ P^1$ on $Q_{\text{r}0}$ is created.
 }  
\label{ADcircs}
\end{figure}

\begin{figure}[h] 
\includegraphics[width=0.4 \textwidth]{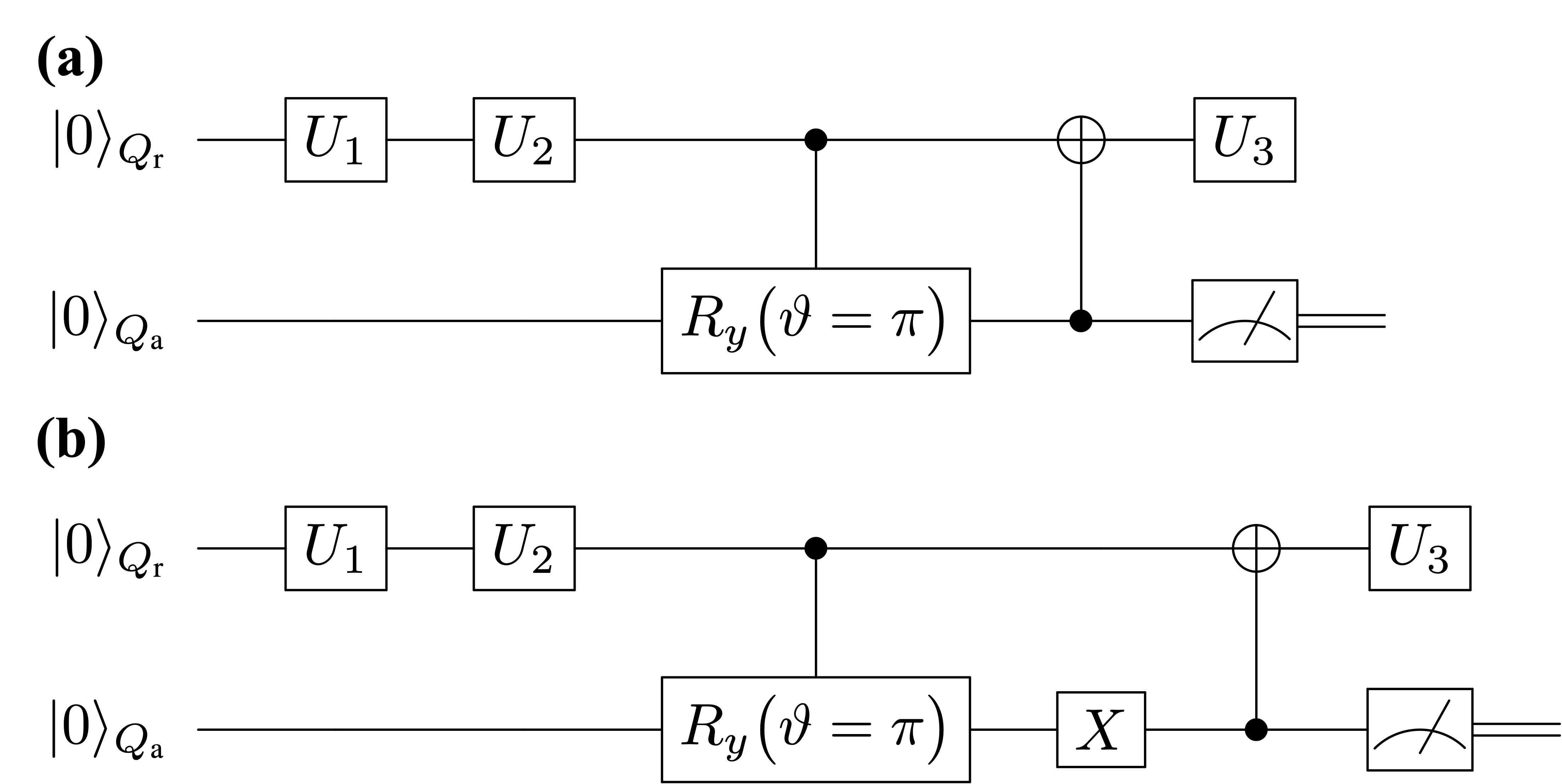}  
\caption{
(a) Schematic of an elementary quantum circuit in the AD-effect circuit group given by the AD-effect circuit A which generates $ \left(\prod_{l=k+1}^d U_l \right) \cdot \left(  \tilde{\sigma}^-  \rho_{k\cdots1}  \tilde{\sigma}^+ \right) \cdot \left(\prod_{l=k+1}^d U_l \right)^\dagger $.
For doing so, we set $\vartheta =\pi $ and post-select the measurement result of the quantum state of $Q_\text{a}$ to be  $|0 \rangle _{Q_\text{a}}$.
(b) Schematic of an elementary quantum circuit in the AD-effect circuit group given by the AD-effect circuit B. When we post-select the measurement outcome of $Q_\text{a}$ to be  $|1 \rangle _{Q_\text{a}}$, we have $ \left(\prod_{l=k+1}^d U_l \right) \cdot \left(  P^1  \rho_{k\cdots1}   P^1  \right) \cdot \left(\prod_{l=k+1}^d U_l \right)^\dagger $. Both of these  quantum circuits in (a) and (b) are for $k=2, d=3$ with $\vartheta=\pi$.  }  
\label{exspk2d3AD}
\end{figure}

Since we have evaluated the single-qubit-state AD effect,  next we discuss our quantum error mitigation (QEM) scheme. 
We denote the operator of which we are aiming to take an expectation value by $\hat{O}$. 
 When we implement the quantum state $\rho$ on a real device
 what we actually obtain is a quantum state which is different from $\rho$ due to quantum noise effects: Note again that hereinafter we only consider the AD effect. Let us write it by  $\rho^{\text{real}}$.   
 We represent the density matrix $\rho^{\text{real}}$ in terms of $\rho$ (ideal state) as  $\rho^{\text{real}}= \rho + \delta^{\text{AD}} \rho$, where $\delta^{\text{AD}} \rho$ represents the deviation from $ \rho$ owing to the AD effect on a real device. 
 Note that we use the symbol $\delta^{\text{AD}}$ to describe the AD effect on a real device while we use $\Delta^{\text{AD}}$ to describe the theoretically-estimated AD effect like Eq.  \eqref{ADDM}. 
 Namely, a quantum computational error occurs due to the deviation $\delta ^{\text{AD}} \rho$. 
 QEM is a prescription for mitigating the error coming from the deviation $\delta ^{\text{AD}} \rho$.  
 Mathematically, this is a task to make the value of Tr$(\hat{O}\delta ^{\text{AD}} \rho)$ as small as possible. 
 In our scheme, we mitigate the error Tr$(\hat{O} \delta ^{\text{AD}} \rho)$ by perturbatively treating the deviation $\delta ^{\text{AD}} \rho$ with respect to $\tau$ and using the theoretically-estimated AD effect $\Delta^{\text{AD}}_p \rho$.
 In the following we show such a perturbative analysis up to the first order in $\tau$. The extension of QEM scheme to higher-order AD effect is discussed in Sec. I in the Supplementary Material.
  The key procedure of our QEM scheme is to construct quantum circuits for computing the quantity Tr$(\hat{O}  \Delta^{\text{AD}}_1 \rho_{d\cdots1})$,
  which describes the theoretically-estimated quantum computational error of the expectation value Tr$(\hat{O} \rho)$ in the first order of $\tau$.
  For doing this, there are two difficulties: 
  (i) the generation of  the anti-commutator term  $ \big{\{} P^1_{j}  ,  \rho_{k\cdots1}   \big{\}} $ in Eq. \eqref{ADdeviationDM2}
  and (ii) the implementation of the non-unitary operators $\tilde{\sigma}^-_j$ and $P^1_j$ .
  Let us discuss from our solution to the difficulty (i). 
  We denote some sort of quantum-computational operation (gate operation or measurement) by $\mathcal{A}$. 
  When the operation $\mathcal{A}$ acts on the quantum state $ \rho_{k\cdots1} $ the output state we have is $ \rho_{k\cdots1} \ \to \ \mathcal{A}   \rho_{k\cdots1}  \mathcal{A}^\dagger$.
  The anti-commutator term  $ \big{\{} P^1_{j}  ,  \rho_{k\cdots1}   \big{\}} $, in contrast, is not represented in this way, and thus, it is not clear how to generate such a term by the quantum-computational operations.
  We solve this in the following way. To make our argument simple, here let focus on the single-register-bit system $(N_q=1)$; the generalization to $N_q \geq2$ is straightforward and is discussed later.  
  First, we rewrite $\tilde{\rho}^{\text{AD}}_{k\cdots1}$ in Eq. \eqref{ADdeviationDM2} as 
  \begin{align}
\tilde{\rho}^{\text{AD}}_{k\cdots1} = -\frac{ \rho_{k\cdots1}}{4} + \frac{ Z  \rho_{k\cdots1}  Z }{4} + \tilde{\sigma}^-  \rho_{k\cdots1}  \tilde{\sigma}^+ - P^1  \rho_{k\cdots1}  P^1.
   \label{ADdeviationDM3}
\end{align}
In the above way, all the four terms in Eq.  \eqref{ADdeviationDM3} are written in the form $ \mathcal{A}   \rho_{k\cdots1}  \mathcal{A}^\dagger$, and thus, we have solved the difficulty (i).
Let us analyze the mathematical structure of the right-hand side of Eq.  \eqref{ADdeviationDM3}.
The quantum circuit for creating the first term is straightforward because it is obtained by the quantum circuit composed of $U^{ \text{QC}}$ (the quantum algorithm under consideration).
The implementation of the quantum circuit for the second term $ \frac{ Z  \rho_{k\cdots1}  Z }{4} $ is also straightforward because we just apply the $Z$ gate after the operation of $U_k.$
 The unclear part is to find ways to construct the quantum circuits for generating the third and fourth terms given by the non-unitary operators $ \tilde{\sigma}^- $ and $P^1$
  and this is nothing but the difficulty (ii).  
 We solve this  by using an ancilla bit and a measurement on it \cite{QCQINandC}. 
 For the creating the operation $\tilde{\sigma}^-,$ we use the quantum circuit presented in Fig. \ref{ADcircs}(a) (AD-effect circuit A) while for the operation of $P^1$ we use the one in Fig. \ref{ADcircs}(b) (AD-effect circuit B). 
In both quantum circuits, we regard the ancilla bit $ Q_\text{a} $ as the environment which induces the AD effect on the register bit $Q_{\text{r}0}$. 
  The interactions between these two qubits are represented by the controlled-rotational gate $U_{CR_y}[Q_{\text{r}0};Q_\text{a}](\vartheta)$  and the controlled-not gate $U_{CX}[Q_\text{a};Q_{\text{r}0}]$. 
  The controlled-rotational gate $U_{CR_y}[Q_{\text{r}0};Q_\text{a}](\vartheta)$ describes the rotation about $y$ axis by the rotational angle $\vartheta$ and it is composed of the control bit  $Q_{\text{r}0}$ and the target bit $Q_\text{a}$. 
  On the other hand, for the gate operation  $U_{CX}[Q_\text{a};Q_{\text{r}0}]$ the ancilla bit $Q_\text{a}$ is the control bit while the register bit $Q_{\text{r}0}$ is the target bit.
 We have used the notation such that the control bit(s) comes before the semicolon while the target bit(s) comes after.
    Let us explain the output states generated by the AD-effect circuits A and B. 
    For both quantum circuits, the initial quantum states of $ Q_{\text{r}0} $ and $ Q_\text{a} $ are the same and it is
  $\rho_{Q_{\text{r}0}Q_\text{a}}(0) = \rho_{Q_{\text{r}0}} (0) \otimes \rho_{Q_\text{a}}(0)$ with $\rho_{Q_{\text{r}0}} (0) = |0\rangle_{Q_{\text{r}0}} \langle0|$ and $ \rho_{Q_\text{a}}(0)= |0\rangle_{Q_\text{a}} \langle0|$.
  The AD-effect circuit  A is given by the unitary operation $U_{\text{ADA}} =U_{CX}[Q_\text{a};Q_{\text{r}0}] \cdot U_{CR_y}[Q_{\text{r}0};Q_\text{a}](\vartheta)$ while 
  the AD-effect circuit  B is given by $U_{\text{ADB}} =U_{CX}[Q_\text{a};Q_{\text{r}0}] \cdot (\boldsymbol{1}_{2\times2} \otimes X_{ Q_\text{a} })\cdot U_{CR_y}[Q_{\text{r}0};Q_\text{a}](\vartheta)$, where $\boldsymbol{1}_{2\times2}$ is the two by two identity operator.
  Owing to these unitary operations, the quantum state generated by the AD-effect circuit A is given by $ \rho_{\text{ADA},Q_{\text{r}0}Q_\text{a}} (\vartheta) =
  U_{\text{ADA}}\cdot \rho_{Q_{\text{r}0}Q_\text{a}}(0) \cdot(U_{\text{ADA}})^\dagger  $ while the quantum state 
   created by the AD-effect circuit B is $\rho_{\text{ADB},Q_{\text{r}0}} (\vartheta)  = U_{\text{ADB}}\cdot \rho_{Q_{\text{r}0}Q_\text{a}}(0) \cdot(U_{\text{ADB}})^\dagger  $.
   At the end, we measure the ancilla bit $Q_\text{a}.$ Then the quantum states of $Q_{\text{r}0}$ (reduced density matrices)  are described by the Kraus operators \cite{QCQINandC,KrausRepresentationref}.  
\begin{align} 
  \mathcal{K} ^{\text{ADA}} _{0}   & = {}_{Q_\text{a}} \langle 0|  U_{\text{ADA}} |0 \rangle _{Q_\text{a}}  
=   P^0 + \cos \left( \frac{ \vartheta }{2} \right)P^1  \notag\\
	 \mathcal{K} ^{\text{ADA}} _{1}   & = {}_{Q_\text{a}} \langle 1|  U_{\text{ADA}} |0 \rangle _{Q_\text{a}}  
= \sin \left( \frac{ \vartheta }{2} \right)\tilde{\sigma}^-, \notag\\
 	 \mathcal{K} ^{\text{ADB}} _{0}   & = {}_{Q_\text{a}} \langle 0|  U_{\text{ADB}} |0 \rangle _{Q_\text{a}}  
= \sin \left( \frac{ \vartheta }{2} \right)P^1 , \notag\\
\mathcal{K} ^{\text{ADB}} _{1}   & = {}_{Q_\text{a}} \langle 1|  U_{\text{ADB}} |0 \rangle _{Q_\text{a}}  
=   \tilde{\sigma}^+ + \cos \left( \frac{ \vartheta }{2} \right)\tilde{\sigma}^- . \label{Kraus1} 
\end{align} 
For the AD-effect circuit A (B) the Kraus operators $ \mathcal{K} ^{\text{ADA}} _{0} $ ($\mathcal{K} ^{\text{ADB}} _{0}$) acts
on the register bit $Q_{\text{r}0}$ when the measurement outcome of the quantum state of the ancilla bit $Q_\text{a}$ is 
$|0 \rangle _{Q_\text{a}}$ while $ \mathcal{K} ^{\text{ADA}} _{1} $ $(\mathcal{K} ^{\text{ADB}} _{1}) $ operates when the measurement outcome is $|1 \rangle _{Q_\text{a}}$. 
 When we average these two outcomes, 
 the quantum state of $Q_{\text{r}0}$ created by the AD-effect circuit A is given by$ \rho_{\text{ADA},Q_{\text{r}0}} (\vartheta)  = \text{Tr}_{Q_\text{a}}\big{[} \rho_{\text{ADA},Q_{\text{r}0}Q_\text{a}} (\vartheta)  \big{]}=
\sum_{\mu=0,1}  \mathcal{K} ^{\text{ADA}} _{\mu} \cdot  \rho_{Q_{\text{r}0}} (0) \cdot ( \mathcal{K} ^{\text{ADA}} _{\mu})^\dagger$, 
where the symbol $\text{Tr}_{Q_\text{a}}$ denotes the partial trace with respect to $Q_\text{a}$ degrees of freedom.   
Similarly, for the AD-effect circuit B we have $ \rho_{\text{ADB},Q_{\text{r}0}} (\vartheta)  = 
 \sum_{\mu=0,1}  \mathcal{K} ^{\text{ADB}} _{\mu} \cdot  \rho_{Q_{\text{r}0}} (0) \cdot ( \mathcal{K} ^{\text{ADB}} _{\mu})^\dagger$. 
 In particular, for the AD-effect circuit A when we take $\vartheta$ to be $\vartheta_t$ such that  $ \cos^2\big{(} \frac{\vartheta_t}{2}\big{)}  = e^{-\gamma t} $ \cite{QCQINandC}, 
 the matrix representation of  $\rho_{\text{ADA},Q_{\text{r}0}} (\vartheta_t)$ is given by
\begin{widetext}  \begin{align}
\rho_{\text{ADA},Q_{\text{r}0}} (\vartheta_t)   =
 \left (
		\begin{array}{cc} 
		 \left[  \rho_{\text{ADA},Q_{\text{r}0}}  (0) \right]_{00}  +  \left[ \rho_{\text{ADA},Q_{\text{r}0}}  (0)    \right]_{11} (1- e^{-\gamma t} )&  \left[ \rho_{\text{ADA},Q_{\text{r}0}}  (0)   \right]_{01} e^{- \frac{\gamma t}{2}}  \\
		 \left[ \rho_{\text{ADA},Q_{\text{r}0}}  (0)   \right]_{10} \cdot    e^{- \frac{\gamma t}{2}} &  \left[ \rho_{\text{ADA},Q_{\text{r}0}}  (0)    \right]_{11}  e^{-\gamma t}
		\end{array}
	\right ).
\label{outRDMQr0}  
\end{align}\end{widetext}
  The matrix element $ \left[  \rho_{\text{ADA},Q_{\text{r}0}}  (0) \right]_{nn^\prime} $ $(n,n^\prime=0,1)$ denotes the $(n,n^\prime)$-element of $ \rho_{\text{ADA},Q_{\text{r}0}}  (0).$
  The reduced density matrix $\rho_{\text{ADA},Q_{\text{r}0}} (\vartheta_t)$ in Eq. \eqref{outRDMQr0}  is nothing but the solution of the quantum master equation \eqref{ADQME}. 
 Further, when we take $\vartheta_t \to \pi$, the Kraus operators in Eq.  \eqref{Kraus1} becomes 
 $\big{\{}\mathcal{K} ^{\text{ADA}} _{0} ,\mathcal{K} ^{\text{ADA}} _{1}        \big{\}} \to \big{\{} P^0 , \tilde{\sigma}^-       \big{\}}, 
 \big{\{}\mathcal{K} ^{\text{ADB}} _{0} ,\mathcal{K} ^{\text{ADB}} _{1}        \big{\}} \to \big{\{}    \tilde{\sigma}^+,P^1        \big{\}}. $
 Therefore, for the case of the AD-effect circuit A by using the measurement result of $Q_\text{a}$ such that we post-select the output state of $Q_\text{a}$ to be $|1 \rangle _{Q_\text{a}}$ 
 we can realize the operation of $\tilde{\sigma}^- $ on $Q_{\text{r}0}$.
 On the other hand, for the AD-effect circuit B by post-selecting the output state of  $Q_\text{a}$ to be $|0 \rangle _{Q_\text{a}}$ we realize the operation of $P^1 $ on $Q_{\text{r}0}$.
 To show the above things concretely, let us present the examples of the quantum circuits for the generation of 
 $ \left(\prod_{l=k+1}^d U_l \right) \cdot \left(  \tilde{\sigma}^-  \rho_{k\cdots1}  \tilde{\sigma}^+ \right) \cdot \left(\prod_{l=k+1}^d U_l \right)^\dagger $ 
 and  $  \left(\prod_{l=k+1}^d U_l \right) \cdot \left(  P^1  \rho_{k\cdots1}   P^1  \right) \cdot \left(\prod_{l=k+1}^d U_l \right)^\dagger $  for $k=2, d=3$, and we show them in Figs. \ref{exspk2d3AD}(a)  and \ref{exspk2d3AD}(b), respectively.
As a result, by using the AD-effect circuits A and B we can perform the actions of $\tilde{\sigma}^- $ and $P^1 $ as described by the third and four terms in Eq. \eqref{ADdeviationDM3},
 and thus, we have solved the second difficulty (ii). 

 Since the difficulties (i) and (ii)  have been solved, we are now ready to establish our QEM scheme.
 To compute the quantity Tr$(\hat{O} \cdot \Delta^{\text{AD}}_1 \rho_{d\cdots1})$  we need four types of quantum circuits:
the original circuit given by $U^{\text{QC}}$, the quantum circuit where the additional $Z$ gate is performed, and the AD-effect circuits A and B.
The latter three quantum-circuit ensembles composed of the additional $Z$-gate, $\tilde{\sigma}^-, $ and $P^1$ operations form the quantum-noise-effect circuit group for the AD effect, i.e, AD-effect circuit group. 
 Hereinafter, let us write the trace of the product between the operator $\hat{O}$ and $\rho$ by $ \text{Tr}(\hat{O} \rho) = \langle \hat{O}  \rangle_\rho$.
 We perturbatively express  the quantum states $ \rho ^{\text{real}} _{d\cdots1} $ with respect to $\tau$ as 
 $ \rho ^{\text{real}} _{d\cdots1} = \rho  _{d\cdots1} + \delta^{\text{AD}} ( \rho  _{d\cdots1} ) $ with 
 $ \delta^{\text{AD}} ( \rho  _{d\cdots1} ) = \sum_{p=1}^\infty \frac{ \tau^p }{p!} \cdot \delta^{\text{AD}}_p ( \rho  _{d\cdots1} )$.
Furthermore, we write the quantity which is obtained by the implementation of $\Delta^{\text{AD}}_1 \rho_{d\cdots1}$ on a real device by $ (\Delta^{\text{AD}}_1 \rho_{d\cdots1})^{\text{real}} $.
With similar to $ \rho ^{\text{real}} _{d\cdots1}$, we perturbatively express $ (\Delta^{\text{AD}}_1 \rho_{d\cdots1})^{\text{real}} $ in terms of 
$ \Delta^{\text{AD}}_1 \rho_{d\cdots1} $ and $\tau$ as
$ (\Delta^{\text{AD}}_1 \rho_{d\cdots1})^{\text{real}} = \Delta^{\text{AD}}_1 \rho_{d\cdots1} + \delta^{\text{AD}}( \Delta^{\text{AD}}_1 \rho_{d\cdots1})$  
with $ \delta^{\text{AD}} ( \Delta^{\text{AD}}_1 \rho_{d\cdots1} ) = \sum_{p=1}^\infty  \frac{ \tau^p }{p!} \cdot \delta^{\text{AD}}_p (\Delta^{\text{AD}}_1 \rho_{d\cdots1} ). $
 Then, by using Eqs. \eqref{ADDM},  \eqref{ADdeviationDM1},  \eqref{ADdeviationDM2}, and  \eqref{ADdeviationDM3} we obtain
 the quantum-error-mitigated expectation value of $\hat{O}$ given by
   \begin{align} 
 \langle \hat{O}  \rangle^{\text{QEM}}_{\rho  _{d\cdots1}} & \equiv 
\langle \hat{O}  \rangle_{\rho^{\text{real}}  _{d\cdots1}} - \tau   \langle \hat{O}  \rangle_{ (\Delta^{\text{AD}}_1 \rho_{d\cdots1})^{\text{real}} } \notag\\
& =   \langle \hat{O}  \rangle_{\rho _{d\cdots1}} 
+ \tau  \Big{(}   \langle \hat{O}  \rangle _{ \delta^{\text{AD}}_1 ( \rho  _{d\cdots1} )} - \langle \hat{O}  \rangle_{ \Delta^{\text{AD}}_1 \rho_{d\cdots1} }   \Big{)} \notag\\ 
&+ \mathcal{O}(\tau^2). \label{QEMformula1}
\end{align} 
The idea of our QEM scheme is clearly represented in the second line of Eq. \eqref{QEMformula1}. 
 The first term is the ideal expectation value while the second term represents the conduction of our QEM scheme. 
It is described as the subtraction between $\langle \hat{O}  \rangle _{ \delta^{\text{AD}}_1 ( \rho  _{d\cdots1} )}$ (the quantum computational error occurring on a real device) and
 $ \langle \hat{O}  \rangle_{ \Delta^{\text{AD}}_1 \rho_{d\cdots1} } $ (theoretically-evaluated quantum computational error), which is computed by the AD-effect circuit group.
 The heart of the idea for doing this is that we have considered that the real noise effect $\delta^{\text{AD}}_1 ( \rho  _{d\cdots1} )$ is (approximately) equivalent to the theoretically-estimated noise effect $\Delta^{\text{AD}}_1 ( \rho  _{d\cdots1} )$.
 When the second term in the second line of Eq. \eqref{QEMformula1} becomes small enough,
 we consider that we have accomplished in mitigating the error of quantum computation on a real device.
Note that the quantum noise effects coming from $ \delta^{\text{AD}} ( \Delta^{\text{AD}} \rho_{d\cdots1} )$ are suppressed by multiplying $ \langle \hat{O}  \rangle_{ (\Delta^{\text{AD}} \rho_{d\cdots1})^{\text{real}} } $
by $\tau$ (the second term in the right-hand side of the first line of Eq.  \eqref{QEMformula1}). 
This is because the lowest-order error AD effect on the implementation of $ \Delta^{\text{AD}}_1 \rho_{d\cdots1}$, which is $\delta^{\text{AD}}_1 (\Delta^{\text{AD}}_1 \rho_{d\cdots1} ),$ becomes $\mathcal{O}(\tau^2)$ due to the multiplication by $\tau$:
$ \delta^{\text{AD}} ( \Delta^{\text{AD}}_1 \rho_{d\cdots1} ) \to 
\tau \cdot   \delta^{\text{AD}} ( \Delta^{\text{AD}}_1 \rho_{d\cdots1} ) = \sum_{p=1}^\infty  \tau^{p+1} \cdot \delta^{\text{AD}}_p (\Delta^{\text{AD}}_1 \rho_{d\cdots1} ). $
Consequently, in the first-order perturbation theory with respect to $\tau$ we have established our QEM scheme by the usage of the AD-effect circuit group and is expressed by the formula given by Eq. \eqref{QEMformula1}.
The above argument on QEM-scheme derivation can be straightforwardly generalized to  register-bit systems for $N_q \geq2.$
In this case, $\tilde{\rho}^{\text{AD}}_{k\cdots1} $ in Eq. \eqref{ADdeviationDM1} is represented as
 \begin{align}
\tilde{\rho}^{\text{AD}}_{k\cdots1} & =  \mathcal{L} _{ \text{AD}} [ \rho_{k\cdots1}] \notag\\
& = \sum_{j=0}^{N_q-1}  -\frac{ \rho_{k\cdots1}}{4} + \frac{ Z_j  \rho_{k\cdots1}  Z_j }{4} + \tilde{\sigma}^-_j  \rho_{k\cdots1}  \tilde{\sigma}^+_j \notag\\
& - P^1_j  \rho_{k\cdots1}  P^1_j.
   \label{ADdeviationDM4}
\end{align}
We can apply our QEM scheme to the $N_q$ register-bit system in the following way. 
We prepare $N_q$  register bits and one ancilla bit $\big{\{} Q_{\text{r}0}, Q_{\text{r}1}, \ldots, Q_{\text{r}N_q-1}, Q_\text{a} \big{\}}$. 
Then we create an ensemble of quantum circuits composed of the $j$-th register bit $Q_{\text{r}j}$ $(j=0,1, \ldots,N_q-1)$ and the ancilla bit $Q_\text{a}$ which describes that 
$Q_{\text{r}j}$ is subject to the AD effect induced by the ancilla bit $Q_\text{a}$. 
Namely, we create the ensemble of four types of quantum circuits composed of $Q_{\text{r}j}$ and $Q_\text{a}$, the original quantum circuits given by $U^{ \text{QC}}$, the quantum circuits with additional $Z$-gate operations, and the AD-effect circuits A and B.
By summing up all these quantum circuits, we obtain the AD-effect circuit group which enables us to perform QEM for $N_q$-register-bit system under the AD effect. The total number of quantum circuits which compose the AD-effect circuit group is $3dN_q+1,$ and thus, it scales polynomial in $dN_q$, which is not so high-cost computational performance.

Before ending this section,  let us explain two ways to extend our QEM formalism. Firstly, we can extend into cases of other quantum noise channels including generalized amplitude damping (GAD), phase damping (PD), their composite channels, 
 and stochastic Pauli noise models such as bit flip, phase flip, bit-phase flip, and depolarizing channel. Secondly, we can create quantum-noise-effect circuit groups which enable us to perform QEM for higher-order quantum noise effects. 
We present  arguments on these two cases in Sec. I in the Supplementary Material.  

\section{Numerical Simulations}\label{nss}   
In this section, we numerically demonstrate our QEM scheme for four types of algorithms.
For the quantum noise effect we choose AD effect. In Sec. \ref{QEMpre}, as a preliminary of our QEM demonstration,
we present the results of two algorithms: the algorithm composed of the initial $X$-gate operation and the repetition of the Hadamard gate acting on a single register bit  
and that composed of the initial $X\otimes X$ operation and the controlled-Hadamard gate acting on two register bits. 
In Sec. \ref{QEMQAAalg}, we show the results of QEM for a long-term quantum algorithm and here we choose quantum amplitude amplification (QAA) algorithm (Grover's search algorithm).
 In Sec. \ref{QEMqaoa}. we show the results of recently developed NISQ quantum algorithms, Quantum Approximate Optimization Algorithm (QAOA).

In the following, let us explain the formalism of our noisy quantum simulations (numerical simulations of running quantum algorithms with real quantum devices performed by classical computers).
As shown in Fig. \ref{QCunderAD1}, every time we apply an unitary (gate) operation, the $N_q$ register bits experience  AD effects.  
Suppose that at a time $t_0$ the quantum states of the register bits were given by the density matrix $\rho(t_0)$. According to the quantum master equation \eqref{ADQME},
when the unitary gate $U$ has been applied to the register bits within the time interval $\Delta t$ the quantum state of the register bits at $t=t_0 + \Delta t$
is expressed by the density matrix
\begin{widetext}
\begin{align}
\rho^{\text{AD}} (t_0 + \Delta t) =  \mathcal{E}^{\text{AD}} [U\rho(t_0)U^\dagger   ],
\label{noisyQsimformula1}
\end{align}
where $\mathcal{E}^{\text{AD}}[\cdots]$ is the superoperator which describes the AD effect on the $N_q$ register bits and it is given by \cite{QCchemistryRMP2020,EMPRL2017,hybridQCalgorithmJPSJ2021,CandSQEMPRAp2021,OttenGrayQEM1,Qiskit}
\begin{align}
& \mathcal{E}^{\text{AD}} [\rho] = \sum_{ n_{Q_{\text{r}0}},  \ldots, n_{Q_{\text{r}N_q-1}}  } \mathcal{M}_{ n_{Q_{\text{r}0}},  \ldots, n_{Q_{\text{r}N_q-1}}  } \cdot
\rho \cdot
\mathcal{M}^\dagger_{ n_{Q_{\text{r}0}},  \ldots, n_{Q_{\text{r}N_q-1}}  }, \notag\\
& \mathcal{M}_{ n_{Q_{\text{r}0}},  \ldots, n_{Q_{\text{r}N_q-1}}  } = \mathcal{M}^{\text{AD}}_{ n_{Q_{\text{r}0}}  }  \otimes \cdots \otimes \mathcal{M}^{\text{AD}}_{ n_{Q_{\text{r}N_q-1}} }, \notag\\ 
& \mathcal{M}  ^{\text{AD}} _{0}   =    \left [
		\begin{array}{cc} 
		 0 & \sin\left( \frac{\vartheta_\tau}{2}  \right) \\
		  & 0
		\end{array}
	\right ], \quad
	 \mathcal{M} ^{\text{AD}} _{1} =  \left [
		\begin{array}{cc} 
		 1   & 0 \\
		0  & \cos\left( \frac{\vartheta_\tau}{2}  \right)
		\end{array}
	\right ],
\label{noisyQsimformula2} 
\end{align}\end{widetext}
where $n_{Q_{\text{r}0}},  \ldots, n_{Q_{\text{r}N_q-1}} =0,1$, and  $ \cos^2\left( \frac{\vartheta_\tau}{2}  \right) = e^{-\tau}$. 
The operators $ \mathcal{M}  ^{\text{AD}} _{0}$ and $ \mathcal{M}  ^{\text{AD}} _{1}$ are the Kraus operators acting on single-qubit states and describe the influence of the AD effect on a single register bit
during the time interval $\Delta t$. 
Here we assume that the $N_q$ register bits homogeneously experience 
 the single-qubit-state AD effect as described in Eq. \eqref{noisyQsimformula2}.
For later convenience, let us introduce the notation which describes the operation of the unitary operator $U$ on the quantum state $\rho$ by 
$\mathcal{T}[\rho,U] (=U\rho U^\dagger)$.
Let us write the quantum state generated by the unitary transformation $U^{ \text{QC}}$ under the AD effect by $\rho^{\text{AD}}_{d\cdots1}$. 
By using the superoperator $\mathcal{E}^{\text{AD}}[\cdots]$, the output state $\rho^{\text{AD}}_{d\cdots1}$ is represented as
\begin{widetext}
\begin{align}
\rho^{\text{AD}}_{d\cdots1} =           \mathcal{E}^{\text{AD}}\big{[}        \mathcal{T}[   \cdots      \mathcal{E}^{\text{AD}}\big{[} \mathcal{T}[ \mathcal{E}^{\text{AD}}\big{[}   \mathcal{T}[ \rho(0), U_1] \big{]}, U_2 ] \big{]}, \cdots U_d ] \big{]}.
\label{noisyQsimformula3}
\end{align}\end{widetext}
Eqs. \eqref{noisyQsimformula2} and \eqref{noisyQsimformula3} are the basic equations of our noisy quantum simulations. 
Namely, we perform our noisy quantum simulations by identifying the quantum state $\rho^{\text{AD}}_{d\cdots1} $ in the above equation with $\rho^{\text{real}}_{d\cdots1}$, which is the AD-affected quantum state generated by the unitary transformation  $U^{ \text{QC}}$ on real devices.
We conduct QEM described by Eq. \eqref{QEMformula1} for various values of $\tau$ by tuning the value of $\vartheta_\tau$.
When we execute the AD circutis A and B, we run quantum simulations of the $N_q+1$ qubit systems. The right-hand side of Eq. (11) becomes $\sum_{ n_{Q_{\text{r}0}},  \ldots, n_{Q_{\text{r}N_q-1}},n_{Q_\text{a}}  } 
\mathcal{M}_{ n_{Q_{\text{r}0}},  \ldots, n_{Q_{\text{r}N_q-1}},n_{Q_\text{a}}  } \cdot \rho \cdot
\mathcal{M}^\dagger_{ n_{Q_{\text{r}0}},  \ldots, n_{Q_{\text{r}N_q-1}},n_{Q_\text{a}}  }$, where $\mathcal{M}_{  n_{Q_{\text{r}0}},  \ldots, n_{Q_{\text{r}N_q-1}}, n_{Q_\text{a}}  } =
  \mathcal{M}^{\text{AD}}_{ n_{Q_{\text{r}0}}  }  \otimes \cdots \otimes \mathcal{M}^{\text{AD}}_{ n_{Q_{\text{r}N_q-1}} }  \otimes\mathcal{M}^{\text{AD}}_{ n_{Q_{\text{a}}}} $. 
Note that the ancilla bit $Q_\text{a}$ is treated similarly as the other $N_q$ register bits such that  $Q_\text{a}$ is subject to the same quantum noise effect described by the Kraus operators in Eq. (11) as the other $N_q$ register bits do. Correspondingly, we mitigate the quantum noise effects influencing both $Q_\text{a}$ and the $N_q$ register bits via Eq.  \eqref{QEMformula1}. 

For performing our noisy quantum simulations, we have created two types of numerical codes. 
The first one is our original numerical code and the second one is the numerical code created by Qiskit \cite{Qiskit}.
In our numerical code, we simply compute the matrix products of the  density matrices,   the unitary operations $U_k$, the Kraus operators, and the additional operations such as $Z,\tilde{\sigma}^\pm,P^1$. Furthermore, we compute the trace operations between the density matrices and the physical operators $O.$ Namely, our original numerical code is a density-matrix simulator. On the other hand, the Qiskit code is   programmed by the two numbers, $N_\text{QC}$ and  $N_\text{samp}$.
To understand them concretely, first let us discuss an example of  quantum computing of a single-qubit system.    
We execute the given quantum circuit and we obtain an output state which is either $|0\rangle$ or $|1\rangle$. 
We repeat this process $N_\text{QC}$ times and say that we obtained $|0\rangle$ for $N_{|0\rangle}$ times as the output state while we obtained  $|1\rangle$ for $N_{|1\rangle}$ times. Then, the probability weight of $|0\rangle$ is $\frac{N_{|0\rangle}}{N_\text{QC}}$ while that of $|1\rangle$ is $\frac{N_{|1\rangle}}{N_\text{QC}}$. Namely, the number $N_\text{QC}$ is the repetition number of quantum computing executed by the given quantum circuit,
and the numerical simulations in our original code describes the quantum simulations in the limit $N_\text{QC}\to \infty.$  
Next let us explain what $N_\text{samp}$ is. 
It is the repetition number of ``the execution of the quantum computation for $N_\text{QC}$ times under the given (fixed) quantum circuit". 
By introducing such a number, we effectively  perform the quantum computation under the given quantum circuit with the repetition number $N_\text{QC} \times N_\text{samp}$ in our Qiskit code. In contrast to $N_\text{QC}$, the repetition number $N_\text{samp}$ is not coming from foundations of quantum mechanics and we have introduced this quantity owing to the following two reasons. First, due to our survey there is an upper limit on $N_\text{QC}$ in the Qiskit program and is $10^7$. By introducing $N_\text{samp}$, we become able to  effectively perform quantum computations with repetition numbers greater than the upper limit of $N_\text{QC}$ ($=10^7$) in the Qiskit program. Second, we consider that it is not enough to just show a single data point of ``the quantum computational result obtained by the fixed quantum circuit and  $N_\text{QC}$" to see how largely it deviates from the true quantum computational result (quantum computation in the limit $N_\text{QC}\to \infty$). To present how largely the quantum computational results for fixed and finite $N_\text{QC}$ deviate (finite-size effects of $N_\text{QC}$) from the true ones, we introduce $N_\text{samp}$ and show visually such deviations.        
In order to describe the second reason more mathematically, let write a binary which labels a quantum state of qubit $Q_\alpha$ ($\alpha=0,\ldots,N_q-1$)  
 by $n_{Q_\alpha}$: $|n_{Q_\alpha}\rangle$ with $n_{Q_\alpha} =0,1$ and an output state is described by the computational basis states $ |n_{Q_{N_q-1}}\rangle_{Q_{N_q-1}} \otimes  
\cdots |n_{Q_0} \rangle_{Q_0}  =  |n_{Q_{N_q-1}}  \cdots n_{Q_0} \rangle$. 
Let us say that we focus on the specific quantum state  $|\hat{n}_{Q_{N_q-1}}  \cdots \hat{n}_{Q_0}  \rangle$ and consider how many times it is  obtained as the output state for given $N_\text{QC}$. 
By saying that  $|\hat{n}_{Q_{N_q-1}}  \cdots \hat{n}_{Q_0}  \rangle$ has been obtained as the output state for $N_{\hat{n}_{Q_{N_q-1}}  \cdots \hat{n}_{Q_0} }(N_\text{QC})$ times, the probability of an output state being 
$|\hat{n}_{Q_{N_q-1}  \cdots \hat{n}_{Q_0} } \rangle$ is $ w_{\hat{n}_{Q_{N_q-1}}  \cdots \hat{n}_{Q_0} }(N_\text{QC}) = \frac{N_{\hat{n}_{Q_{N_q-1}}  \cdots \hat{n}_{Q_0} }(N_\text{QC}) }{N_\text{QC}} $.  
Furthermore, we write the probability such that $|\hat{n}_{Q_{N_q-1}}  \cdots \hat{n}_{Q_0} \rangle$ is going to become observed in quantum computing under $N_\text{QC}\to \infty$ as $ p_{\hat{n}_{Q_{N_q-1}}  \cdots \hat{n}_{Q_0} }$.   
We redescribe such a circumstance as the binomial distribution denoted by  $B(p_{\hat{n}_{Q_{N_q-1}}  \cdots \hat{n}_{Q_0} },N_\text{QC})$  and introduce the random variable $X_i$ such that in the $i$th round of quantum computation we take $X_i=1$ when the output state is 
$|\hat{n}_{Q_{N_q-1}}  \cdots \hat{n}_{Q_0} \rangle$, otherwise $X_i=0$. Next, we introduce another random variable $\bar{X} = \frac{\sum_{j=0}^{N_\text{QC}-1}X_j}{N_\text{QC}}$, which is equivalent to   
$ w_{\hat{n}_{Q_{N_q-1}}  \cdots \hat{n}_{Q_0} }(N_\text{QC})  $. Owing to the central limit theorem, we obtain   
$\lim_{N_\text{QC} \to \infty} w_{\hat{n}_{Q_{N_q-1}}  \cdots \hat{n}_{Q_0} }(N_\text{QC}) = p_{\hat{n}_{Q_{N_q-1}}  \cdots \hat{n}_{Q_0} }$.  
In other words, the binomial distribution $B(p,N_\text{QC})$ approaches to the Gaussian distribution function given by the mean $p_{\hat{n}_{Q_{N_q-1}}  \cdots \hat{n}_{Q_0} }$ 
and the standard deviation $\sqrt{\frac{p_{\hat{n}_{Q_{N_q-1}}  \cdots \hat{n}_{Q_0} }(1-p_{\hat{n}_{Q_{N_q-1}}  \cdots \hat{n}_{Q_0} })}{N_\text{QC}}}$, 
namely $N\left(p_{\hat{n}_{Q_{N_q-1}}  \cdots \hat{n}_{Q_0} },  \frac{p_{\hat{n}_{Q_{N_q-1}}  \cdots \hat{n}_{Q_0} }(1-p_{\hat{n}_{Q_{N_q-1}}  \cdots \hat{n}_{Q_0} })}{N_\text{QC}}   \right)$.  
Let us say that we perform quantum computing with accuracy $\epsilon.$  
From the standard deviation of  $N\left(p_{\hat{n}_{Q_{N_q-1}}  \cdots \hat{n}_{Q_0} },  \frac{p_{\hat{n}_{Q_{N_q-1}}  \cdots \hat{n}_{Q_0} }(1-p_{\hat{n}_{Q_{N_q-1}}  \cdots \hat{n}_{Q_0} })}{N_\text{QC}}   \right)$, 
we can estimate the lower bound of $N_\text{QC}$ for doing this and is equal  to  
$\frac{p_{\hat{n}_{Q_{N_q-1}}  \cdots \hat{n}_{Q_0} }(1-p_{\hat{n}_{Q_{N_q-1}}  \cdots \hat{n}_{Q_0} })}{\epsilon^2} $.   
Note that in order to perform QEM with the accuracy $\epsilon$ the total repetition number of quantum computing  gets larger with respect to $d$ and $N_q$  due to the quantum-noise-effect circuit group and this issue is addressed later.
The number $N_\text{QC}$ describes the repetition number of quantum computation owing to the given quantum circuit whereas the number $N_\text{samp}$ describes how many times you conduct the sampling for the expectation values of physical operators obtained by the $N_\text{QC}$-repeated quantum computation. 
Owing to this sampling,  the repetitive number of the quantum computations effectively becomes $N_\text{QC} \times N_\text{samp}$, and our simulation results become more trustable.
In our simulations we take $N_\text{QC} = 2^{10}$ and $ N_\text{samp}=100$ for both the original quantum circuit and each elementary circuit of quantum-noise-effect circuit group.
Our original numerical code, on the other hand, is the code for a noisy quantum simulation in the limit $N_{\text{QC}} \to \infty$, and basically, it performs pure linear algebraical computations such as matrix-product and trace operations.
We note that when we create the numerical codes with Qiskit, we need to be careful with how controlled-unitary operators are implemented.
We have examined that on Qiskit program  the controlled-unitary operators are implemented as the decomposition of  $U_{CX}$ gates and single-qubit unitary gates.  
For example, the control-$R_y$ gate $U_{CR_{y}(\vartheta)} [Q_{\text{r}0};Q_{\text{r}1}]$
is decomposed as $U_{CR_{y}(\vartheta)} [Q_{\text{r}0};Q_{\text{r}1}] = \left( {\boldsymbol{1}_{2\times2}}_{Q_{\text{r}0}} \otimes R_{y}(\frac{\vartheta}{2})_{Q_{\text{r}1}} \right) \cdot U_{CX} [Q_{\text{r}1};Q_{\text{r}0}] \cdot \left( {\boldsymbol{1}_{2\times2}}_{Q_{\text{r}0}} \otimes R_{y}(-\frac{\vartheta}{2})_{Q_{\text{r}1}} \right) \cdot U_{CX}[Q_{\text{r}0};Q_{\text{r}1}]$. 
Therefore, when we simulate QAA for three-qubit systems and QAOA with our original code we implement $U_{CR_{y}(\vartheta)} [Q_{\text{r}0};Q_{\text{r}1}]$ in the same way as Qiskit program does.

In order to quantitatively describe the validity of our QEM scheme  we introduce the measure defined by 
\begin{align}
\text{RT}_{\text{QEM}} = \frac{| \langle \hat{O}  \rangle_{\rho^{\text{real}}  _{d\cdots1}}  - \langle \hat{O}  \rangle_{\rho  _{d\cdots1}} |}{| \langle \hat{O}  \rangle_{\rho  _{d\cdots1}} - \langle \hat{O}  \rangle_{\rho^{\text{QEM}}  _{d\cdots1}}  |}.
   \label{QEMratio}
\end{align}
The numerator of $\text{RT}_{\text{QEM}}$ in Eq. \eqref{QEMratio} describes the absolute of the difference between the expectation value owing to the noisy quantum simulation $\langle \hat{O}  \rangle_{\rho^{\text{real}}  _{d\cdots1}}$ (no QEM) and the ideal expectation value $\langle \hat{O}  \rangle_{\rho^{\text{ideal}}  _{d\cdots1}}$ .
On the other hand, the denominator represents the absolute of the difference between the expectation value obtained by our QEM scheme $\langle \hat{O}  \rangle_{\rho^{\text{QEM}} _{d\cdots1}}$ (see Eq. \eqref{QEMformula1}) and the ideal expectation value. 
In other words, the measure $\text{RT}_{\text{QEM}}$ in Eq. \eqref{QEMratio} is the ratio between the absolute of the error without QEM and the one with QEM.
Thus, when $\text{RT}_{\text{QEM}} >1$ the expectation value $\langle \hat{O}  \rangle_{\rho^{\text{QEM}}  _{d\cdots1}}$ is closer to the ideal value than $\langle \hat{O}  \rangle_{\rho^{\text{real}}  _{d\cdots1}}$,
which means that our QEM scheme is working.  
In addition to the ratio $\text{RT}_{\text{QEM}} $ in Eq. \eqref{QEMratio}, we display the results of the expectation values obtained by the ideal simulations, the noisy simulations without QEM, and the noisy simulations with QEM, and show explicitly the validity of our QEM scheme.
Note that for our original code we take the noise-strength parameter to be $\vartheta_\tau = i_\text{o} \times 0.01$ 
with $i_\text{o}=0,1,\ldots,50$ while for Qiskit code we take $\vartheta_\tau = i_\text{Q} \times 0.05$ 
with $i_\text{Q}=0,1,\ldots,10$. For computing the expectation values we include the case $i_\text{o}=0$ and $i_\text{Q}=0$
whereas for computing the ratio $\text{RT}_{\text{QEM}}$ we omit $i_\text{o}=0$ and $i_\text{Q}=0$.
This is because in this case we have 
$ \langle \hat{O}  \rangle_{\rho  _{d\cdots1}} = \langle \hat{O}  \rangle_{\rho^{\text{real}}  _{d\cdots1}} = \langle \hat{O}  \rangle_{\rho^{\text{QEM}}  _{d\cdots1}} $ and we encounter in the indefinite $\frac{0}{0}$.  
In the following, we create a subsection for each quantum algorithm and discuss the results in detail.
\begin{figure}[!h] 
\includegraphics[width=0.3 \textwidth]{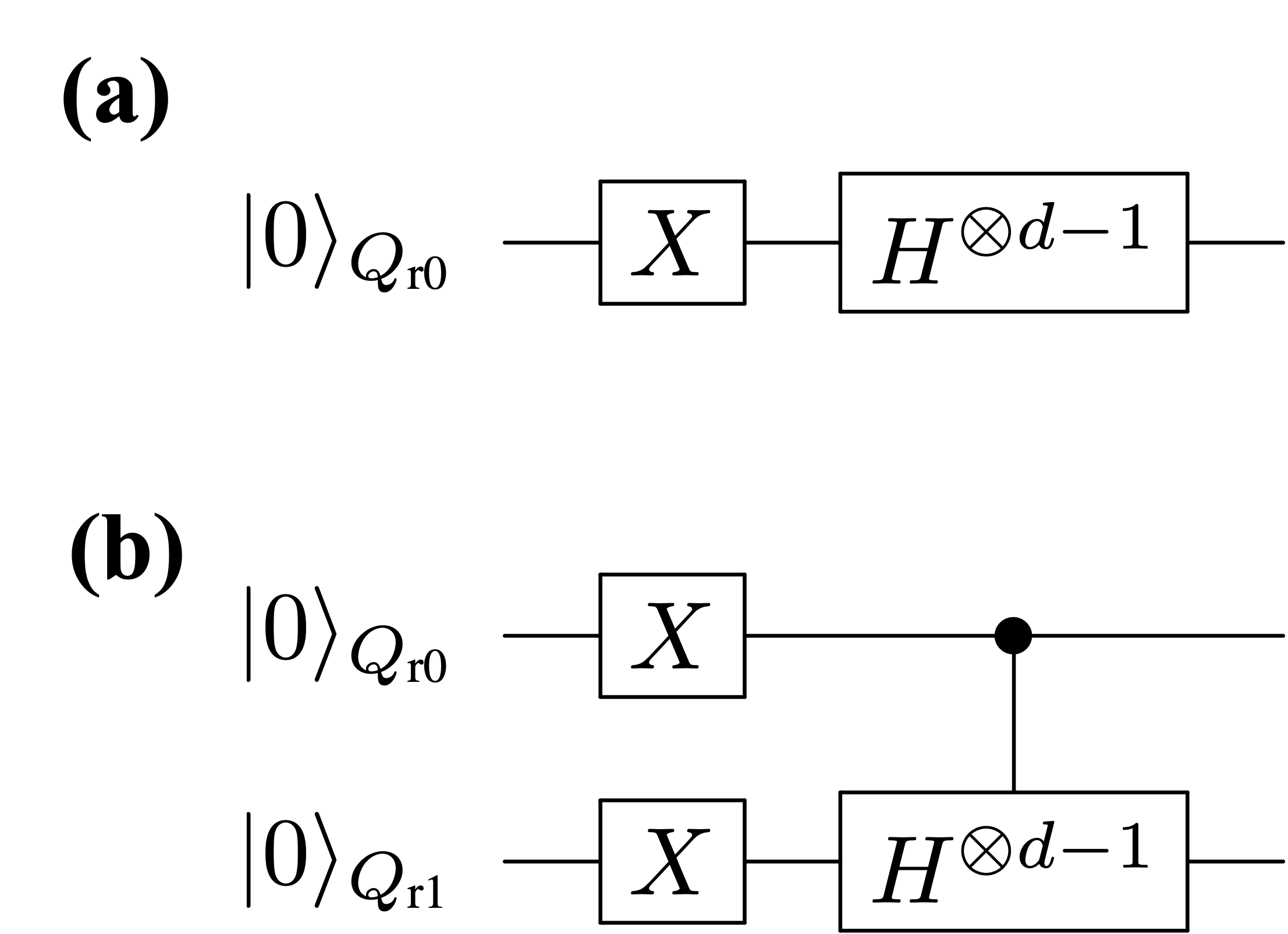}
\caption{
Quantum circuits for (a) $U^{ \text{QC}}_{ \text{pre}1}=H^{ \otimes d-1} \cdot  X  $ and 
(b) $U^{ \text{QC}}_{ \text{pre}2}=(U_{\text{C}H}[Q_{\text{r}0};Q_{\text{r}1}] )^{ \otimes d-1} ) \cdot X_{Q_{\text{r}0}} \otimes X_{Q_{\text{r}1}}$.
 }  
\label{XHXXCHqcircs}
\end{figure} 
\begin{figure*}[!ht] 
\centering
\includegraphics[width=0.9 \textwidth ]{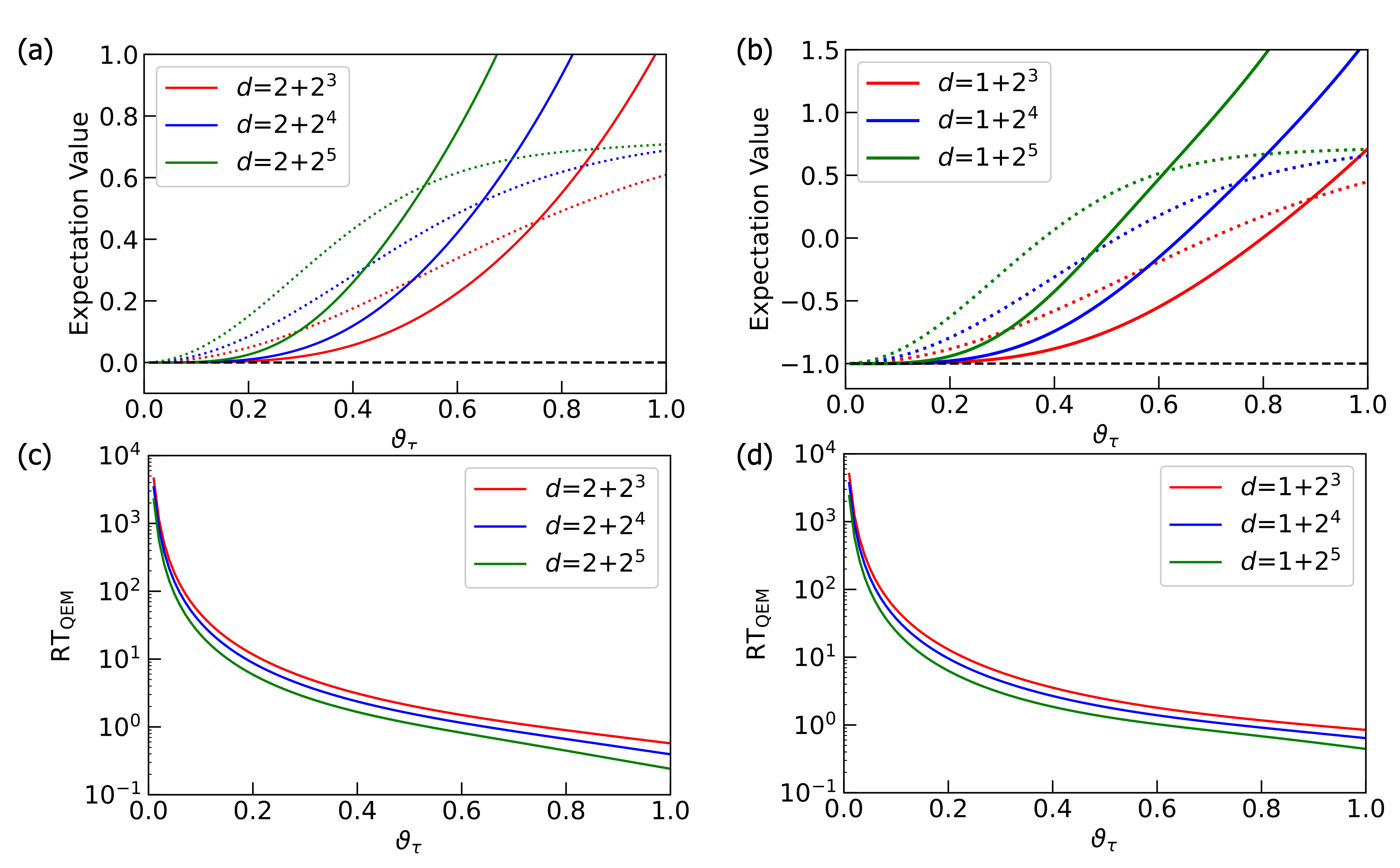}
\caption{ Quantum simulations for the quantum algorithm $U^{ \text{QC}}_{ \text{pre}1}=H^{ \otimes d-1} \cdot  X$.
Plots  in (a) and (b) are the results of QEM for the expectation value of $X$ and $Z$,  respectively. 
The dotted lines are the expectation values without QEM while the solid lines represent the expectation values with QEM. 
The black dotted lines are the results of the ideal simulations. In (c) and (d), we show the ratio $\text{RT}_{\text{QEM}} $ for the expectation values of $X$ and $Z$
respectively. The red, blue, and green plots are the results for $d = 1+2^3,d = 1+2^4,$ and $d = 1+2^5$, respectively.
}
\label{XHnoisyQsims}
\end{figure*} 
\begin{figure*}[!htb] 
\centering
\includegraphics[width=0.8 \textwidth]{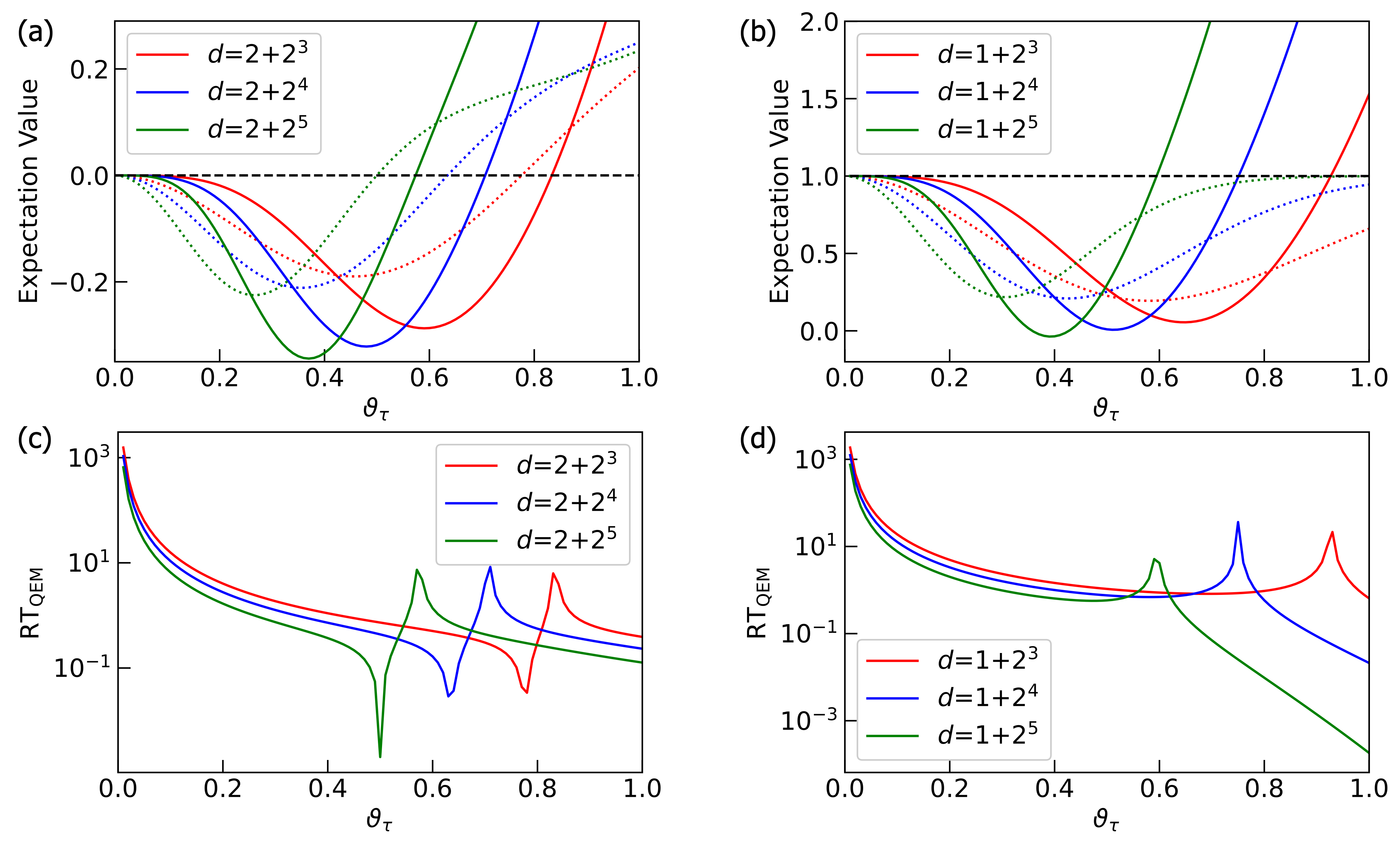}
\caption{ Quantum simulations for the quantum algorithm $U^{ \text{QC}}_{ \text{pre}2}=(U_{\text{C}H}[Q_{\text{r}0};Q_{\text{r}1}] )^{ \otimes d-1} ) \cdot X_{Q_{\text{r}0}} \otimes X_{Q_{\text{r}1}}$.
Plots in (a) and (b) are the results of QEM for the expectation value of $ZX$ and $ZZ$, respectively.
The dotted lines are the expectation values without QEM while the solid lines represent the expectation values with QEM. 
The black dotted lines are the ideal simulation results. 
We plot the ratio $\text{RT}_{\text{QEM}}$ for expectation value of $ZX$ and $ZZ$ in (c) and (d), respectively. The red, blue, and green plots are the results for $d = 1+2^3,d = 1+2^4,$ and $d = 1+2^5$, respectively.}
\label{XXCHnoisyQsims}
\end{figure*} 
\subsection{Preliminary}\label{QEMpre}  
As a preliminary, let us conduct the noisy quantum simulations for two simple algorithms.  
The first algorithm is given by the unitary operation $U^{ \text{QC}}_{ \text{pre}1}=H^{ \otimes d-1} \cdot  X  $, where $H$ denotes the Hadamard gate. 
The second one is given by $U^{ \text{QC}}_{ \text{pre}2}=(U_{\text{C}H}[Q_{\text{r}0};Q_{\text{r}1}] )^{ \otimes d-1} ) \cdot X_{Q_{\text{r}0}} \otimes X_{Q_{\text{r}1}}$.
The unitary operator $U_{CH}[Q_{\text{r}0};Q_{\text{r}1}]$ is the controlled-Hadamard gate composed of the control bit $Q_{\text{r}0}$ and the target bit $Q_{\text{r}1}$. 
Here we take the circuit depth $d$ to be $d = 1 + 2^{n}$ with $n$ being positive integers.  
Let us show the quantum circuits for these two algorithms in Fig. \ref{XHXXCHqcircs}.
By changing the values of $d$ and $\tau$, we numerically analyze how well our QEM scheme works in terms of these two parameters.  
Let us discuss from the results of noisy quantum simulations conducted by the quantum circuit in Fig. \ref{XHXXCHqcircs}(a) and show them in Fig. \ref{XHnoisyQsims}.
We have taken the physical operators $\hat{O}$ as $\hat{O} = X,Z$.
The horizontal axis represents $\vartheta_\tau$, which can be regarded as the strength of AD effect.
On the other hand, the vertical axis in Figs. \ref{XHnoisyQsims}(a) and \ref{XHnoisyQsims}(b) denote the expectation values of $\hat{O}$ while those 
in Figs. \ref{XHnoisyQsims}(c) and \ref{XHnoisyQsims}(d) describe the ratio $\text{RT}_{\text{QEM}} $ given in Eq. \eqref{QEMratio}: 
Figs. \ref{XHnoisyQsims}(a) and \ref{XHnoisyQsims}(c) are the results for $ \hat{O}=X$ while Figs. \ref{XHnoisyQsims}(b) and \ref{XHnoisyQsims}(d) are those for $ \hat{O}=Z$. 
The dotted lines in Figs. \ref{XHnoisyQsims}(a) and \ref{XHnoisyQsims}(b)
describe the expectation values without QEM being performed whereas the solid lines represent the expectation values with QEM being performed.
For both the dotted and solid lines the red, blue, and green plots in Figs. \ref{XHnoisyQsims}(a) and \ref{XHnoisyQsims}(b) are the computational results of $\langle X \rangle$ and $\langle Z \rangle$ 
for $d=2^3 +1$, $2^4 +1$, and $2^5 +1$, respectively. The black dotted lines are the results of the ideal simulations. 
We have computed $ \langle \hat{X}  \rangle_{\rho  _{d\cdots1}},\langle \hat{X}  \rangle_{\rho^{\text{real}}  _{d\cdots1}},$ and
$\langle \hat{X}  \rangle_{\rho^{\text{QEM}} _{d\cdots1}}$ not as the expectation values of $X$ with respect to the quantum states generated by $U^{ \text{QC}}_{ \text{pre}1}$
but as those of $Z$ with respect to the quantum states generated by $U^{ \text{QC}}_{ \text{pre}1}$ and the subsequent operation of $H$, i.e., 
we have switched the basis vectors for the measurement from the computational basis to $\{ |+\rangle, |-\rangle     \}$,  where  
$|+\rangle = H| 0\rangle,|-\rangle = H| 1\rangle.$  Let us analyze our simulation results by comparing the behaviors of the expectation values and the ratio $\text{RT}_{\text{QEM}} $ in Eq. \eqref{QEMratio}
as functions of $\vartheta_\tau$. In this way, we can clearly see whether our QEM scheme is working or not, and for this purpose in the following we rewrite  $\text{RT}_{\text{QEM}}$ as $\text{RT}_{\text{QEM}}(\vartheta_\tau)$ to emphasize that they are the functions of $\vartheta_\tau$. 
Furthermore, we introduce the angle $\vartheta^{\text{c}}_\tau$ such that $\text{RT}_{\text{QEM}} (\vartheta_\tau =\vartheta^{\text{c}}_\tau)=1$, which indicates that the point $ \vartheta_\tau =  \vartheta^{\text{c}}_\tau $ is the critical point of our QEM to become failed.  
Let us look from the results shown in Figs. \ref{XHnoisyQsims}(a) and \ref{XHnoisyQsims}(c) by focusing on how the behaviors of $\langle \hat{O}  \rangle_{\rho _{d\cdots1}}, \langle \hat{O}  \rangle_{\rho^{\text{real}}  _{d\cdots1}}, \langle \hat{O}  \rangle_{\rho^{\text{QEM}}  _{d\cdots1}},$ and $\text{RT}_{\text{QEM}}$ change by increasing $\vartheta_\tau$.
In Fig. \ref{XHnoisyQsims}(a), as the definition of $ \vartheta^{\text{c}}_\tau $ we certainly see that in the range $0< \vartheta_\tau \leq \vartheta^{\text{c}}_\tau$ 
the absolute  $| \langle \hat{O}  \rangle_{\rho^{\text{real}}  _{d\cdots1}}  - \langle \hat{O}  \rangle_{\rho  _{d\cdots1}} |$ is bigger than 
 $| \langle \hat{O}  \rangle_{\rho^{\text{QEM}}  _{d\cdots1}}  - \langle \hat{O}  \rangle_{\rho  _{d\cdots1}} |$, which implies that  $\langle \hat{O}  \rangle_{\rho^{\text{QEM}}  _{d\cdots1}} $ is numerically closer to $\langle \hat{O}  \rangle_{\rho  _{d\cdots1}} $ than $\langle \hat{O}  \rangle_{\rho^{\text{real}}  _{d\cdots1}}$, and correspondingly, in Fig. \ref{XHnoisyQsims}(c) we see $\text{RT}_{\text{QEM}} (\vartheta_\tau) \geq 1$. 
As we increase the value of $\vartheta_\tau$ from $ \vartheta^{\text{c}}_\tau $, the absolute $| \langle \hat{O}  \rangle_{\rho^{\text{real}}  _{d\cdots1}}  - \langle \hat{O}  \rangle_{\rho  _{d\cdots1}} |$ becomes smaller than $| \langle \hat{O}  \rangle_{\rho^{\text{QEM}}  _{d\cdots1}}  - \langle \hat{O}  \rangle_{\rho  _{d\cdots1}} |$, and correspondingly, the ratio $\text{RT}_{\text{QEM}}(\vartheta_\tau)$ decreases monotonically from one.  
For the region $ \vartheta_\tau \geq \vartheta^{\text{c}}_\tau $ to improve the quality of our QEM we need take into account higher-order AD effects and establish QEM schemes for mitigating them and we expect the value of $\vartheta^{\text{c}}_\tau$
to become larger. Next, let us analyze how the quality of our QEM becomes when we vary the circuit depth $d$. 
We see that for every $\vartheta_\tau$ both $| \langle \hat{O}  \rangle_{\rho^{\text{real}}  _{d\cdots1}}  - \langle \hat{O}  \rangle_{\rho  _{d\cdots1}} |$ and $| \langle \hat{O}  \rangle_{\rho^{\text{QEM}}  _{d\cdots1}}  - \langle \hat{O}  \rangle_{\rho  _{d\cdots1}} |$ become bigger and $\text{RT}_{\text{QEM}}(\vartheta_\tau)$ decreases as we increase $d$. This is reasonable because when $d$ gets larger the amount of error gets bigger. 
For the results in Figs. \ref{XHnoisyQsims}(b) and \ref{XHnoisyQsims}(d), 
basically we see that both the expectation values of $Z$ and $\text{RT}_{\text{QEM}}(\vartheta_\tau)$ show the similar behaviors as those for $ \hat{O}=X$:
(I) the validity of QEM ($\text{RT}_{\text{QEM}} (\vartheta_\tau) \geq 1$) in the range $0< \vartheta_\tau \leq \vartheta^{\text{c}}_\tau$ and monotonic decrease of $\text{RT}_{\text{QEM}}(\vartheta_\tau)$ for $ \vartheta_\tau > \vartheta^{\text{c}}_\tau$, and
(II) worsening of the quality of our QEM for large $d$.
In contrast to the above characteristics of $\langle X \rangle$  and $\langle Z \rangle$, we have numerically verified that  the expectation value of $Y $  takes zero for any $ \vartheta_\tau$. 
This is because when the density matrix $\rho$ is a real matrix the expectation value $\langle Y \rangle $ is  zero. 
Since both the quantum algorithm and the AD effect are described by real numbers (see also Eq. \eqref{outRDMQr0} or the Kraus operators in Eq. \eqref{noisyQsimformula2}) the density matrix generated by these two things is real and we have $\langle Y \rangle =0$.  

Let us discuss the results in Fig. \ref{XXCHnoisyQsims}. 
They are the noisy simulation results of the quantum algorithm given by $U^{ \text{QC}}_{ \text{pre}2}$ (see the quantum circuit in Fig. \ref{XHXXCHqcircs}(b)) and we have taken $d = 1 + 2^{n}$ as in the case of simulations for  $U^{ \text{QC}}_{ \text{pre}1}$. 
Here we have simulated $\text{RT}_{\text{QEM}} (\vartheta_\tau)$ for the expectation values of the operators  $\hat{O} = ZX, ZZ$.
As similar to the computations of $ \langle \hat{X}  \rangle_{\rho  _{d\cdots1}},\langle \hat{X}  \rangle_{\rho^{\text{real}}  _{d\cdots1}},$ and
$\langle \hat{X}  \rangle_{\rho^{\text{QEM}} _{d\cdots1}}$,
we have computed $ \langle \hat{ZX}  \rangle_{\rho  _{d\cdots1}},\langle \hat{ZX}  \rangle_{\rho^{\text{real}}  _{d\cdots1}},$ and
$\langle \hat{ZX}  \rangle_{\rho^{\text{QEM}} _{d\cdots1}}$
not as the expectation values of $ZX$ with respect to the quantum states generated by $U^{ \text{QC}}_{ \text{pre}2}$
but as those of $ZZ$ with respect to the quantum states generated by $U^{ \text{QC}}_{ \text{pre}2}$ and the subsequent operation of $H$ on $Q_{\text{r}1}$.
 Overall, we see the same characteristics with the cases of $ \hat{O}=X,Z$: the characteristics (I) and (II) mentioned above. 
For any  $\vartheta_\tau$, the ratio  $\text{RT}_{\text{QEM}} (\vartheta_\tau)$ for the noisy simulations of $U^{ \text{QC}}_{ \text{pre}2}$ are smaller than those of $U^{ \text{QC}}_{ \text{pre}1}$.
This is because $U^{ \text{QC}}_{ \text{pre}1}$ is solely comprised of the single-qubit gates ($X$ and $H$) while $U^{ \text{QC}}_{ \text{pre}2}$ is constructed by 
$n$-operation of the controlled-Hadamard gate (two-qubit gate), and thus, the bigger amount of errors are accumulated in the latter case.  
The difference between the characteristics of noisy simulations for $U^{ \text{QC}}_{ \text{pre}1}$ and those for $U^{ \text{QC}}_{ \text{pre}2}$, 
although it is not an essential point for the validity of our QEM, is that we see both one minima and one maxima in each plot for $\text{RT}_{\text{QEM}} (\vartheta_\tau)$ in Fig. \ref{XXCHnoisyQsims}(c) while only one maxima appears in Fig. \ref{XXCHnoisyQsims}(d). 
Let us denote the point where $\text{RT}_{\text{QEM}} (\vartheta_\tau)$ takes the minimum (maximum) by 
$\vartheta^{\text{min}}_\tau$ ($\vartheta^{\text{max}}_\tau$): note that these values depend on $d$.  
We can understand why these points emerge by looking at Figs. \ref{XXCHnoisyQsims}(a) and \ref{XXCHnoisyQsims}(b). 
Let us explain from the origins of the minima and the maxima in Fig. \ref{XXCHnoisyQsims}(c) by looking at the plots in Fig. \ref{XXCHnoisyQsims}(a).   
In the range  $0< \vartheta_\tau \leq \vartheta^{\text{min}}_\tau$ we have $\langle \hat{O}  \rangle_{\rho^{\text{real}}  _{d\cdots1}} - \langle \hat{O}  \rangle_{\rho _{d\cdots1}}<0 $ and $\langle \hat{O}  \rangle_{\rho^{\text{QEM}}  _{d\cdots1}} - \langle \hat{O}  \rangle_{\rho _{d\cdots1}}<0 $
 whereas in the range  $\vartheta^{\text{min}}_\tau < \vartheta_\tau \leq \vartheta^{\text{max}}_\tau$ we have 
 $\langle \hat{O}  \rangle_{\rho^{\text{real}}  _{d\cdots1}} - \langle \hat{O}  \rangle_{\rho _{d\cdots1}}>0 $ and $\langle \hat{O}  \rangle_{\rho^{\text{QEM}}  _{d\cdots1}} - \langle \hat{O}  \rangle_{\rho _{d\cdots1}}<0 $. 
  Then, in the range $\vartheta^{\text{max}}_\tau < \vartheta_\tau$ we have $\langle \hat{O}  \rangle_{\rho^{\text{real}}  _{d\cdots1}} - \langle \hat{O}  \rangle_{\rho _{d\cdots1}}>0 $ and $\langle \hat{O}  \rangle_{\rho^{\text{QEM}}  _{d\cdots1}} - \langle \hat{O}  \rangle_{\rho _{d\cdots1}}>0 $.
 As a result,  the minima appears at $ \vartheta_\tau = \vartheta^{\text{min}}_\tau$  whereas the maxima emerges at $ \vartheta_\tau = \vartheta^{\text{max}}_\tau$ in Fig. \ref{XXCHnoisyQsims}(c).  
 The origin of the maxima in Fig. \ref{XXCHnoisyQsims}(d) can be similarly explained  by looking at the plots in Fig. \ref{XXCHnoisyQsims}(b).
 In the range $0< \vartheta_\tau \leq \vartheta^{\text{max}}_\tau$ we have $\langle \hat{O}  \rangle_{\rho^{\text{real}}  _{d\cdots1}} - \langle \hat{O}  \rangle_{\rho _{d\cdots1}}<0 $ and $\langle \hat{O}  \rangle_{\rho^{\text{QEM}}  _{d\cdots1}} - \langle \hat{O}  \rangle_{\rho _{d\cdots1}}<0 $, while in $\vartheta_\tau >  \vartheta^{\text{max}}_\tau$ we have $\langle \hat{O}  \rangle_{\rho^{\text{real}}  _{d\cdots1}} - \langle \hat{O}  \rangle_{\rho _{d\cdots1}}<0 $ and $\langle \hat{O}  \rangle_{\rho^{\text{QEM}}  _{d\cdots1}} - \langle \hat{O}  \rangle_{\rho _{d\cdots1}}>0 $,
 and consequently, the maxima appears at $\vartheta_\tau = \vartheta^{\text{max}}_\tau$.
Like the case of the noisy simulations of $U^{ \text{QC}}_{ \text{pre}1}$, the density matrices are generated as real matrices (the unitary transformation $U^{ \text{QC}}_{ \text{pre}2}$ as well as the AD effects are described by real numbers), 
and the expectation values of the Pauli operators, $ IY,   X Y, YI, YX, YZ, ZY$ are zero for both ideal and noisy simulations. Here we have rewritten $\boldsymbol{1}_{2\times2} $ as $I$ for convenience. 
Note that the expectation value of the identity operator ($= \boldsymbol{1}_{4\times4}:$ four by four identity operator) is one for any quantum state including noise-affected quantum states since the trace of density matrix is one for any quantum state.
 In other words, it is unnecessary to do QEM for the expectation value of the identity operator.  Note, however, that when leakage occurs the trace preservation is not held anymore and we need to consider QEM for the error induced by the leakage.  
\begin{figure*}[!htb] 
\centering
\includegraphics[width=0.7 \textwidth]{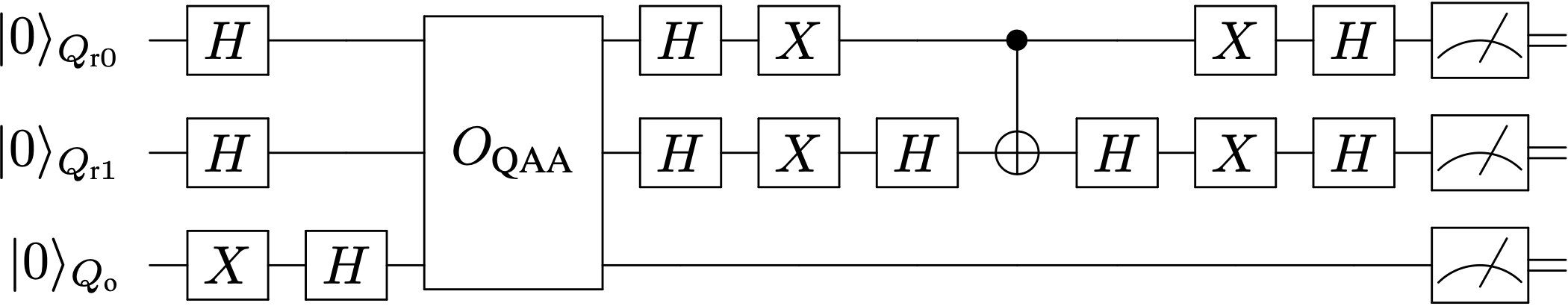}
\caption{ Quantum circuits for $ U^{ \text{QAA}}$ in Eq. \eqref{UnitaryQAA}. }
\label{QAAQcirc} 
\end{figure*} 
\begin{figure*}[!htb] 
\centering
\includegraphics[width=0.8 \textwidth]{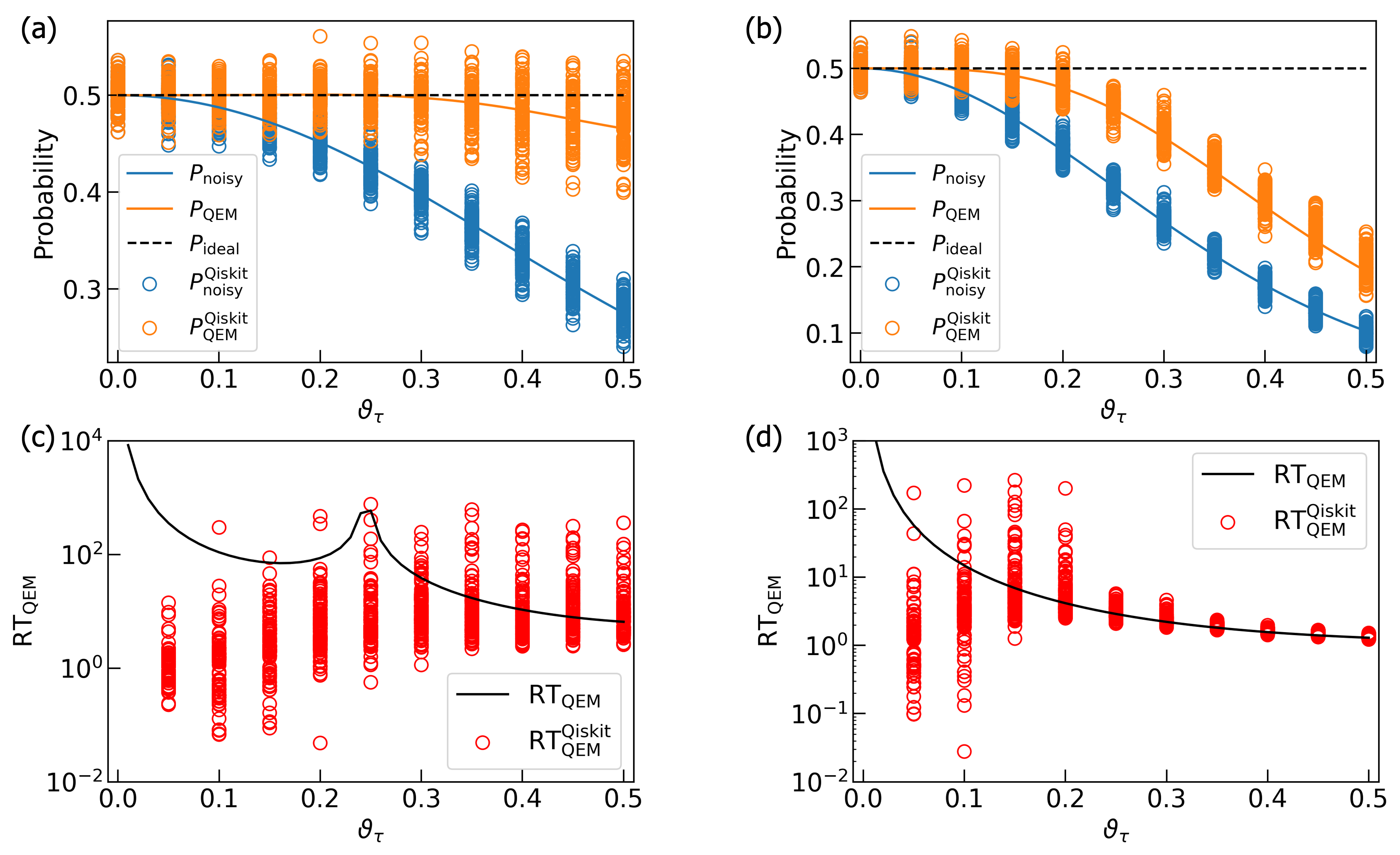}
\caption{ Quantum simulation results for the quantum algorithm $ U^{ \text{QAA}}$ in Eq. \eqref{UnitaryQAA}.
Plots in (a) and (b) are the results of the probabilities $P_{110}$ (probability of $| 110 \rangle$) and $P_{111}$ (probability of $| 111 \rangle$), respectively.
The dotted black lines are the ideal simulation results whereas the blue and orange curves are the noisy simulation results without QEM and the ones with QEM, respectively. All of them are obtained by our original code.
The blue and orange circles are the noisy simulation results without QEM and the ones with QEM, respectively, and they are obtained by our Qiskit code.
Plots in (c) and (d) are the results of the ratio $\text{RT}_{\text{QEM}}$ for $P_{110}$ and $P_{111}$, respectively. The black curves are obtained by our original code while the red circles by our Qiskit code.
For each  $\vartheta_\tau$ we have plotted 100 circles in (a) - (d), i.e., $N_{\text{samp}}=100.$
}
\label{groverq3probsandratios}
\end{figure*} 
\begin{figure*}[!htb] 
\centering
\includegraphics[width=0.85 \textwidth]{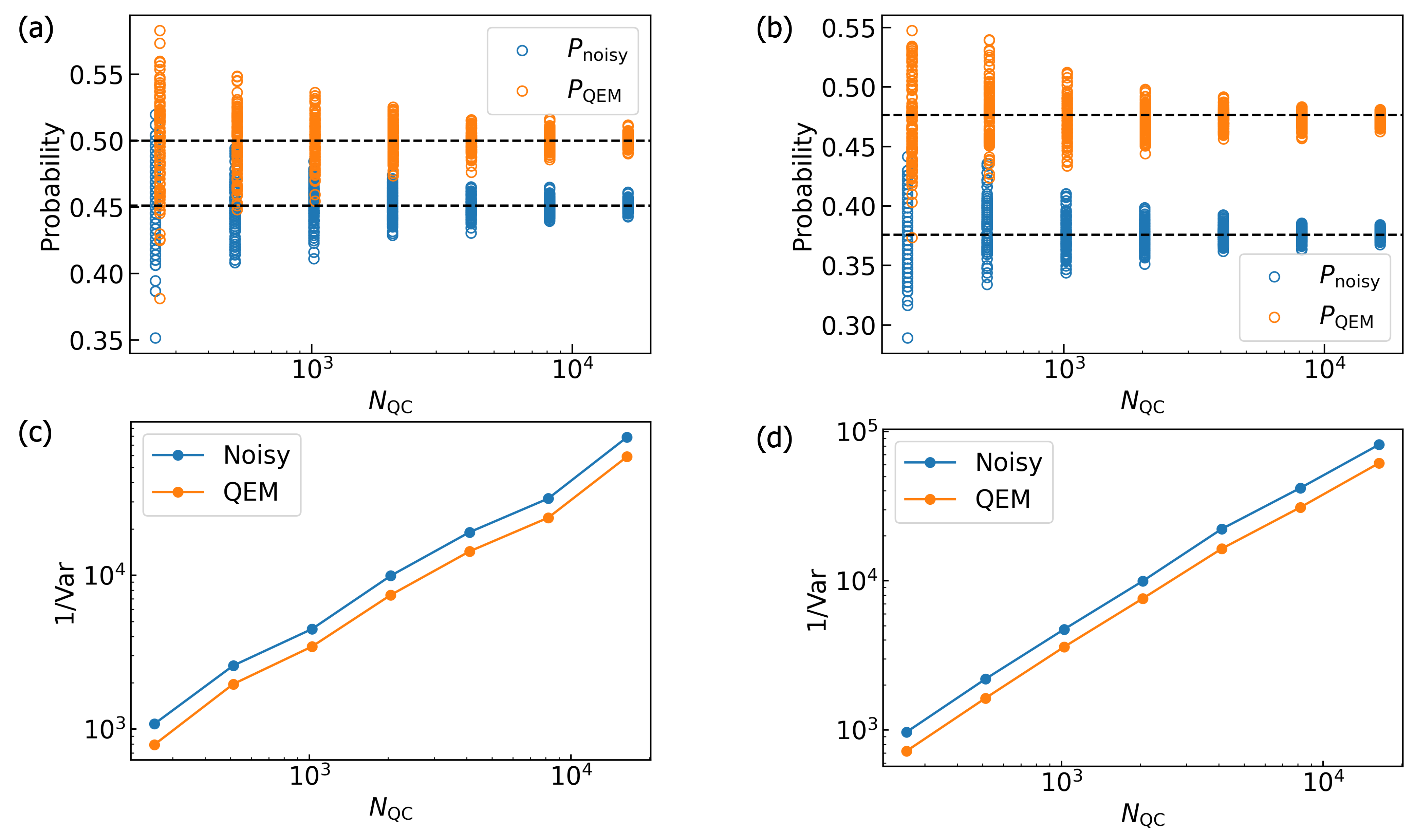}
\caption{ Numerical results of  $\langle O  \rangle_\text{noisy}^\text{Qiskit}(i,N_\text{QC})$ and $\langle O  \rangle_\text{QEM}^\text{Qiskit}(i,N_\text{QC})$
for (a) $P_{110}$ and (b) $P_{111}$. The plots in (c) and (d)  are  $\left( \sigma^{2}_\text{noisy,\text{Qiskit}}[O,N_\text{QC},N_\text{samp}] \right)^{-1} $ (blue) and $\left[ \sigma^{2}_\text{QEM,\text{Qiskit}}[O,N_\text{QC},N_\text{samp},N_q,d] \right]^{-1}$ (orange)
for $P_{110}$ and $P_{111}$, respectively. We take $N_\text{QC}=2^{n_\text{QC}}$ with $n_\text{QC}=8,9,\ldots,14$, $N_\text{samp}=100$, and $\vartheta_\tau=0.2$. 
}
\label{GroverNqcdependence}
\end{figure*} 
\subsection{Quantum Amplitude Amplification}\label{QEMQAAalg}  
By taking account of the previous analysis, let us apply our QEM scheme to Quantum Amplitude Amplification (QAA) \cite{TIsing3} for three-qubit systems: two-register bits and one oracle bit.
One of the important application of QAA is the database retrieval  and the quantum algorithms for this is called the Grover's search algorithm \cite{TIsing3,GroverQA,Grover1997,montanaro2016quantum,QAZoo}.   
Let us denote the (classical) oracle function by $f$ and a binary by $x$ which takes $``00",``01",``10",``11".$  
We consider that we have only one solution of $f$ and write it by $x^\ast$, which satisfy  $f(x^\ast)=1$, and assume $x^\ast=``11"$:  for $x= ``00",``01",``10"$ we have $f(x)=0$.   
The oracle operator $O_{\text{QAA}}$ can be implemented on a quantum circuit by using one oracle bit $Q_\text{o}$  such that $O_{\text{QAA}}\left[ | x \rangle_{Q_{\text{r}0}Q_{\text{r}1}}\otimes \left( \frac{ |0 \rangle_ {Q_\text{o}} - |1 \rangle_{Q_\text{o}} } {\sqrt{2} } \right) \right]
= (-1)^{f(x)} \left[ | x \rangle_{Q_{\text{r}0}Q_{\text{r}1}}\otimes \left( \frac{ |0 \rangle_ {Q_\text{o}} - |1 \rangle_{Q_\text{o}} } {\sqrt{2} } \right) \right],$
where the superposition state $\frac{ |0 \rangle_ {Q_\text{o}} - |1 \rangle_{Q_\text{o}} } {\sqrt{2} }$ is created by applying $H\cdot X$ on the oracle-bit state $|0 \rangle_ {Q_\text{o}}.$
In our case, $(-1)^{f(x)} =-1$ when $ | x \rangle_{Q_{\text{r}0}Q_{\text{r}1}} = | 11 \rangle_{Q_{\text{r}0}Q_{\text{r}1}}$, and the oracle operator $O_{\text{QAA}}$ is equivalent to the Toffoli gate comprised of the two controlled bits $Q_{\text{r}0}$ and $Q_{\text{r}1}$ and the target bit $Q_\text{o}$ \cite{TIsing3}, and write it by  $U_{\text{C}X}[Q_{\text{r}0}Q_{\text{r}1}; Q_\text{o}]$. 
To construct QAA we need one more unitary transformation and that is $U_\psi = \left(\boldsymbol{1}_{4\times4}-2| \psi \rangle\langle \psi|\right) \otimes  \boldsymbol{1}_{2\times2}$, where $|\psi \rangle =H^{\otimes 2} | 00\rangle_{Q_{\text{r}0}Q_{\text{r}1}}$. 
By introducing $U_{\text{init}}= H^{\otimes 2} \otimes (H\cdot X)_{ Q_\text{o} } $, QAA is given by the unitary operation \cite{TIsing3}
\begin{widetext}
\begin{align}
    U^{ \text{QAA}}  &  =  ( U_{\text{G}}  )^k \cdot  U_{\text{init}},  \notag\\
 U^{ \text{G}} & = - U_\psi \cdot O_{\text{QAA}}, \notag\\
  U_\psi & =     \big{(}  (H \cdot X)^{\otimes 2} \otimes \boldsymbol{1}_{2\times2, Q_\text{o}}  \big{)} 
  \cdot
   \big{(}  \boldsymbol{1}_{2\times2, Q_{\text{r}0}} \otimes H_{ Q_{\text{r}1} }   \otimes \boldsymbol{1}_{2\times2, Q_\text{o}}  \big{)} 
  \cdot    \big{(} U_{\text{C}X}[Q_{\text{r}0};Q_{\text{r}1}] \otimes \boldsymbol{1}_{2\times2, Q_\text{o}}  \big{)}                  \notag\\
   &\cdot  \big{(}  \boldsymbol{1}_{2\times2, Q_{\text{r}0}} \otimes H_{ Q_{\text{r}1} }   \otimes \boldsymbol{1}_{2\times2, Q_\text{o}}  \big{)}
 \cdot \big{(}  (X \cdot H)^{\otimes 2} \otimes \boldsymbol{1}_{2\times2, Q_\text{o}}  \big{)} 
   \label{UnitaryQAA}
\end{align} \end{widetext}
\begin{figure*}[!htb] 
\centering
\includegraphics[width=1.0 \textwidth]{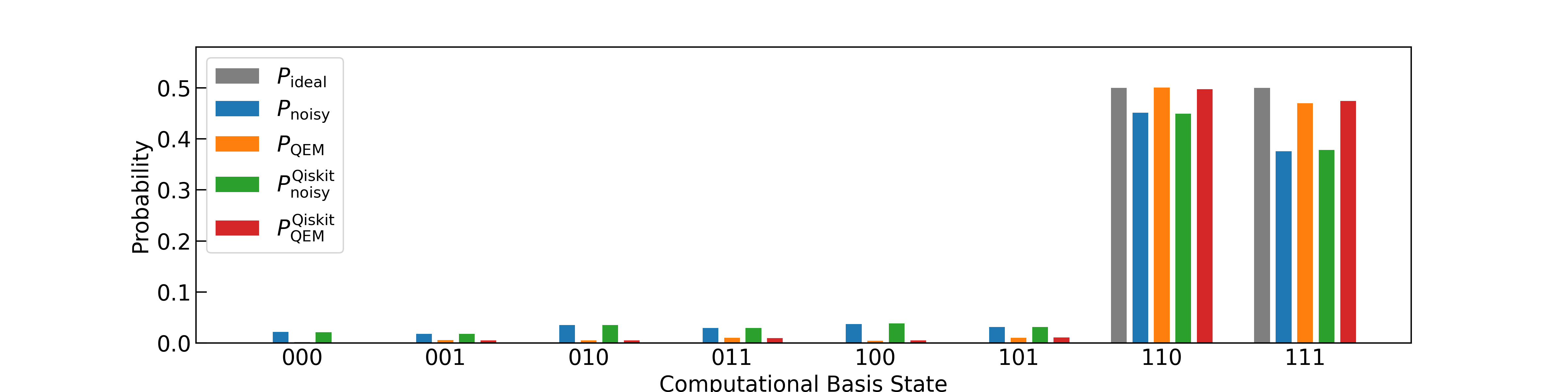}
\caption{ Histogram of the probability distribution of the computational basis states for QAA simulations given by $ U^{ \text{QAA}}$ in Eq. \eqref{UnitaryQAA}.
 Here we have set $\vartheta_\tau = 0.2$.
}
\label{groverq3histogram} 
\end{figure*} 
We show the quantum circuit for the unitary operation $U^{ \text{QAA}} $ in Fig. \ref{QAAQcirc} \cite{TIsing3}: the quantum circuit for the oracle operator  $O_{\text{QAA}}$ ($=U_{\text{C}X}[Q_{\text{r}0}Q_{\text{r}1}; Q_\text{o}]$) is shown in Sec. II in the Supplementary Material. We note that on the quantum circuit in Fig. \ref{QAAQcirc}, what is actually implemented is $-U^{ \text{G}} =  U_\psi \cdot O_{\text{QAA}}$ and the 
global phase factor $(-1)$ does not affect our result. 
The unitary operation $U_{\text{G}} $ is called the Grover operator and $k$ is the repetitive number number of its operation 
and here we have $k=1.$ After the operation of $U^{ \text{QAA}} $ in Eq. \eqref{UnitaryQAA}, ideally 
both of the probability weights of $|110\rangle$ and $|111\rangle$ are $\frac{1}{2}$, and thus, the probability of obtaining the quantum state $|11\rangle$
as the output state is $\frac{1}{2}\times 2=1$, which implies the success of searching the solution $x^\ast=``11"$.
By taking account of the above theoretical framework, we examine whether our QEM scheme works or not for QAA given by  $U^{ \text{QAA}} $ in Eq. \eqref{UnitaryQAA} 
by computing the probability weights of $|110\rangle$ and $|111\rangle$ which we name as $P_{110}$ and $P_{111}$, respectively, 
and show these results in Fig. \ref{groverq3probsandratios}. 
Solid lines in Figs. \ref{groverq3probsandratios}(a) and \ref{groverq3probsandratios}(b) describe the probability weights obtained by our original numerical code and we have denoted $\langle P_{110(111)} \rangle_{\rho^{\text{ideal}}  _{d\cdots1}}, \langle P_{110(111)} \rangle_{\rho^{\text{real}}  _{d\cdots1}},$ and $\langle P_{110(111)} \rangle_{\rho^{\text{QEM}}  _{d\cdots1}}$ by $ P_\text{ideal},  P_\text{real}$, and $ P_\text{QEM} $, respectively. 
On the other hand, the blue and orange circles are calculated by our Qiskit code and we have denoted $\langle P_{110(111)} \rangle_{\rho^{\text{real}}  _{d\cdots1}}$ and $\langle P_{110(111)} \rangle_{\rho^{\text{QEM}}  _{d\cdots1}}$ by $P_{\mathrm{noisy}}^{\mathrm{Qiskit}}$ and $P_{\mathrm{QEM}}^{\mathrm{Qiskit}}$, respectively.
Similarly, in Figs. \ref{groverq3probsandratios}(c) and \ref{groverq3probsandratios}(d), we have denoted the ratio $\text{RT}_{\text{QEM}}(\vartheta_\tau)$ calculated by our Qiskit code by $\text{RT}_{\text{QEM}}^{\text{Qiskit}}$: 
for the results obtained by our original code we have just used the notation $\text{RT}_{\text{QEM}}$ for describing them.
Let us look from the simulation results of $P_{110}$ and the associated ratio $\text{RT}_{\text{QEM}}(\vartheta_\tau)$ given by Figs. \ref{groverq3probsandratios}(a) and \ref{groverq3probsandratios}(c), respectively. 
In the range $0 \leq \vartheta_\tau \leq 0.5$, overall  the simulation results with QEM are numerically closer to the ideal values than the noisy simulation results without QEM, and correspondingly, we have $\text{RT}_{\text{QEM}}(\vartheta_\tau) >1$. 
The similar characteristics can be seen in  Figs.  \ref{groverq3probsandratios}(b) (simulation results of $P_{111}$) and  \ref{groverq3probsandratios}(d) ($\text{RT}_{\text{QEM}}(\vartheta_\tau) $ for $P_{111}$).  
For the results obtained by our Qiskit code, the deviation between $\text{RT}_{\text{QEM}}$ and $\text{RT}_{\text{QEM}}^{\text{Qiskit}}$ becomes prominent in the small $\vartheta_\tau$ region and we consider this as follows.
When noise strength $\vartheta_\tau$ is weak enough, on the Qiskit code the difference between the noisy value and the ideal value is very tiny such that our QEM becomes invalid and $\text{RT}_{\text{QEM}}^{\text{Qiskit}}$
gets lower than one. On the other hand, we see that some red points are above $\text{RT}_{\text{QEM}}(\vartheta_\tau)=1.$
We consider that by greatly increasing $N_\text{QC}$, we expect that $\text{RT}_{\text{QEM}}^{\text{Qiskit}}$
approaches to $\text{RT}_{\text{QEM}}$. 
As a result, our QEM works for the noisy simulations of both $P_{110}$ and $P_{111}$. 
To show clearly that the simulation results obtained by our Qiskit code approach to those obtained by our original code by increasing $N_\text{QC}$, in Figs. \ref{GroverNqcdependence}(a) and \ref{GroverNqcdependence}(b) we show the $N_\text{QC}$ dependencies of $P_{110}$ and $P_{111}$, respectively. In these plots the horizontal axes represent $N_\text{QC}$ whereas the vertical axis in Fig. \ref{GroverNqcdependence}(a) represents the numerical values of $P_{110}$ and that in Fig. \ref{GroverNqcdependence}(b) represents those of $P_{111}$. 
In the following let us write a noisy expectation value obtained in the $i$th round of quantum computing by $\langle O  \rangle_\text{noisy}^\text{Qiskit}(i,N_\text{QC})$ ($O=P_{110}, P_{111}$ and $i=1,\ldots, N_\text{samp}$)
and similarly an expectation value with QEM by  $\langle O  \rangle_\text{QEM}^\text{Qiskit}(i,N_\text{QC})$
with taking $N_\text{QC}=2^{n_\text{QC}}$ with $n_\text{QC}=8,9,\ldots,14$, $N_\text{samp}=100$, and $\vartheta_\tau=0.2$.  
From these two figures, we clearly see that both $\langle O  \rangle_\text{noisy}^\text{Qiskit}(i,N_\text{QC})$ and $\langle O  \rangle_\text{QEM}^\text{Qiskit}(i,N_\text{QC})$
approach to $\langle O  \rangle_\text{noisy}$ and $\langle O  \rangle_\text{QEM}$, respectively,  i.e.,  the deviations of $\langle O  \rangle_\text{noisy}^\text{Qiskit}(i,N_\text{QC})$ from $\langle O  \rangle_\text{noisy}$ and those of
$\langle O  \rangle_\text{QEM}^\text{Qiskit}(i,N_\text{QC})$ from $\langle O  \rangle_\text{QEM}$ get smaller for larger $N_\text{QC}=2^{n_\text{QC}}$. 
To evaluate these deviations numerically, in Figs. \ref{GroverNqcdependence}(c)  ($O= P_{110}$) and \ref{GroverNqcdependence}(d)  ($O= P_{111}$ ) we plot inverse variances defined by  $\left( \sigma^{2}_\text{noisy,\text{Qiskit}}[O,N_\text{QC},N_\text{samp}] \right)^{-1} = \left[ \sum_{i=1}^{N_\text{samp}} \frac{1}{N_\text{samp}} \left( \langle O  \rangle_\text{noisy}^\text{Qiskit}(i,N_\text{QC}) -  \bar{\langle O  \rangle}_\text{noisy}^\text{Qiskit}(N_\text{QC})        \right)^2 \right]^{-1}  $, where $ \bar{\langle O  \rangle}_\text{noisy}^\text{Qiskit}(N_\text{QC},N_\text{samp}) =   \sum_{i=1}^{N_\text{samp}} \frac{\langle O  \rangle_\text{noisy}^\text{Qiskit}(i,N_\text{QC})}{N_\text{samp}} $. 
We approximate $\left( \sigma^{2}_\text{noisy,\text{Qiskit}}[O,N_\text{QC},N_\text{samp}] \right)^{-1}$ as a linear function of $N_\text{QC}$ as 
$\left( \sigma^{2}_\text{noisy,\text{Qiskit}}[O,N_\text{QC},N_\text{samp}] \right)^{-1}= \alpha^\text{Qiskit}_\text{noisy}[O,N_\text{samp}] N_\text{QC}$ and we have $\alpha^\text{Qiskit}_\text{noisy}[O,N_\text{samp}] = 4.62$ and $5.05$ for  $P_{110}$ and $P_{111}$, respectively. On the other hand, the variances of $\langle O  \rangle_\text{noisy}^\text{Qiskit}(i,N_\text{QC})$ can be analytically evaluated (see also the description in page 7) as
$\left(\sigma^2_\text{noisy}[O,N_\text{QC}]\right)^{-1} = \left( \frac{ \langle O  \rangle_\text{noisy}(1-\langle O  \rangle_\text{noisy})}{N_\text{QC}} \right)^{-1}= \alpha_\text{noisy}[O]N_\text{QC}$, and
 we have  $ \alpha_\text{noisy}[O]  \simeq  4.04$ for $P_{110}$ and $ \alpha_\text{noisy}[O]  \simeq  4.26$ for $P_{111}$. Therefore, the two variances $\sigma^2_\text{noisy}[O,N_\text{QC}]$ and $\sigma^{2}_\text{noisy,\text{Qiskit}}[O,N_\text{QC}N_\text{samp}] $ have good numerical agreements.  
In addition to these two variances, we also plot the quantity defined by $\left[ \sigma^{2}_\text{QEM,\text{Qiskit}}[O,N_\text{QC},N_\text{samp},N_q,d] \right]^{-1} = \left[
 \sum_{i=1}^{N_\text{samp}} \frac{1}{N_\text{samp}} \left( \langle O  \rangle_\text{QEM}^\text{Qiskit}(i,N_\text{QC}) -  
\bar{\langle O  \rangle}_\text{QEM}^\text{Qiskit}(N_\text{QC},N_\text{samp})        \right)^2 \right]= \alpha^\text{Qiskit}_\text{QEM}(O,N_\text{samp},N_q,d) N_\text{QC} $, where $ \bar{\langle O  \rangle}_\text{QEM}^\text{Qiskit}(N_\text{QC},N_\text{samp}) =   \sum_{i=1}^{N_\text{samp}} \frac{\langle O  \rangle_\text{QEM}^\text{Qiskit}(i,N_\text{QC})}{N_\text{samp}} $. 
Such a quantity describes the deviations of $\langle O  \rangle_\text{QEM}^\text{Qiskit}(i,N_\text{QC})$ from  $\langle O  \rangle_\text{QEM}$ owing to the finite effect of $N_\text{QC}$ and plays the role of variance. Like $\left( \sigma^2_\text{noisy}[O,N_\text{QC}] \right)^{-1}$ and $\left[ \sigma^{2}_\text{noisy,\text{Qiskit}}[O,N_\text{QC},N_\text{samp}]] \right]^{-1}$,
we take it as the linear function of $N_\text{QC}$ given by the coefficient $ \alpha^\text{Qiskit}_\text{QEM}(O,N_\text{samp},N_q,d) $. 
Note that the dependency of the coefficient $ \alpha^\text{Qiskit}_\text{QEM}(O,N_\text{samp},N_q,d)$ in terms of $N_q$ and $d$ originates in the size of the quantum-noise-effect circuit group $3N_qd$.
Owing to our simulation, we obtain  $ \alpha^\text{Qiskit}_\text{QEM}(O,N_\text{samp},N_q,d)=3.45$ for $O=P_{110}$ and $ \alpha^\text{Qiskit}_\text{QEM}(O,N_\text{samp},N_q,d)=3.77$ for $O=P_{111}$.
The coefficients $ \alpha^\text{Qiskit}_\text{QEM}(O,N_\text{samp},N_q,d)$ are smaller than $\alpha^\text{Qiskit}_\text{noisy}[O,N_\text{samp}] $, as expected, since the deviations get larger owing to the usage of the quantum-noise-effect circuit group (additional computational resource for QEM). The ratio $\frac{\alpha^\text{Qiskit}_\text{noisy}[O,N_\text{samp}] }{ \alpha^\text{Qiskit}_\text{QEM}(O,N_\text{samp},N_q,d)}$, however, is about 1.34 for both cases which implies that the broadening of the deviations is not so big. We leave the detailed mathematical analysis on  $\left( \sigma^2_\text{QEM}[O,N_\text{QC}] \right)^{-1} $ and $\left[ \sigma^{2}_\text{QEM,\text{Qiskit}}[O,N_\text{QC},N_\text{samp},N_q,d] \right]^{-1} $ 
as well as the coefficient $ \alpha_\text{QEM}(O,N_\text{samp},N_q,d)$  as our future work. 
In addition to $N_\text{samp}=100$,  we also perform the simulations for $N_\text{samp}=1000$,  and as a result, we obtain 
$\alpha^\text{Qiskit}_\text{noisy}[O,N_\text{samp}] = 4.25$ and $ \alpha^\text{Qiskit}_\text{QEM}(O,N_\text{samp},N_q,d)=3.18$ for  $P_{110}$ and 
$\alpha^\text{Qiskit}_\text{noisy}[O,N_\text{samp}] = 4.47$ and $ \alpha^\text{Qiskit}_\text{QEM}(O,N_\text{samp},N_q,d)=3.34$  for $P_{111}$.  
As a result, by increasing $N_\text{samp}$ 
both  $\langle O  \rangle_\text{noisy}^\text{Qiskit}(i,N_\text{QC})$ and $\langle O  \rangle_\text{QEM}^\text{Qiskit}(i,N_\text{QC})$ approach to $\langle O  \rangle_\text{noisy}$ 
and $\langle O  \rangle_\text{QEM}$, respectively, owing to the law of large numbers. 
Let us also show the simulation results of the rest of the six probabilities of the computational basis states for $\vartheta_\tau = 0.2$ as the histogram in Fig. \ref{groverq3histogram}, which also includes $P_{110}$ and $P_{111}$.
The ideal values of these six probabilities are all zero and we see that the noisy simulation results with QEM are numerically close to them compared with those without QEM, which indicates that our QEM scheme also works for the other six probabilities.
\begin{figure*}[!htb]  
\centering
\includegraphics[width=0.8 \textwidth]{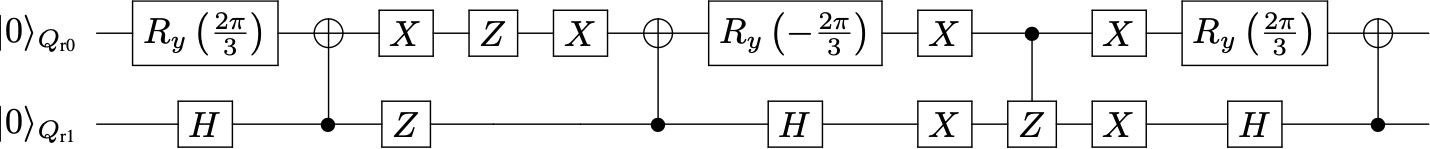}
\caption{ Quantum circuit for QAA given by $V^{ \text{QAA}} $ with $k=1$.}
\label{circuitQAA}
\end{figure*} 
\begin{figure*}[!htb] 
\centering
\includegraphics[width=0.8 \textwidth]{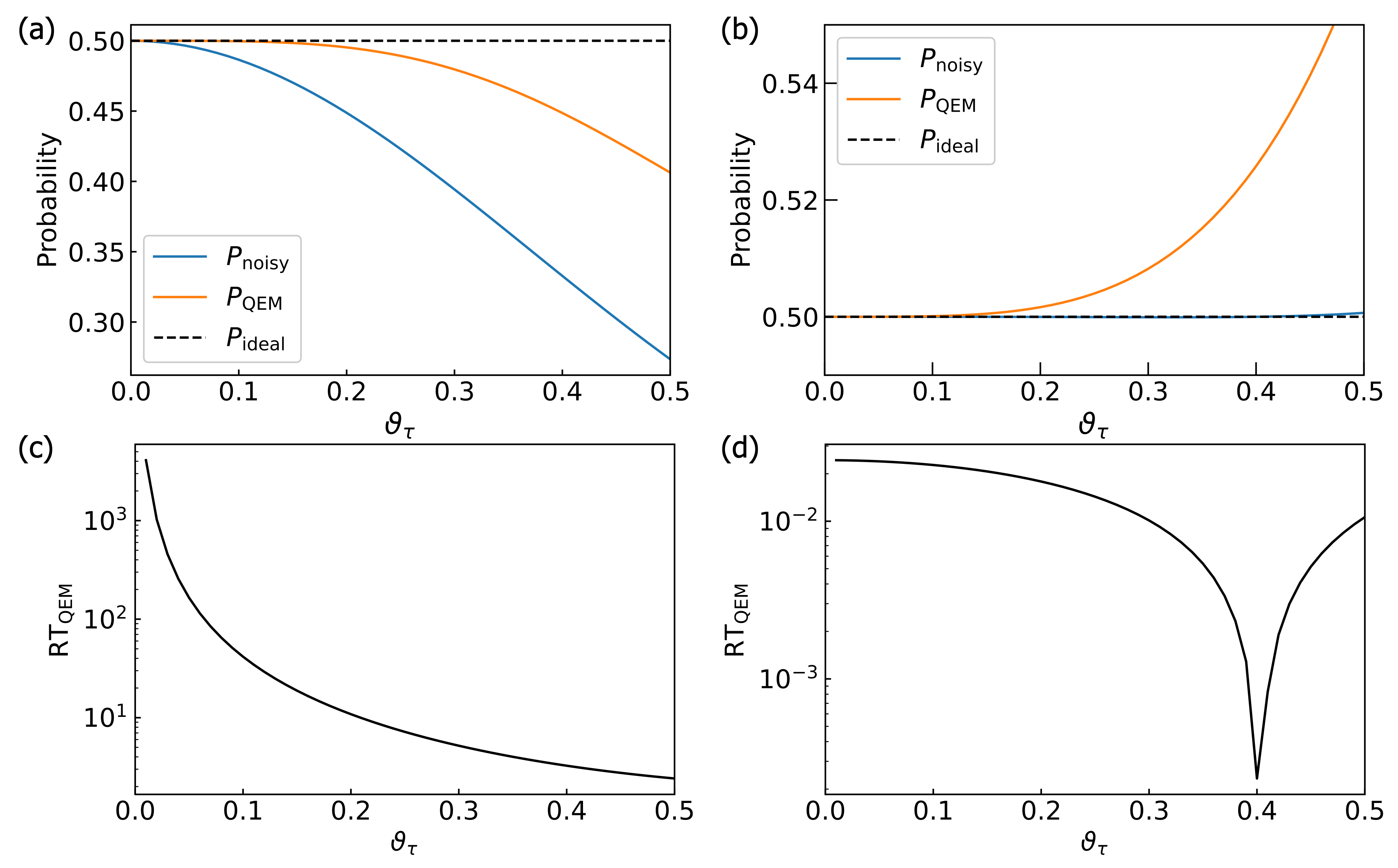}
\caption{ Noisy quantum simulations for QAA given by $V^{ \text{QAA}}$ for $\hat{O}=P_{00}, P_{11}$. 
All these results are obtained by our numerical code.
(a)  Plots of the results of $P_{\mathrm{ideal}} = \langle P_{11}  \rangle_{\rho _{d\cdots1}}, P_{\mathrm{noisy}} = \langle P_{11}  \rangle_{\rho^{\text{real}}  _{d\cdots1}},$ and  
$P_{\mathrm{QEM}} = \langle P_{11}  \rangle_{\rho^{\text{QEM}}  _{d\cdots1}}$
presented by the black dashed line, the blue solid line, and the orange solid line, respectively.
(b)  Plots of the results of $P_{\mathrm{ideal}} = \langle P_{00}  \rangle_{\rho _{d\cdots1}}, P_{\mathrm{noisy}} = \langle P_{00}  \rangle_{\rho^{\text{real}}  _{d\cdots1}},$ and $P_{\mathrm{QEM}} = \langle P_{00}  \rangle_{\rho^{\text{QEM}}  _{d\cdots1}}$
presented by the black dashed line, the blue solid line, and the orange solid line, respectively.
(c) The ratio $\text{RT}_{\text{QEM}} (\vartheta_\tau)$ for $P_{11}$. 
(d) The ratio $\text{RT}_{\text{QEM}} (\vartheta_\tau)$ for $P_{00}$. 
}
\label{GrovernoisyQsims} 
\end{figure*} %
\begin{figure}[!htb] 
\centering
\includegraphics[width=0.45 \textwidth]{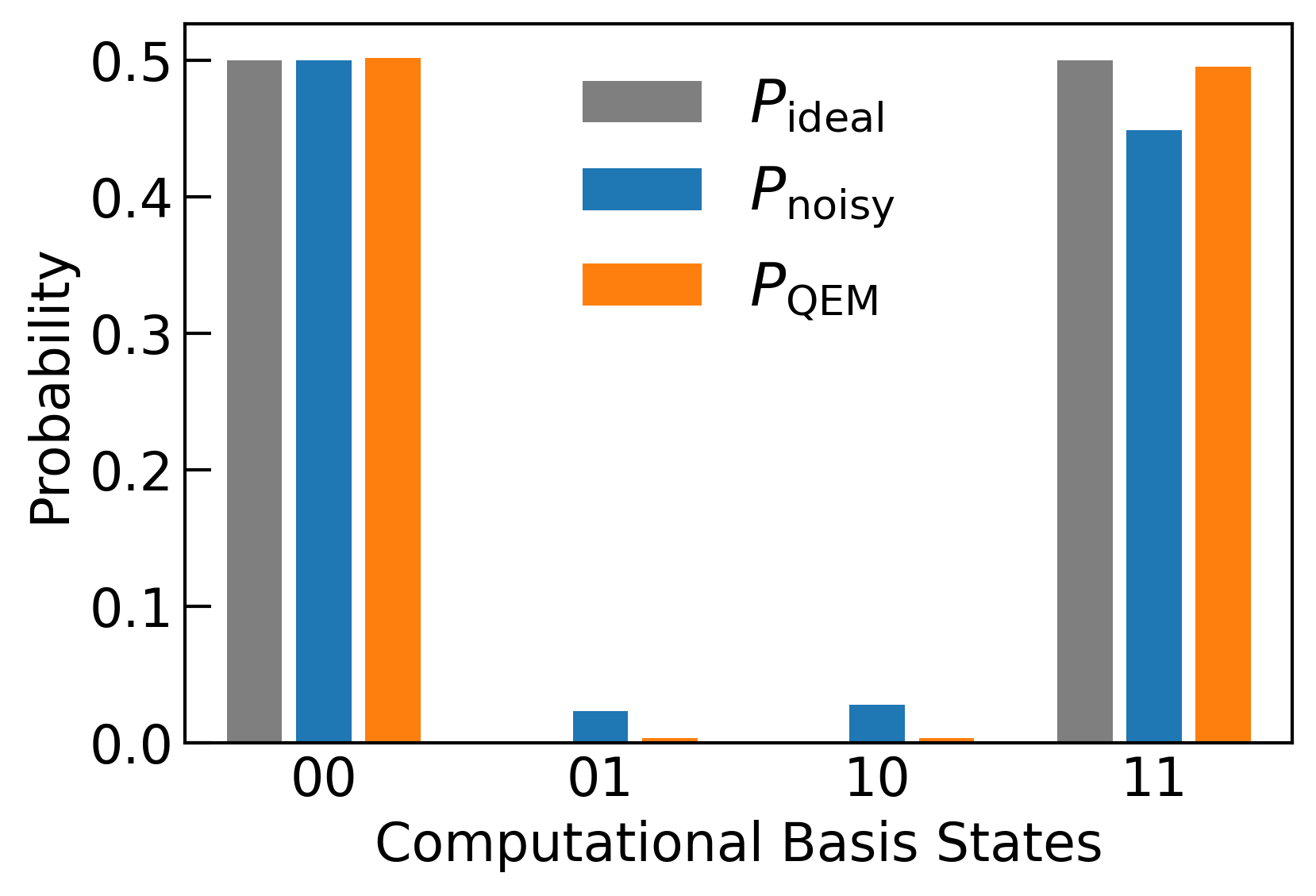}
\caption{ Histogram of the probability distribution of the computational basis states for two-qubit-system QAA simulations. We have set $\vartheta_\tau = 0.2$.}
\label{QAAhistogram}
\end{figure} 

In addition to the above simulation, let us present the simplified version of QAA for the two-qubit systems \cite{2021ibm}.
In this problem setting, we consider the effective two-dimensional space spanned by the two Bell states $| \Psi^+ \rangle = \frac{|01\rangle + |10\rangle}{\sqrt{2}} $ 
and $| \Phi^+ \rangle = \frac{|00\rangle + |11\rangle}{\sqrt{2}} $, and write their superposition state by 
$|\Sigma \rangle = c_{\Psi^+} | \Psi^+ \rangle + c_{\Phi^+} | \Phi^+ \rangle $,  where the complex coefficients $ c_{\Psi^+} ,  c_{\Phi^+} $
satisfy $|c_{\Psi^+}|^2 + |c_{\Phi^+}|^2=1.$
Our goal is to amplify the probability amplitude  $c_{\Phi^+} .$ 
Here we take the initialization operator to be  $V_\text{init} = U_{CX}[Q_{\text{r}1};Q_{\text{r}0}]  \cdot ( R_y\left(\frac{2\pi}{3}\right)_{Q_{\text{r}0}}  \otimes H_{Q_{\text{r}1}}  )$ 
while we take the Grover operator  to be $V_\text{G} = (V_\text{s}  V_\omega)^k$, where
$ V_\omega =    ( X \cdot Z \cdot X )_{Q_{r0}} \otimes Z_{Q_{r1}}$ and 
$V_\text{s}= V_\text{init} \cdot( 2|0\rangle^{\otimes 2}\langle 0| - \boldsymbol{1}_{4\times4}) \cdot V^\dagger_\text{init}  \equiv -\tilde{V}_\text{s}$ 
with $\tilde{V}_\text{s} =   V_\text{init} \cdot( X\otimes X \cdot U_{CZ}[Q_{r0};Q_{r1}]  \cdot X\otimes X) \cdot V^\dagger_\text{init}$. 
 $k$ is the number of $V_\text{G}$ to be applied.  In total, the unitary operation for running QAA  is given by $V^{ \text{QAA}}  =  ( V_{\text{G}}  )^k \cdot  V_{\text{init}}.$    
We present the corresponding quantum circuit in Fig. \ref{circuitQAA}.  
As similar to the above case,  on the quantum circuit we implement $\tilde{V}_{\text{s}}$ instead of  $V_{\text{s}}$.
In the following, let us take a look at the meanings of the three unitary operations  $V_{\text{init}}, V_{\text{s}}$, and $V_{\omega}$. 
First, the unitary operation $V_{\text{init}}$ generates the superposition of  $| \Phi^+ \rangle  $ and $| \Psi^+ \rangle  $ as  
$|00\rangle \to |\text{s}\rangle = V_{\text{init}}|00\rangle =
  \cos \frac{\vartheta_V}{2}|\Psi^{+}\rangle + \sin \frac{\vartheta_V}{2}|\Phi^{+}\rangle$ with $\vartheta_V=\pi/3$. 
Second, the unitary operation $V_{\omega} $ is the oracle operator and when it is applied to the initial  state $| \text{s} \rangle$ 
we have $V_{\omega}| \text{s} \rangle = \cos\frac{\vartheta_V}{2}| \Psi^+\rangle - \sin\frac{\vartheta_V}{2}| \Phi^+\rangle$, i.e., the oracle operator $V_{\omega} $ is the operator such that it reverses the sign of the Bell state $ | \Phi^+\rangle$.
Third, $V_{\text{s}}$ is the reflection of the vector $V_{\omega}|\text{s} \rangle$ with respect to the vector  $|\text{s}\rangle$.
When we operate $V_\text{G}  $  on  
$|\text{s}\rangle$ for $k$ times we have  
 $(V_{\text{G}})^k |s\rangle = \cos\frac{(2k+1)\vartheta_V}{2}| \Psi^+\rangle +  \sin\frac{(2k+1)\vartheta_V}{2}| \Phi^+\rangle,$ and since $\vartheta_V=\pi/3$ we have $k=1.$
 We can understand the geometrical meaning of the operation $V_{\text{G}} $  as follows.
 Let us consider the effective three-dimensional space spanned by the two Bell states $| \Phi^+ \rangle$ and $| \Psi^+ \rangle$ and the vector $|\xi \rangle$ which is
 perpendicular to both $| \Phi^+ \rangle$ and $| \Psi^+ \rangle$. 
 Furthermore, we call the axis which is parallel to $|\xi \rangle$ (perpendicular to the two-dimensional plane spanned by $| \Phi^+ \rangle$ and $| \Psi^+ \rangle$) as $\xi$-axis. 
  The unitary operation $V_{\text{G}} $ is the rotation about  $\xi$-axis by the angle $\vartheta_V$ in this effective three-dimensional space.  
 We can verify whether $| \Phi^+\rangle$ has been generated as the output state or not by measuring the probabilities of  $|00\rangle$ state and $|11\rangle$ state,
 which are denoted by  $p_{00}$ and $p_{11}$, respectively. 
 In other words, the probabilities $p_{00}$ and $p_{11}$ are the expectation values of the projection operators 
 $P_{00}$ and $P_{11}$, respectively, where $P_{00}$ ($P_{11}$) is the projection operator of $|00\rangle$ ($|11\rangle$) state.
 Namely, we run noisy quantum simulation for $\hat{O}=P_{00}, P_{11}$ and perform QEM on them: Note that in the case of the ideal simulation we obtain $p_{00} = p_{11} = \frac{1}{2}$. 
We plot the numerical results of $p_{11}$ and $p_{00}$ for the range $0 \leq \vartheta_\tau \leq 0.5$ in Figs. \ref{GrovernoisyQsims}(a) and \ref{GrovernoisyQsims}(b), respectively, and in Figs. \ref{GrovernoisyQsims}(c) and \ref{GrovernoisyQsims}(d) we plot $\text{RT}_{\text{QEM}} (\vartheta_\tau)$
for $p_{11}$ and $p_{00}$, respectively. 
All these results shown here are obtained by our original numerical code and we have denoted $\langle P_{00(11)}  \rangle_{\rho^{\text{ideal}}  _{d\cdots1}}, \langle P_{00(11)} \rangle_{\rho^{\text{real}}  _{d\cdots1}},$ and $\langle P_{00(11)} \rangle_{\rho^{\text{QEM}}  _{d\cdots1}}$
by $ P_\text{ideal},  P_\text{real}$, and $ P_\text{QEM} $, respectively.                           
Let us look from the results of the probability $p_{11}$. 
In Fig. \ref{GrovernoisyQsims}(a) we see that for any $\vartheta_\tau$ the absolute of the deviation 
$\big{|} \langle P_{11}  \rangle_{\rho^{\text{real}}  _{d\cdots1}} - \langle P_{11}  \rangle_{\rho  _{d\cdots1}} \big{|}$ is bigger than $\big{|} \langle P_{11}  \rangle_{\rho^{\text{QEM}}  _{d\cdots1}} - \langle P_{11}  \rangle_{\rho  _{d\cdots1}} \big{|}$, 
and correspondingly, as we see in Fig. \ref{GrovernoisyQsims}(c)
 the ratio $\text{RT}_{\text{QEM}} (\vartheta_\tau)$ is greater than one. Therefore, our QEM scheme works well for the noisy simulation of $p_{11}$.
In contrast, in Figs. \ref{GrovernoisyQsims}(b) and \ref{GrovernoisyQsims}(d)  we see that the probability $p_{00}$ shows essentially a different behavior.
 That is the expectation value without QEM $ \langle P_{00}  \rangle_{\rho^{\text{real}}  _{d\cdots1}}$ is numerically close to the ideal value $\langle P_{00}  \rangle_{\rho  _{d\cdots1}}$ compared with the QEM-performed expectation value $ \langle P_{00}  \rangle_{\rho^{\text{QEM}}  _{d\cdots1}}$, and correspondingly,
 the ratio $\text{RT}_{\text{QEM}} (\vartheta_\tau)$ is lower than one. 
 Such a characteristic is understood as follows.  
 Firstly, we have analytically examined that 
 the expectation value $ \langle P_{00}  \rangle_{\rho^{\text{real}}  _{d\cdots1}}$ does not include first-order term in $\tau$, i.e., $ \langle P_{00}  \rangle_{\Delta^{\text{AD}}_1 \rho _{d\cdots1}} =0$.
The lowest-order term included in the numerator of $\text{RT}_{\text{QEM}} (\vartheta_\tau)$  is  $\mathcal{O}(\tau^2)$.
 Secondly, due to our QEM the lowest order of the denominator of $\text{RT}_{\text{QEM}} (\vartheta_\tau)$ is also $\mathcal{O}(\tau^2)$. 
As a result, the ratio $\text{RT}_{\text{QEM}} (\vartheta_\tau)$ becomes lower than one, which indicates that it is not appropriate to adopt our QEM scheme.  
We consider that this is because our QEM scheme described by Eq. \eqref{QEMformula1} is the scheme for mitigating the first-order AD effect. 
 In the limit of $\vartheta_\tau \to 0$, the ratio $\text{RT}_{\text{QEM}}(\vartheta_\tau) $ takes finite value, and analytically it is the ratio between the absolute of the coefficient of $ \langle P_{00}  \rangle_{\delta^{\text{AD}}_2 \rho _{d\cdots1}} $ and that of $ \langle P_{00}  \rangle_{\delta_1(\Delta^{\text{AD}}_1 \rho _{d\cdots1})}.$ 
We note that we have performed two types of simulations with our original code. 
In the first one we have directly implemented $U_{CZ}[Q_{\text{r}1};Q_{\text{r}0}] $ gate while in the second one we have implemented it via 
the decomposition $\left({\boldsymbol{1}_{2\times2}}_{Q_{\text{r}0}} \otimes H_{Q_{\text{r}1}} \right) \cdot U_{CX}[Q_{\text{r}1};Q_{\text{r}0}]  \cdot \left({\boldsymbol{1}_{2\times2}}_{Q_{\text{r}0}} \otimes H_{Q_{\text{r}1}} \right)$.
According to the results of these two simulations, we have analyzed that on the Qiskit code $U_{CZ}[Q_{\text{r}1};Q_{\text{r}0}] $ gate
is automatically implemented by the above decomposition since the result of the second simulation with our original code shows better matching with that obtained by our Qiskit code. In such a case, the first-order term in $\tau$ appears for $ \langle P_{00}  \rangle_{\rho^{\text{real}}  _{d\cdots1}}$ and our QEM works well.
Besides $p_{00}$ and $p_{11}$, let us briefly discuss the noisy simulation results of the probability weights of $|01\rangle$ and $|10\rangle$ and write them by $p_{01}$ and $p_{10},$ respectively.
We show them in the histogram in Fig. \ref{QAAhistogram} which describes the probability distribution of the computational basis states of the two register qubits $Q_{\text{r}0}$ and $Q_{\text{r}1}$.
Here we have taken  $\vartheta_\tau=0.2$. 
We see that like the probability weight of $|11\rangle$, 
our QEM scheme works for the probability weights of $|01\rangle$ and $|10\rangle$.

Let us end this subsection by giving the following comment.
In the previous subsection, we have seen that our QEM scheme becomes meaningless in the cases when the expectation values of the ideal simulations are equivalent to those of noisy simulations such as the simulation for the expectation value $\langle Y \rangle$. Besides these cases, our QEM scheme represented by the formula in Eq. \eqref{QEMformula1} does not work when noisy expectation values do not include the first-order term in $\tau$ like the noisy simulation for the probability $p_{00}$ discussed above. In other words, if we construct the QEM formula which describes the mitigation for a higher-order quantum noise effect, which is discussed in Sec. IA in the Supplementary Material, by using it we become able to accomplish the noisy quantum simulation obtaining $\text{RT}_{\text{QEM}}(\vartheta_\tau) >1.$ 
 In practice, however, when we run quantum algorithms on real quantum devices we cannot compute $\text{RT}_{\text{QEM}}(\vartheta_\tau)$ since we cannot compute ideal expectation values. We can check whether noisy expectation values include the first-order terms in $\tau$ or not, for example, in the following way. We perform two types of QEM, QEM of both the first- and second order quantum noise effects and the one of only the second-order effect. 
Let us denote the density matrices obtained by the former QEM and the latter one by
$\rho^\text{QEM}_\text{2nd}$ and $\rho^\text{QEM}_\text{2nd,only}$, respectively. 
Next, we introduce the measure $M_\text{1st/2nd}=\Big{|} 
\text{Tr}\Big{(}\big{(}  \rho^\text{QEM}_\text{2nd}-\rho^\text{QEM}_\text{2nd,only} \big{)}O\Big{)}\times\tau^{-1}\Big{|}$, where $O$ is a physical operator.
If the first-order QEM fails (noisy expectation values do not include the first-order terms in $\tau$) then the measure $M_\text{1st/2nd}$ is $\mathcal{O}(\tau^2)$.
On the other hand, if the first-order QEM succeeds (noisy expectation values include the first-order terms in $\tau$ and is mitigated) then $M_\text{1st/2nd}$ is $\mathcal{O}(1)$.
By extending this approach we can examine whether noisy expectation values include 
higher-order quantum noise effects or not. We consider, however, that the failure of the first-order QEM for the noisy quantum computation when the linear order in $\tau$ does not appear is not so crucial compared with the case when we have failed in mitigating the linear-order quantum noise effects included in noisy expectation values, and indeed we can see this by looking at Fig. \ref{QAAhistogram}. For the probability weight of |00> state, the numerical differences among the three expectation values, $ \langle \hat{O}  \rangle_{\rho  _{d\cdots1}},  \langle \hat{O}  \rangle_{\rho^{\text{real}} _{d\cdots1}},  \langle \hat{O}  \rangle_{\rho^{\text{QEM}}  _{d\cdots1}} $, are small. On the other hand,
 for the probability weight of |11> state, $ \langle \hat{O}  \rangle_{\rho  _{d\cdots1}}$ and  $  \langle \hat{O}  \rangle_{\rho^{\text{QEM}}  _{d\cdots1}} $ is close enough while $ \langle \hat{O}  \rangle_{\rho  _{d\cdots1}}$ and  $\langle\hat{O}  \rangle_{\rho^{\text{real}} _{d\cdots1}}$ are quite separated.
\begin{figure}[b] 
\centering
\includegraphics[width=0.35 \textwidth]{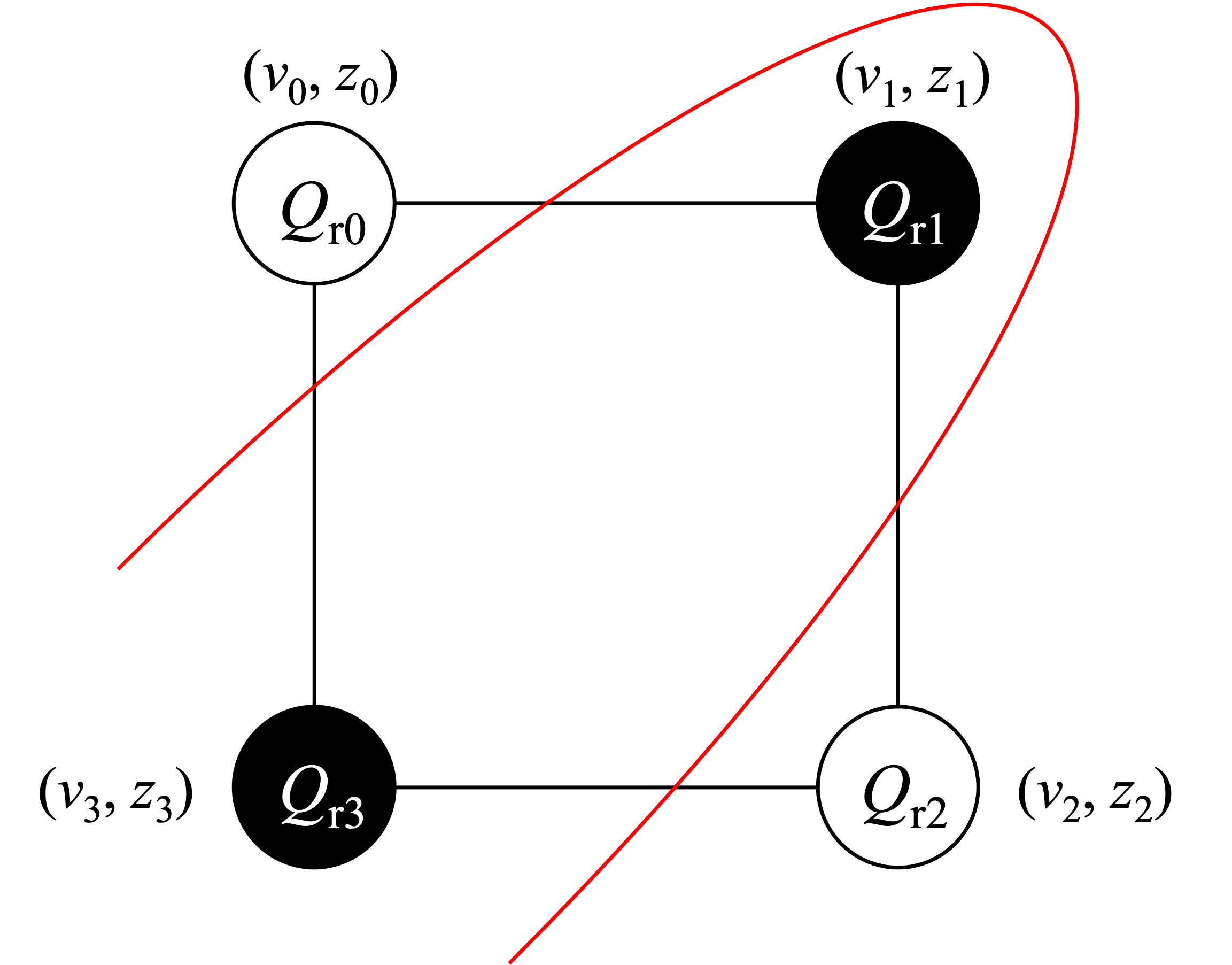}
\caption{ Structure of the graph $G = (V,E)$. It is the square composed of the four vertices $v_0, v_1, v_2,$ and $v_3$ 
  and the four edges $ \langle v_0 v_{1} \rangle,  \langle v_1 v_2 \rangle,  \langle v_2 v_3 \rangle,$ and $ \langle v_3 v_0 \rangle$. 
  For each vertex $v_i$ ($i=0,1,2,3$) the binary value $z_i=\pm1$ is assigned and the set $(v_i,z_i)$ is encoded in the qubit $Q_{\text{r}i}$ in the QAOA simulation.} 
\label{QAOAgraph}
\end{figure} 
\begin{figure}[b] 
\centering
\includegraphics[width=0.475 \textwidth]{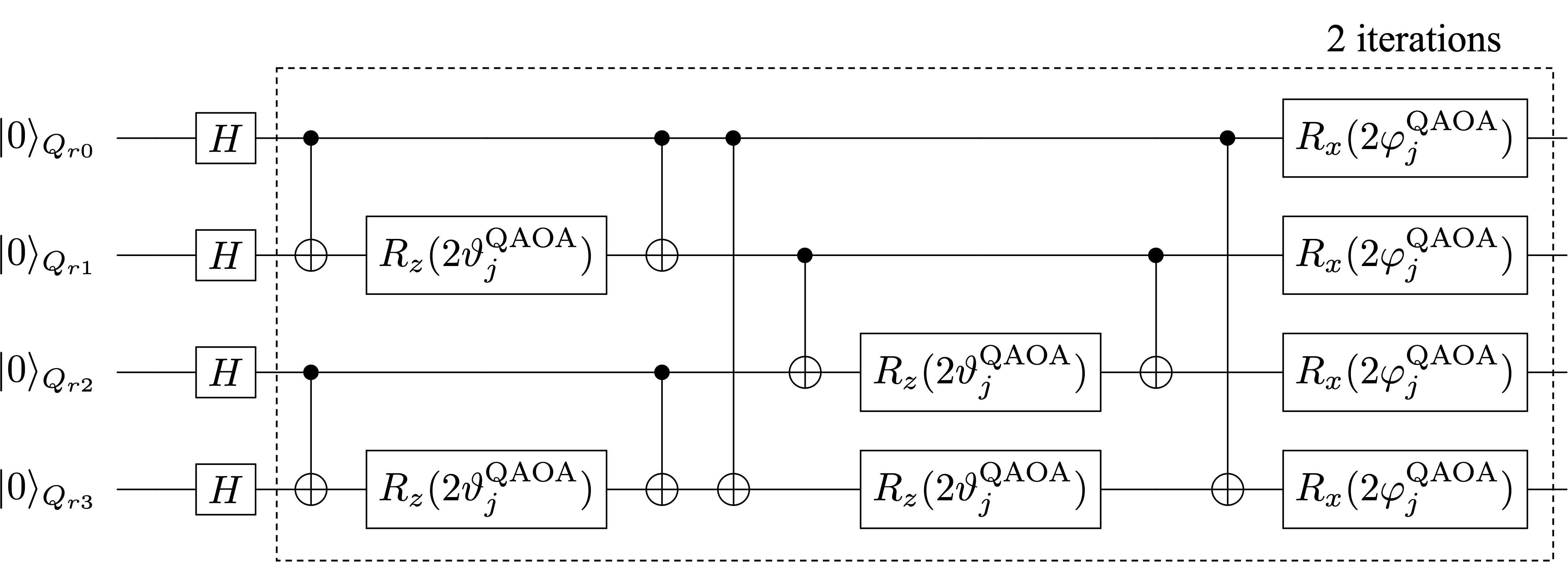} 
\caption{ Quantum circuit for QAOA.}
\label{QAOAqcirc}
\end{figure} 
\subsection{ QAOA}\label{QEMqaoa}  
As a final example, let us apply our QEM scheme to the noisy simulation of the variational quantum algorithm called 
Quantum Approximate Optimization Algorithm (QAOA)  \cite{hybridQCalgorithmJPSJ2021,QAOA2014,crooks2018performance,wang2018quantum,shaydulin2019evaluating,zhou2020quantum,TIsing3}. In the following, we analyze QAOA for a max-cut problem which is to divide vertices (nodes) of a given graph into two groups 
 so that the number of edges connecting two vertices belonging to the different groups is maximized and is a NP (Non-deterministic Polynomial time)-hard problem. 
 
 First, we discuss from a theoretical framework of a classical approximate optimization. 
  We express the given graph $G$ as $G = (V,E)$, where $V = \{ v_0, \ldots, v_i, \ldots, v_{N_\text{V}-1} \}$ is the set of vertices with $N_\text{V}$ denoting their total number and for each vertex $v_i$ the binary value $z_i = \pm 1$ is assigned. 
  $E = \Big{\{} \big{\{} \langle v_0 v_1 \rangle, C_{0,1} \big{\}}, \ldots,  \big{\{} \langle v_i v_{i+1} \rangle, C_{i,i+1} \big{\}}, \ldots,$  $\big{\{} \langle v_{N_\text{V}-2} v_{N_\text{V}-1} \rangle, C_{N_\text{V}-2,N_\text{V}-1} \big{\}} \Big{\}}$
  is the set of the edges with $\langle v_i v_{i+1} \rangle$ denoting the edge connected by the vertices $v_i$ and $v_{i+1} $.
  The quantity $C_{ij}$ ($i,j=0,\ldots, N_\text{V}-1$ with $i\neq j$) is the adjacency matrix element (weight) for the edge $\langle v_i v_j \rangle$ which is semi-positive.
Let us write the $N_\text{V}$ strings of  $z_i$ by $\boldsymbol{z} = (z_0,\ldots, z_{N_\text{V}-1})$. The goal of a classical approximate optimization is to minimize the cost function 
\begin{align}
C(\boldsymbol{z})= \frac{1}{2}\sum_{ (v_i, v_j)} C_{ij} \big{(} z_iz_j   -1   \big{)} ,  \label{AOAcostfunction} 
\end{align}
  or equivalently to maximize the ratio $r_\text{CAO}$ ($\leq1$) which satisfies
\begin{align}
\frac { C(\boldsymbol{z}) }{C_\text{min}} \geq  r_\text{CAO},  \label{AOAcostfunction2}
\end{align}
  where $C_\text{min}$ is the minimum value of $ C(\boldsymbol{z})$.
  In our simulation,  as illustrated in Fig. \ref{QAOAgraph} we adopt the square graph given by the four vertices $v_0, v_1, v_2,$ and $v_3,$ 
  and the edges are $ \langle v_0 v_{1} \rangle,  \langle v_1 v_2 \rangle,  \langle v_2 v_3 \rangle,$ and $ \langle v_3 v_0 \rangle$,  and take $C_{ij}=1$ for any edge $\langle v_i v_{j} \rangle$.  
  Next, we discuss the theoretical framework for QAOA. The four vertices $v_0, v_1, v_2,$ and $v_3$  are encoded in  four qubits $Q_{ \text{r} 0}, Q_{ \text{r} 1}, Q_{ \text{r} 2},$ and $Q_{ \text{r} 3}$, respectively,
  and the values   $z_i z_j $  in the expectation values of the operators $Z_{Q_i} \otimes Z_{Q_j}$. 
 The cost function $C(\boldsymbol{z})$ in Eq. \eqref{AOAcostfunction} is given by the expectation of the Hamiltonian    
\begin{align}
	H_{\text{C}} = \frac{1}{2}\sum_{(i,j)  } \big{(} Z_{Q_{\text{r}i}} \otimes Z_{ Q_{\text{r}j} }   -1   \big{)}  , \label{QAOAHamiltonian}
\end{align} 
where the symbol $(i, j)$ ($i, j =0,1,2,3$) denotes the summation for the edges connected by the qubits $Q_{ \text{r} i}$ and $Q_{ \text{r} j}$ under the square-graph structure in Fig. \ref{QAOAgraph}.
In this simulation the physical operator $\hat{O}$ is the Hamiltonian $H_{\text{C}} $ in Eq. \eqref{QAOAHamiltonian}.
 The unitary operation for running QAOA, which we denote by $U^{ \text{QAOA}}$, consists of three elements. 
The first one is the unitary operation for creating the reference state and is given by the Hadamard-gate operation on all four qubits, $ U_\text{int} = H^{\otimes 4}$.  
The other two are the unitary operations $U_{ \text{C}}(\vartheta^{ \text{QAOA}}_j)$ and $U_{ X}(\varphi^{ \text{QAOA}}_j)$ which are generated 
by the Hamiltonian $H_{\text{C}} $ in Eq. \eqref{QAOAHamiltonian} with the angle $\vartheta^{ \text{QAOA}}_j$ and the term $H_X = \sum_{i} X_{Q_{\text{r}i}}$
 with the angle $\varphi^{ \text{QAOA}}_j$, respectively: in QAOA $H_X$  is called the transverse-field (mixing or driving) term.
The two types of angles $\vartheta^{ \text{QAOA}}_j$ and $\varphi^{ \text{QAOA}}_j$ ($j=1,\ldots, p$) are the variational parameters
and $p$ is the repetition number of applying the unitary operation $U_{ X}(\varphi^{ \text{QAOA}}_j) \cdot U_{ \text{C}}(\vartheta^{ \text{QAOA}}_j) $ (the number of iteration), which determines the accuracy of QAOA.
 In total, $U^{ \text{QAOA}}  $ is given by 
\begin{align}
&	U^{ \text{QAOA}}  
    = 
    \left[ \prod_{j=1}^{p} \left( U_X (\varphi^{\text{QAOA}}_j)  \cdot U_C (\vartheta^{\text{QAOA}}_j) \right) \right]
    \cdot
    \left[\bigotimes_{a=0}^3 H_{Q_{ra}} \right], \notag\\
&  U_C (\vartheta^{\text{QAOA}}_j)   = e^{-i \vartheta^{\text{QAOA}}_j H_C}, \quad  U_X (\varphi^{\text{QAOA}}_j)   =  e^{-i \varphi^{\text{QAOA}}_j H_X}.
    \label{UQAOA1}
 \end{align} 
 The quantum circuit for $U^{ \text{QAOA}} $ in Eq. \eqref{UQAOA1} is presented in Fig. \ref{QAOAqcirc}. 
 In our simulation we set $p=2$  and the circuit depth is $d=15$.
 The unitary operation $U_X (\varphi^{\text{QAOA}}_j)$ is implemented by the $R_x$ gate (rotation about $x$ axis) with the angle $2\varphi^{\text{QAOA}}_j$.
  Meanwhile, the quantum circuit for $U_C (\vartheta^{\text{QAOA}}_j)$ is composed of the sets of the quantum gates $\big{[} U_{CX}[Q_{\text{r}i}; Q_{\text{r}i} ],  R_z (2\vartheta^{\text{QAOA}}_j)  \big{]}$,
  where $R_z$ denotes the rotational gate about $z$ axis and the associated angle is  $2\vartheta^{\text{QAOA}}_j$.
 Corresponding to Eq. \eqref{AOAcostfunction2}, the goal of QAOA simulation is to compute and minimize the expectation value 
 \begin{align} 
C\big{(}\boldsymbol{\vartheta}^{\text{QAOA}}, \boldsymbol{\varphi}^{\text{QAOA}}\big{)} =  \langle \boldsymbol{\vartheta}^{\text{QAOA}}, \boldsymbol{\varphi}^{\text{QAOA}}|   H_{\text{C}} 
  | \boldsymbol{\vartheta}^{\text{QAOA}}, \boldsymbol{\varphi}^{\text{QAOA}} \rangle  ,
 \label{QAOAcostfunction}
\end{align}
\begin{figure}[!htb] 
\centering
\includegraphics[width=0.5 \textwidth]{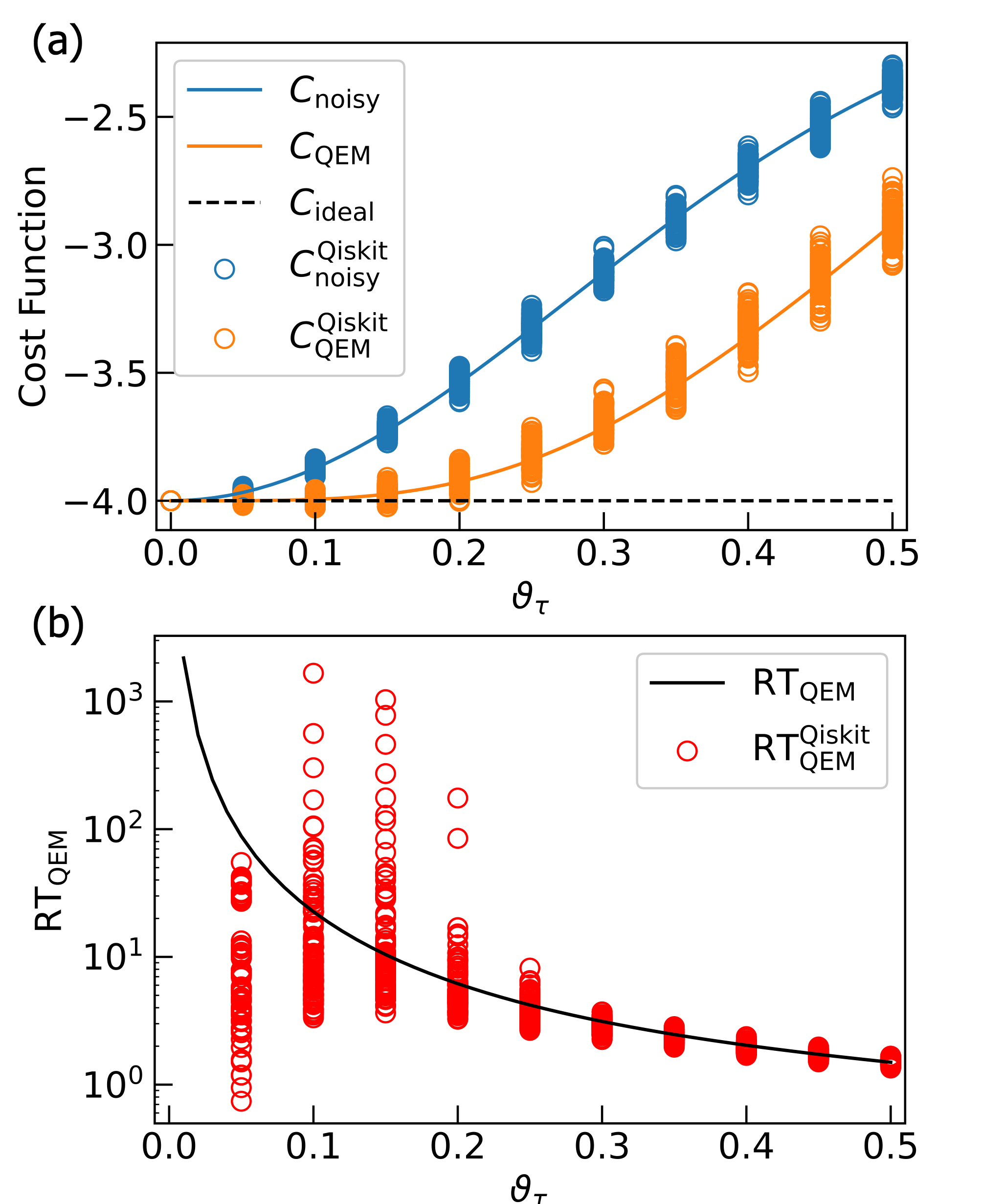}
\caption{ Quantum simulations for QAOA. 
In (a) we show the results of QEM for the cost function $ C\big{(}\boldsymbol{\vartheta}^{\text{QAOA}}, \boldsymbol{\varphi}^{\text{QAOA}}\big{)} $.
The blue and orange solid lines are $C_\text{noisy} = \langle H_{C}  \rangle_{\rho^{\text{real}}  _{d\cdots1}}$ and $C_\text{QEM} = \langle H_{C}  \rangle_{\rho^{\text{QEM}}  _{d\cdots1}}$, respectively, and they obtained by our original code.  
The blue and orange circles are the results of $ \langle H_{C}  \rangle_{\rho^{\text{real}}  _{d\cdots1}}$ and  $\langle H_{C}  \rangle_{\rho^{\text{QEM}}  _{d\cdots1}}$ obtained by our Qiskit code, respectively. 
 For this simulation, we have described  $\langle H_{C}  \rangle_{\rho^{\text{real}}  _{d\cdots1}}$ and $ \langle H_{C}  \rangle_{\rho^{\text{QEM}}  _{d\cdots1}} $
 as $ C_\text{noisy}^{\text{Qiskit}}$ and $ C_\text{QEM}^{\text{Qiskit}}$, respectively. 
The dotted black line is the ideal expectation value $C_\text{ideal} = \langle H_{C}  \rangle_{\rho  _{d\cdots1}}$. 
In (b) we have plotted the results of the ratio $\text{RT}_{\text{QEM}} (\vartheta_\tau)$. The black curve is the result obtained by our original code whereas the red circles are the one obtained by our Qiskit code.
For each $\vartheta_\tau$, we have plotted 100 circles.   }
\label{QAOAnoisyQsims}
\end{figure}
\begin{figure*}[!htb] 
\centering
\includegraphics[width=0.85 \textwidth]{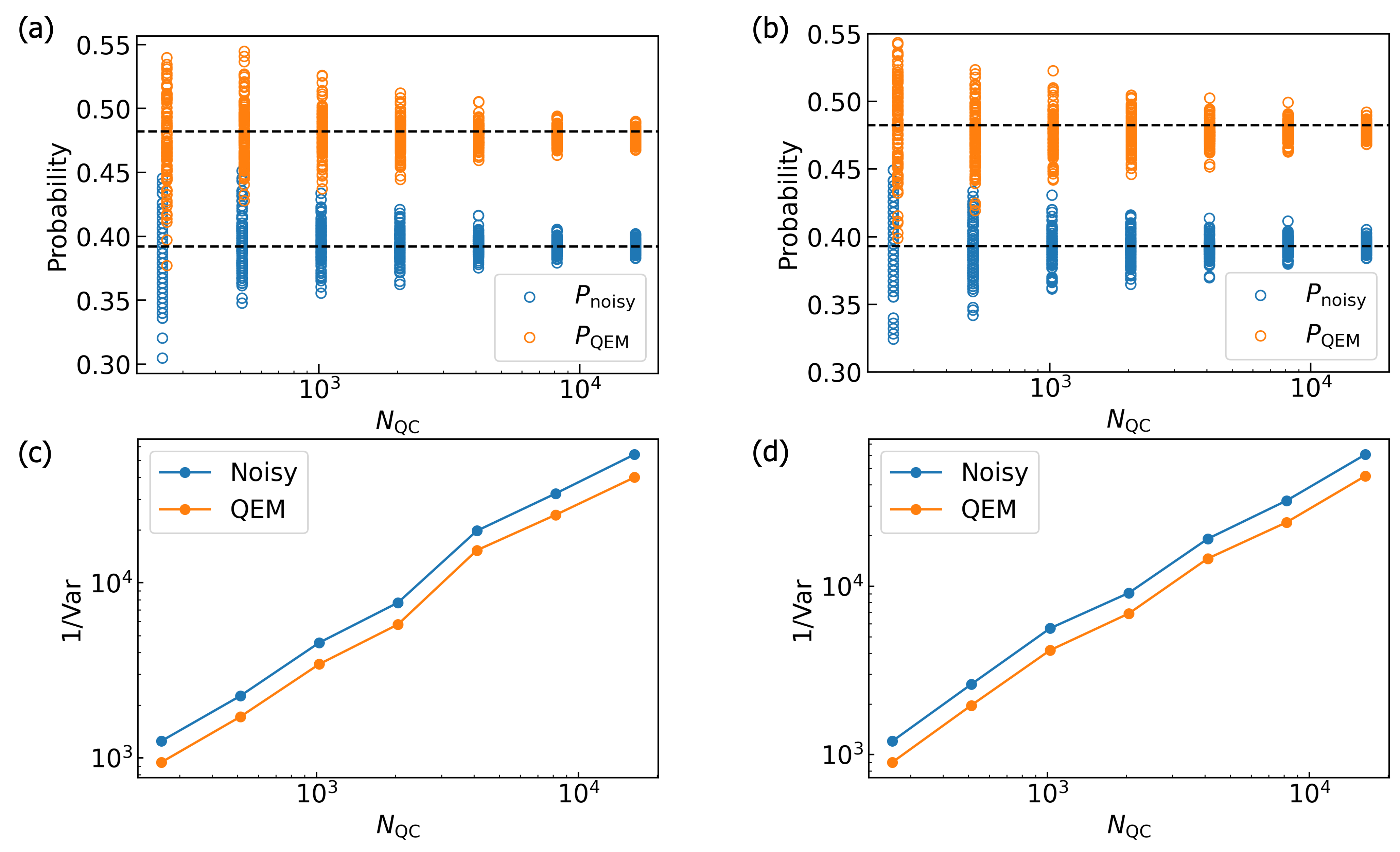}
\caption{ Numerical results  of  $\langle O  \rangle_\text{noisy}^\text{Qiskit}(i,N_\text{QC})$ and $\langle O  \rangle_\text{QEM}^\text{Qiskit}(i,N_\text{QC})$
for (a) $P_{0101}$ and (b) $P_{1010}$. We plot $\left( \sigma^{2}_\text{noisy,\text{Qiskit}}[O,N_\text{QC},N_\text{samp}] \right)^{-1} $ (blue) and $\left[ \sigma^{2}_\text{QEM,\text{Qiskit}}[O,N_\text{QC},N_\text{samp},N_q,d] \right]^{-1}$ (orange)
for $P_{0101}$ and $P_{1010}$ in (c) and (d), respectively. We take $N_\text{QC}=2^{n_\text{QC}}$ with $n_\text{QC}=8,9,\ldots,14$, $N_\text{samp}=100$, and $\vartheta_\tau=0.2$. 
}
\label{QAOANqcdependence}
\end{figure*} 
\begin{figure*}[!htb] 
\centering
\includegraphics[width=1.0 \textwidth]{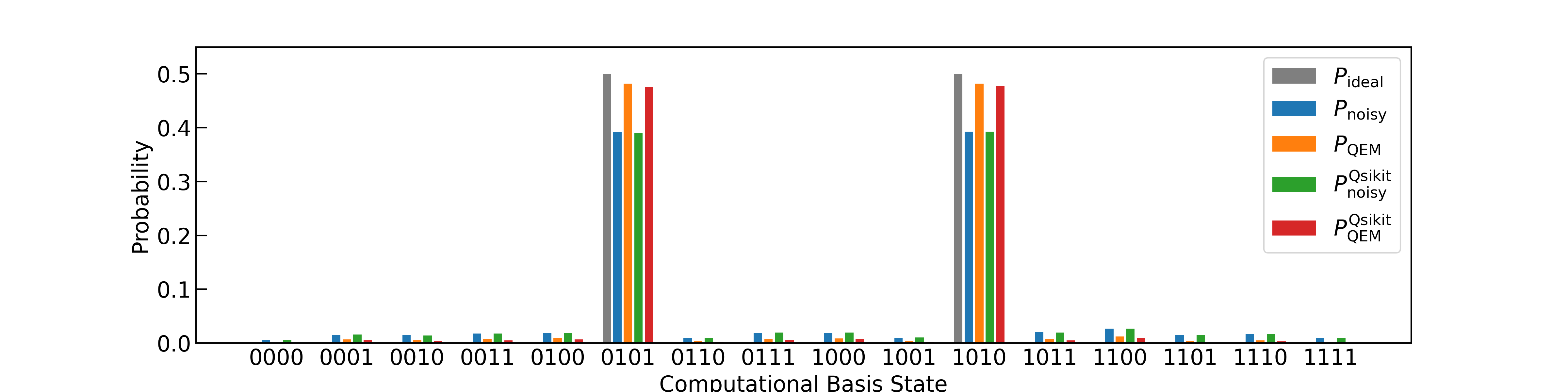}
\caption{ Histogram of the probability distribution of the computational basis states for QAOA simulations. We have set  $\vartheta_\tau = 0.2$. 
 $P_\text{ideal}, P_\text{noisy},$ and $P_\text{QEM} $ are obtained by our original code while $P_\text{noisy}^\text{Qiskit}$ and $P_\text{QEM}^\text{Qiskit} $ are obtained by the Qiskit code.}
\label{QAOAhistogram}
\end{figure*} 
 where $ | \boldsymbol{\vartheta}^\text{QAOA}, \boldsymbol{\varphi}^\text{QAOA} \rangle =  U^\text{QAOA}     |0\rangle^{\otimes 4}$
 with $ \boldsymbol{\vartheta}^\text{QAOA} = (\vartheta^\text{QAOA}_1, \vartheta^\text{QAOA}_2)$ and 
 $ \boldsymbol{\varphi}^\text{QAOA} = (\varphi^\text{QAOA}_1, \varphi^\text{QAOA}_2)$.
Let us also call the expectation value $ C\big{(}\boldsymbol{\vartheta}^\text{QAOA}, \boldsymbol{\varphi}^\text{QAOA}\big{)} $ in Eq. \eqref{QAOAcostfunction} as the cost function 
and its minimization is equivalent to the optimization of the variational parameters $\vartheta_j^\text{QAOA}, \varphi_j^\text{QAOA} $,
and write them by $\vartheta_j^\text{QAOA,opt}, \varphi_j^\text{QAOA,opt} $. We show their values in Sec. III in the Supplementary Material.
Let us now discuss our simulation results shown in Figs. \ref{QAOAnoisyQsims}-\ref{QAOAhistogram}. 
All these results are obtained by computing the probability distribution of the computational basis states since the Hamiltonian $H_{\text{C}} $ in Eq. \eqref{QAOAHamiltonian} is given by the $Z$ gate operations.  
First, let us take a look at the simulation results in Fig. \ref{QAOAnoisyQsims}.  
In Fig.  \ref{QAOAnoisyQsims}(a), we have plotted the results of the cost function $C\big{(}\boldsymbol{\vartheta}^{\text{QAOA}}, \boldsymbol{\varphi}^{\text{QAOA}}\big{)} $
for the ideal simulation ($C_\text{ideal}  = \langle H_{C}  \rangle_{\rho  _{d\cdots1}}$), the noisy simulation ($C_\text{noisy} = \langle H_{C}  \rangle_{\rho^{\text{real}}  _{d\cdots1}}$), 
and the simulation with QEM  ($C_\text{QEM} = \langle H_{C}  \rangle_{\rho^{\text{QEM}}  _{d\cdots1}}$).
The dashed black line, the blue solid line, and the orange solid line are  $C_\text{ideal}, C_\text{noisy}$, and $C_\text{QEM}$, respectively, and they are all obtained by our original code.
On the other hand, the blue and orange circles are $ \langle H_{C}  \rangle_{\rho^{\text{real}}  _{d\cdots1}}$ and $ \langle H_{C}  \rangle_{\rho^{\text{QEM}}  _{d\cdots1}}$, respectively, and they have been calculated by our Qiskit code. 
We have denoted $\langle H_{C}  \rangle_{\rho^{\text{real}}  _{d\cdots1}}$ and $ \langle H_{C}  \rangle_{\rho^{\text{QEM}}  _{d\cdots1}}$ by $C_{\mathrm{noisy}}^{\mathrm{Qiskit}}$ and $C_{\mathrm{QEM}}^{\mathrm{Qiskit}}$, respectively.
We have plotted 100 circles for each $\vartheta_\tau$.  We see that in the range  $0.1 \leq\vartheta_\tau \leq 0.5$ our QEM works well.  
In Fig.  \ref{QAOAnoisyQsims}(b), we have plotted the ratio $\text{RT}_{\text{QEM}} (\vartheta_\tau)$.
The black curve is the result obtained by our original code while the red circles are those obtained by our Qiskit code and 100 circles are plotted for each $\vartheta_\tau$.
Corresponding to the result shown in Fig.  \ref{QAOAnoisyQsims}(a), the ratio satisfies $\text{RT}_{\text{QEM}}(\vartheta_\tau) >1$.  
Like the results in Fig. \ref{GroverNqcdependence}, we present the $N_\text{QC}$ dependencies of the probabilities  $P_{0101}$ and $P_{1010}$
 in Figs. \ref{QAOANqcdependence}(a) and \ref{QAOANqcdependence}(b), respectively. We take $N_\text{QC}=2^{n_\text{QC}}$ with $n_\text{QC}=8,9,\ldots,14$, $N_\text{samp}=100$, and $\vartheta_\tau=0.2$. 
We see that as we increase $N_\text{QC}$
both $\langle P_{0101} \rangle^\text{Qiskit}_\text{noisy}(i,N_\text{QC})$ and $\langle P_{1010} \rangle^\text{Qiskit}_\text{noisy}(i,N_\text{QC})$ approach to  
$\langle P_{0101} \rangle_\text{noisy}$ and $\langle P_{1010} \rangle_\text{noisy}$, respectively,
and similarly,  $\langle P_{0101} \rangle^\text{Qiskit}_\text{QEM}(i,N_\text{QC})$ and $\langle P_{1010} \rangle^\text{Qiskit}_\text{QEM}(i,N_\text{QC})$ approach to  
$\langle P_{0101} \rangle_\text{QEM}$ and $\langle P_{1010} \rangle_\text{QEM}$, respectively.  
In Figs. \ref{QAOANqcdependence}(c) and \ref{QAOANqcdependence}(d)   we plot $\left( \sigma^{2}_\text{noisy,\text{Qiskit}}[O,N_\text{QC},N_\text{samp}] \right)^{-1},$  and $\left[ \sigma^{2}_\text{QEM,\text{Qiskit}}[O,N_\text{QC},N_\text{samp},N_q,d] \right]^{-1}$ for $P_{0101}$ and $P_{1010}$, respectively.
For $P_{0101}$ we obtain $\alpha_\text{noisy}[O]  \simeq4.20, \alpha^\text{Qiskit}_\text{noisy}[O,N_\text{samp}]\simeq3.50, \alpha_\text{QEM}[O,N_\text{samp},N_q,d]\simeq2.61$
and for $P_{1010}$ $\alpha_\text{noisy}[O]  \simeq4.19, \alpha^\text{Qiskit}_\text{noisy}[O,N_\text{samp}]\simeq3.80, \alpha_\text{QEM}[O,N_\text{samp},N_q,d]\simeq2.84$.
For both $P_{0101}$ and $P_{1010}$ the ratio $\frac{\alpha^\text{Qiskit}_\text{noisy}[O,N_\text{samp}] }{ \alpha^\text{Qiskit}_\text{QEM}(O,N_\text{samp},N_q,d)}$ is 1.34, which indicates that the broadening of the deviations of $\langle O \rangle^\text{Qiskit}_\text{QEM}(i,N_\text{QC})$ from $\langle O \rangle_\text{QEM}$ due to our QEM method is not so large (not so high computational cost). 
We have also performed our simulations for $N_\text{samp}=1000$ and we obtain $ \alpha^\text{Qiskit}_\text{noisy}[O,N_\text{samp}]\simeq4.04, \alpha^\text{Qiskit}_\text{QEM}[O,N_\text{samp},N_q,d]\simeq3.02$
for $P_{0101}$ and $ \alpha^\text{Qiskit}_\text{noisy}[O,N_\text{samp}]\simeq4.33, \alpha^\text{Qiskit}_\text{QEM}[O,N_\text{samp},N_q,d]\simeq3.27$
for $P_{1010}$. Therefore, compared with the results for $N_\text{samp}=100$ the coefficients $ \alpha^\text{Qiskit}_\text{noisy}[O,N_\text{samp}]$ get bigger and become closer to $\alpha_\text{noisy}[O] $  and 
$\alpha^\text{Qiskit}_\text{QEM}[O,N_\text{samp},N_q,d]$. 
Finally, let us explain the results in Fig. \ref{QAOAhistogram}.  
Here we have presented the histogram of the probability distribution of the computational basis states for the ideal case $(P_\text{ideal})$, the noisy case $(P_\text{noisy}, P_\text{noisy}^\text{Qiskit})$, and the case with QEM being performed 
$(P_\text{QEM}, P_\text{QEM}^\text{Qiskit})$,
with setting $\vartheta_\tau=0.2$.
We see that ideally the probabilities of the two quantum states $|0101\rangle$ and $|1010\rangle$ are both equal to 0.5. 
This implies that under the optimized variational parameters the cost function $C\big{(}\boldsymbol{\vartheta}^{\text{QAOA}}, \boldsymbol{\varphi}^{\text{QAOA}}\big{)} $ 
becomes minimized such that the four qubits $Q_{\text{r}i}$ are partitioned into the two groups $\big{[}Q_{\text{r}0},Q_{\text{r}2}\big{]}$ and $\big{[}Q_{\text{r}1},Q_{\text{r}3}\big{]}$, and it implies that 
all the edges of the square are to be cut, i.e.,  the maximum number of edges to be cut is four. 
Correspondingly, as shown in Fig.  \ref{QAOAnoisyQsims}(a) the ideal minimum value of the cost function is $-4.0,$ and we obtain the maximum cut number four by multiplying minus one.
Consequently, our QEM scheme works for the noisy QAOA simulation. 

\section{ QEM Scheme Implementation }\label{QEMimplementation} 
In this section, we demonstrate our QEM scheme using  two IBM Q Experience processors, ibmq$\_$belem and ibm$\_$perth \cite{IBMQExp}. 
In Fig. 4 in Sec. V of the Supplementary Material, we show schematics of spatial configurations for qubits in these machines: Fig. 4(a) is the illustration for ibmq$\_$belem whereas Fig. 4(b) is that for ibm$\_$perth.   As similar to the quantum simulations demonstrated in Sec. \ref{nss}, we examine the efficacy of our QEM scheme for the real quantum devices by varying the depths of the quantum algorithms to be run.  
We do this for two quantum algorithms, $U^{ \text{QC}}_{ \text{pre}1}$ in Fig. \ref{XHXXCHqcircs}(a) and 
a quantum algorithm for a two-qubit system defined by 
$U^\text{QC}_\text{imp,2} =  (U_{CZ}[Q_{\text{r}i};Q_{\text{r}j}])^{n^\text{rep}} \cdot X_iX_j$, where $n^\text{rep}$ is an integer.
For the usage of ibmq$\_$belem, we choose qubits $Q_\text{1}$ and $Q_\text{3}$ as register qubits whereas we use $Q_\text{2}$ ($Q_\text{4}$) as an ancilla bit for mitigating the quantum noise effect on $Q_\text{r1}$  ($Q_\text{r3}$). On the other hand, we use $Q_\text{1}$ and $Q_\text{3}$ as register qubits while we use an ancilla bit $Q_\text{2}$ ($Q_\text{5}$) for QEM of the quantum noise effect on $Q_\text{1}$  ($Q_\text{3}$) for the usage of ibm$\_$perth. In the following, we rewrite $Q_\text{1}$  and $Q_\text{3}$ as $Q_\text{r1}$  and $Q_\text{r3}$, respectively, to emphasize that they are used as register bits.  
Our QEM scheme can be applied provided that noise parameters (strengths) are given a priori. 
Thus, we start  from the acquisition of the quantum noise strengths of these devices (noise characterization).  Then, we discuss the results of our QEM method obtained with these machines (implementation).   
\subsection{Noise Characterization}\label{noisecharacterization}
Let us first discuss from how to obtain the AD strength $\tau$ ($=\gamma \Delta t$) or the $T_1$ time. The relation between the decay rate $\gamma$ and the $T_1$ time is
$T_1=\frac{1}{\gamma(2\bar{n}+1)}$, where $\bar{n}$ is the Bose-Einstein distribution function. 
For superconducting qubit systems, qubit frequencies and temperatures are about 5.0 GHz (see also Table II, Table IV,  and Table V in the Supplementary Material) and 10 mK \cite{SCQNISQ20191,TsaigroupSCCQC2021}, respectively. As a result,  the Bose-Einstein distribution function $\bar{n}$ is estimated to be $\bar{n}\sim10^{-11}$ and we approximate $\bar{n}$ to be zero.  The value of $T_1$ can be obtained as follows. 
First, we prepare a single qubit and apply the $X$ gate so that the qubit is in the $|1\rangle$ state (excited state).  
Next, we let the qubit relax until certain time $t^\text{relax}_a$ ($a=1,2,\ldots,N_\text{m}$ and $0<t_1<t_2<\cdots<t_{N_\text{m}}$), 
measure the output state of the qubit which is either $|0\rangle$ or $|1\rangle$, and repeat this process to obtain the probability weights of the $|0\rangle$ and $|1\rangle$ states. 
The total exposure time of the qubit is $t^{\text{tot},T_1}_a =\Delta t^{X}+ t^\text{relax}_a$, where $ \Delta t^{X}$  denotes a $X$-gate operation time.  
This is equivalent to measuring the dynamics of the expectation value $\langle P^1 \rangle(t^{\text{tot},T_1}_a)$ (the dynamics of the probability such that $|1\rangle$ is to be measured as the output state),  
and by doing an exponential fitting on the $\langle P^1 \rangle(t^{\text{tot},T_1}_a)$ plots we can extract the value of $T_1$. 
We can also extract a $T_2$ time and here let us consider a $T_2^\ast$ time which is obtained by the Ramsey measurement explained in the following and hereinafter we just denote $T_2^\ast$ time as $T_2$ time. 
Basically, what we do is first we  apply the Hadamard gate, let the qubit relax for $t^\text{relax}_a$, apply the second Hadamard gate, measure the output state, and repeat this process:   
The actual gate operations implemented on these experiments are not the two Hadamard gates owing to the issue of transmon qubit systems and its details is described in Sec. IV in the Supplementary Material. 
The above procedure is essentially equivalent to measuring the dynamics of $\langle P^1 \rangle(t^{\text{tot},T_2}_a)$, where $t^{\text{tot},T_2}_a=  2\Delta t^{H} + t^\text{relax}_a$  
with $\Delta t^{H}$ denoting Hadamard-gate operation time. From the $\langle P^1 \rangle(t^{\text{tot},T_2}_a)$ curve we obtain  a $T_2$ time. In the Supplementary Material IV, we explain how to estimate the $T_1$ and $T_2$ times with presenting the experimental data plots of the expectation values: Fig. 2 and Fig. 3 show the data plots of  the $T_1$ and $T_2$ times, respectively, and Table I lists the values of them for ibmq$\_$belem.

\begin{figure*}[!t] 
\centering
\includegraphics[width=0.85 \textwidth]{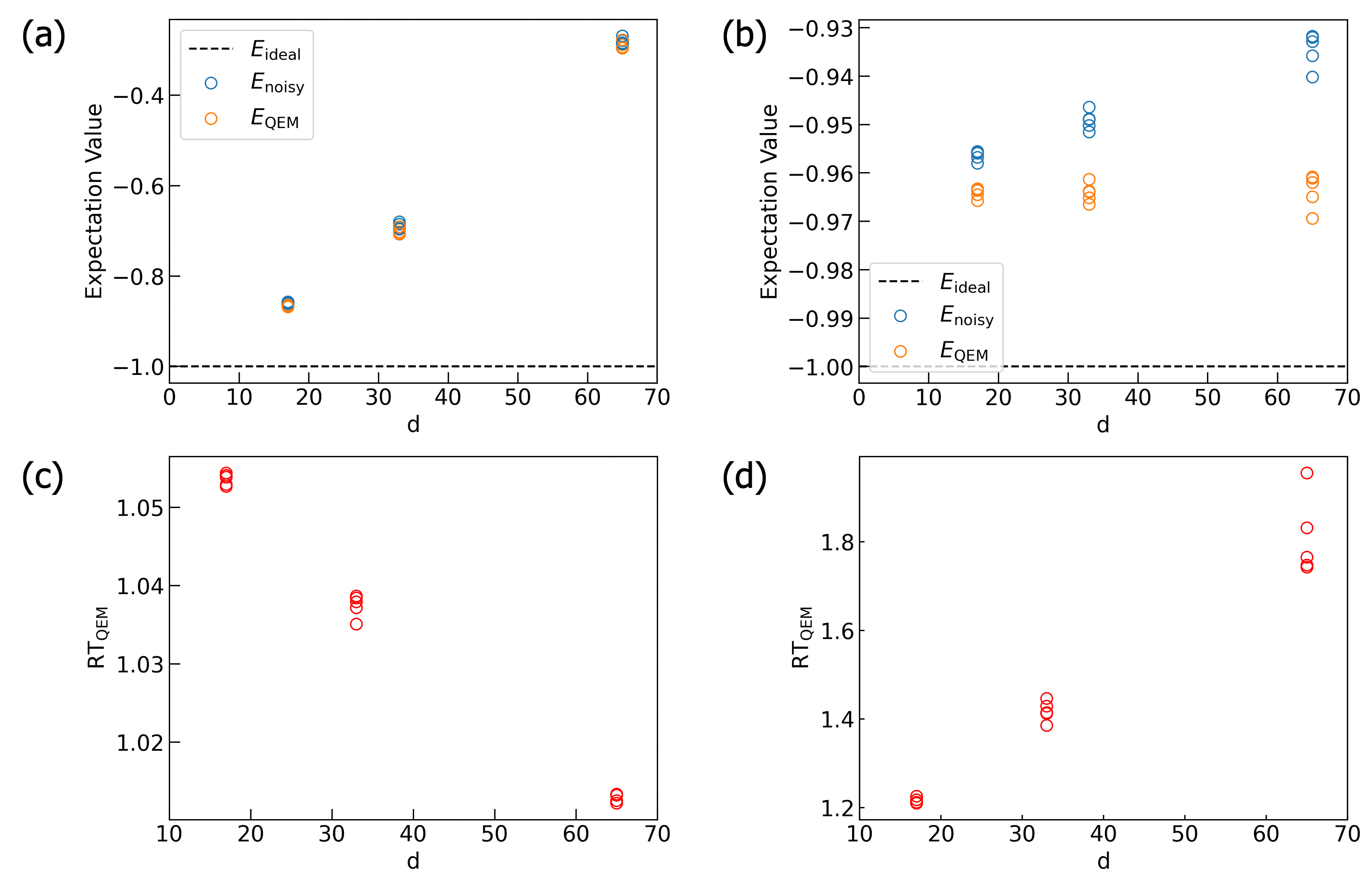}
\caption{Plots in (a) and (c) are
quantum computation results obtained by ibm$\_$perth
and (b) and (d) are quantum simulation results.
The quantum algorithm which has been run is $U^{ \text{QC}}_{ \text{pre}1}$. 
Both (a) and (b) are the results of expectation values of $Z_1$ and (c) and (d) are the ones of RT$_\text{QEM}$. 
All the quantum circuits have been executed under $N_\text{QC}=2^{14},N_\text{samp}=5.$
The horizontal axes represent $d=n^\text{rep}+1$, where $n^\text{rep}=16,32,64$. The quantum computation has been conducted at 07:17, 09/26/2023.}
\label{ibmperthZexpandRT}
\end{figure*} 
In Sec. V in the Supplementary Material, we present the physical properties (single-qubit and two-qubit gate times, qubit frequencies) of ibmq$\_$belem and ibm$\_$perth:
 Table II and Table III (Table IV-Table VII) list the physical properties of ibmq$\_$belem  (ibm$\_$perth).
Here let us briefly explain the gate properties. The native gates of these machines are the following,  identity gate, virtual Z gate, $\sqrt{X}$ gate, 
$X$ gate, CNOT, and reset operation (reset into $|0\rangle$).  The Hadamard gate is decomposed into a series of the native gates as $H=R_z\left(\frac{\pi}{2}\right)\cdot \sqrt{X}  \cdot R_z\left(\frac{\pi}{2}\right).$ 
\begin{figure*}[!t] 
\centering
\includegraphics[width=0.85 \textwidth]{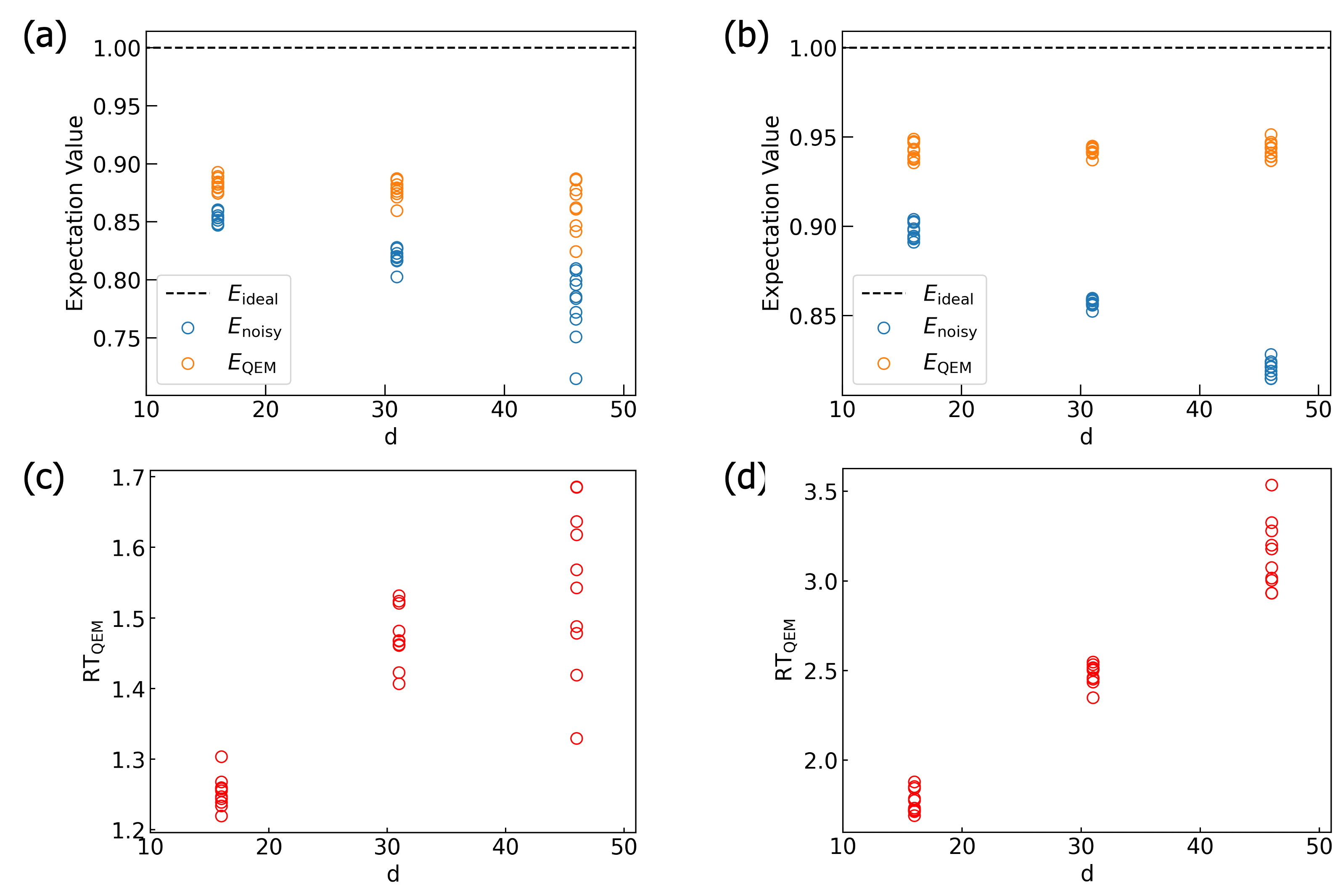}
\caption{Quantum computation and simulation results for  $U^\text{QC}_\text{imp,2} $.
(a) ((b)) and (c) ((d)) are the quantum computation (simulation) results of expectation values $\langle Z_1Z_3 \rangle$ and  ratios RT$_\text{QEM}$, respectively. 
 All the horizontal axes represent $d=3n^\text{rep}+1$ with $n^\text{rep}=5,10,15$.   
 The quantum computations have been conducted via ibm$\_$perth at 06:55, 09/21/2023 under $N_\text{QC}=2^{14},N_\text{samp}=10$. }
\label{ibmperthZZexpandRT}
\end{figure*} 
By using the data in Sec. IV and Sec. V in the Supplementary Material we are now able to implement our QEM scheme.  Before showing the results, let us comment five things about our experiment.
Firstly, the values of  $T_1$ and $T_2$ times differ from qubit to qubit on a real device due to imperfection (inhomogeneities
of $T_1$ and $T_2$). To take these inhomogeneities into account and extract $T_1$ and $T_2$ times, we need to perform the above two procedures independently on each qubit.
Secondly, a gate time $\Delta t$ is actually gate dependent as shown in Table II-Table VII in the Supplementary Material.   
By taking  the inhomogeneities of both the $T_1$ and $T_2$ times and the gate times into account, instead of $\tau$, which does not depend on both the qubits and the quantum gates, we perform our QEM  scheme by using a perturbative parameter $\tau_{jk}=\gamma_j \Delta t_k$, where $j(=0,\ldots,N_q-1)$ is the index for qubit numbering whereas $k(=1,\ldots,d)$ is that for labeling the gates.  
In the following, we denote the $T_1$ and $T_2$ times of the qubit $Q_j$ by $T_{1,j}$ and $T_{2,j}$, respectively.
Thirdly, $T_1$ and $T_2$ times fluctuate temporally on a real device. To run our QEM scheme, it is necessary to record the data of $T_1$ and $T_2$ times day-by-day and use these values.
Thus, we indicate the time and the date in the coordinated universal time (UCT) when we list the data of $T_1$ and $T_2$ times.  
Fourthly, the $CZ$ gate operation is decomposed into the form $U_{CZ}[Q_{\text{r}1};Q_{\text{r}3}] =  \left(R_z\left(\frac{\pi}{2}\right)\cdot \sqrt{X}  \cdot R_z\left(\frac{\pi}{2}\right)\right)_{3} \cdot
U_{CX}[Q_{\text{r}1};Q_{\text{r}3}] 
\cdot \left(R_z\left(\frac{\pi}{2}\right)\cdot \sqrt{X}  \cdot R_z\left(\frac{\pi}{2}\right)\right)_{3}$,
and
according to the transpilation of IBM Q Experience programming $(U_{CZ}[Q_{\text{r}1};Q_{\text{r}3}])^2 $ has been processed as     
$(U_{CZ}[Q_{\text{r}1};Q_{\text{r}3}])^2 =  \left(R_z\left(\frac{\pi}{2}\right)\cdot \sqrt{X} \cdot R_z\left(\frac{\pi}{2}\right) \right) 
\cdot U_{CX}[Q_{\text{r}1};Q_{\text{r}3}] \cdot \boldsymbol{1} \cdot \boldsymbol{1} \cdot U_{CX}[Q_{\text{r}1};Q_{\text{r}3}] \cdot
\left(R_z\left(\frac{\pi}{2}\right)\cdot \sqrt{X} \cdot R_z\left(\frac{\pi}{2}\right) \right), $
and $U^\text{QC}_\text{imp,2} $ has been executed as $U^\text{QC}_\text{imp,2} = 
 \left(R_z\left(\frac{\pi}{2}\right)\cdot \sqrt{X}  \cdot R_z\left(\frac{\pi}{2}\right)\right)_{3} \cdot U_{CX}[Q_{\text{r}1};Q_{\text{r}3}] \cdot
(\boldsymbol{1} \cdot \boldsymbol{1} \cdot U_{CX}[Q_{\text{r}1};Q_{\text{r}3}] )
^{n^\text{rep}-1} \cdot  \left(R_z\left(\frac{\pi}{2}\right)\cdot \sqrt{X}  \cdot R_z\left(\frac{\pi}{2}\right)\right)_{3} \cdot X_1X_3$  \cite{IBMQExp}.
Note that $\boldsymbol{1}$ denotes the identity operator.
Fifthly, the time when the quantum computation has been conducted (see the captions of Fig. \ref{ibmperthZexpandRT}, Fig. \ref{ibmperthZZexpandRT}, and Fig. \ref{ibmqbelemZZexpandRT}) indicates the time when the original circuit and the quantum-noise-effect-circuit group have been executed.
Finally, the measured (experimental) value of $\tau$, which we denote by $\tau_\text{exp}$,
can be differ from the true value of $\tau$ due to, for instance, imperfection of experimental apparatuses. Although that is the case, the conduction of our QEM method is said to be a success provided that the condition $\text{RT}_\text{QEM}>1$ is satisfied. 
\begin{figure*}[!t] 
\centering
\includegraphics[width=0.85 \textwidth]{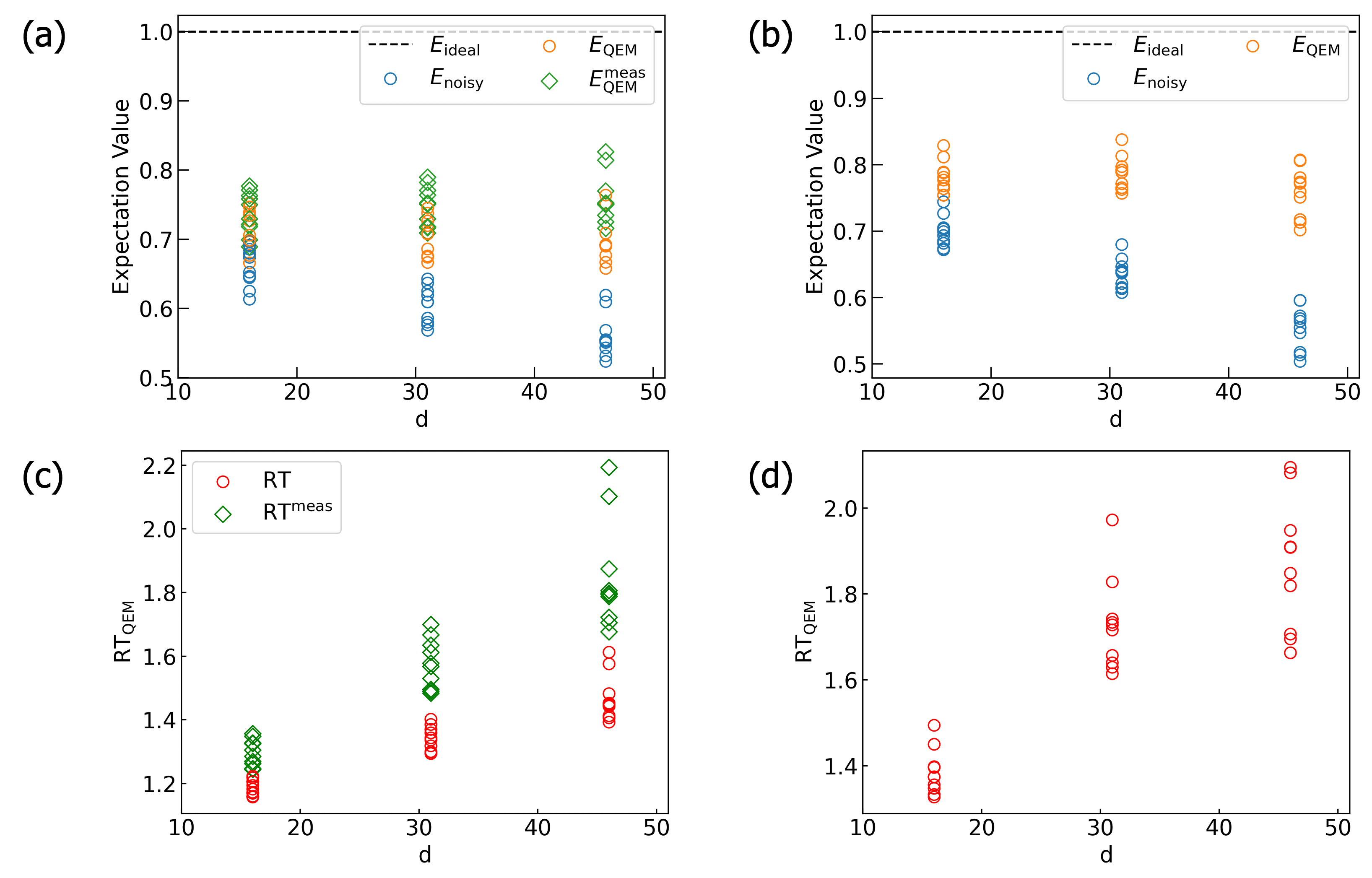}
\caption{Quantum computation ((a) and (c)) and simulation results ((b) and (d)) for  $U^\text{QC}_\text{imp,2} $.
(a) and (b) are the plots of $\langle Z_1Z_3 \rangle$ and (c) and (d) are those of RT$_\text{QEM}$. All the horizontal axes represent $d=3n^\text{rep}+1$ with $n^\text{rep}=5,10,15$.   
 The quantum computations have been conducted via ibmq$\_$belem at 11:43, 09/11/2023 under $N_\text{QC}=2^{10},N_\text{samp}=10$. }
\label{ibmqbelemZZexpandRT}
\end{figure*} 
\subsection{Implementation}\label{implementation}
As mentioned previously, in real quantum devices both  $T_1$ and $T_2$ times and  gate operation times are inhomogeneous, and moreover, both  AD and PD effects exist.
By taking these elements into account, we perform our QEM scheme by improving the formulas given by Eqs. \eqref{ADdeviationDM1} and \eqref{QEMformula1} as
\begin{widetext}
\begin{align}
 \langle \hat{O}  \rangle^{\text{QEM}}_{\rho  _{d\cdots1}} & \equiv 
\langle \hat{O}  \rangle_{\rho^{\text{real}}  _{d\cdots1}} - \sum_{j=0}^{N_q-1}  \sum_{k=1}^{d}   \langle \hat{O}  \rangle_{ (\Delta^{\text{AD}\&\text{PD}}_{1,jk} \rho_{d\cdots1})^{\text{real}} } , \notag\\
\Delta^{\text{AD}\&\text{PD}}_{1,jk} \rho_{d\cdots1}&=    \left(\prod_{l=k+1}^d U_l  \right) \cdot \left[
-\frac{\Delta t_k}{2T_{2,j}}\left( \rho_{k\cdots1} - Z_j \rho_{k\cdots1} Z_j  \right) + \frac{\Delta t_k}{T_{1,j}}\left( \tilde{\sigma}^-_j\rho_{k\cdots1}\tilde{\sigma}^+_j - P^1_j \rho_{k\cdots1} P^1_j  \right)
\right]      \cdot \left(\prod_{l=k+1}^d U_l \right)^\dagger .
 \label{improvedQEM}
\end{align} \end{widetext}
We note that the quantum circuits such that the additional operations $\{Z, \tilde{\sigma}^-, P^1 \}$ are inserted after the virtual $Z$ gates are not needed for (or do not contribute to)
the calculation of Eq. \eqref{improvedQEM} because the operation time of the virtual $Z$ gate is zero.

First, let us discuss from the results for $U^\text{QC}_\text{pre1}$ which are shown in Fig. \ref{ibmperthZexpandRT}. Here all the horizontal axes represent $d=n^\text{rep}+1$, where the repetition number $n^\text{rep}$ is taken to be 16,32, and 64. 
Fig. \ref{ibmperthZexpandRT}(a) presents a quantum computation result of expectation values $\langle Z_1\rangle$ obtained by ibm$\_$perth 
whereas the plots in Fig. \ref{ibmperthZexpandRT}(b) is a quantum simulation result of  $\langle Z_1\rangle$. On the other hand,
Fig. \ref{ibmperthZexpandRT}(c) displays a quantum computation result of RT$_\text{QEM}$ with ibm$\_$perth and Fig. \ref{ibmperthZexpandRT}(d) plots a quantum simulation result of RT$_\text{QEM}$. 
 In Fig. \ref{ibmperthZexpandRT}(a), we observe that both the expectation values with and without QEM get farther from the ideal expectation value (black dashed line) rapidly as we increase  $n^\text{rep}$, and correspondingly, the monotonic decreasing of RT$_\text{QEM}$ is exhibited in Fig. \ref{ibmperthZexpandRT}(c), which are similar to the characteristics  in Figs. \ref{XHnoisyQsims}(a) and  \ref{XHnoisyQsims}(c). Compared with these results, however,  RT$_\text{QEM}$ decreases gradually since not only the noisy expectation values but also the expectation values with QEM increase rapidly.
 In contrast, in Fig. \ref{ibmperthZexpandRT}(b), we see that while the expectation values without QEM or noisy expectation values (blue open circles) increase gradually as $d$ (or $n^\text{rep}$) gets larger the expectation values with QEM (orange open circles) take almost the same value. In Fig. \ref{ibmperthZexpandRT}(d), the monotonic increasing of RT$_\text{QEM}$ is exhibited. 
 In the experiment of the implementation of $U^{ \text{QC}}_{ \text{pre}1}$, both the  expectation values and  RT$_\text{QEM}$ obtained by the quantum computation and those by the quantum simulation show qualitatively different behaviors. On the other hand, all the ratios RT$_\text{QEM}$ are greater than one. As a result, our QEM scheme has worked, however, not effectively.

Next, let us explain the results of the implementation which are shown in Fig. \ref{ibmperthZZexpandRT}.
Figs. \ref{ibmperthZZexpandRT}(a) and \ref{ibmperthZZexpandRT}(c) display quantum computation results via ibm$\_$perth
and Figs. \ref{ibmperthZZexpandRT}(b) and \ref{ibmperthZZexpandRT}(d) present quantum simulation results:
Figs. \ref{ibmperthZZexpandRT}(a) and  \ref{ibmperthZZexpandRT}(b) are the results of expectation values $\langle Z_1Z_3 \rangle$ while Figs. \ref{ibmperthZZexpandRT}(c) and \ref{ibmperthZZexpandRT}(d) are those of RT$_\text{QEM}$. All the horizontal axes describe $d=3n^\text{rep}+1$ with $n^\text{rep}=5,10,15$: the factor three represents the implementation of one CNOT gate $U_{CX}[Q_{\text{r}1};Q_{\text{r}3}]$ and two identity operations acting on both $Q_{\text{r}1}$ and $Q_{\text{r}3}$. Such  identity operations are coming from the operation $\left(R_z\left(\frac{\pi}{2}\right)\cdot \sqrt{X}  \cdot R_z\left(\frac{\pi}{2}\right)\right)^2 = \boldsymbol{1}\cdot\boldsymbol{1}$: see also the fourth comment in Sec. \ref{noisecharacterization}.   Note that the number of the implementation of the virtual $Z$ gate is not included in the depths of the quantum algorithms.  
Let us explain from the results in Figs. \ref{ibmperthZZexpandRT}(a) and  \ref{ibmperthZZexpandRT}(b). We observe that both of these results show qualitatively the same behavior, i.e.,
while the noisy expectation values (blue open circles) gradually get farther from the ideal expectation value (black dashed line) as $n^\text{rep}$ increases, the expectation values with QEM (orange open circles) approximately take the same value.  
 On the other hand, in Figs.  \ref{ibmperthZZexpandRT}(c) and  \ref{ibmperthZZexpandRT}(d)
 the ratios RT$_\text{QEM}$ exhibit the monotonic increasing, and moreover  the ratio RT$_\text{QEM}$ is greater than one for every $d$: RT$_\text{QEM}$ shown in Fig.  \ref{ibmperthZZexpandRT}(c)  are larger than those in Fig. \ref{ibmperthZexpandRT}(c) for every $d$. 
 In addition to the quantum computation with  ibm$\_$perth, we have also conducted a quantum computation using  ibmq$\_$belem and a quantum simulation for $U^\text{QC}_\text{imp,2} $: Figs. \ref{ibmqbelemZZexpandRT}(a) and  \ref{ibmqbelemZZexpandRT}(c) are the quantum computation results and Figs. \ref{ibmqbelemZZexpandRT}(b) and  \ref{ibmqbelemZZexpandRT}(d) are the quantum simulation results.
 As similar to the labeling in Fig. \ref{ibmperthZZexpandRT}, 
Figs. \ref{ibmqbelemZZexpandRT}(a) and  \ref{ibmqbelemZZexpandRT}(b) (Figs. \ref{ibmqbelemZZexpandRT}(c) and  \ref{ibmqbelemZZexpandRT}(d))
are the results of $\langle Z_1Z_3 \rangle$ (RT$_\text{QEM}$),
and all the vertical axes represent $d=3n^\text{rep}+1$ with $n^\text{rep}=5,10,15$.
Overall, the characteristics of these results are similar to those for ibm$\_$perth.
Here we plot two types of expectation values with QEM which are indicated by the orange open circles and the green squares and two types of RT$_\text{QEM}$ by the red open circles and the green squares. The quantities plotted by the orange and red open circles have been calculated by using the data of the $T_1$ and $T_2$ times obtained at 05:08 whereas the others obtained by the data taken at 12:33: see Table I in the Supplementary Material.
The time when the latter data has been taken (12:33) is closer to the time when the quantum computation has been conducted (11:43), and thus we consider that the latter RT$_\text{QEM}$ are greater than the former ones: see also the third comment in Sec. \ref{noisecharacterization}. As a result, we observe RT$_\text{QEM}>1$ for every quantum computation in our experiment, and we consider that our QEM works for the IBM Q Experience processors.   

To summarize our experiment of the implementation of our QEM scheme,  the quantum computations for $U^\text{QC}_\text{pre1}$ and $U^\text{QC}_\text{imp,2} $ show the different behaviors although the same machine has been used: the former case exhibits the different behavior from the simulation result while the latter case shows qualitatively the similar behavior. Moreover, the ratios  RT$_\text{QEM}$ for $U^\text{QC}_\text{imp,2} $ are larger than those for $U^\text{QC}_\text{pre1}$,
which are in contrast to the results in  Fig. \ref{XHnoisyQsims} and Fig. \ref{XXCHnoisyQsims}: On the whole,  
 RT$_\text{QEM}$ for $U^\text{QC}_\text{pre2}$ are basically smaller than those for $U^\text{QC}_\text{pre1}$.
One way to interpret the characteristics in Fig. \ref{ibmperthZZexpandRT}(c)  and Fig. \ref{ibmqbelemZZexpandRT}(c), the increasing behavior of RT$_\text{QEM}$ with respect to $n^\text{rep}$ , is as follows. When the noise strength is too small the noisy expectation values become sufficiently close to  the ideal ones and in such circumstances the conduction of our QEM scheme can give rise to negative effects since computers can treat up to certain digits. Indeed, such a characteristic has been observed in the quantum simulation results in Figs. \ref{groverq3probsandratios}(c) and \ref{groverq3probsandratios}(d) and Fig. \ref{QAOAnoisyQsims}(b) in the range $0 < \vartheta_\tau \leq 0.1$: for a superconducting qubit system $T_1\approx100$ $\mu$sec and $\Delta t\approx100$ nsec and $\vartheta_\tau$ is estimated to be
$\vartheta_\tau \approx 0.063$. Thus, in order to utilize our QEM method effectively we need to use it under circumstances with moderate quantum noise strengths or for running moderately long quantum algorithms under weak quantum noise effects: 
such a way of interpretation, however, cannot be adapted to understand the characteristics in Fig. \ref{ibmperthZexpandRT}(c). Consequently,  the interpretation of the discrepancy between the quantum simulation result and the quantum computational result for $U^\text{QC}_\text{pre1}$ and the discrepancy between the characteristic of RT$_\text{QEM}$ for $U^\text{QC}_\text{imp,2} $ and that for $U^\text{QC}_\text{pre1}$ remain unresolved for this experiment. 

We note that our experiment  has been conducted under restricted conditions such as the time for which we could have used the IBM Q Experience processors and the machines which have been available.
Provided that we have no such restrictions, let us give several comments on how to improve our results. 
The first way is to increase the value of $N_\text{samp}$. 
As indicated in Fig. \ref{GroverNqcdependence} and Fig. \ref{QAOANqcdependence}, by increasing the value of $N_\text{samp}$ we expect that the expectation values with QEM approach to unique values
and we become clearer to see whether our QEM scheme is working or not.  
The second way is to mitigate other types of errors. By combining our QEM scheme with QEM methods for other errors such as state preparation and measurement errors or errors according to other quantum noise channels such as crosstalk  \cite{sarovar2020detecting},
we anticipate that the value  of RT$_\text{QEM}$ becomes larger.  

In addition to the above discussion, let us consider the effectiveness of the implementation of our QEM scheme on  different quantum hardware and here we choose ion trap qubit systems.
Ion trap qubit systems are engineered, for instance, as linear chains and a two-qubit gate operation can be exploited such that it can be performed on any pair of qubits  \cite{wright2019benchmarking}:
In contrast, in order to implement a two-qubit gate on two separated  qubits, say $Q_a$ and $Q_b$, on the superconducting quantum devices which have been used in this experiment we need to insert SWAP operations acting on qubits which locate between $Q_a$ and $Q_b$  \cite{IBMQExp}.   
Therefore, all quantum algorithms as well as quantum-noise-effect circuit groups are able to be implemented as indicated by the quantum circuits for them. In other words, we can harness our QEM scheme on  ion trap qubit devices without reformulating the quantum circuits.
Next, let us discuss quantum noise in  ion trap qubit systems. The quantum noise occur in, for instance, hyperfine-state type ion-trap qubit systems are considered to be the phase damping and  
$T_2$ times are about 10 sec while single-qubit gate times are around 10 microseconds and 
two-qubit gate times are about 100 $\mu$sec \cite{linke2017experimental,trappedionNISQ2019}. 
By setting $T_2 = (2\gamma_\text{p})^{-1}$, we have $\tau_\text{p} = \gamma_\text{p} \Delta t \approx 5\times 10^{-6}$, where we have set $\Delta t=100$ $\mu$sec.
$\tau_\text{p}$ is sufficiently small and therefore, we expect that our QEM scheme also works  for ion-trap NISQ devices.
From these ingredients, we expect that  the implementation our QEM scheme works more effectively and is more suitable for ion trap qubit systems compared with the superconducting qubit systems.

\section{Comparison with other methods}\label{sec:comparison_with_other_methods}
We make comparisons between our method and other methods.  
Although many types of QEM methods have been proposed up to now \cite{EMPRL2017,EMNature2019,EMPRX2017,EMPRX2018,EMarxiv2018,PhysRevA.98.062339,mcardle2019error,jattana2020general,xiong2020sampling,zlokapa2020deep,EMPRA2021,CandSQEMPRAp2021,QCchemistryRMP2020,hybridQCalgorithmJPSJ2021,OttenGrayQEM1,OttenGrayQEM2,QSEQEM, CliffordQEM,LearningBasedQEM,VirtualDistillationQEM,koczor2021exponential,PRXQuantum.2.010316,piveteau2021error,lostaglio2021error,suzuki2022quantum,piveteau2022quasiprobability, pascuzzi2022computationally, takagi2021optimal, larose2022mitiq, koczor2021dominant,cai2022quantum}, here we focus on the following three methods, probabilistic error cancellation (PEC)  \cite{QCchemistryRMP2020,hybridQCalgorithmJPSJ2021,EMPRL2017,EMPRX2018,xiong2020sampling,CandSQEMPRAp2021,LearningBasedQEM,piveteau2021error,piveteau2022quasiprobability,takagi2021optimal,larose2022mitiq,cai2022quantum}, zero noise extrapolation (ZNE) \cite{QCchemistryRMP2020,hybridQCalgorithmJPSJ2021,EMPRL2017,EMNature2019,EMPRX2017,EMPRX2018,pascuzzi2022computationally,larose2022mitiq,cai2022quantum}, 
and error suppression by dearangements (ESD) \cite{koczor2021exponential,koczor2021dominant,cai2022quantum} and virtual distillation (VD) \cite{VirtualDistillationQEM,cai2022quantum}. In Table \ref{QEMcomparison}, we summarize and present the comparison between our method and the others in terms of the number of ancilla bits  $N_a$ and the number of additional circuits $N_{\text{AQC}}$. We choose PEC from these three methods and numerically compare with our method and the reason we do this is the following. 
Both methods are theoretically similar such that they evaluate quantum noise effects on quantum states by quantum computational operations
and perform QEM which are represented as
sums of expectation values yielded by ensembles of quantum circuits including original circuits and quantum circuits containing additional operations. 
\begin{table}[htb]
\centering 
  \begin{tabular}{lcc} \\ \hline
    Method & $N_a$ & $N_\text{AQC}$ \\ \hline
    Ours & $k$ & $\mathcal{O}\left((N_qd)^k\right)$  \\
    PEC & 0 & $\mathcal{O}\left(\exp(2b\epsilon_\text{err} N_q d)\right)$\\
    ZNE & 0 & 0 \\ 
   ESD/VD & $(n-1)N_q+1$ \ $\text{or}$ \ $(n-1)N_q$ & 0 \\ \hline
   \color{black}
     \end{tabular}
    \caption{Comparison between our QEM method in $k$th-order perturbation theory and the other three methods. Here $b$ is a positive constant and $\epsilon_\text{err}$ is a gate error rate. The number of ancilla bits $N_a$ is equal to $(n-1)N_q$ for VD without ancilla bits.  } 
   \label{QEMcomparison}
\end{table}  

\subsection{Comparison with PEC}
\begin{figure*}[!htb]  
\centering
\includegraphics[width=0.85 \textwidth]{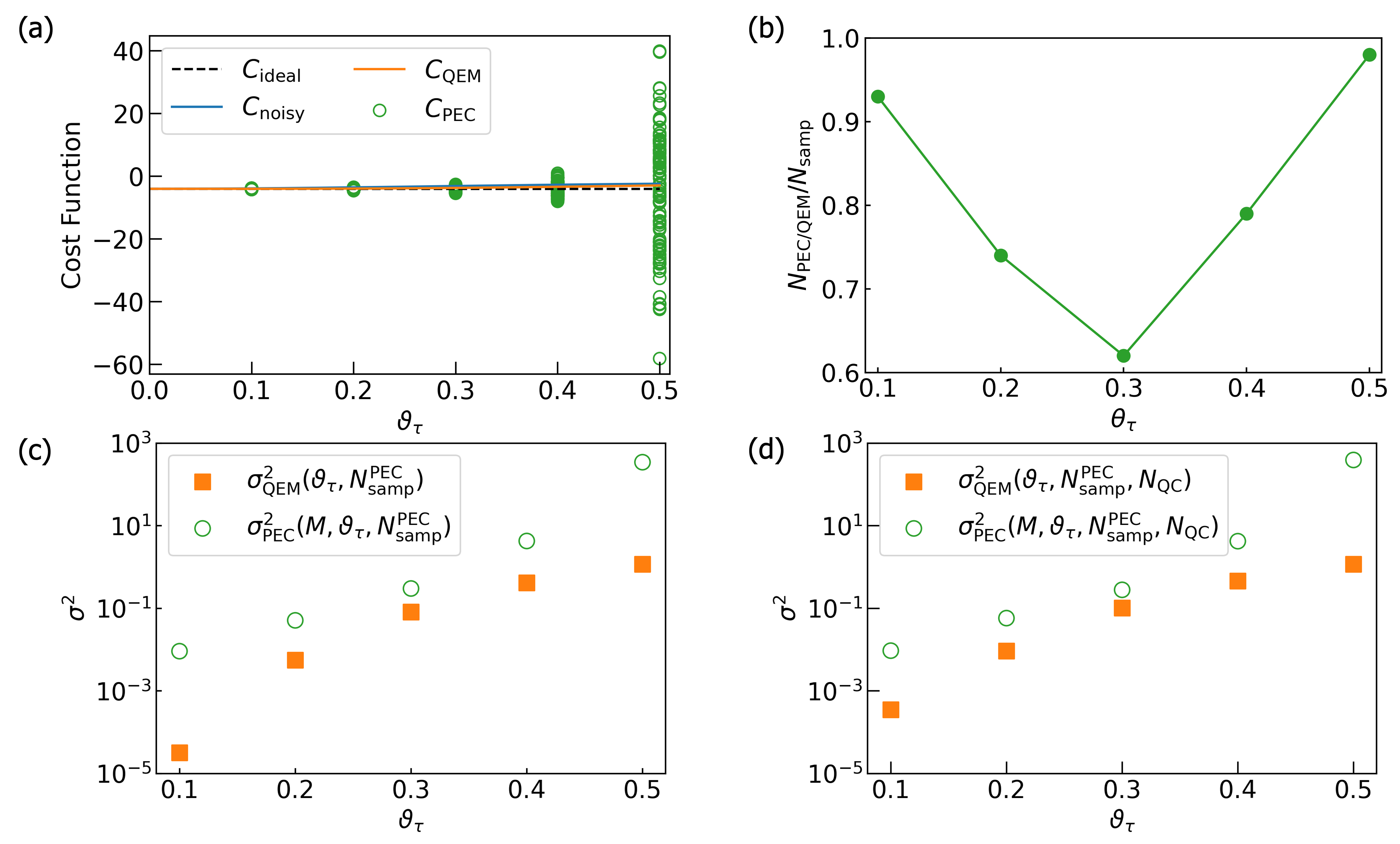}
\caption{ Numerical comparisons between our QEM method and PEC. We take $\vartheta_\tau= 0.1\times a$ with $a=1,2,\ldots,5$.
The results in (a)-(c) are the ones for $N_\text{QC}\to \infty$.
(a) Plots of the ideal cost function
$C_\text{ideal}$ (black dotted line), the cost function without QEM $C_\text{noisy}$ (blue curve), the cost function with QEM $C_\text{QEM}$ (orange curve), and the cost function with PEC $C_\text{PEC}$ which are plotted by 100 ($N^\text{PEC}_\text{samp}=100$)
green circles for each $\vartheta_\tau$. We take $M=181$. (b) The ratio $N_{\text{PEC/QEM}}/N_\text{samp}$ which satisfy  $\text{RT}_{\text{PEC/QEM}}>1$. 
(c) Plots of the variances $\sigma^2_\text{QEM}(\vartheta_\tau,N^\text{PEC}_\text{samp})$ and  $\sigma^2_\text{PEC}(M,\vartheta_\tau,N^\text{PEC}_\text{samp}) $ in Eq.  \eqref{QEMvariances_PECQEM}. 
We take $M=181$ and  $N_\text{samp}=100$.
(d) Plots of the variances $\sigma^2_\text{QEM}(\vartheta_\tau,N^\text{PEC}_\text{samp} , N_{\text{QC}})$ and  $\sigma^2_\text{PEC}(M,\vartheta_\tau,N^\text{PEC}_\text{samp}, N_{\text{QC}}) $ by taking $N_\text{QC}=2^{10}$.
}
\label{QEMandPECplots}
\end{figure*}

First, we make a comparison between our method and PEC \cite{QCchemistryRMP2020,hybridQCalgorithmJPSJ2021,EMPRL2017,EMPRX2018,xiong2020sampling,CandSQEMPRAp2021,LearningBasedQEM,piveteau2021error,piveteau2022quasiprobability,takagi2021optimal,larose2022mitiq,cai2022quantum}. 
These two methods have a theoretical similarity such that they
use additional quantum computational operations to mitigate quantum noise effects, however, their treatments are technically different. 
In PEC, quantum noise effects are mitigated by constructing  inverse
processes of quantum noises comprised of the sixteen basis 
operations and quasiprobabilities called recovery operations. In other words, both original circuits and additional quantum circuits are probabilistically generated  owing to  quasiprobabilities. 
Suppose that the recovery operation for the quantum noise under consideration, which acts on a single-qubit state, is composed of $N_\text{quasp}$ nonzero quasiprobabilities. 
The elementary operations of the recovery operation (the terms composing the recovery operation) is inserted after each operation of $U_l$ and the maximum number of  circuits for performing PEC is $N_\text{quasp}^{N_qd}$.
Once all these circuits are run obeying the quasibrobabilities the quantum noise effects are completely mitigated (non-perturbavtive method with respect to the noise strength).
We mathematically describe  the recovery operation as follows.
By writing the non-zero quasiprobabilities and the associated basis operations as $\eta_\alpha$ and  $\mathcal{B}_\alpha$ ($\alpha=1,\ldots,N_\text{quasp}$), respectively, 
the recovery operation executed after the operation of $U_l$ is described in terms of these quantities as $\mathcal{E}^{-1}_{\text{QN},l} = \bigotimes_{j_l=1}^{N^\text{PEC}_{q,l} } \left(\sum_{\alpha_{j_l}=1}^{N_\text{quasp}} \eta_{\alpha_{j_l}}  \mathcal{B}_{\alpha_{j_l}}\right),$ where $N^\text{PEC}_{q,l}$ is the number of qubits on which the recovery operations in the $l$th layer act,   $\mathcal{B}_{\alpha_{j_l}}$ is the basis operation acting on the $j$th qubit $Q_j$, and  $ \eta_{\alpha_{j_l}} $ is the associated quasiprobability of $\mathcal{B}_{\alpha_{j_l}}$:
note that $ \eta_{\alpha_{j_l}} $ include both  positive and negative values.
Furthermore, let us say that we compute an expectation value $\langle O \rangle$ with a repetition number $N_\text{QC}$ and accuracy $\epsilon$ and write an associated quantum mechanical variance by 
$\Delta^2_O[N_\text{QC}, \epsilon]$. When we perform PEC under the same repetition number $N_\text{QC}$ and the accuracy $\epsilon$ the variance of the expectation value with PEC being performed, which we denote by 
$\Delta^2_{O,\text{PEC}}[N_\text{QC}, \epsilon]$,  becomes larger than the original variance $\Delta^2_O[N_\text{QC}, \epsilon]$ as $\Delta^2_{O,\text{PEC}}[N_\text{QC}, \epsilon] = \mathcal{C}^2_\text{PEC}\Delta^2_O[N_\text{QC}, \epsilon]$,
where $\mathcal{C}^2_\text{PEC}= \prod_{l=1}^d \sum_{\alpha_=1}^{N_\text{quasp}^{N^\text{PEC}_{q,l}}} |\eta_\alpha| = \exp(2b\epsilon_\text{err}d)$ with $b$ a positive constant and $\epsilon_\text{err}$ is a gate error rate which is assumed to be gate independent
(or a typical gate error rate) \cite{hybridQCalgorithmJPSJ2021,EMPRL2017,EMPRX2018,CandSQEMPRAp2021}.
 Let us say that we generate $M$ quantum circuits (we call $M$ as PEC sampling number) obeying the quasiprobability and an elementary circuit of the $M$ quantum circuits can be either the original circuit or one of the additional circuits.
 By writing the set of quantum circuits for PEC which is composed of $N_\text{quas}^{N_qd}$ quantum circutis by $\mathcal{S}^\text{PEC}$ and
 the $i$th generated circuit by $C^\text{PEC}_i$ ($\in\mathcal{S}^\text{PEC}$ ), the exact error cancellation owing to PEC is described by 
 $ \lim_{M\to\infty}  \langle O \rangle_{\text{PEC},M}= \lim_{M\to\infty} \sum_{i=1}^M \langle O \rangle_{C^\text{PEC}_i} = \langle O \rangle_\text{ideal} $,
 where $\langle O \rangle_{C^\text{PEC}_i}$ is the expectation value of $O$ obtained by the quantum circuit $C^\text{PEC}_i$, which is generated with the probability $\frac{1}{M}$ \cite{EMPRL2017}. How many times a quantum circuit, say $\mathcal{C}_a$, appears in the $M$ sampling depends on its quasiprobability.
The computational resource, however, is finite in practice and in order to implement PEC in a real circumstance we take the sampling number $M$ such that $\big{|}  \langle O \rangle_{\text{PEC},M}-   \langle O \rangle_\text{ideal}     \big{|} \leq \epsilon$.
The sampling number $M$ scales in $\epsilon$ as $\mathcal{O}\left(\epsilon^{-2}\right)$ \cite{EMPRL2017,EMPRX2018,CandSQEMPRAp2021,takagi2021optimal,larose2022mitiq,cai2022quantum}.
On the other hand,  our method QEM is conducted by the estimation of the $k$th-order quantum noise effect $\Delta^\text{AD}_k \rho_{d\cdots1}$
and we subtract it from the noisy expectation value. In contrast to the quasiprobabilities, the coefficients which compose $\Delta^\text{AD}_k \rho_{d\cdots1}$ such as $\pm1$ and $\pm\frac{1}{4}$ are not probabilistic 
but deterministic (see Eq. \eqref{ADdeviationDM4} and the argument in Sec. I in the Supplementary Material, and furthermore, to evaluate $\Delta^\text{AD}_k \rho_{d\cdots1}$ we deterministically prepare and execute the quantum-noise-effect circuit group whose size (the number of elementary circuits composing it) is $\mathcal{O}\left( (dN_q)^k \right)$.  
The quality of our QEM method gets better as we increase $k$ and this corresponds to the increasement of $M$ in PEC.
Let us make numerical comparisons between these two methods for QAOA discussed in Sec. \ref{QEMqaoa} and show them in Fig. \ref{QEMandPECplots}. 
We perform the quantum simulations by taking the AD strength $\vartheta_\tau= 0.1\times a$ with $a=1,2,\ldots,5$.
We take $k=1$ (first-order perturbation theory) for our QEM method and for PEC we take 
$M=3dN_q+1,$ which is the number of quantum circuits we need to perform our first-order QEM scheme, and in this case $d=15,N_q=4$ and $M=181.$ The reason why we take $M=3dN_q+1$ is the following.  
Let us say that we conduct each QEM method with the same amount of quantum computational resource (under the same condition) and this can be regarded as the number of quantum circuits to be used for QEM. This is because, as mentioned previously, both methods are described as sums of expectation values generated by ensembles of quantum circuits consist of original circuits and quantum circuits including additional operations. By taking account  this, next let us introduce an error defined by $\delta_\text{QEM}= \Big{|}      \langle \hat{O}  \rangle_{\rho  _{d\cdots1}} - \langle \hat{O}  \rangle_{\rho^{\text{QEM}}  _{d\cdots1}} \Big{|}$, 
which represents the absolute of the difference between the ideal expectation value and the expectation value with our QEM method being performed. 
On the other hand, 
we introduce an error $\delta_\text{PEC}= \Big{|} \langle \hat{O}  \rangle_{\rho  _{d\cdots1}} - \langle O \rangle_{\text{PEC},M} \Big{|}$
with setting $M=3dN_q+1$. 
The numerical comparison between our method and PEC can be redescribed such that we perform each method using 
the common quantum computational resource which is the $3dN_q+1$ quantum circuits and examine whether $\delta_\text{QEM}$ is smaller than $\delta_\text{PEC}$ or not: the method having smaller error can be represented as the better QEM method.  
Hereinafter,  let us rewrite the expectation value with QEM and that with PEC as  $ \langle \hat{O}  \rangle_{\rho^{\text{QEM}}  _{d\cdots1}} \to  \langle \hat{O}  \rangle_\text{QEM}(\vartheta_\tau)$ and 
$\langle O \rangle_{\text{PEC},M} \to \langle O \rangle_\text{PEC}(\vartheta_\tau,M)$, respectively, to express the noise-strength dependencies. Furthermore, we rewrite the ideal expectation value and the noisy expectation value as  $\langle \hat{O}  \rangle_{\rho  _{d\cdots1}} \to \langle \hat{O}  \rangle_\text{ideal} $ and
$\langle \hat{O}  \rangle_{\rho ^\text{real} _{d\cdots1}} \to \langle \hat{O}  \rangle_\text{noisy}(\vartheta_\tau) $, respectively.
Note that the operator $O$ is chosen to be $H_C$ in Eq. \eqref{QAOAHamiltonian}.    
The recovery operation of AD acting on single-qubit states after the operation of $U_l$ is described by \cite{takagi2021optimal}
\begin{align}
\mathcal{E}^{-1}_{\text{AD},l} &= \bigotimes_{j_l=1}^{5}   \frac{1+\sqrt{1-\epsilon^\text{AD}(\vartheta_\tau)}}{2(1-\epsilon^\text{AD}(\vartheta_\tau))} \boldsymbol{1}_{j_l}+
 \frac{1-\sqrt{1-\epsilon^\text{AD}(\vartheta_\tau)}}{2(1-\epsilon^\text{AD}(\vartheta_\tau))} \boldsymbol{\mathcal{Z}}_{j_l} \notag\\
&- \frac{\epsilon^\text{AD}(\vartheta_\tau)}{1-\epsilon^\text{AD}(\vartheta_\tau)} (\boldsymbol{\mathcal{P}_{|0\rangle}})_{j_l}, \label{ADrecoveryoperation}
\end{align}   
where $ \boldsymbol{1}_{j_l}$ is the identity operator acting on $Q_j$, $ \boldsymbol{\mathcal{Z}}_{j_l}[\rho] = Z_j \rho Z_j,$  
 $(\boldsymbol{\mathcal{P}_{|0\rangle}})_{j_l}$ is the reset operator acting on $Q_j$, 
and  $\epsilon^\text{AD}(\vartheta_\tau)$ is the error rate given by $\epsilon^\text{AD}(\vartheta_\tau)= \sin^2\left( \frac{\vartheta_\tau}{2}  \right)=1-e^{-\tau}.$ Let us explain from the result in Fig. \ref{QEMandPECplots}(a). 
Here we plot four types of quantities, the ideal cost function $\langle \hat{O}  \rangle_\text{ideal}$  (black dotted line), the noisy (no QEM) cost function $\langle \hat{O}  \rangle_\text{noisy}(\vartheta_\tau)$ (blue curve), 
the cost function with QEM $\langle \hat{O}  \rangle_\text{QEM}(\vartheta_\tau)$ (orange curve), and the cost function with PEC $\langle \hat{O}  \rangle_\text{PEC}(M,\vartheta_\tau,l)$ with $l=1,\ldots,N^\text{PEC}_\text{samp}$,
where $N^\text{PEC}_\text{samp}$ is the repetition number of the PEC simulation under the same condition in terms of $M$ and $\vartheta_\tau$ and $\langle \hat{O}  \rangle_\text{PEC}(M,\vartheta_\tau,l)$ is the cost function acquired in the $l$th round of the PEC simulation.
We note that $N_\text{samp}^\text{PEC}$ is different from $N_\text{samp}$ introduced in Sec. \ref{nss}.
In contrast to our QEM scheme, PEC is a probabilistic theory and when we perform a PEC simulation for $N^\text{PEC}_\text{samp}$ times and obtain $N^\text{PEC}_\text{samp}$ data of the cost function 
$\langle \hat{O}  \rangle_\text{PEC}(M,\vartheta_\tau,l)$,
 in general $\langle \hat{O}  \rangle_\text{PEC}(M,\vartheta_\tau,l_1) \neq \langle \hat{O}  \rangle_\text{PEC}(M,\vartheta_\tau,l_2)$ for $l_1\neq l_2$. 
This is why we introduce the repetition number $N^\text{PEC}_\text{samp}$
for the PEC simulation.
On the other hand, $\langle \hat{O}  \rangle_\text{QEM}(\vartheta_\tau,l_1) = \langle \hat{O}  \rangle_\text{QEM}(\vartheta_\tau,l_2)=\langle \hat{O}  \rangle_\text{QEM}(\vartheta_\tau)$ for $l_1\neq l_2$  and the repetition number  $N^\text{PEC}_\text{samp}$  is unnecessary for our QEM scheme.
 We compute $\langle \hat{O}  \rangle_\text{ideal},\langle \hat{O}  \rangle_\text{noisy}(\vartheta_\tau),\langle \hat{O}  \rangle_\text{QEM}(\vartheta_\tau)$  with our original code while we compute $\langle \hat{O}  \rangle_\text{PEC}(M,\vartheta_\tau,l)$ with the software package Mitiq \cite{larose2022mitiq} and Cirq \cite{cirq_developers_2022_6599601},
and the PEC simulations are done under the condition  $M=181$ and $N^\text{PEC}_\text{samp}=100$ and the data points of $\langle \hat{O}  \rangle_\text{PEC}(M,\vartheta_\tau,l) $ are plotted with green circles for each $\vartheta_\tau$.  
To quantify which of the two cost functions, $\langle \hat{O}  \rangle_\text{QEM}(\vartheta_\tau)$ and $\langle \hat{O}  \rangle_\text{PEC}(M,\vartheta_\tau,l) $, is closer to $\langle \hat{O}  \rangle_\text{ideal}$, we introduce a measure defined by  
\begin{align}
\text{RT}_{\text{PEC/QEM}}(M,\vartheta_\tau,l) = \frac{|\langle \hat{O}  \rangle_\text{PEC}(M,\vartheta_\tau,l) -\langle \hat{O}  \rangle_\text{ideal} |}{| \langle \hat{O}  \rangle_\text{QEM}(\vartheta_\tau)-\langle \hat{O}  \rangle_\text{ideal}  |}.
   \label{QEMratio_PECQEM}
\end{align} 
Like $\text{RT}_{\text{QEM}}$ in Eq. \eqref{QEMratio}, the ratio $\text{RT}_{\text{PEC/QEM}}(M,\vartheta_\tau,l)$  becoming greater than one implies that our QEM scheme is working better than PEC. 
By denoting the number of $\langle \hat{O}  \rangle_\text{PEC}(M,\vartheta_\tau,l)$ data points satisfying $\text{RT}_{\text{PEC/QEM}}(M,\vartheta_\tau,l)>1$ as $N_{\text{PEC/QEM}}$, in Fig. \ref{QEMandPECplots}(b) we show the ratio $N_{\text{PEC/QEM}}/N_\text{samp}^\text{PEC}$.
We see that the ratio $\text{RT}_{\text{PEC/QEM}}(M,\vartheta_\tau,l)$ is greater than 50 percent for every $\vartheta_\tau$.
We also obtain the ratio $N_{\text{PEC/QEM}}/N_\text{samp}=0.99$ for the realistic noise strength $\vartheta_{\tau}=0.045$ in the current superconducting qubit devices, as discussed in Sec. \ref{discussionQEMNISQ}.

In addition to the ratio $\text{RT}_{\text{QEM}}(M,\vartheta_\tau,l)$ in Eq. \eqref{QEMratio_PECQEM}, we compute  mean squared errors defined by
\footnotesize\begin{align}
\sigma^2_\text{PEC}(M,\vartheta_\tau,N^\text{PEC}_\text{samp}) &= \sum_{l=1}^{N^\text{PEC}_\text{samp}} 
\frac{1}{N^\text{PEC}_\text{samp}}  \left(\langle \hat{O}  \rangle_\text{PEC}(M,\vartheta_\tau,l) - \langle \hat{O}  \rangle_\text{ideal} \right)^2, \notag\\
\sigma^2_\text{QEM}(\vartheta_\tau,N^\text{PEC}_\text{samp}) &= \sum_{l=1}^{N^\text{PEC}_\text{samp}}
 \frac{1}{N^\text{PEC}_\text{samp}} \left(\langle \hat{O}  \rangle_\text{QEM}(\vartheta_\tau,l) - \langle \hat{O}  \rangle_\text{ideal}  \right)^2.
 \label{QEMvariances_PECQEM}
\end{align}\normalsize
 Note that $\sigma^2_\text{QEM}(\vartheta_\tau,N^\text{PEC}_\text{samp})=\sigma^2_\text{QEM}(\vartheta_\tau)$ (no $N^\text{PEC}_\text{samp}$ dependency).
 By using the variances in Eq. \eqref{QEMvariances_PECQEM} we can re-describe the comparison between the quality of our QEM and that of PEC as the comparison of the magnitudes of the two variances, i.e.,
 the method exhibiting smaller variance has the better QEM quality. 
 We show this numerical result in Fig. \ref{QEMandPECplots}(c)  and we obtain 
 $\sigma^2_\text{QEM}(\vartheta_\tau) <\sigma^2_\text{PEC}(\vartheta_\tau,N^\text{PEC}_\text{samp}) $ for every $\vartheta_\tau$.
  The results in Figs. \ref{QEMandPECplots}(a)-\ref{QEMandPECplots}(c) are the ones for $N_\text{QC}\to \infty$. 
 In Fig. \ref{QEMandPECplots}(d), we compute the variances $\sigma^2_\text{QEM}(\vartheta_\tau, N^\text{PEC}_\text{samp})$ and $\sigma^2_\text{PEC}(M, \vartheta_\tau,N^\text{PEC}_\text{samp})$ for $N_\text{QC}=2^{10}$ (each circuit is executed by taking $N_\text{QC}=2^{10}$) and denote them as $\sigma^2_\text{QEM}(\vartheta_\tau,N^\text{PEC}_\text{samp}, N_\text{QC})$ and $\sigma^2_\text{PEC}(M, \vartheta_\tau,N^\text{PEC}_\text{samp}, N_\text{QC})$, respectively. 
 Note that we have used $N^\text{PEC}_\text{samp} N_\text{QC}$ data of  expectation values $\langle \hat{O}  \rangle_\text{PEC}(M,\vartheta_\tau,N_\text{QC},l)$ to compute $\sigma^2_\text{PEC}(M, \vartheta_\tau,N^\text{PEC}_\text{samp}, N_\text{QC})$.
 We do not observe any significant difference in the variance of PEC between the finite and infinite $N_\text{QC}$ cases, as can be seen by comparing Figs. \ref{QEMandPECplots}(c) and \ref{QEMandPECplots}(d). 
 However, due to the finite $N_\text{QC}=2^{10}$, we observe a slight increase in the variance of our QEM in the range $0.1 \leq \vartheta_{\tau} \leq 0.3$.
 Nonetheless, our QEM still exhibits $\sigma^2_\text{QEM}(\vartheta_\tau,N^\text{PEC}_\text{samp}, N_\text{QC}) <\sigma^2_\text{PEC}(M, \vartheta_\tau,N^\text{PEC}_\text{samp},N_\text{QC}) $ for every $\vartheta_{\tau}$.

As a result, the quality of our QEM method outperforms that of PEC under such conditions.
Let us examine for realistic cases when $M$ is sufficiently larger than 181 and we consider that there are two types of effects. The first one is that, owing to the concept of PEC,  $ \langle O \rangle_\text{PEC}(\vartheta_\tau,M)$  gets closer to $ \langle O \rangle_\text{ideal}$  owing to an increasement of $M$, 
which is a positive effect. The second one, which is a negative effect, is that the total amount of errors associated with the additional operations gets bigger by increasing $M$ which makes $ \langle O \rangle_\text{PEC}(\vartheta_\tau,M)$ to become farther from $ \langle O \rangle_\text{ideal}$.  
At some point, say $M=M_\text{c}$, we consider that the second (negative) effect gets larger than the first (positive) effect 
because the second effect is not to be mitigated and gets bigger. 
As a result, $ \langle O \rangle_\text{PEC}(\vartheta_\tau,M)$  becomes farther from $ \langle O \rangle_\text{ideal}$ for $M >M_\text{c}$.    
 It has been shown in \cite{CandSQEMPRAp2021} that the errors associated with the additional operations can be mitigated by combining with ZNE. 
 The necessity of ZNE, however, implies that we need an additional computational resource to mitigate such errors.  
In contrast, our QEM method is conducted self-consistently such that the quantum noise effects on both the original circuit and the quantum-noise-effect circuit groups are mitigated.
In other words, we do not need additional resources to mitigate the quantum noise effects on the quantum-noise-effect circuit groups.
We consider that such a self-consistency is the advantage of our method compared with PEC. 
Furthermore,  the size of the quantum-noise-effect circuit group (the computational cost) is practically polynomial in $N_qd$ while the number of quantum circuits for performing PEC is $N_\text{quasp}^{N_qd}$ (exponential in $N_qd$), 
and thus the practical computational cost of our QEM method is lower than that of PEC.
In total, we consider that our QEM method is superior to PEC even for sufficiently large $M.$
\subsection{Comparison with ZNE}
Next, let us make a comparison between ZNE \cite{QCchemistryRMP2020,hybridQCalgorithmJPSJ2021,EMPRL2017,EMNature2019,EMPRX2017,EMPRX2018,pascuzzi2022computationally,larose2022mitiq,cai2022quantum} and our method.    
The similarity between ZNE and our method (as well as PEC) is that both of these methods require additional quantum circuits. 
In ZNE, first a quantum mechanical expectation of an physical observable, say $O_\text{phys}$, is measured which includes a quantum computational error given by a noise strength (or an error rate) $\gamma_{0}$. Then, we create extra quantum circuits which ideally yield the same expectation value of $O_\text{phys}$ as the original quantum circuit does but with noise strengths larger than $\gamma_{0}$. 
Such boostings of the noise strengths can be done, for instance, by insertions of identity gate operations \cite{EMNature2019,hybridQCalgorithmJPSJ2021,pascuzzi2022computationally}. Let us denote the original circuit by $C_\text{org}$, which is subject to the quantum noise with the strength $\gamma_{0}$, while we denote the extra circuits subject to noises with strengths $\gamma_j$ by $C_j$ with $j=1,2,\ldots,N_\text{ZNE}$, i.e., for ZNE $N_\text{AQC}=N_\text{ZNE}$. Here we labeled the extra circuits $C_j$ so that $\gamma_0<\gamma_1<\cdots\gamma_{N_\text{ZNE}}.$    
The highest order of the quantum noise effects which can be mitigated is $N_\text{ZNE}$ and this is one of the powerful advantage of ZNE. 
The difference between our method and ZNE, on the other hand, is that in our method (as well as in PEC and  ESD and VD) QEM tasks are performed by quantum computational operations (gate operations and measurements) whereas in ZNE quantum computers are used for calculating expectation values while the QEM tasks are done by classical computers.  
The advantages of ZNE are that it does not require ancilla bits and QEM can be performed without knowledge of quantum noises. 
There is a drawback, however, that its quality becomes poor when $\gamma_{0}$ is too big. 
Furthermore, in experiments we need to obtain the values $\gamma_0,\gamma_1,\cdots,\gamma_\text{ZNE}$ with high precision. 
On the other hand, our method requires both the ancilla bits and knowledge of  quantum noises but
 can be applied to quantum noises for arbitrary noise strengths although higher computational cost is demanded for doing higher-order perturbation and both the values of decay rates and gate operation times are needed with high precision.   
Next let us compare the two methods with respect to the depth of a quantum algorithm and a coherence time with which we identify as a $T_1$ time.
Here we assume that all single-qubit and two-qubit gate times are identical and denote the single-qubit time and the two-qubit gate time by 
$t^\text{g}_1$ and $t^\text{g}_2$, respectively. Moreover, we assume that all $T_1$ times are identical with respect to qubits.
Let us first consider from the conduction of Richardson-extrapolation-based  ZNE and for simplicity we make an approximation $t^\text{g}_1 \ll t^\text{g}_2$ so that the depth of a quantum algorithm $d$ can be identified with the depth of (the number of the layers of) two-qubit gates to be implemented.    
Let us say that we enhance the noise strength $\gamma_i$ by the factor $c_i = r^i$
with taking  $r_i=1+0.1\times i$ with $i=1,2,\ldots.$. Such a situation can be created by inserting an identity operation for the time $0.1\times i\times t^\text{g}_2$ after the operation of each $U_k$. 
In this way, we can effectively make the operation time of $U_k$ to be $r_i t^\text{g}_2$ or we can effectively make the decay rate to be $r_i\gamma_i$ while the operation time of $U_k$ to be $t^\text{g}_2$.
The operation time to execute the quantum algorithm $\prod_{k=1}^d  U_{ k} =U_{ d} \cdot U_{ d-1} \cdots U_{ 2} \cdot U_{ 1}$ plus the additional identity operations is $r_i dt^\text{g}_2$ and it must satisfy the condition $r_i dt^\text{g}_2<T_1$.
For superconducting qubits  $t^\text{g}_2 \approx 100$ nsec and $T_1 \approx 100$ $\mu$sec \cite{SCQARCMP2020,ZhugroupSQC2020,Qiskit} and under such conditions we obtain $r_i d < 1000$.
For $d=800$,  we have $ r_1d = 880, r_2d=960, r_3d>1000$, which means that we are able to perform QEM up to second order: 
For  $d=900$  we have $ r_1d = 990, r_2d>1000$, which means that we are able to perform QEM up to linear order,
and for $d>910$ we become unable perform QEM. 
For larger $c_i $ the upper limit of $d$ such that QEM is valid gets smaller. From these considerations, we see that we can perform high-order QEM for small $d$ while for large $d$ we can only perform low-order QEM and for sufficiently large $d$ we cannot apply QEM.   
Such a characteristics comes from the $d$ dependence of the total additional identity operation time. Furthermore, the exponential extrapolation also becomes invalid for large $d$ since it is only effective for small error rate $\epsilon_\text{err}$ such that $\mathcal{\epsilon}_\text{err}N_qd=\mathcal{O}(1)$ \cite{hybridQCalgorithmJPSJ2021}: with the same reason for the order estimation of $\mathcal{C}^2_\text{PEC}$ we consider that the condition of the exponential extrapolation to become invalid is not $\mathcal{\epsilon}_\text{err}N_g=\mathcal{O}(1)$ but $\mathcal{\epsilon}_\text{err}N_qd=\mathcal{O}(1)$. 
Next, let us consider the conduction of our QEM. The time for performing our $k$th-order QEM method is at most 
$dt^\text{g}_2+ (2t^\text{g}_2+t^\text{g}_1)k \simeq dt^\text{g}_2+ 2.1kt^\text{g}_2$ (we assume that all additional operations are $P^1$), where we have set $ t^\text{g}_1 \simeq 0.1t^\text{g}_2$  \cite{EMNature2019,Qiskit}. 
Thus, the operational time for implementing the $k$ additional operations, which is $2.1kt^\text{g}_2$, does not depend on $d$. For $d=800, 900,$ and 910 the highest orders of quantum noise effects which we can mitigate are $95,47,$ and 42, respectively.  
Consequently, we consider that for small $d$ satisfying  $r_\text{ZNE} dt^\text{g}_2<T_1$ the quality of ZNE is better while for large $d$ our method has a better quality and such a  tendency does not change even $T_1$ times are extended and the gate times $t^\text{g}_2$ get shorter.  At the end, we mention that ZNE is not applicable to time-dependent noise \cite{hybridQCalgorithmJPSJ2021,EMPRL2017,EMNature2019} whereas our method is by reformulating $\tau$ as a function of time. 
 \subsection{Comparison with ESD and VD}  
Finally, we make a comparison between our method and ESD \cite{koczor2021exponential,koczor2021dominant,cai2022quantum} and VD \cite{VirtualDistillationQEM,cai2022quantum}. 
Since these two are similar approaches we bring them together and abbreviate it as ESD/VD.  
Like PEC and our method, ESD/VD is composed of gate operations and/or quantum measurement on an ancilla bit. 
ESD/VD has two advantages, it can be applied to various types of quantum noise and additional quantum circuits are unneeded.   
The procedure of ESD/VD is comprised of three parts, a preparation of $n$ copies of an original circuit which generate $(\rho^\text{real}_{d\cdots1})^{\otimes n}$, a performance of derangement operation, which is a generalization of SWAP operation (in \cite{VirtualDistillationQEM} it is called cyclic shift operator), and an operation of controlled-$O$ operator with $O$ denoting the physical operator of which we take an expectation value \cite{koczor2021exponential,koczor2021dominant}: In \cite{VirtualDistillationQEM}, the scheme which does not use an ancilla-assisted measurement has been presented and in this case $N_a=(n-1)N_q$ (see Table \ref{QEMcomparison}).
Let us express $\rho^\text{real}_{d\cdots1}$ in a spectral decomposition form denoted by 
$\rho^\text{real}_{d\cdots1}= \lambda_\text{dom}|\psi_\text{dom}\rangle\langle\psi_\text{dom}|+\sum_{a=1}^{2^{N_q}-1} \lambda_a|\psi_a\rangle\langle\psi_a|$, where the eigenvalues satisfy the descending order $\lambda_\text{dom}>\lambda_1\cdots>\lambda_{2^{N_q}-1}.$ Due to such a process, we obtain the expectation value $\langle\psi_\text{dom}|O|\psi_\text{dom}\rangle.$ 
In ESD/VD, there are two problems which need to be handled with, the mismatch between the ideal state  $\rho_{d\cdots1}$ and the dominant eigenvector state $\rho^\text{dom} = |\psi_\text{dom}\rangle\langle\psi_\text{dom}|$ which is called coherent mismatch or noise floor, and the mitigation of the quantum noise effects associated with the operations of the derangement operator and the controlled-$O$ operator. It was argued in \cite{koczor2021exponential} that the quantum computational error coming from the coherent mismatch and that due to the quantum noise effects associated with the operations constructing the quantum circuit for ESD/VD, which we call ESD/VD circuit, can be mitigated by combining with ZNE.
As argued in the previous discussion for ZNE, 
we consider, however, that such a hybrid scheme works provided that the composite circuit of the original circuit and the ESD/VD circuit can be executed within a coherence time of a qubit.
Furthermore, we need large numbers of qubits to perform ESD/VD and this limits sizes of overhead. 
On the other hand, our method needs the additional quantum circuits (quantum-noise-effect circuit group) but both the errors associated with the operations of $U_k$ and those associated with the insertions of the additional operations implemented on the quantum-noise-effect circuit group are self-consistently mitigated and saves an additional computational resource for mitigating the errors associated with the additional operations since our scheme does not need to be combined with other methods for doing this. Moreover, the number of ancilla bits which we need to perform $k$-th order QEM is $k$, which does not enlarge the overhead that much. Such properties, in principle, enable us to perform QEM for both short-term (NISQ algorithms) and long-term algorithms.

\section{ Conclusion and Outlook}\label{conclusion} 
In this paper, we have established our QEM scheme for reducing the quantum noise (decoherence) effects on the single-qubit states which occur during the gate operations.
We have formulated it as the perturbation theory with respect to the noise strength (in the case of AD effect it is $\tau$), which are evaluated by the gate time and decay rate ($T_1$ time and/or $T_2$ time),
and is represented by the ensemble of quantum circuits,  namely the quantum-noise-effect circuit groups.
The numbers of quantum circuits composing the quantum-noise-effect circuit groups are polynomial with respect to the product of the depth of the quantum algorithm under consideration  
and the number of register bits, which can be considered that the conduction of our QEM scheme is not so high-cost computational performance.  
To demonstrate the validity of our QEM scheme, we have performed the noisy quantum simulations of the qubits under the AD effects for four types of quantum algorithms based on the linear-order perturbation theory. 
It is to be noted that before we conduct our QEM scheme, we need to be careful with if the expectation values on which we are aiming to perform QEM do not include the linear-order term in $\tau$ or
if they are equivalent to the ideal simulation results like the computation of the expectation value of the identity operator.  
As long as these two are not the cases then our linear-order QEM scheme works as we have demonstrated in Sec. \ref{nss}, and it is valid in a broad region of $\tau$, which implies its effectiveness and powerfulness.  
Our QEM scheme can be generalized to error mitigation for other types of quantum noises including the generalized amplitude damping, the phase damping, the composition of these two, and the stochastic Pauli noises like the depolarizing channel. Furthermore, it can be extended to cases of error mitigation for higher-order quantum noise effects and once this is established we expect that we become able to perform quantum computations in high accuracy even for long-depth quantum algorithms.
In Sec. \ref{QEMimplementation}, we have discussed the experimental results of the implementation of our QEM scheme.
Consequently, we have observed RT$_\text{QEM}>1$ for every quantum computation and our QEM scheme has worked for the IBM Q Experience processors. 
In this work, we have focused on the decoherence effects acting on the single-qubit states and established the QEM scheme for them. In real quantum devices, however, there exist many types of errors and complex quantum noise channels. We expect that by conducting an elaborate noise (device) characterization we become able to improve the quality of QEM. 
For instance, by combining our QEM scheme with other error-mitigation techniques such as those for state  preparation (initialization) and measurement and imperfections of gate operations or those for other quantum noise channels like crosstalk we anticipate to realize QEM with higher quality.

Our QEM scheme is solely conducted by gate operations and measurements on ancilla bits and can be applied to any type of quantum algorithm.
Such a characteristic enables us to programmably perform high-accurate quantum computing solely by the quantum-computational operations (software manipulations).    
Furthermore, our QEM scheme can be performed with any type of quantum hardware such as solid-state systems and atomic-molecule and optical systems and with quantum devices of any generation, and both the computational cost whose order is polynomial in $N_qd$ and its accuracy can be coherently controlled. These three characteristics are the big advantages of our scheme.

One of the important outlooks of this work is quantum computations by large-scale quantum devices or future (next-generation) quantum devices using various types of quantum hardware such as superconducting circuits and ion-trap qubit systems.
In such a case, we consider that we also need to take into account quantum noises acting on many-body quantum states such as collective quantum noises \cite{carmichaeltxb,Agarwalltxb,opendynamicstext,superradiance1,GH82,DuancollectivedecohePRA1998,ViolagroupNJP2002} and correlated noises \cite{EMPRA2021,EMarxiv2018,PhysRevA.65.050301,TCADPRA2003,Cnoisepra2005,NoviasprlCnoise,KitaevgroupCnoiseprl2006}.
Another important problem is the establishment of QEM schemes for time-dependent quantum noises including non-Markovian quantum noises  \cite{ban2005decoherence,yu2010entanglement}.

Our QEM scheme can be extended to mitigation of these complex quantum noise effects provided that they are formulated as groups of quantum circuits.
When such formalisms are being constructed, we expect that we become able to realize QEM scheme which mitigates various types of  quantum noise effects.   
We expect that this leads to conduction of quantum computing for big-size problems with high-quality results being obtained. 
We believe that this paves the way to realize high-quality quantum computing for application to problems in many branches of science and engineering
including material science, quantum chemistry, combinatorial optimization problems, and machine learning using large-scale quantum computers.

\acknowledgements
 We thank all the other members of Quemix Inc. for giving us the fruitful comments and reading this manuscript carefully.  
 This work was supported by MEXT as ”Program for Promoting Researches on the Supercomputer Fugaku” (JP-MXP1020200205) and JSPS KAKENHI as ”Grant-in- Aid for Scientific Research(A)” Grant Number 21H04553. This study was carried out using the TSUBAME3.0 supercomputer at Tokyo Institute of Technology.
 
\section{Author Contributions}
Y.H. and H.N. accomplished the theoretical analysis. 
Y.H. and H.N. developed the codes, performed the numerical simulations, and verify the results. All authors contributed to the manuscript preparation and presentation of results. 
\section{Competing Interests}
The authors declare no competing interests. 
\section{Data Availability}
The datasets generated and/or analyzed during the current study are available from
the corresponding author on reasonable request.

\section{References}
\bibliographystyle{apsrev4-2} 
\bibliography{cleanQEMlistref.bib}

\begin{thebibliography}{110}%
\makeatletter
\providecommand \@ifxundefined [1]{%
 \@ifx{#1\undefined}
}%
\providecommand \@ifnum [1]{%
 \ifnum #1\expandafter \@firstoftwo
 \else \expandafter \@secondoftwo
 \fi
}%
\providecommand \@ifx [1]{%
 \ifx #1\expandafter \@firstoftwo
 \else \expandafter \@secondoftwo
 \fi
}%
\providecommand \natexlab [1]{#1}%
\providecommand \enquote  [1]{``#1''}%
\providecommand \bibnamefont  [1]{#1}%
\providecommand \bibfnamefont [1]{#1}%
\providecommand \citenamefont [1]{#1}%
\providecommand \href@noop [0]{\@secondoftwo}%
\providecommand \href [0]{\begingroup \@sanitize@url \@href}%
\providecommand \@href[1]{\@@startlink{#1}\@@href}%
\providecommand \@@href[1]{\endgroup#1\@@endlink}%
\providecommand \@sanitize@url [0]{\catcode `\\12\catcode `\$12\catcode `\&12\catcode `\#12\catcode `\^12\catcode `\_12\catcode `\%12\relax}%
\providecommand \@@startlink[1]{}%
\providecommand \@@endlink[0]{}%
\providecommand \url  [0]{\begingroup\@sanitize@url \@url }%
\providecommand \@url [1]{\endgroup\@href {#1}{\urlprefix }}%
\providecommand \urlprefix  [0]{URL }%
\providecommand \Eprint [0]{\href }%
\providecommand \doibase [0]{https://doi.org/}%
\providecommand \selectlanguage [0]{\@gobble}%
\providecommand \bibinfo  [0]{\@secondoftwo}%
\providecommand \bibfield  [0]{\@secondoftwo}%
\providecommand \translation [1]{[#1]}%
\providecommand \BibitemOpen [0]{}%
\providecommand \bibitemStop [0]{}%
\providecommand \bibitemNoStop [0]{.\EOS\space}%
\providecommand \EOS [0]{\spacefactor3000\relax}%
\providecommand \BibitemShut  [1]{\csname bibitem#1\endcsname}%
\let\auto@bib@innerbib\@empty
\bibitem [{\citenamefont {Feynman}(1982)}]{QCFeynman}%
  \BibitemOpen
  \bibfield  {author} {\bibinfo {author} {\bibfnamefont {R.~P.}\ \bibnamefont {Feynman}},\ }\href@noop {} {\bibfield  {journal} {\bibinfo  {journal} {International Journal of Theoretical Physics}\ }\textbf {\bibinfo {volume} {21}} (\bibinfo {year} {1982})}\BibitemShut {NoStop}%
\bibitem [{\citenamefont {Deutsch}(1985)}]{QCDeutsch}%
  \BibitemOpen
  \bibfield  {author} {\bibinfo {author} {\bibfnamefont {D.}~\bibnamefont {Deutsch}},\ }\href@noop {} {\bibfield  {journal} {\bibinfo  {journal} {Proceedings of the Royal Society of London. A. Mathematical and Physical Sciences}\ }\textbf {\bibinfo {volume} {400}},\ \bibinfo {pages} {97} (\bibinfo {year} {1985})}\BibitemShut {NoStop}%
\bibitem [{\citenamefont {Lloyd}(1996)}]{QCLloyd}%
  \BibitemOpen
  \bibfield  {author} {\bibinfo {author} {\bibfnamefont {S.}~\bibnamefont {Lloyd}},\ }\href@noop {} {\bibfield  {journal} {\bibinfo  {journal} {Science}\ ,\ \bibinfo {pages} {1073}} (\bibinfo {year} {1996})}\BibitemShut {NoStop}%
\bibitem [{\citenamefont {DiVincenzo}(2000)}]{DiVincenzoQC}%
  \BibitemOpen
  \bibfield  {author} {\bibinfo {author} {\bibfnamefont {D.~P.}\ \bibnamefont {DiVincenzo}},\ }\href@noop {} {\bibfield  {journal} {\bibinfo  {journal} {Fortschritte der Physik: Progress of Physics}\ }\textbf {\bibinfo {volume} {48}},\ \bibinfo {pages} {771} (\bibinfo {year} {2000})}\BibitemShut {NoStop}%
\bibitem [{\citenamefont {Nielsen}\ and\ \citenamefont {Chuang}(2002)}]{QCQINandC}%
  \BibitemOpen
  \bibfield  {author} {\bibinfo {author} {\bibfnamefont {M.~A.}\ \bibnamefont {Nielsen}}\ and\ \bibinfo {author} {\bibfnamefont {I.}~\bibnamefont {Chuang}},\ }\href@noop {} {\bibinfo {title} {Quantum computation and quantum information}} (\bibinfo {year} {2002})\BibitemShut {NoStop}%
\bibitem [{\citenamefont {Georgescu}\ \emph {et~al.}(2014)\citenamefont {Georgescu}, \citenamefont {Ashhab},\ and\ \citenamefont {Nori}}]{QSRMP2014}%
  \BibitemOpen
  \bibfield  {author} {\bibinfo {author} {\bibfnamefont {I.~M.}\ \bibnamefont {Georgescu}}, \bibinfo {author} {\bibfnamefont {S.}~\bibnamefont {Ashhab}},\ and\ \bibinfo {author} {\bibfnamefont {F.}~\bibnamefont {Nori}},\ }\href {https://doi.org/10.1103/RevModPhys.86.153} {\bibfield  {journal} {\bibinfo  {journal} {Rev. Mod. Phys.}\ }\textbf {\bibinfo {volume} {86}},\ \bibinfo {pages} {153} (\bibinfo {year} {2014})}\BibitemShut {NoStop}%
\bibitem [{\citenamefont {Linke}\ \emph {et~al.}(2017)\citenamefont {Linke}, \citenamefont {Maslov}, \citenamefont {Roetteler}, \citenamefont {Debnath}, \citenamefont {Figgatt}, \citenamefont {Landsman}, \citenamefont {Wright},\ and\ \citenamefont {Monroe}}]{linke2017experimental}%
  \BibitemOpen
  \bibfield  {author} {\bibinfo {author} {\bibfnamefont {N.~M.}\ \bibnamefont {Linke}}, \bibinfo {author} {\bibfnamefont {D.}~\bibnamefont {Maslov}}, \bibinfo {author} {\bibfnamefont {M.}~\bibnamefont {Roetteler}}, \bibinfo {author} {\bibfnamefont {S.}~\bibnamefont {Debnath}}, \bibinfo {author} {\bibfnamefont {C.}~\bibnamefont {Figgatt}}, \bibinfo {author} {\bibfnamefont {K.~A.}\ \bibnamefont {Landsman}}, \bibinfo {author} {\bibfnamefont {K.}~\bibnamefont {Wright}},\ and\ \bibinfo {author} {\bibfnamefont {C.}~\bibnamefont {Monroe}},\ }\href@noop {} {\bibfield  {journal} {\bibinfo  {journal} {Proceedings of the National Academy of Sciences}\ }\textbf {\bibinfo {volume} {114}},\ \bibinfo {pages} {3305} (\bibinfo {year} {2017})}\BibitemShut {NoStop}%
\bibitem [{\citenamefont {Wendin}(2017)}]{SCQRPP2017}%
  \BibitemOpen
  \bibfield  {author} {\bibinfo {author} {\bibfnamefont {G.}~\bibnamefont {Wendin}},\ }\href@noop {} {\bibfield  {journal} {\bibinfo  {journal} {Reports on Progress in Physics}\ }\textbf {\bibinfo {volume} {80}},\ \bibinfo {pages} {106001} (\bibinfo {year} {2017})}\BibitemShut {NoStop}%
\bibitem [{\citenamefont {Krantz}\ \emph {et~al.}(2019)\citenamefont {Krantz}, \citenamefont {Kjaergaard}, \citenamefont {Yan}, \citenamefont {Orlando}, \citenamefont {Gustavsson},\ and\ \citenamefont {Oliver}}]{SCQNISQ20191}%
  \BibitemOpen
  \bibfield  {author} {\bibinfo {author} {\bibfnamefont {P.}~\bibnamefont {Krantz}}, \bibinfo {author} {\bibfnamefont {M.}~\bibnamefont {Kjaergaard}}, \bibinfo {author} {\bibfnamefont {F.}~\bibnamefont {Yan}}, \bibinfo {author} {\bibfnamefont {T.~P.}\ \bibnamefont {Orlando}}, \bibinfo {author} {\bibfnamefont {S.}~\bibnamefont {Gustavsson}},\ and\ \bibinfo {author} {\bibfnamefont {W.~D.}\ \bibnamefont {Oliver}},\ }\href@noop {} {\bibfield  {journal} {\bibinfo  {journal} {Applied Physics Reviews}\ }\textbf {\bibinfo {volume} {6}},\ \bibinfo {pages} {021318} (\bibinfo {year} {2019})}\BibitemShut {NoStop}%
\bibitem [{\citenamefont {Kjaergaard}\ \emph {et~al.}(2020)\citenamefont {Kjaergaard}, \citenamefont {Schwartz}, \citenamefont {Braum{\"u}ller}, \citenamefont {Krantz}, \citenamefont {Wang}, \citenamefont {Gustavsson},\ and\ \citenamefont {Oliver}}]{SCQARCMP2020}%
  \BibitemOpen
  \bibfield  {author} {\bibinfo {author} {\bibfnamefont {M.}~\bibnamefont {Kjaergaard}}, \bibinfo {author} {\bibfnamefont {M.~E.}\ \bibnamefont {Schwartz}}, \bibinfo {author} {\bibfnamefont {J.}~\bibnamefont {Braum{\"u}ller}}, \bibinfo {author} {\bibfnamefont {P.}~\bibnamefont {Krantz}}, \bibinfo {author} {\bibfnamefont {J.~I.-J.}\ \bibnamefont {Wang}}, \bibinfo {author} {\bibfnamefont {S.}~\bibnamefont {Gustavsson}},\ and\ \bibinfo {author} {\bibfnamefont {W.~D.}\ \bibnamefont {Oliver}},\ }\href@noop {} {\bibfield  {journal} {\bibinfo  {journal} {Annual Review of Condensed Matter Physics}\ }\textbf {\bibinfo {volume} {11}},\ \bibinfo {pages} {369} (\bibinfo {year} {2020})}\BibitemShut {NoStop}%
\bibitem [{\citenamefont {Huang}\ \emph {et~al.}(2020)\citenamefont {Huang}, \citenamefont {Wu}, \citenamefont {Fan},\ and\ \citenamefont {Zhu}}]{ZhugroupSQC2020}%
  \BibitemOpen
  \bibfield  {author} {\bibinfo {author} {\bibfnamefont {H.-L.}\ \bibnamefont {Huang}}, \bibinfo {author} {\bibfnamefont {D.}~\bibnamefont {Wu}}, \bibinfo {author} {\bibfnamefont {D.}~\bibnamefont {Fan}},\ and\ \bibinfo {author} {\bibfnamefont {X.}~\bibnamefont {Zhu}},\ }\href@noop {} {\bibfield  {journal} {\bibinfo  {journal} {Science China Information Sciences}\ }\textbf {\bibinfo {volume} {63}},\ \bibinfo {pages} {1} (\bibinfo {year} {2020})}\BibitemShut {NoStop}%
\bibitem [{\citenamefont {Bruzewicz}\ \emph {et~al.}(2019)\citenamefont {Bruzewicz}, \citenamefont {Chiaverini}, \citenamefont {McConnell},\ and\ \citenamefont {Sage}}]{trappedionNISQ2019}%
  \BibitemOpen
  \bibfield  {author} {\bibinfo {author} {\bibfnamefont {C.~D.}\ \bibnamefont {Bruzewicz}}, \bibinfo {author} {\bibfnamefont {J.}~\bibnamefont {Chiaverini}}, \bibinfo {author} {\bibfnamefont {R.}~\bibnamefont {McConnell}},\ and\ \bibinfo {author} {\bibfnamefont {J.~M.}\ \bibnamefont {Sage}},\ }\href@noop {} {\bibfield  {journal} {\bibinfo  {journal} {Applied Physics Reviews}\ }\textbf {\bibinfo {volume} {6}},\ \bibinfo {pages} {021314} (\bibinfo {year} {2019})}\BibitemShut {NoStop}%
\bibitem [{\citenamefont {Kaushal}\ \emph {et~al.}(2020)\citenamefont {Kaushal}, \citenamefont {Lekitsch}, \citenamefont {Stahl}, \citenamefont {Hilder}, \citenamefont {Pijn}, \citenamefont {Schmiegelow}, \citenamefont {Bermudez}, \citenamefont {M{\"u}ller}, \citenamefont {Schmidt-Kaler},\ and\ \citenamefont {Poschinger}}]{AVSQSiontrapQC2020}%
  \BibitemOpen
  \bibfield  {author} {\bibinfo {author} {\bibfnamefont {V.}~\bibnamefont {Kaushal}}, \bibinfo {author} {\bibfnamefont {B.}~\bibnamefont {Lekitsch}}, \bibinfo {author} {\bibfnamefont {A.}~\bibnamefont {Stahl}}, \bibinfo {author} {\bibfnamefont {J.}~\bibnamefont {Hilder}}, \bibinfo {author} {\bibfnamefont {D.}~\bibnamefont {Pijn}}, \bibinfo {author} {\bibfnamefont {C.}~\bibnamefont {Schmiegelow}}, \bibinfo {author} {\bibfnamefont {A.}~\bibnamefont {Bermudez}}, \bibinfo {author} {\bibfnamefont {M.}~\bibnamefont {M{\"u}ller}}, \bibinfo {author} {\bibfnamefont {F.}~\bibnamefont {Schmidt-Kaler}},\ and\ \bibinfo {author} {\bibfnamefont {U.}~\bibnamefont {Poschinger}},\ }\href@noop {} {\bibfield  {journal} {\bibinfo  {journal} {AVS Quantum Science}\ }\textbf {\bibinfo {volume} {2}},\ \bibinfo {pages} {014101} (\bibinfo {year} {2020})}\BibitemShut {NoStop}%
\bibitem [{\citenamefont {Nakamura}\ \emph {et~al.}(1999)\citenamefont {Nakamura}, \citenamefont {Pashkin},\ and\ \citenamefont {Tsai}}]{NakamuraTsaigroupSC}%
  \BibitemOpen
  \bibfield  {author} {\bibinfo {author} {\bibfnamefont {Y.}~\bibnamefont {Nakamura}}, \bibinfo {author} {\bibfnamefont {Y.~A.}\ \bibnamefont {Pashkin}},\ and\ \bibinfo {author} {\bibfnamefont {J.}~\bibnamefont {Tsai}},\ }\href@noop {} {\bibfield  {journal} {\bibinfo  {journal} {nature}\ }\textbf {\bibinfo {volume} {398}},\ \bibinfo {pages} {786} (\bibinfo {year} {1999})}\BibitemShut {NoStop}%
\bibitem [{\citenamefont {Makhlin}\ \emph {et~al.}(2001)\citenamefont {Makhlin}, \citenamefont {Sch\"on},\ and\ \citenamefont {Shnirman}}]{SCQRMP2001}%
  \BibitemOpen
  \bibfield  {author} {\bibinfo {author} {\bibfnamefont {Y.}~\bibnamefont {Makhlin}}, \bibinfo {author} {\bibfnamefont {G.}~\bibnamefont {Sch\"on}},\ and\ \bibinfo {author} {\bibfnamefont {A.}~\bibnamefont {Shnirman}},\ }\href {https://doi.org/10.1103/RevModPhys.73.357} {\bibfield  {journal} {\bibinfo  {journal} {Rev. Mod. Phys.}\ }\textbf {\bibinfo {volume} {73}},\ \bibinfo {pages} {357} (\bibinfo {year} {2001})}\BibitemShut {NoStop}%
\bibitem [{\citenamefont {Xiang}\ \emph {et~al.}(2013)\citenamefont {Xiang}, \citenamefont {Ashhab}, \citenamefont {You},\ and\ \citenamefont {Nori}}]{circuitqedreview1}%
  \BibitemOpen
  \bibfield  {author} {\bibinfo {author} {\bibfnamefont {Z.-L.}\ \bibnamefont {Xiang}}, \bibinfo {author} {\bibfnamefont {S.}~\bibnamefont {Ashhab}}, \bibinfo {author} {\bibfnamefont {J.~Q.}\ \bibnamefont {You}},\ and\ \bibinfo {author} {\bibfnamefont {F.}~\bibnamefont {Nori}},\ }\href {https://doi.org/10.1103/RevModPhys.85.623} {\bibfield  {journal} {\bibinfo  {journal} {Rev. Mod. Phys.}\ }\textbf {\bibinfo {volume} {85}},\ \bibinfo {pages} {623} (\bibinfo {year} {2013})}\BibitemShut {NoStop}%
\bibitem [{\citenamefont {Kwon}\ \emph {et~al.}(2021)\citenamefont {Kwon}, \citenamefont {Tomonaga}, \citenamefont {Lakshmi~Bhai}, \citenamefont {Devitt},\ and\ \citenamefont {Tsai}}]{TsaigroupSCCQC2021}%
  \BibitemOpen
  \bibfield  {author} {\bibinfo {author} {\bibfnamefont {S.}~\bibnamefont {Kwon}}, \bibinfo {author} {\bibfnamefont {A.}~\bibnamefont {Tomonaga}}, \bibinfo {author} {\bibfnamefont {G.}~\bibnamefont {Lakshmi~Bhai}}, \bibinfo {author} {\bibfnamefont {S.~J.}\ \bibnamefont {Devitt}},\ and\ \bibinfo {author} {\bibfnamefont {J.-S.}\ \bibnamefont {Tsai}},\ }\href@noop {} {\bibfield  {journal} {\bibinfo  {journal} {Journal of Applied Physics}\ }\textbf {\bibinfo {volume} {129}},\ \bibinfo {pages} {041102} (\bibinfo {year} {2021})}\BibitemShut {NoStop}%
\bibitem [{\citenamefont {Cirac}\ and\ \citenamefont {Zoller}(1995)}]{CiracZollergate}%
  \BibitemOpen
  \bibfield  {author} {\bibinfo {author} {\bibfnamefont {J.~I.}\ \bibnamefont {Cirac}}\ and\ \bibinfo {author} {\bibfnamefont {P.}~\bibnamefont {Zoller}},\ }\href {https://doi.org/10.1103/PhysRevLett.74.4091} {\bibfield  {journal} {\bibinfo  {journal} {Phys. Rev. Lett.}\ }\textbf {\bibinfo {volume} {74}},\ \bibinfo {pages} {4091} (\bibinfo {year} {1995})}\BibitemShut {NoStop}%
\bibitem [{\citenamefont {S\o{}rensen}\ and\ \citenamefont {M\o{}lmer}(1999)}]{TionSMgate}%
  \BibitemOpen
  \bibfield  {author} {\bibinfo {author} {\bibfnamefont {A.}~\bibnamefont {S\o{}rensen}}\ and\ \bibinfo {author} {\bibfnamefont {K.}~\bibnamefont {M\o{}lmer}},\ }\href {https://doi.org/10.1103/PhysRevLett.82.1971} {\bibfield  {journal} {\bibinfo  {journal} {Phys. Rev. Lett.}\ }\textbf {\bibinfo {volume} {82}},\ \bibinfo {pages} {1971} (\bibinfo {year} {1999})}\BibitemShut {NoStop}%
\bibitem [{\citenamefont {Leibfried}\ \emph {et~al.}(2003)\citenamefont {Leibfried}, \citenamefont {Blatt}, \citenamefont {Monroe},\ and\ \citenamefont {Wineland}}]{trappedionsreview1A}%
  \BibitemOpen
  \bibfield  {author} {\bibinfo {author} {\bibfnamefont {D.}~\bibnamefont {Leibfried}}, \bibinfo {author} {\bibfnamefont {R.}~\bibnamefont {Blatt}}, \bibinfo {author} {\bibfnamefont {C.}~\bibnamefont {Monroe}},\ and\ \bibinfo {author} {\bibfnamefont {D.}~\bibnamefont {Wineland}},\ }\href {https://doi.org/10.1103/RevModPhys.75.281} {\bibfield  {journal} {\bibinfo  {journal} {Rev. Mod. Phys.}\ }\textbf {\bibinfo {volume} {75}},\ \bibinfo {pages} {281} (\bibinfo {year} {2003})}\BibitemShut {NoStop}%
\bibitem [{\citenamefont {Wineland}(2009)}]{trappedionsreview1B}%
  \BibitemOpen
  \bibfield  {author} {\bibinfo {author} {\bibfnamefont {D.~J.}\ \bibnamefont {Wineland}},\ }\href@noop {} {\bibfield  {journal} {\bibinfo  {journal} {Physica Scripta}\ }\textbf {\bibinfo {volume} {2009}},\ \bibinfo {pages} {014007} (\bibinfo {year} {2009})}\BibitemShut {NoStop}%
\bibitem [{\citenamefont {H{\"a}ffner}\ \emph {et~al.}(2008)\citenamefont {H{\"a}ffner}, \citenamefont {Roos},\ and\ \citenamefont {Blatt}}]{trappedionsreview2}%
  \BibitemOpen
  \bibfield  {author} {\bibinfo {author} {\bibfnamefont {H.}~\bibnamefont {H{\"a}ffner}}, \bibinfo {author} {\bibfnamefont {C.~F.}\ \bibnamefont {Roos}},\ and\ \bibinfo {author} {\bibfnamefont {R.}~\bibnamefont {Blatt}},\ }\href@noop {} {\bibfield  {journal} {\bibinfo  {journal} {Physics reports}\ }\textbf {\bibinfo {volume} {469}},\ \bibinfo {pages} {155} (\bibinfo {year} {2008})}\BibitemShut {NoStop}%
\bibitem [{\citenamefont {Peruzzo}\ \emph {et~al.}(2014)\citenamefont {Peruzzo}, \citenamefont {McClean}, \citenamefont {Shadbolt}, \citenamefont {Yung}, \citenamefont {Zhou}, \citenamefont {Love}, \citenamefont {Aspuru-Guzik},\ and\ \citenamefont {O’brien}}]{VQEnc2014}%
  \BibitemOpen
  \bibfield  {author} {\bibinfo {author} {\bibfnamefont {A.}~\bibnamefont {Peruzzo}}, \bibinfo {author} {\bibfnamefont {J.}~\bibnamefont {McClean}}, \bibinfo {author} {\bibfnamefont {P.}~\bibnamefont {Shadbolt}}, \bibinfo {author} {\bibfnamefont {M.-H.}\ \bibnamefont {Yung}}, \bibinfo {author} {\bibfnamefont {X.-Q.}\ \bibnamefont {Zhou}}, \bibinfo {author} {\bibfnamefont {P.~J.}\ \bibnamefont {Love}}, \bibinfo {author} {\bibfnamefont {A.}~\bibnamefont {Aspuru-Guzik}},\ and\ \bibinfo {author} {\bibfnamefont {J.~L.}\ \bibnamefont {O’brien}},\ }\href@noop {} {\bibfield  {journal} {\bibinfo  {journal} {Nature communications}\ }\textbf {\bibinfo {volume} {5}},\ \bibinfo {pages} {1} (\bibinfo {year} {2014})}\BibitemShut {NoStop}%
\bibitem [{\citenamefont {McClean}\ \emph {et~al.}(2016)\citenamefont {McClean}, \citenamefont {Romero}, \citenamefont {Babbush},\ and\ \citenamefont {Aspuru-Guzik}}]{VQE0}%
  \BibitemOpen
  \bibfield  {author} {\bibinfo {author} {\bibfnamefont {J.~R.}\ \bibnamefont {McClean}}, \bibinfo {author} {\bibfnamefont {J.}~\bibnamefont {Romero}}, \bibinfo {author} {\bibfnamefont {R.}~\bibnamefont {Babbush}},\ and\ \bibinfo {author} {\bibfnamefont {A.}~\bibnamefont {Aspuru-Guzik}},\ }\href@noop {} {\bibfield  {journal} {\bibinfo  {journal} {New Journal of Physics}\ }\textbf {\bibinfo {volume} {18}},\ \bibinfo {pages} {023023} (\bibinfo {year} {2016})}\BibitemShut {NoStop}%
\bibitem [{\citenamefont {Kandala}\ \emph {et~al.}(2017)\citenamefont {Kandala}, \citenamefont {Mezzacapo}, \citenamefont {Temme}, \citenamefont {Takita}, \citenamefont {Brink}, \citenamefont {Chow},\ and\ \citenamefont {Gambetta}}]{VQEnature2017}%
  \BibitemOpen
  \bibfield  {author} {\bibinfo {author} {\bibfnamefont {A.}~\bibnamefont {Kandala}}, \bibinfo {author} {\bibfnamefont {A.}~\bibnamefont {Mezzacapo}}, \bibinfo {author} {\bibfnamefont {K.}~\bibnamefont {Temme}}, \bibinfo {author} {\bibfnamefont {M.}~\bibnamefont {Takita}}, \bibinfo {author} {\bibfnamefont {M.}~\bibnamefont {Brink}}, \bibinfo {author} {\bibfnamefont {J.~M.}\ \bibnamefont {Chow}},\ and\ \bibinfo {author} {\bibfnamefont {J.~M.}\ \bibnamefont {Gambetta}},\ }\href@noop {} {\bibfield  {journal} {\bibinfo  {journal} {Nature}\ }\textbf {\bibinfo {volume} {549}},\ \bibinfo {pages} {242} (\bibinfo {year} {2017})}\BibitemShut {NoStop}%
\bibitem [{\citenamefont {McArdle}\ \emph {et~al.}(2020)\citenamefont {McArdle}, \citenamefont {Endo}, \citenamefont {Aspuru-Guzik}, \citenamefont {Benjamin},\ and\ \citenamefont {Yuan}}]{QCchemistryRMP2020}%
  \BibitemOpen
  \bibfield  {author} {\bibinfo {author} {\bibfnamefont {S.}~\bibnamefont {McArdle}}, \bibinfo {author} {\bibfnamefont {S.}~\bibnamefont {Endo}}, \bibinfo {author} {\bibfnamefont {A.}~\bibnamefont {Aspuru-Guzik}}, \bibinfo {author} {\bibfnamefont {S.~C.}\ \bibnamefont {Benjamin}},\ and\ \bibinfo {author} {\bibfnamefont {X.}~\bibnamefont {Yuan}},\ }\href {https://doi.org/10.1103/RevModPhys.92.015003} {\bibfield  {journal} {\bibinfo  {journal} {Rev. Mod. Phys.}\ }\textbf {\bibinfo {volume} {92}},\ \bibinfo {pages} {015003} (\bibinfo {year} {2020})}\BibitemShut {NoStop}%
\bibitem [{\citenamefont {Endo}\ \emph {et~al.}(2021)\citenamefont {Endo}, \citenamefont {Cai}, \citenamefont {Benjamin},\ and\ \citenamefont {Yuan}}]{hybridQCalgorithmJPSJ2021}%
  \BibitemOpen
  \bibfield  {author} {\bibinfo {author} {\bibfnamefont {S.}~\bibnamefont {Endo}}, \bibinfo {author} {\bibfnamefont {Z.}~\bibnamefont {Cai}}, \bibinfo {author} {\bibfnamefont {S.~C.}\ \bibnamefont {Benjamin}},\ and\ \bibinfo {author} {\bibfnamefont {X.}~\bibnamefont {Yuan}},\ }\href@noop {} {\bibfield  {journal} {\bibinfo  {journal} {Journal of the Physical Society of Japan}\ }\textbf {\bibinfo {volume} {90}},\ \bibinfo {pages} {032001} (\bibinfo {year} {2021})}\BibitemShut {NoStop}%
\bibitem [{\citenamefont {Farhi}\ \emph {et~al.}(2014)\citenamefont {Farhi}, \citenamefont {Goldstone},\ and\ \citenamefont {Gutmann}}]{QAOA2014}%
  \BibitemOpen
  \bibfield  {author} {\bibinfo {author} {\bibfnamefont {E.}~\bibnamefont {Farhi}}, \bibinfo {author} {\bibfnamefont {J.}~\bibnamefont {Goldstone}},\ and\ \bibinfo {author} {\bibfnamefont {S.}~\bibnamefont {Gutmann}},\ }\href@noop {} {\bibfield  {journal} {\bibinfo  {journal} {arXiv preprint arXiv:1411.4028}\ } (\bibinfo {year} {2014})}\BibitemShut {NoStop}%
\bibitem [{\citenamefont {Crooks}(2018)}]{crooks2018performance}%
  \BibitemOpen
  \bibfield  {author} {\bibinfo {author} {\bibfnamefont {G.~E.}\ \bibnamefont {Crooks}},\ }\href@noop {} {\bibfield  {journal} {\bibinfo  {journal} {arXiv preprint arXiv:1811.08419}\ } (\bibinfo {year} {2018})}\BibitemShut {NoStop}%
\bibitem [{\citenamefont {Wang}\ \emph {et~al.}(2018)\citenamefont {Wang}, \citenamefont {Hadfield}, \citenamefont {Jiang},\ and\ \citenamefont {Rieffel}}]{wang2018quantum}%
  \BibitemOpen
  \bibfield  {author} {\bibinfo {author} {\bibfnamefont {Z.}~\bibnamefont {Wang}}, \bibinfo {author} {\bibfnamefont {S.}~\bibnamefont {Hadfield}}, \bibinfo {author} {\bibfnamefont {Z.}~\bibnamefont {Jiang}},\ and\ \bibinfo {author} {\bibfnamefont {E.~G.}\ \bibnamefont {Rieffel}},\ }\href@noop {} {\bibfield  {journal} {\bibinfo  {journal} {Physical Review A}\ }\textbf {\bibinfo {volume} {97}},\ \bibinfo {pages} {022304} (\bibinfo {year} {2018})}\BibitemShut {NoStop}%
\bibitem [{\citenamefont {Shaydulin}\ and\ \citenamefont {Alexeev}(2019)}]{shaydulin2019evaluating}%
  \BibitemOpen
  \bibfield  {author} {\bibinfo {author} {\bibfnamefont {R.}~\bibnamefont {Shaydulin}}\ and\ \bibinfo {author} {\bibfnamefont {Y.}~\bibnamefont {Alexeev}},\ }in\ \href@noop {} {\emph {\bibinfo {booktitle} {2019 tenth international green and sustainable computing conference (IGSC)}}}\ (\bibinfo {organization} {IEEE},\ \bibinfo {year} {2019})\ pp.\ \bibinfo {pages} {1--6}\BibitemShut {NoStop}%
\bibitem [{\citenamefont {Zhou}\ \emph {et~al.}(2020)\citenamefont {Zhou}, \citenamefont {Wang}, \citenamefont {Choi}, \citenamefont {Pichler},\ and\ \citenamefont {Lukin}}]{zhou2020quantum}%
  \BibitemOpen
  \bibfield  {author} {\bibinfo {author} {\bibfnamefont {L.}~\bibnamefont {Zhou}}, \bibinfo {author} {\bibfnamefont {S.-T.}\ \bibnamefont {Wang}}, \bibinfo {author} {\bibfnamefont {S.}~\bibnamefont {Choi}}, \bibinfo {author} {\bibfnamefont {H.}~\bibnamefont {Pichler}},\ and\ \bibinfo {author} {\bibfnamefont {M.~D.}\ \bibnamefont {Lukin}},\ }\href@noop {} {\bibfield  {journal} {\bibinfo  {journal} {Physical Review X}\ }\textbf {\bibinfo {volume} {10}},\ \bibinfo {pages} {021067} (\bibinfo {year} {2020})}\BibitemShut {NoStop}%
\bibitem [{\citenamefont {Abhijith}\ \emph {et~al.}(2018)\citenamefont {Abhijith}, \citenamefont {Adedoyin}, \citenamefont {Ambrosiano}, \citenamefont {Anisimov}, \citenamefont {B{\"a}rtschi}, \citenamefont {Casper}, \citenamefont {Chennupati}, \citenamefont {Coffrin}, \citenamefont {Djidjev}, \citenamefont {Gunter} \emph {et~al.}}]{TIsing3}%
  \BibitemOpen
  \bibfield  {author} {\bibinfo {author} {\bibfnamefont {J.}~\bibnamefont {Abhijith}}, \bibinfo {author} {\bibfnamefont {A.}~\bibnamefont {Adedoyin}}, \bibinfo {author} {\bibfnamefont {J.}~\bibnamefont {Ambrosiano}}, \bibinfo {author} {\bibfnamefont {P.}~\bibnamefont {Anisimov}}, \bibinfo {author} {\bibfnamefont {A.}~\bibnamefont {B{\"a}rtschi}}, \bibinfo {author} {\bibfnamefont {W.}~\bibnamefont {Casper}}, \bibinfo {author} {\bibfnamefont {G.}~\bibnamefont {Chennupati}}, \bibinfo {author} {\bibfnamefont {C.}~\bibnamefont {Coffrin}}, \bibinfo {author} {\bibfnamefont {H.}~\bibnamefont {Djidjev}}, \bibinfo {author} {\bibfnamefont {D.}~\bibnamefont {Gunter}}, \emph {et~al.},\ }\href@noop {} {\bibfield  {journal} {\bibinfo  {journal} {arXiv e-prints}\ ,\ \bibinfo {pages} {arXiv}} (\bibinfo {year} {2018})}\BibitemShut {NoStop}%
\bibitem [{\citenamefont {Schuld}\ and\ \citenamefont {Petruccione}(2018)}]{SchuldandPetruccionegroupQML}%
  \BibitemOpen
  \bibfield  {author} {\bibinfo {author} {\bibfnamefont {M.}~\bibnamefont {Schuld}}\ and\ \bibinfo {author} {\bibfnamefont {F.}~\bibnamefont {Petruccione}},\ }\href@noop {} {\emph {\bibinfo {title} {Supervised learning with quantum computers}}},\ Vol.~\bibinfo {volume} {17}\ (\bibinfo  {publisher} {Springer},\ \bibinfo {year} {2018})\BibitemShut {NoStop}%
\bibitem [{\citenamefont {Mitarai}\ \emph {et~al.}(2018)\citenamefont {Mitarai}, \citenamefont {Negoro}, \citenamefont {Kitagawa},\ and\ \citenamefont {Fujii}}]{MitaraietalQML}%
  \BibitemOpen
  \bibfield  {author} {\bibinfo {author} {\bibfnamefont {K.}~\bibnamefont {Mitarai}}, \bibinfo {author} {\bibfnamefont {M.}~\bibnamefont {Negoro}}, \bibinfo {author} {\bibfnamefont {M.}~\bibnamefont {Kitagawa}},\ and\ \bibinfo {author} {\bibfnamefont {K.}~\bibnamefont {Fujii}},\ }\href {https://doi.org/10.1103/PhysRevA.98.032309} {\bibfield  {journal} {\bibinfo  {journal} {Phys. Rev. A}\ }\textbf {\bibinfo {volume} {98}},\ \bibinfo {pages} {032309} (\bibinfo {year} {2018})}\BibitemShut {NoStop}%
\bibitem [{\citenamefont {Benedetti}\ \emph {et~al.}(2019)\citenamefont {Benedetti}, \citenamefont {Lloyd}, \citenamefont {Sack},\ and\ \citenamefont {Fiorentini}}]{QSTQML2019}%
  \BibitemOpen
  \bibfield  {author} {\bibinfo {author} {\bibfnamefont {M.}~\bibnamefont {Benedetti}}, \bibinfo {author} {\bibfnamefont {E.}~\bibnamefont {Lloyd}}, \bibinfo {author} {\bibfnamefont {S.}~\bibnamefont {Sack}},\ and\ \bibinfo {author} {\bibfnamefont {M.}~\bibnamefont {Fiorentini}},\ }\href@noop {} {\bibfield  {journal} {\bibinfo  {journal} {Quantum Science and Technology}\ }\textbf {\bibinfo {volume} {5}},\ \bibinfo {pages} {019601} (\bibinfo {year} {2019})}\BibitemShut {NoStop}%
\bibitem [{\citenamefont {Arute}\ \emph {et~al.}(2019)\citenamefont {Arute}, \citenamefont {Arya}, \citenamefont {Babbush}, \citenamefont {Bacon}, \citenamefont {Bardin}, \citenamefont {Barends}, \citenamefont {Biswas}, \citenamefont {Boixo}, \citenamefont {Brandao}, \citenamefont {Buell} \emph {et~al.}}]{Qsup}%
  \BibitemOpen
  \bibfield  {author} {\bibinfo {author} {\bibfnamefont {F.}~\bibnamefont {Arute}}, \bibinfo {author} {\bibfnamefont {K.}~\bibnamefont {Arya}}, \bibinfo {author} {\bibfnamefont {R.}~\bibnamefont {Babbush}}, \bibinfo {author} {\bibfnamefont {D.}~\bibnamefont {Bacon}}, \bibinfo {author} {\bibfnamefont {J.~C.}\ \bibnamefont {Bardin}}, \bibinfo {author} {\bibfnamefont {R.}~\bibnamefont {Barends}}, \bibinfo {author} {\bibfnamefont {R.}~\bibnamefont {Biswas}}, \bibinfo {author} {\bibfnamefont {S.}~\bibnamefont {Boixo}}, \bibinfo {author} {\bibfnamefont {F.~G.}\ \bibnamefont {Brandao}}, \bibinfo {author} {\bibfnamefont {D.~A.}\ \bibnamefont {Buell}}, \emph {et~al.},\ }\href@noop {} {\bibfield  {journal} {\bibinfo  {journal} {Nature}\ }\textbf {\bibinfo {volume} {574}},\ \bibinfo {pages} {505} (\bibinfo {year} {2019})}\BibitemShut {NoStop}%
\bibitem [{\citenamefont {Preskill}(2018)}]{PreskillNISQ2018}%
  \BibitemOpen
  \bibfield  {author} {\bibinfo {author} {\bibfnamefont {J.}~\bibnamefont {Preskill}},\ }\href@noop {} {\bibfield  {journal} {\bibinfo  {journal} {Quantum}\ }\textbf {\bibinfo {volume} {2}},\ \bibinfo {pages} {79} (\bibinfo {year} {2018})}\BibitemShut {NoStop}%
\bibitem [{\citenamefont {Palma}\ and\ \citenamefont {Suominen}(1996)}]{EkertgroupQCdissipation}%
  \BibitemOpen
  \bibfield  {author} {\bibinfo {author} {\bibfnamefont {G.}~\bibnamefont {Palma}}\ and\ \bibinfo {author} {\bibfnamefont {K.}~\bibnamefont {Suominen}},\ }\href@noop {} {\bibfield  {journal} {\bibinfo  {journal} {London A}\ }\textbf {\bibinfo {volume} {452}},\ \bibinfo {pages} {567} (\bibinfo {year} {1996})}\BibitemShut {NoStop}%
\bibitem [{\citenamefont {Resch}\ and\ \citenamefont {Karpuzcu}(2021)}]{resch2021benchmarking}%
  \BibitemOpen
  \bibfield  {author} {\bibinfo {author} {\bibfnamefont {S.}~\bibnamefont {Resch}}\ and\ \bibinfo {author} {\bibfnamefont {U.~R.}\ \bibnamefont {Karpuzcu}},\ }\href@noop {} {\bibfield  {journal} {\bibinfo  {journal} {ACM Computing Surveys (CSUR)}\ }\textbf {\bibinfo {volume} {54}},\ \bibinfo {pages} {1} (\bibinfo {year} {2021})}\BibitemShut {NoStop}%
\bibitem [{\citenamefont {Shor}(1995)}]{ShorPRAQEC1995}%
  \BibitemOpen
  \bibfield  {author} {\bibinfo {author} {\bibfnamefont {P.~W.}\ \bibnamefont {Shor}},\ }\href {https://doi.org/10.1103/PhysRevA.52.R2493} {\bibfield  {journal} {\bibinfo  {journal} {Phys. Rev. A}\ }\textbf {\bibinfo {volume} {52}},\ \bibinfo {pages} {R2493} (\bibinfo {year} {1995})}\BibitemShut {NoStop}%
\bibitem [{\citenamefont {Devitt}\ \emph {et~al.}(2013)\citenamefont {Devitt}, \citenamefont {Munro},\ and\ \citenamefont {Nemoto}}]{NemotogroupQEC}%
  \BibitemOpen
  \bibfield  {author} {\bibinfo {author} {\bibfnamefont {S.~J.}\ \bibnamefont {Devitt}}, \bibinfo {author} {\bibfnamefont {W.~J.}\ \bibnamefont {Munro}},\ and\ \bibinfo {author} {\bibfnamefont {K.}~\bibnamefont {Nemoto}},\ }\href@noop {} {\bibfield  {journal} {\bibinfo  {journal} {Reports on Progress in Physics}\ }\textbf {\bibinfo {volume} {76}},\ \bibinfo {pages} {076001} (\bibinfo {year} {2013})}\BibitemShut {NoStop}%
\bibitem [{\citenamefont {Lidar}\ and\ \citenamefont {Brun}(2013)}]{lidar2013quantum}%
  \BibitemOpen
  \bibfield  {author} {\bibinfo {author} {\bibfnamefont {D.~A.}\ \bibnamefont {Lidar}}\ and\ \bibinfo {author} {\bibfnamefont {T.~A.}\ \bibnamefont {Brun}},\ }\href@noop {} {\emph {\bibinfo {title} {Quantum error correction}}}\ (\bibinfo  {publisher} {Cambridge university press},\ \bibinfo {year} {2013})\BibitemShut {NoStop}%
\bibitem [{\citenamefont {Roffe}(2019)}]{QECRoffe}%
  \BibitemOpen
  \bibfield  {author} {\bibinfo {author} {\bibfnamefont {J.}~\bibnamefont {Roffe}},\ }\href@noop {} {\bibfield  {journal} {\bibinfo  {journal} {Contemporary Physics}\ }\textbf {\bibinfo {volume} {60}},\ \bibinfo {pages} {226} (\bibinfo {year} {2019})}\BibitemShut {NoStop}%
\bibitem [{\citenamefont {Viola}\ and\ \citenamefont {Lloyd}(1998)}]{viola1998dynamical}%
  \BibitemOpen
  \bibfield  {author} {\bibinfo {author} {\bibfnamefont {L.}~\bibnamefont {Viola}}\ and\ \bibinfo {author} {\bibfnamefont {S.}~\bibnamefont {Lloyd}},\ }\href@noop {} {\bibfield  {journal} {\bibinfo  {journal} {Physical Review A}\ }\textbf {\bibinfo {volume} {58}},\ \bibinfo {pages} {2733} (\bibinfo {year} {1998})}\BibitemShut {NoStop}%
\bibitem [{\citenamefont {Viola}\ \emph {et~al.}(1999)\citenamefont {Viola}, \citenamefont {Knill},\ and\ \citenamefont {Lloyd}}]{viola1999dynamical}%
  \BibitemOpen
  \bibfield  {author} {\bibinfo {author} {\bibfnamefont {L.}~\bibnamefont {Viola}}, \bibinfo {author} {\bibfnamefont {E.}~\bibnamefont {Knill}},\ and\ \bibinfo {author} {\bibfnamefont {S.}~\bibnamefont {Lloyd}},\ }\href@noop {} {\bibfield  {journal} {\bibinfo  {journal} {Physical Review Letters}\ }\textbf {\bibinfo {volume} {82}},\ \bibinfo {pages} {2417} (\bibinfo {year} {1999})}\BibitemShut {NoStop}%
\bibitem [{\citenamefont {Khodjasteh}\ and\ \citenamefont {Lidar}(2005)}]{khodjasteh2005fault}%
  \BibitemOpen
  \bibfield  {author} {\bibinfo {author} {\bibfnamefont {K.}~\bibnamefont {Khodjasteh}}\ and\ \bibinfo {author} {\bibfnamefont {D.~A.}\ \bibnamefont {Lidar}},\ }\href@noop {} {\bibfield  {journal} {\bibinfo  {journal} {Physical review letters}\ }\textbf {\bibinfo {volume} {95}},\ \bibinfo {pages} {180501} (\bibinfo {year} {2005})}\BibitemShut {NoStop}%
\bibitem [{\citenamefont {Masuyama}\ \emph {et~al.}(2018)\citenamefont {Masuyama}, \citenamefont {Mizuno}, \citenamefont {Ozawa}, \citenamefont {Ishiwata}, \citenamefont {Hatano}, \citenamefont {Ohshima}, \citenamefont {Iwasaki},\ and\ \citenamefont {Hatano}}]{masuyama2018extending}%
  \BibitemOpen
  \bibfield  {author} {\bibinfo {author} {\bibfnamefont {Y.}~\bibnamefont {Masuyama}}, \bibinfo {author} {\bibfnamefont {K.}~\bibnamefont {Mizuno}}, \bibinfo {author} {\bibfnamefont {H.}~\bibnamefont {Ozawa}}, \bibinfo {author} {\bibfnamefont {H.}~\bibnamefont {Ishiwata}}, \bibinfo {author} {\bibfnamefont {Y.}~\bibnamefont {Hatano}}, \bibinfo {author} {\bibfnamefont {T.}~\bibnamefont {Ohshima}}, \bibinfo {author} {\bibfnamefont {T.}~\bibnamefont {Iwasaki}},\ and\ \bibinfo {author} {\bibfnamefont {M.}~\bibnamefont {Hatano}},\ }\href@noop {} {\bibfield  {journal} {\bibinfo  {journal} {Review of Scientific Instruments}\ }\textbf {\bibinfo {volume} {89}},\ \bibinfo {pages} {125007} (\bibinfo {year} {2018})}\BibitemShut {NoStop}%
\bibitem [{\citenamefont {Temme}\ \emph {et~al.}(2017)\citenamefont {Temme}, \citenamefont {Bravyi},\ and\ \citenamefont {Gambetta}}]{EMPRL2017}%
  \BibitemOpen
  \bibfield  {author} {\bibinfo {author} {\bibfnamefont {K.}~\bibnamefont {Temme}}, \bibinfo {author} {\bibfnamefont {S.}~\bibnamefont {Bravyi}},\ and\ \bibinfo {author} {\bibfnamefont {J.~M.}\ \bibnamefont {Gambetta}},\ }\href@noop {} {\bibfield  {journal} {\bibinfo  {journal} {Physical review letters}\ }\textbf {\bibinfo {volume} {119}},\ \bibinfo {pages} {180509} (\bibinfo {year} {2017})}\BibitemShut {NoStop}%
\bibitem [{\citenamefont {Kandala}\ \emph {et~al.}(2019)\citenamefont {Kandala}, \citenamefont {Temme}, \citenamefont {C{\'o}rcoles}, \citenamefont {Mezzacapo}, \citenamefont {Chow},\ and\ \citenamefont {Gambetta}}]{EMNature2019}%
  \BibitemOpen
  \bibfield  {author} {\bibinfo {author} {\bibfnamefont {A.}~\bibnamefont {Kandala}}, \bibinfo {author} {\bibfnamefont {K.}~\bibnamefont {Temme}}, \bibinfo {author} {\bibfnamefont {A.~D.}\ \bibnamefont {C{\'o}rcoles}}, \bibinfo {author} {\bibfnamefont {A.}~\bibnamefont {Mezzacapo}}, \bibinfo {author} {\bibfnamefont {J.~M.}\ \bibnamefont {Chow}},\ and\ \bibinfo {author} {\bibfnamefont {J.~M.}\ \bibnamefont {Gambetta}},\ }\href@noop {} {\bibfield  {journal} {\bibinfo  {journal} {Nature}\ }\textbf {\bibinfo {volume} {567}},\ \bibinfo {pages} {491} (\bibinfo {year} {2019})}\BibitemShut {NoStop}%
\bibitem [{\citenamefont {Li}\ and\ \citenamefont {Benjamin}(2017)}]{EMPRX2017}%
  \BibitemOpen
  \bibfield  {author} {\bibinfo {author} {\bibfnamefont {Y.}~\bibnamefont {Li}}\ and\ \bibinfo {author} {\bibfnamefont {S.~C.}\ \bibnamefont {Benjamin}},\ }\href@noop {} {\bibfield  {journal} {\bibinfo  {journal} {Physical Review X}\ }\textbf {\bibinfo {volume} {7}},\ \bibinfo {pages} {021050} (\bibinfo {year} {2017})}\BibitemShut {NoStop}%
\bibitem [{\citenamefont {Endo}\ \emph {et~al.}(2018)\citenamefont {Endo}, \citenamefont {Benjamin},\ and\ \citenamefont {Li}}]{EMPRX2018}%
  \BibitemOpen
  \bibfield  {author} {\bibinfo {author} {\bibfnamefont {S.}~\bibnamefont {Endo}}, \bibinfo {author} {\bibfnamefont {S.~C.}\ \bibnamefont {Benjamin}},\ and\ \bibinfo {author} {\bibfnamefont {Y.}~\bibnamefont {Li}},\ }\href@noop {} {\bibfield  {journal} {\bibinfo  {journal} {Physical Review X}\ }\textbf {\bibinfo {volume} {8}},\ \bibinfo {pages} {031027} (\bibinfo {year} {2018})}\BibitemShut {NoStop}%
\bibitem [{\citenamefont {Premakumar}\ and\ \citenamefont {Joynt}(2018)}]{EMarxiv2018}%
  \BibitemOpen
  \bibfield  {author} {\bibinfo {author} {\bibfnamefont {V.~N.}\ \bibnamefont {Premakumar}}\ and\ \bibinfo {author} {\bibfnamefont {R.}~\bibnamefont {Joynt}},\ }\href@noop {} {\bibfield  {journal} {\bibinfo  {journal} {arXiv preprint arXiv:1812.07076}\ } (\bibinfo {year} {2018})}\BibitemShut {NoStop}%
\bibitem [{\citenamefont {Bonet-Monroig}\ \emph {et~al.}(2018)\citenamefont {Bonet-Monroig}, \citenamefont {Sagastizabal}, \citenamefont {Singh},\ and\ \citenamefont {O'Brien}}]{PhysRevA.98.062339}%
  \BibitemOpen
  \bibfield  {author} {\bibinfo {author} {\bibfnamefont {X.}~\bibnamefont {Bonet-Monroig}}, \bibinfo {author} {\bibfnamefont {R.}~\bibnamefont {Sagastizabal}}, \bibinfo {author} {\bibfnamefont {M.}~\bibnamefont {Singh}},\ and\ \bibinfo {author} {\bibfnamefont {T.~E.}\ \bibnamefont {O'Brien}},\ }\href {https://doi.org/10.1103/PhysRevA.98.062339} {\bibfield  {journal} {\bibinfo  {journal} {Phys. Rev. A}\ }\textbf {\bibinfo {volume} {98}},\ \bibinfo {pages} {062339} (\bibinfo {year} {2018})}\BibitemShut {NoStop}%
\bibitem [{\citenamefont {McArdle}\ \emph {et~al.}(2019)\citenamefont {McArdle}, \citenamefont {Yuan},\ and\ \citenamefont {Benjamin}}]{mcardle2019error}%
  \BibitemOpen
  \bibfield  {author} {\bibinfo {author} {\bibfnamefont {S.}~\bibnamefont {McArdle}}, \bibinfo {author} {\bibfnamefont {X.}~\bibnamefont {Yuan}},\ and\ \bibinfo {author} {\bibfnamefont {S.}~\bibnamefont {Benjamin}},\ }\href@noop {} {\bibfield  {journal} {\bibinfo  {journal} {Physical review letters}\ }\textbf {\bibinfo {volume} {122}},\ \bibinfo {pages} {180501} (\bibinfo {year} {2019})}\BibitemShut {NoStop}%
\bibitem [{\citenamefont {Jattana}\ \emph {et~al.}(2020)\citenamefont {Jattana}, \citenamefont {Jin}, \citenamefont {De~Raedt},\ and\ \citenamefont {Michielsen}}]{jattana2020general}%
  \BibitemOpen
  \bibfield  {author} {\bibinfo {author} {\bibfnamefont {M.~S.}\ \bibnamefont {Jattana}}, \bibinfo {author} {\bibfnamefont {F.}~\bibnamefont {Jin}}, \bibinfo {author} {\bibfnamefont {H.}~\bibnamefont {De~Raedt}},\ and\ \bibinfo {author} {\bibfnamefont {K.}~\bibnamefont {Michielsen}},\ }\href@noop {} {\bibfield  {journal} {\bibinfo  {journal} {Quantum Information Processing}\ }\textbf {\bibinfo {volume} {19}},\ \bibinfo {pages} {1} (\bibinfo {year} {2020})}\BibitemShut {NoStop}%
\bibitem [{\citenamefont {Xiong}\ \emph {et~al.}(2020)\citenamefont {Xiong}, \citenamefont {Chandra}, \citenamefont {Ng},\ and\ \citenamefont {Hanzo}}]{xiong2020sampling}%
  \BibitemOpen
  \bibfield  {author} {\bibinfo {author} {\bibfnamefont {Y.}~\bibnamefont {Xiong}}, \bibinfo {author} {\bibfnamefont {D.}~\bibnamefont {Chandra}}, \bibinfo {author} {\bibfnamefont {S.~X.}\ \bibnamefont {Ng}},\ and\ \bibinfo {author} {\bibfnamefont {L.}~\bibnamefont {Hanzo}},\ }\href@noop {} {\bibfield  {journal} {\bibinfo  {journal} {IEEE Access}\ }\textbf {\bibinfo {volume} {8}},\ \bibinfo {pages} {228967} (\bibinfo {year} {2020})}\BibitemShut {NoStop}%
\bibitem [{\citenamefont {Zlokapa}\ and\ \citenamefont {Gheorghiu}(2020)}]{zlokapa2020deep}%
  \BibitemOpen
  \bibfield  {author} {\bibinfo {author} {\bibfnamefont {A.}~\bibnamefont {Zlokapa}}\ and\ \bibinfo {author} {\bibfnamefont {A.}~\bibnamefont {Gheorghiu}},\ }\href@noop {} {\bibfield  {journal} {\bibinfo  {journal} {arXiv preprint arXiv:2005.10811}\ } (\bibinfo {year} {2020})}\BibitemShut {NoStop}%
\bibitem [{\citenamefont {Bravyi}\ \emph {et~al.}(2021)\citenamefont {Bravyi}, \citenamefont {Sheldon}, \citenamefont {Kandala}, \citenamefont {Mckay},\ and\ \citenamefont {Gambetta}}]{EMPRA2021}%
  \BibitemOpen
  \bibfield  {author} {\bibinfo {author} {\bibfnamefont {S.}~\bibnamefont {Bravyi}}, \bibinfo {author} {\bibfnamefont {S.}~\bibnamefont {Sheldon}}, \bibinfo {author} {\bibfnamefont {A.}~\bibnamefont {Kandala}}, \bibinfo {author} {\bibfnamefont {D.~C.}\ \bibnamefont {Mckay}},\ and\ \bibinfo {author} {\bibfnamefont {J.~M.}\ \bibnamefont {Gambetta}},\ }\href {https://doi.org/10.1103/PhysRevA.103.042605} {\bibfield  {journal} {\bibinfo  {journal} {Phys. Rev. A}\ }\textbf {\bibinfo {volume} {103}},\ \bibinfo {pages} {042605} (\bibinfo {year} {2021})}\BibitemShut {NoStop}%
\bibitem [{\citenamefont {Sun}\ \emph {et~al.}(2021)\citenamefont {Sun}, \citenamefont {Yuan}, \citenamefont {Tsunoda}, \citenamefont {Vedral}, \citenamefont {Benjamin},\ and\ \citenamefont {Endo}}]{CandSQEMPRAp2021}%
  \BibitemOpen
  \bibfield  {author} {\bibinfo {author} {\bibfnamefont {J.}~\bibnamefont {Sun}}, \bibinfo {author} {\bibfnamefont {X.}~\bibnamefont {Yuan}}, \bibinfo {author} {\bibfnamefont {T.}~\bibnamefont {Tsunoda}}, \bibinfo {author} {\bibfnamefont {V.}~\bibnamefont {Vedral}}, \bibinfo {author} {\bibfnamefont {S.~C.}\ \bibnamefont {Benjamin}},\ and\ \bibinfo {author} {\bibfnamefont {S.}~\bibnamefont {Endo}},\ }\href {https://doi.org/10.1103/PhysRevApplied.15.034026} {\bibfield  {journal} {\bibinfo  {journal} {Phys. Rev. Applied}\ }\textbf {\bibinfo {volume} {15}},\ \bibinfo {pages} {034026} (\bibinfo {year} {2021})}\BibitemShut {NoStop}%
\bibitem [{\citenamefont {Otten}\ and\ \citenamefont {Gray}(2019{\natexlab{a}})}]{OttenGrayQEM1}%
  \BibitemOpen
  \bibfield  {author} {\bibinfo {author} {\bibfnamefont {M.}~\bibnamefont {Otten}}\ and\ \bibinfo {author} {\bibfnamefont {S.~K.}\ \bibnamefont {Gray}},\ }\href {https://doi.org/https://doi.org/10.1038/s41534-019-0125-3} {\bibfield  {journal} {\bibinfo  {journal} {npj Quantum Information}\ }\textbf {\bibinfo {volume} {5}},\ \bibinfo {pages} {1} (\bibinfo {year} {2019}{\natexlab{a}})}\BibitemShut {NoStop}%
\bibitem [{\citenamefont {Otten}\ and\ \citenamefont {Gray}(2019{\natexlab{b}})}]{OttenGrayQEM2}%
  \BibitemOpen
  \bibfield  {author} {\bibinfo {author} {\bibfnamefont {M.}~\bibnamefont {Otten}}\ and\ \bibinfo {author} {\bibfnamefont {S.~K.}\ \bibnamefont {Gray}},\ }\href {https://doi.org/10.1103/PhysRevA.99.012338} {\bibfield  {journal} {\bibinfo  {journal} {Phys. Rev. A}\ }\textbf {\bibinfo {volume} {99}},\ \bibinfo {pages} {012338} (\bibinfo {year} {2019}{\natexlab{b}})}\BibitemShut {NoStop}%
\bibitem [{\citenamefont {McClean}\ \emph {et~al.}(2017)\citenamefont {McClean}, \citenamefont {Kimchi-Schwartz}, \citenamefont {Carter},\ and\ \citenamefont {de~Jong}}]{QSEQEM}%
  \BibitemOpen
  \bibfield  {author} {\bibinfo {author} {\bibfnamefont {J.~R.}\ \bibnamefont {McClean}}, \bibinfo {author} {\bibfnamefont {M.~E.}\ \bibnamefont {Kimchi-Schwartz}}, \bibinfo {author} {\bibfnamefont {J.}~\bibnamefont {Carter}},\ and\ \bibinfo {author} {\bibfnamefont {W.~A.}\ \bibnamefont {de~Jong}},\ }\href {https://doi.org/10.1103/PhysRevA.95.042308} {\bibfield  {journal} {\bibinfo  {journal} {Phys. Rev. A}\ }\textbf {\bibinfo {volume} {95}},\ \bibinfo {pages} {042308} (\bibinfo {year} {2017})}\BibitemShut {NoStop}%
\bibitem [{\citenamefont {Czarnik}\ \emph {et~al.}(2021)\citenamefont {Czarnik}, \citenamefont {Arrasmith}, \citenamefont {Coles},\ and\ \citenamefont {Cincio}}]{CliffordQEM}%
  \BibitemOpen
  \bibfield  {author} {\bibinfo {author} {\bibfnamefont {P.}~\bibnamefont {Czarnik}}, \bibinfo {author} {\bibfnamefont {A.}~\bibnamefont {Arrasmith}}, \bibinfo {author} {\bibfnamefont {P.~J.}\ \bibnamefont {Coles}},\ and\ \bibinfo {author} {\bibfnamefont {L.}~\bibnamefont {Cincio}},\ }\href {https://doi.org/10.22331/q-2021-11-26-592} {\bibfield  {journal} {\bibinfo  {journal} {{Quantum}}\ }\textbf {\bibinfo {volume} {5}},\ \bibinfo {pages} {592} (\bibinfo {year} {2021})}\BibitemShut {NoStop}%
\bibitem [{\citenamefont {Strikis}\ \emph {et~al.}(2021)\citenamefont {Strikis}, \citenamefont {Qin}, \citenamefont {Chen}, \citenamefont {Benjamin},\ and\ \citenamefont {Li}}]{LearningBasedQEM}%
  \BibitemOpen
  \bibfield  {author} {\bibinfo {author} {\bibfnamefont {A.}~\bibnamefont {Strikis}}, \bibinfo {author} {\bibfnamefont {D.}~\bibnamefont {Qin}}, \bibinfo {author} {\bibfnamefont {Y.}~\bibnamefont {Chen}}, \bibinfo {author} {\bibfnamefont {S.~C.}\ \bibnamefont {Benjamin}},\ and\ \bibinfo {author} {\bibfnamefont {Y.}~\bibnamefont {Li}},\ }\href {https://doi.org/10.1103/PRXQuantum.2.040330} {\bibfield  {journal} {\bibinfo  {journal} {PRX Quantum}\ }\textbf {\bibinfo {volume} {2}},\ \bibinfo {pages} {040330} (\bibinfo {year} {2021})}\BibitemShut {NoStop}%
\bibitem [{\citenamefont {Huggins}\ \emph {et~al.}(2021)\citenamefont {Huggins}, \citenamefont {McArdle}, \citenamefont {O'Brien}, \citenamefont {Lee}, \citenamefont {Rubin}, \citenamefont {Boixo}, \citenamefont {Whaley}, \citenamefont {Babbush},\ and\ \citenamefont {McClean}}]{VirtualDistillationQEM}%
  \BibitemOpen
  \bibfield  {author} {\bibinfo {author} {\bibfnamefont {W.~J.}\ \bibnamefont {Huggins}}, \bibinfo {author} {\bibfnamefont {S.}~\bibnamefont {McArdle}}, \bibinfo {author} {\bibfnamefont {T.~E.}\ \bibnamefont {O'Brien}}, \bibinfo {author} {\bibfnamefont {J.}~\bibnamefont {Lee}}, \bibinfo {author} {\bibfnamefont {N.~C.}\ \bibnamefont {Rubin}}, \bibinfo {author} {\bibfnamefont {S.}~\bibnamefont {Boixo}}, \bibinfo {author} {\bibfnamefont {K.~B.}\ \bibnamefont {Whaley}}, \bibinfo {author} {\bibfnamefont {R.}~\bibnamefont {Babbush}},\ and\ \bibinfo {author} {\bibfnamefont {J.~R.}\ \bibnamefont {McClean}},\ }\href {https://doi.org/10.1103/PhysRevX.11.041036} {\bibfield  {journal} {\bibinfo  {journal} {Phys. Rev. X}\ }\textbf {\bibinfo {volume} {11}},\ \bibinfo {pages} {041036} (\bibinfo {year} {2021})}\BibitemShut {NoStop}%
\bibitem [{\citenamefont {Koczor}(2021{\natexlab{a}})}]{koczor2021exponential}%
  \BibitemOpen
  \bibfield  {author} {\bibinfo {author} {\bibfnamefont {B.}~\bibnamefont {Koczor}},\ }\href@noop {} {\bibfield  {journal} {\bibinfo  {journal} {Physical Review X}\ }\textbf {\bibinfo {volume} {11}},\ \bibinfo {pages} {031057} (\bibinfo {year} {2021}{\natexlab{a}})}\BibitemShut {NoStop}%
\bibitem [{\citenamefont {Wise}\ \emph {et~al.}(2021)\citenamefont {Wise}, \citenamefont {Morton},\ and\ \citenamefont {Dhomkar}}]{PRXQuantum.2.010316}%
  \BibitemOpen
  \bibfield  {author} {\bibinfo {author} {\bibfnamefont {D.~F.}\ \bibnamefont {Wise}}, \bibinfo {author} {\bibfnamefont {J.~J.}\ \bibnamefont {Morton}},\ and\ \bibinfo {author} {\bibfnamefont {S.}~\bibnamefont {Dhomkar}},\ }\href {https://doi.org/10.1103/PRXQuantum.2.010316} {\bibfield  {journal} {\bibinfo  {journal} {PRX Quantum}\ }\textbf {\bibinfo {volume} {2}},\ \bibinfo {pages} {010316} (\bibinfo {year} {2021})}\BibitemShut {NoStop}%
\bibitem [{\citenamefont {Piveteau}\ \emph {et~al.}(2021)\citenamefont {Piveteau}, \citenamefont {Sutter}, \citenamefont {Bravyi}, \citenamefont {Gambetta},\ and\ \citenamefont {Temme}}]{piveteau2021error}%
  \BibitemOpen
  \bibfield  {author} {\bibinfo {author} {\bibfnamefont {C.}~\bibnamefont {Piveteau}}, \bibinfo {author} {\bibfnamefont {D.}~\bibnamefont {Sutter}}, \bibinfo {author} {\bibfnamefont {S.}~\bibnamefont {Bravyi}}, \bibinfo {author} {\bibfnamefont {J.~M.}\ \bibnamefont {Gambetta}},\ and\ \bibinfo {author} {\bibfnamefont {K.}~\bibnamefont {Temme}},\ }\href@noop {} {\bibfield  {journal} {\bibinfo  {journal} {Physical review letters}\ }\textbf {\bibinfo {volume} {127}},\ \bibinfo {pages} {200505} (\bibinfo {year} {2021})}\BibitemShut {NoStop}%
\bibitem [{\citenamefont {Lostaglio}\ and\ \citenamefont {Ciani}(2021)}]{lostaglio2021error}%
  \BibitemOpen
  \bibfield  {author} {\bibinfo {author} {\bibfnamefont {M.}~\bibnamefont {Lostaglio}}\ and\ \bibinfo {author} {\bibfnamefont {A.}~\bibnamefont {Ciani}},\ }\href@noop {} {\bibfield  {journal} {\bibinfo  {journal} {Physical review letters}\ }\textbf {\bibinfo {volume} {127}},\ \bibinfo {pages} {200506} (\bibinfo {year} {2021})}\BibitemShut {NoStop}%
\bibitem [{\citenamefont {Suzuki}\ \emph {et~al.}(2022)\citenamefont {Suzuki}, \citenamefont {Endo}, \citenamefont {Fujii},\ and\ \citenamefont {Tokunaga}}]{suzuki2022quantum}%
  \BibitemOpen
  \bibfield  {author} {\bibinfo {author} {\bibfnamefont {Y.}~\bibnamefont {Suzuki}}, \bibinfo {author} {\bibfnamefont {S.}~\bibnamefont {Endo}}, \bibinfo {author} {\bibfnamefont {K.}~\bibnamefont {Fujii}},\ and\ \bibinfo {author} {\bibfnamefont {Y.}~\bibnamefont {Tokunaga}},\ }\href {https://doi.org/10.1103/PRXQuantum.3.010345} {\bibfield  {journal} {\bibinfo  {journal} {PRX Quantum}\ }\textbf {\bibinfo {volume} {3}},\ \bibinfo {pages} {010345} (\bibinfo {year} {2022})}\BibitemShut {NoStop}%
\bibitem [{\citenamefont {Piveteau}\ \emph {et~al.}(2022)\citenamefont {Piveteau}, \citenamefont {Sutter},\ and\ \citenamefont {Woerner}}]{piveteau2022quasiprobability}%
  \BibitemOpen
  \bibfield  {author} {\bibinfo {author} {\bibfnamefont {C.}~\bibnamefont {Piveteau}}, \bibinfo {author} {\bibfnamefont {D.}~\bibnamefont {Sutter}},\ and\ \bibinfo {author} {\bibfnamefont {S.}~\bibnamefont {Woerner}},\ }\href@noop {} {\bibfield  {journal} {\bibinfo  {journal} {npj Quantum Information}\ }\textbf {\bibinfo {volume} {8}},\ \bibinfo {pages} {1} (\bibinfo {year} {2022})}\BibitemShut {NoStop}%
\bibitem [{\citenamefont {Pascuzzi}\ \emph {et~al.}(2022)\citenamefont {Pascuzzi}, \citenamefont {He}, \citenamefont {Bauer}, \citenamefont {de~Jong},\ and\ \citenamefont {Nachman}}]{pascuzzi2022computationally}%
  \BibitemOpen
  \bibfield  {author} {\bibinfo {author} {\bibfnamefont {V.~R.}\ \bibnamefont {Pascuzzi}}, \bibinfo {author} {\bibfnamefont {A.}~\bibnamefont {He}}, \bibinfo {author} {\bibfnamefont {C.~W.}\ \bibnamefont {Bauer}}, \bibinfo {author} {\bibfnamefont {W.~A.}\ \bibnamefont {de~Jong}},\ and\ \bibinfo {author} {\bibfnamefont {B.}~\bibnamefont {Nachman}},\ }\href {https://doi.org/10.1103/PhysRevA.105.042406} {\bibfield  {journal} {\bibinfo  {journal} {Phys. Rev. A}\ }\textbf {\bibinfo {volume} {105}},\ \bibinfo {pages} {042406} (\bibinfo {year} {2022})}\BibitemShut {NoStop}%
\bibitem [{\citenamefont {Takagi}(2021)}]{takagi2021optimal}%
  \BibitemOpen
  \bibfield  {author} {\bibinfo {author} {\bibfnamefont {R.}~\bibnamefont {Takagi}},\ }\href {https://doi.org/10.1103/PhysRevResearch.3.033178} {\bibfield  {journal} {\bibinfo  {journal} {Phys. Rev. Research}\ }\textbf {\bibinfo {volume} {3}},\ \bibinfo {pages} {033178} (\bibinfo {year} {2021})}\BibitemShut {NoStop}%
\bibitem [{\citenamefont {LaRose}\ \emph {et~al.}(2022)\citenamefont {LaRose}, \citenamefont {Mari}, \citenamefont {Kaiser}, \citenamefont {Karalekas}, \citenamefont {Alves}, \citenamefont {Czarnik}, \citenamefont {El~Mandouh}, \citenamefont {Gordon}, \citenamefont {Hindy}, \citenamefont {Robertson} \emph {et~al.}}]{larose2022mitiq}%
  \BibitemOpen
  \bibfield  {author} {\bibinfo {author} {\bibfnamefont {R.}~\bibnamefont {LaRose}}, \bibinfo {author} {\bibfnamefont {A.}~\bibnamefont {Mari}}, \bibinfo {author} {\bibfnamefont {S.}~\bibnamefont {Kaiser}}, \bibinfo {author} {\bibfnamefont {P.~J.}\ \bibnamefont {Karalekas}}, \bibinfo {author} {\bibfnamefont {A.~A.}\ \bibnamefont {Alves}}, \bibinfo {author} {\bibfnamefont {P.}~\bibnamefont {Czarnik}}, \bibinfo {author} {\bibfnamefont {M.}~\bibnamefont {El~Mandouh}}, \bibinfo {author} {\bibfnamefont {M.~H.}\ \bibnamefont {Gordon}}, \bibinfo {author} {\bibfnamefont {Y.}~\bibnamefont {Hindy}}, \bibinfo {author} {\bibfnamefont {A.}~\bibnamefont {Robertson}}, \emph {et~al.},\ }\href {https://doi.org/https://doi.org/10.22331/q-2022-08-11-774} {\bibfield  {journal} {\bibinfo  {journal} {Quantum}\ }\textbf {\bibinfo {volume} {6}},\ \bibinfo {pages} {774} (\bibinfo {year} {2022})}\BibitemShut {NoStop}%
\bibitem [{\citenamefont {Koczor}(2021{\natexlab{b}})}]{koczor2021dominant}%
  \BibitemOpen
  \bibfield  {author} {\bibinfo {author} {\bibfnamefont {B.}~\bibnamefont {Koczor}},\ }\href {https://doi.org/10.1088/1367-2630/ac37ae} {\bibfield  {journal} {\bibinfo  {journal} {New Journal of Physics}\ }\textbf {\bibinfo {volume} {23}},\ \bibinfo {pages} {123047} (\bibinfo {year} {2021}{\natexlab{b}})}\BibitemShut {NoStop}%
\bibitem [{\citenamefont {Cai}\ \emph {et~al.}(2022)\citenamefont {Cai}, \citenamefont {Babbush}, \citenamefont {Benjamin}, \citenamefont {Endo}, \citenamefont {Huggins}, \citenamefont {Li}, \citenamefont {McClean},\ and\ \citenamefont {O'Brien}}]{cai2022quantum}%
  \BibitemOpen
  \bibfield  {author} {\bibinfo {author} {\bibfnamefont {Z.}~\bibnamefont {Cai}}, \bibinfo {author} {\bibfnamefont {R.}~\bibnamefont {Babbush}}, \bibinfo {author} {\bibfnamefont {S.~C.}\ \bibnamefont {Benjamin}}, \bibinfo {author} {\bibfnamefont {S.}~\bibnamefont {Endo}}, \bibinfo {author} {\bibfnamefont {W.~J.}\ \bibnamefont {Huggins}}, \bibinfo {author} {\bibfnamefont {Y.}~\bibnamefont {Li}}, \bibinfo {author} {\bibfnamefont {J.~R.}\ \bibnamefont {McClean}},\ and\ \bibinfo {author} {\bibfnamefont {T.~E.}\ \bibnamefont {O'Brien}},\ }\bibfield  {journal} {\bibinfo  {journal} {arXiv preprint arXiv:2210.00921}\ }\href {https://doi.org/10.48550/ARXIV.2210.00921} {10.48550/ARXIV.2210.00921} (\bibinfo {year} {2022})\BibitemShut {NoStop}%
\bibitem [{\citenamefont {Wang}\ \emph {et~al.}(2011)\citenamefont {Wang}, \citenamefont {Ashhab},\ and\ \citenamefont {Nori}}]{NorigroupNEQPRA2011}%
  \BibitemOpen
  \bibfield  {author} {\bibinfo {author} {\bibfnamefont {H.}~\bibnamefont {Wang}}, \bibinfo {author} {\bibfnamefont {S.}~\bibnamefont {Ashhab}},\ and\ \bibinfo {author} {\bibfnamefont {F.}~\bibnamefont {Nori}},\ }\href {https://doi.org/10.1103/PhysRevA.83.062317} {\bibfield  {journal} {\bibinfo  {journal} {Phys. Rev. A}\ }\textbf {\bibinfo {volume} {83}},\ \bibinfo {pages} {062317} (\bibinfo {year} {2011})}\BibitemShut {NoStop}%
\bibitem [{\citenamefont {Hu}\ \emph {et~al.}(2020)\citenamefont {Hu}, \citenamefont {Xia},\ and\ \citenamefont {Kais}}]{KaisgroupOpenQ2020}%
  \BibitemOpen
  \bibfield  {author} {\bibinfo {author} {\bibfnamefont {Z.}~\bibnamefont {Hu}}, \bibinfo {author} {\bibfnamefont {R.}~\bibnamefont {Xia}},\ and\ \bibinfo {author} {\bibfnamefont {S.}~\bibnamefont {Kais}},\ }\href@noop {} {\bibfield  {journal} {\bibinfo  {journal} {Scientific reports}\ }\textbf {\bibinfo {volume} {10}},\ \bibinfo {pages} {1} (\bibinfo {year} {2020})}\BibitemShut {NoStop}%
\bibitem [{\citenamefont {Tornow}\ \emph {et~al.}(2020)\citenamefont {Tornow}, \citenamefont {Gehrke},\ and\ \citenamefont {Helmbrecht}}]{openQHubbard}%
  \BibitemOpen
  \bibfield  {author} {\bibinfo {author} {\bibfnamefont {S.}~\bibnamefont {Tornow}}, \bibinfo {author} {\bibfnamefont {W.}~\bibnamefont {Gehrke}},\ and\ \bibinfo {author} {\bibfnamefont {U.}~\bibnamefont {Helmbrecht}},\ }\href@noop {} {\bibfield  {journal} {\bibinfo  {journal} {arXiv preprint arXiv:2011.11059}\ } (\bibinfo {year} {2020})}\BibitemShut {NoStop}%
\bibitem [{\citenamefont {Garc{\'\i}a-P{\'e}rez}\ \emph {et~al.}(2020)\citenamefont {Garc{\'\i}a-P{\'e}rez}, \citenamefont {Rossi},\ and\ \citenamefont {Maniscalco}}]{openQsimnpj2020}%
  \BibitemOpen
  \bibfield  {author} {\bibinfo {author} {\bibfnamefont {G.}~\bibnamefont {Garc{\'\i}a-P{\'e}rez}}, \bibinfo {author} {\bibfnamefont {M.~A.}\ \bibnamefont {Rossi}},\ and\ \bibinfo {author} {\bibfnamefont {S.}~\bibnamefont {Maniscalco}},\ }\href@noop {} {\bibfield  {journal} {\bibinfo  {journal} {npj Quantum Information}\ }\textbf {\bibinfo {volume} {6}},\ \bibinfo {pages} {1} (\bibinfo {year} {2020})}\BibitemShut {NoStop}%
\bibitem [{\citenamefont {Del~Re}\ \emph {et~al.}(2020)\citenamefont {Del~Re}, \citenamefont {Rost}, \citenamefont {Kemper},\ and\ \citenamefont {Freericks}}]{drivendisspativePRB2020}%
  \BibitemOpen
  \bibfield  {author} {\bibinfo {author} {\bibfnamefont {L.}~\bibnamefont {Del~Re}}, \bibinfo {author} {\bibfnamefont {B.}~\bibnamefont {Rost}}, \bibinfo {author} {\bibfnamefont {A.~F.}\ \bibnamefont {Kemper}},\ and\ \bibinfo {author} {\bibfnamefont {J.~K.}\ \bibnamefont {Freericks}},\ }\href {https://doi.org/10.1103/PhysRevB.102.125112} {\bibfield  {journal} {\bibinfo  {journal} {Phys. Rev. B}\ }\textbf {\bibinfo {volume} {102}},\ \bibinfo {pages} {125112} (\bibinfo {year} {2020})}\BibitemShut {NoStop}%
\bibitem [{\citenamefont {Koppenh{\"o}fer}\ \emph {et~al.}(2020)\citenamefont {Koppenh{\"o}fer}, \citenamefont {Bruder},\ and\ \citenamefont {Roulet}}]{koppenhofer2020quantum}%
  \BibitemOpen
  \bibfield  {author} {\bibinfo {author} {\bibfnamefont {M.}~\bibnamefont {Koppenh{\"o}fer}}, \bibinfo {author} {\bibfnamefont {C.}~\bibnamefont {Bruder}},\ and\ \bibinfo {author} {\bibfnamefont {A.}~\bibnamefont {Roulet}},\ }\href@noop {} {\bibfield  {journal} {\bibinfo  {journal} {Physical Review Research}\ }\textbf {\bibinfo {volume} {2}},\ \bibinfo {pages} {023026} (\bibinfo {year} {2020})}\BibitemShut {NoStop}%
\bibitem [{\citenamefont {de~Jong}\ \emph {et~al.}(2021)\citenamefont {de~Jong}, \citenamefont {Metcalf}, \citenamefont {Mulligan}, \citenamefont {P{\l}osko{\'n}}, \citenamefont {Ringer},\ and\ \citenamefont {Yao}}]{de2021quantum}%
  \BibitemOpen
  \bibfield  {author} {\bibinfo {author} {\bibfnamefont {W.~A.}\ \bibnamefont {de~Jong}}, \bibinfo {author} {\bibfnamefont {M.}~\bibnamefont {Metcalf}}, \bibinfo {author} {\bibfnamefont {J.}~\bibnamefont {Mulligan}}, \bibinfo {author} {\bibfnamefont {M.}~\bibnamefont {P{\l}osko{\'n}}}, \bibinfo {author} {\bibfnamefont {F.}~\bibnamefont {Ringer}},\ and\ \bibinfo {author} {\bibfnamefont {X.}~\bibnamefont {Yao}},\ }\href@noop {} {\bibfield  {journal} {\bibinfo  {journal} {Physical Review D}\ }\textbf {\bibinfo {volume} {104}},\ \bibinfo {pages} {L051501} (\bibinfo {year} {2021})}\BibitemShut {NoStop}%
\bibitem [{\citenamefont {Hama}(2020)}]{hama2020quantum}%
  \BibitemOpen
  \bibfield  {author} {\bibinfo {author} {\bibfnamefont {Y.}~\bibnamefont {Hama}},\ }\href {https://doi.org/10.48550/ARXIV.2012.02410} {\bibinfo {title} {Quantum circuits for collective amplitude damping in two-qubit systems}} (\bibinfo {year} {2020})\BibitemShut {NoStop}%
\bibitem [{\citenamefont {Abraham}\ \emph {et~al.}(2019)\citenamefont {Abraham}, \citenamefont {AduOffei}, \citenamefont {Agarwal}, \citenamefont {Akhalwaya}, \citenamefont {Aleksandrowicz}, \citenamefont {Alexander}, \citenamefont {Amy}, \citenamefont {Arbel}, \citenamefont {Arijit02}, \citenamefont {Asfaw}, \citenamefont {Avkhadiev}, \citenamefont {Azaustre}, \citenamefont {AzizNgoueya}, \citenamefont {Banerjee}, \citenamefont {Bansal}, \citenamefont {Barkoutsos}, \citenamefont {Barnawal}, \citenamefont {Barron}, \citenamefont {Barron}, \citenamefont {Bello}, \citenamefont {Ben-Haim}, \citenamefont {Bevenius}, \citenamefont {Bhobe}, \citenamefont {Bishop}, \citenamefont {Blank}, \citenamefont {Bolos}, \citenamefont {Bosch}, \citenamefont {Brandon}, \citenamefont {Bravyi}, \citenamefont {Bryce-Fuller}, \citenamefont {Bucher}, \citenamefont {Burov}, \citenamefont {Cabrera}, \citenamefont {Calpin}, \citenamefont {Capelluto}, \citenamefont {Carballo}, \citenamefont {Carrascal}, \citenamefont {Chen},
  \citenamefont {Chen}, \citenamefont {Chen}, \citenamefont {Chen}, \citenamefont {Chen}, \citenamefont {Chow}, \citenamefont {Churchill}, \citenamefont {Claus}, \citenamefont {Clauss}, \citenamefont {Cocking}, \citenamefont {Correa}, \citenamefont {Cross}, \citenamefont {Cross}, \citenamefont {Cross}, \citenamefont {Cruz-Benito}, \citenamefont {Culver}, \citenamefont {C{\'o}rcoles-Gonzales}, \citenamefont {Dague}, \citenamefont {Dandachi}, \citenamefont {Daniels}, \citenamefont {Dartiailh}, \citenamefont {DavideFrr}, \citenamefont {Davila}, \citenamefont {Dekusar}, \citenamefont {Ding}, \citenamefont {Doi}, \citenamefont {Drechsler}, \citenamefont {Drew}, \citenamefont {Dumitrescu}, \citenamefont {Dumon}, \citenamefont {Duran}, \citenamefont {EL-Safty}, \citenamefont {Eastman}, \citenamefont {Eberle}, \citenamefont {Eendebak}, \citenamefont {Egger}, \citenamefont {Everitt}, \citenamefont {Fern{\'a}ndez}, \citenamefont {Ferrera}, \citenamefont {Fouilland}, \citenamefont {FranckChevallier}, \citenamefont
  {Frisch}, \citenamefont {Fuhrer}, \citenamefont {Fuller}, \citenamefont {GEORGE}, \citenamefont {Gacon}, \citenamefont {Gago}, \citenamefont {Gambella}, \citenamefont {Gambetta}, \citenamefont {Gammanpila}, \citenamefont {Garcia}, \citenamefont {Garg}, \citenamefont {Garion}, \citenamefont {Gilliam}, \citenamefont {Giridharan}, \citenamefont {Gomez-Mosquera}, \citenamefont {Gonzalo}, \citenamefont {de~la Puente~Gonz{\'a}lez}, \citenamefont {Gorzinski}, \citenamefont {Gould}, \citenamefont {Greenberg}, \citenamefont {Grinko}, \citenamefont {Guan}, \citenamefont {Gunnels}, \citenamefont {Haglund}, \citenamefont {Haide}, \citenamefont {Hamamura}, \citenamefont {Hamido}, \citenamefont {Harkins}, \citenamefont {Havlicek}, \citenamefont {Hellmers}, \citenamefont {Herok}, \citenamefont {Hillmich}, \citenamefont {Horii}, \citenamefont {Howington}, \citenamefont {Hu}, \citenamefont {Hu}, \citenamefont {Huang}, \citenamefont {Huisman}, \citenamefont {Imai}, \citenamefont {Imamichi}, \citenamefont {Ishizaki},
  \citenamefont {Iten}, \citenamefont {Itoko}, \citenamefont {JamesSeaward}, \citenamefont {Javadi}, \citenamefont {Javadi-Abhari}, \citenamefont {Javed}, \citenamefont {Jessica}, \citenamefont {Jivrajani}, \citenamefont {Johns}, \citenamefont {Johnstun}, \citenamefont {Jonathan-Shoemaker}, \citenamefont {K}, \citenamefont {Kachmann}, \citenamefont {Kale}, \citenamefont {Kanazawa}, \citenamefont {Kang-Bae}, \citenamefont {Karazeev}, \citenamefont {Kassebaum}, \citenamefont {Kelso}, \citenamefont {King}, \citenamefont {Knabberjoe}, \citenamefont {Kobayashi}, \citenamefont {Kovyrshin}, \citenamefont {Krishnakumar}, \citenamefont {Krishnan}, \citenamefont {Krsulich}, \citenamefont {Kumkar}, \citenamefont {Kus}, \citenamefont {LaRose}, \citenamefont {Lacal}, \citenamefont {Lambert}, \citenamefont {Lapeyre}, \citenamefont {Latone}, \citenamefont {Lawrence}, \citenamefont {Lee}, \citenamefont {Li}, \citenamefont {Liu}, \citenamefont {Liu}, \citenamefont {Maeng}, \citenamefont {Majmudar}, \citenamefont {Malyshev},
  \citenamefont {Manela}, \citenamefont {Marecek}, \citenamefont {Marques}, \citenamefont {Maslov}, \citenamefont {Mathews}, \citenamefont {Matsuo}, \citenamefont {McClure}, \citenamefont {McGarry}, \citenamefont {McKay}, \citenamefont {McPherson}, \citenamefont {Meesala}, \citenamefont {Metcalfe}, \citenamefont {Mevissen}, \citenamefont {Meyer}, \citenamefont {Mezzacapo}, \citenamefont {Midha}, \citenamefont {Minev}, \citenamefont {Mitchell}, \citenamefont {Moll}, \citenamefont {Montanez}, \citenamefont {Monteiro}, \citenamefont {Mooring}, \citenamefont {Morales}, \citenamefont {Moran}, \citenamefont {Motta}, \citenamefont {MrF}, \citenamefont {Murali}, \citenamefont {M{\"u}ggenburg}, \citenamefont {Nadlinger}, \citenamefont {Nakanishi}, \citenamefont {Nannicini}, \citenamefont {Nation}, \citenamefont {Navarro}, \citenamefont {Naveh}, \citenamefont {Neagle}, \citenamefont {Neuweiler}, \citenamefont {Nicander}, \citenamefont {Niroula}, \citenamefont {Norlen}, \citenamefont {NuoWenLei}, \citenamefont
  {O'Riordan}, \citenamefont {Ogunbayo}, \citenamefont {Ollitrault}, \citenamefont {Otaolea}, \citenamefont {Oud}, \citenamefont {Padilha}, \citenamefont {Paik}, \citenamefont {Pal}, \citenamefont {Pang}, \citenamefont {Pascuzzi}, \citenamefont {Perriello}, \citenamefont {Phan}, \citenamefont {Piro}, \citenamefont {Pistoia}, \citenamefont {Piveteau}, \citenamefont {Pocreau}, \citenamefont {Pozas-Kerstjens}, \citenamefont {Prokop}, \citenamefont {Prutyanov}, \citenamefont {Puzzuoli}, \citenamefont {P{\'e}rez}, \citenamefont {Quintiii}, \citenamefont {Rahman}, \citenamefont {Raja}, \citenamefont {Ramagiri}, \citenamefont {Rao}, \citenamefont {Raymond}, \citenamefont {Redondo}, \citenamefont {Reuter}, \citenamefont {Rice}, \citenamefont {Riedemann}, \citenamefont {Rocca}, \citenamefont {Rodr{\'\i}guez}, \citenamefont {RohithKarur}, \citenamefont {Rossmannek}, \citenamefont {Ryu}, \citenamefont {SAPV}, \citenamefont {SamFerracin}, \citenamefont {Sandberg}, \citenamefont {Sandesara}, \citenamefont {Sapra},
  \citenamefont {Sargsyan}, \citenamefont {Sarkar}, \citenamefont {Sathaye}, \citenamefont {Schmitt}, \citenamefont {Schnabel}, \citenamefont {Schoenfeld}, \citenamefont {Scholten}, \citenamefont {Schoute}, \citenamefont {Schwarm}, \citenamefont {Sertage}, \citenamefont {Setia}, \citenamefont {Shammah}, \citenamefont {Shi}, \citenamefont {Silva}, \citenamefont {Simonetto}, \citenamefont {Singstock}, \citenamefont {Siraichi}, \citenamefont {Sitdikov}, \citenamefont {Sivarajah}, \citenamefont {Sletfjerding}, \citenamefont {Smolin}, \citenamefont {Soeken}, \citenamefont {Sokolov}, \citenamefont {Sokolov}, \citenamefont {SooluThomas}, \citenamefont {Starfish}, \citenamefont {Steenken}, \citenamefont {Stypulkoski}, \citenamefont {Sun}, \citenamefont {Sung}, \citenamefont {Takahashi}, \citenamefont {Takawale}, \citenamefont {Tavernelli}, \citenamefont {Taylor}, \citenamefont {Taylour}, \citenamefont {Thomas}, \citenamefont {Tillet}, \citenamefont {Tod}, \citenamefont {Tomasik}, \citenamefont {de~la Torre},
  \citenamefont {Trabing}, \citenamefont {Treinish}, \citenamefont {TrishaPe}, \citenamefont {Tulsi}, \citenamefont {Turner}, \citenamefont {Vaknin}, \citenamefont {Valcarce}, \citenamefont {Varchon}, \citenamefont {Vazquez}, \citenamefont {Villar}, \citenamefont {Vogt-Lee}, \citenamefont {Vuillot}, \citenamefont {Weaver}, \citenamefont {Weidenfeller}, \citenamefont {Wieczorek}, \citenamefont {Wildstrom}, \citenamefont {Winston}, \citenamefont {Woehr}, \citenamefont {Woerner}, \citenamefont {Woo}, \citenamefont {Wood}, \citenamefont {Wood}, \citenamefont {Wood}, \citenamefont {Wood}, \citenamefont {Wootton}, \citenamefont {Yeralin}, \citenamefont {Yonge-Mallo}, \citenamefont {Young}, \citenamefont {Yu}, \citenamefont {Zachow}, \citenamefont {Zdanski}, \citenamefont {Zhang}, \citenamefont {Zoufal}, \citenamefont {Zoufalc}, \citenamefont {a~kapila}, \citenamefont {a~matsuo}, \citenamefont {bcamorrison}, \citenamefont {brandhsn}, \citenamefont {nick bronn}, \citenamefont {brosand}, \citenamefont {chlorophyll
  zz}, \citenamefont {csseifms}, \citenamefont {dekel.meirom}, \citenamefont {dekelmeirom}, \citenamefont {dekool}, \citenamefont {dime10}, \citenamefont {drholmie}, \citenamefont {dtrenev}, \citenamefont {ehchen}, \citenamefont {elfrocampeador}, \citenamefont {faisaldebouni}, \citenamefont {fanizzamarco}, \citenamefont {gabrieleagl}, \citenamefont {gadial}, \citenamefont {galeinston}, \citenamefont {georgios ts}, \citenamefont {gruu}, \citenamefont {hhorii}, \citenamefont {hykavitha}, \citenamefont {jagunther}, \citenamefont {jliu45}, \citenamefont {jscott2}, \citenamefont {kanejess}, \citenamefont {klinvill}, \citenamefont {krutik2966}, \citenamefont {kurarrr}, \citenamefont {lerongil}, \citenamefont {ma5x}, \citenamefont {merav aharoni}, \citenamefont {michelle4654}, \citenamefont {ordmoj}, \citenamefont {sagar pahwa}, \citenamefont {rmoyard}, \citenamefont {saswati qiskit}, \citenamefont {scottkelso}, \citenamefont {sethmerkel}, \citenamefont {shaashwat}, \citenamefont {sternparky}, \citenamefont
  {strickroman}, \citenamefont {sumitpuri}, \citenamefont {tigerjack}, \citenamefont {toural}, \citenamefont {tsura crisaldo}, \citenamefont {vvilpas}, \citenamefont {welien}, \citenamefont {willhbang}, \citenamefont {yang.luh}, \citenamefont {yotamvakninibm},\ and\ \citenamefont {{\v{C}}epulkovskis}}]{Qiskit}%
  \BibitemOpen
  \bibfield  {author} {\bibinfo {author} {\bibfnamefont {H.}~\bibnamefont {Abraham}}, \bibinfo {author} {\bibnamefont {AduOffei}}, \bibinfo {author} {\bibfnamefont {R.}~\bibnamefont {Agarwal}}, \bibinfo {author} {\bibfnamefont {I.~Y.}\ \bibnamefont {Akhalwaya}}, \bibinfo {author} {\bibfnamefont {G.}~\bibnamefont {Aleksandrowicz}}, \bibinfo {author} {\bibfnamefont {T.}~\bibnamefont {Alexander}}, \bibinfo {author} {\bibfnamefont {M.}~\bibnamefont {Amy}}, \bibinfo {author} {\bibfnamefont {E.}~\bibnamefont {Arbel}}, \bibinfo {author} {\bibnamefont {Arijit02}}, \bibinfo {author} {\bibfnamefont {A.}~\bibnamefont {Asfaw}}, \bibinfo {author} {\bibfnamefont {A.}~\bibnamefont {Avkhadiev}}, \bibinfo {author} {\bibfnamefont {C.}~\bibnamefont {Azaustre}}, \bibinfo {author} {\bibnamefont {AzizNgoueya}}, \bibinfo {author} {\bibfnamefont {A.}~\bibnamefont {Banerjee}}, \bibinfo {author} {\bibfnamefont {A.}~\bibnamefont {Bansal}}, \bibinfo {author} {\bibfnamefont {P.}~\bibnamefont {Barkoutsos}}, \bibinfo {author}
  {\bibfnamefont {A.}~\bibnamefont {Barnawal}}, \bibinfo {author} {\bibfnamefont {G.}~\bibnamefont {Barron}}, \bibinfo {author} {\bibfnamefont {G.~S.}\ \bibnamefont {Barron}}, \bibinfo {author} {\bibfnamefont {L.}~\bibnamefont {Bello}}, \bibinfo {author} {\bibfnamefont {Y.}~\bibnamefont {Ben-Haim}}, \bibinfo {author} {\bibfnamefont {D.}~\bibnamefont {Bevenius}}, \bibinfo {author} {\bibfnamefont {A.}~\bibnamefont {Bhobe}}, \bibinfo {author} {\bibfnamefont {L.~S.}\ \bibnamefont {Bishop}}, \bibinfo {author} {\bibfnamefont {C.}~\bibnamefont {Blank}}, \bibinfo {author} {\bibfnamefont {S.}~\bibnamefont {Bolos}}, \bibinfo {author} {\bibfnamefont {S.}~\bibnamefont {Bosch}}, \bibinfo {author} {\bibnamefont {Brandon}}, \bibinfo {author} {\bibfnamefont {S.}~\bibnamefont {Bravyi}}, \bibinfo {author} {\bibnamefont {Bryce-Fuller}}, \bibinfo {author} {\bibfnamefont {D.}~\bibnamefont {Bucher}}, \bibinfo {author} {\bibfnamefont {A.}~\bibnamefont {Burov}}, \bibinfo {author} {\bibfnamefont {F.}~\bibnamefont {Cabrera}}, \bibinfo
  {author} {\bibfnamefont {P.}~\bibnamefont {Calpin}}, \bibinfo {author} {\bibfnamefont {L.}~\bibnamefont {Capelluto}}, \bibinfo {author} {\bibfnamefont {J.}~\bibnamefont {Carballo}}, \bibinfo {author} {\bibfnamefont {G.}~\bibnamefont {Carrascal}}, \bibinfo {author} {\bibfnamefont {A.}~\bibnamefont {Chen}}, \bibinfo {author} {\bibfnamefont {C.-F.}\ \bibnamefont {Chen}}, \bibinfo {author} {\bibfnamefont {E.}~\bibnamefont {Chen}}, \bibinfo {author} {\bibfnamefont {J.~C.}\ \bibnamefont {Chen}}, \bibinfo {author} {\bibfnamefont {R.}~\bibnamefont {Chen}}, \bibinfo {author} {\bibfnamefont {J.~M.}\ \bibnamefont {Chow}}, \bibinfo {author} {\bibfnamefont {S.}~\bibnamefont {Churchill}}, \bibinfo {author} {\bibfnamefont {C.}~\bibnamefont {Claus}}, \bibinfo {author} {\bibfnamefont {C.}~\bibnamefont {Clauss}}, \bibinfo {author} {\bibfnamefont {R.}~\bibnamefont {Cocking}}, \bibinfo {author} {\bibfnamefont {F.}~\bibnamefont {Correa}}, \bibinfo {author} {\bibfnamefont {A.~J.}\ \bibnamefont {Cross}}, \bibinfo {author}
  {\bibfnamefont {A.~W.}\ \bibnamefont {Cross}}, \bibinfo {author} {\bibfnamefont {S.}~\bibnamefont {Cross}}, \bibinfo {author} {\bibfnamefont {J.}~\bibnamefont {Cruz-Benito}}, \bibinfo {author} {\bibfnamefont {C.}~\bibnamefont {Culver}}, \bibinfo {author} {\bibfnamefont {A.~D.}\ \bibnamefont {C{\'o}rcoles-Gonzales}}, \bibinfo {author} {\bibfnamefont {S.}~\bibnamefont {Dague}}, \bibinfo {author} {\bibfnamefont {T.~E.}\ \bibnamefont {Dandachi}}, \bibinfo {author} {\bibfnamefont {M.}~\bibnamefont {Daniels}}, \bibinfo {author} {\bibfnamefont {M.}~\bibnamefont {Dartiailh}}, \bibinfo {author} {\bibnamefont {DavideFrr}}, \bibinfo {author} {\bibfnamefont {A.~R.}\ \bibnamefont {Davila}}, \bibinfo {author} {\bibfnamefont {A.}~\bibnamefont {Dekusar}}, \bibinfo {author} {\bibfnamefont {D.}~\bibnamefont {Ding}}, \bibinfo {author} {\bibfnamefont {J.}~\bibnamefont {Doi}}, \bibinfo {author} {\bibfnamefont {E.}~\bibnamefont {Drechsler}}, \bibinfo {author} {\bibnamefont {Drew}}, \bibinfo {author} {\bibfnamefont
  {E.}~\bibnamefont {Dumitrescu}}, \bibinfo {author} {\bibfnamefont {K.}~\bibnamefont {Dumon}}, \bibinfo {author} {\bibfnamefont {I.}~\bibnamefont {Duran}}, \bibinfo {author} {\bibfnamefont {K.}~\bibnamefont {EL-Safty}}, \bibinfo {author} {\bibfnamefont {E.}~\bibnamefont {Eastman}}, \bibinfo {author} {\bibfnamefont {G.}~\bibnamefont {Eberle}}, \bibinfo {author} {\bibfnamefont {P.}~\bibnamefont {Eendebak}}, \bibinfo {author} {\bibfnamefont {D.}~\bibnamefont {Egger}}, \bibinfo {author} {\bibfnamefont {M.}~\bibnamefont {Everitt}}, \bibinfo {author} {\bibfnamefont {P.~M.}\ \bibnamefont {Fern{\'a}ndez}}, \bibinfo {author} {\bibfnamefont {A.~H.}\ \bibnamefont {Ferrera}}, \bibinfo {author} {\bibfnamefont {R.}~\bibnamefont {Fouilland}}, \bibinfo {author} {\bibnamefont {FranckChevallier}}, \bibinfo {author} {\bibfnamefont {A.}~\bibnamefont {Frisch}}, \bibinfo {author} {\bibfnamefont {A.}~\bibnamefont {Fuhrer}}, \bibinfo {author} {\bibfnamefont {B.}~\bibnamefont {Fuller}}, \bibinfo {author} {\bibfnamefont
  {M.}~\bibnamefont {GEORGE}}, \bibinfo {author} {\bibfnamefont {J.}~\bibnamefont {Gacon}}, \bibinfo {author} {\bibfnamefont {B.~G.}\ \bibnamefont {Gago}}, \bibinfo {author} {\bibfnamefont {C.}~\bibnamefont {Gambella}}, \bibinfo {author} {\bibfnamefont {J.~M.}\ \bibnamefont {Gambetta}}, \bibinfo {author} {\bibfnamefont {A.}~\bibnamefont {Gammanpila}}, \bibinfo {author} {\bibfnamefont {L.}~\bibnamefont {Garcia}}, \bibinfo {author} {\bibfnamefont {T.}~\bibnamefont {Garg}}, \bibinfo {author} {\bibfnamefont {S.}~\bibnamefont {Garion}}, \bibinfo {author} {\bibfnamefont {A.}~\bibnamefont {Gilliam}}, \bibinfo {author} {\bibfnamefont {A.}~\bibnamefont {Giridharan}}, \bibinfo {author} {\bibfnamefont {J.}~\bibnamefont {Gomez-Mosquera}}, \bibinfo {author} {\bibnamefont {Gonzalo}}, \bibinfo {author} {\bibfnamefont {S.}~\bibnamefont {de~la Puente~Gonz{\'a}lez}}, \bibinfo {author} {\bibfnamefont {J.}~\bibnamefont {Gorzinski}}, \bibinfo {author} {\bibfnamefont {I.}~\bibnamefont {Gould}}, \bibinfo {author} {\bibfnamefont
  {D.}~\bibnamefont {Greenberg}}, \bibinfo {author} {\bibfnamefont {D.}~\bibnamefont {Grinko}}, \bibinfo {author} {\bibfnamefont {W.}~\bibnamefont {Guan}}, \bibinfo {author} {\bibfnamefont {J.~A.}\ \bibnamefont {Gunnels}}, \bibinfo {author} {\bibfnamefont {M.}~\bibnamefont {Haglund}}, \bibinfo {author} {\bibfnamefont {I.}~\bibnamefont {Haide}}, \bibinfo {author} {\bibfnamefont {I.}~\bibnamefont {Hamamura}}, \bibinfo {author} {\bibfnamefont {O.~C.}\ \bibnamefont {Hamido}}, \bibinfo {author} {\bibfnamefont {F.}~\bibnamefont {Harkins}}, \bibinfo {author} {\bibfnamefont {V.}~\bibnamefont {Havlicek}}, \bibinfo {author} {\bibfnamefont {J.}~\bibnamefont {Hellmers}}, \bibinfo {author} {\bibfnamefont {{\L}.}~\bibnamefont {Herok}}, \bibinfo {author} {\bibfnamefont {S.}~\bibnamefont {Hillmich}}, \bibinfo {author} {\bibfnamefont {H.}~\bibnamefont {Horii}}, \bibinfo {author} {\bibfnamefont {C.}~\bibnamefont {Howington}}, \bibinfo {author} {\bibfnamefont {S.}~\bibnamefont {Hu}}, \bibinfo {author} {\bibfnamefont
  {W.}~\bibnamefont {Hu}}, \bibinfo {author} {\bibfnamefont {J.}~\bibnamefont {Huang}}, \bibinfo {author} {\bibfnamefont {R.}~\bibnamefont {Huisman}}, \bibinfo {author} {\bibfnamefont {H.}~\bibnamefont {Imai}}, \bibinfo {author} {\bibfnamefont {T.}~\bibnamefont {Imamichi}}, \bibinfo {author} {\bibfnamefont {K.}~\bibnamefont {Ishizaki}}, \bibinfo {author} {\bibfnamefont {R.}~\bibnamefont {Iten}}, \bibinfo {author} {\bibfnamefont {T.}~\bibnamefont {Itoko}}, \bibinfo {author} {\bibnamefont {JamesSeaward}}, \bibinfo {author} {\bibfnamefont {A.}~\bibnamefont {Javadi}}, \bibinfo {author} {\bibfnamefont {A.}~\bibnamefont {Javadi-Abhari}}, \bibinfo {author} {\bibfnamefont {W.}~\bibnamefont {Javed}}, \bibinfo {author} {\bibnamefont {Jessica}}, \bibinfo {author} {\bibfnamefont {M.}~\bibnamefont {Jivrajani}}, \bibinfo {author} {\bibfnamefont {K.}~\bibnamefont {Johns}}, \bibinfo {author} {\bibfnamefont {S.}~\bibnamefont {Johnstun}}, \bibinfo {author} {\bibnamefont {Jonathan-Shoemaker}}, \bibinfo {author} {\bibfnamefont
  {V.}~\bibnamefont {K}}, \bibinfo {author} {\bibfnamefont {T.}~\bibnamefont {Kachmann}}, \bibinfo {author} {\bibfnamefont {A.}~\bibnamefont {Kale}}, \bibinfo {author} {\bibfnamefont {N.}~\bibnamefont {Kanazawa}}, \bibinfo {author} {\bibnamefont {Kang-Bae}}, \bibinfo {author} {\bibfnamefont {A.}~\bibnamefont {Karazeev}}, \bibinfo {author} {\bibfnamefont {P.}~\bibnamefont {Kassebaum}}, \bibinfo {author} {\bibfnamefont {J.}~\bibnamefont {Kelso}}, \bibinfo {author} {\bibfnamefont {S.}~\bibnamefont {King}}, \bibinfo {author} {\bibnamefont {Knabberjoe}}, \bibinfo {author} {\bibfnamefont {Y.}~\bibnamefont {Kobayashi}}, \bibinfo {author} {\bibfnamefont {A.}~\bibnamefont {Kovyrshin}}, \bibinfo {author} {\bibfnamefont {R.}~\bibnamefont {Krishnakumar}}, \bibinfo {author} {\bibfnamefont {V.}~\bibnamefont {Krishnan}}, \bibinfo {author} {\bibfnamefont {K.}~\bibnamefont {Krsulich}}, \bibinfo {author} {\bibfnamefont {P.}~\bibnamefont {Kumkar}}, \bibinfo {author} {\bibfnamefont {G.}~\bibnamefont {Kus}}, \bibinfo {author}
  {\bibfnamefont {R.}~\bibnamefont {LaRose}}, \bibinfo {author} {\bibfnamefont {E.}~\bibnamefont {Lacal}}, \bibinfo {author} {\bibfnamefont {R.}~\bibnamefont {Lambert}}, \bibinfo {author} {\bibfnamefont {J.}~\bibnamefont {Lapeyre}}, \bibinfo {author} {\bibfnamefont {J.}~\bibnamefont {Latone}}, \bibinfo {author} {\bibfnamefont {S.}~\bibnamefont {Lawrence}}, \bibinfo {author} {\bibfnamefont {C.}~\bibnamefont {Lee}}, \bibinfo {author} {\bibfnamefont {G.}~\bibnamefont {Li}}, \bibinfo {author} {\bibfnamefont {D.}~\bibnamefont {Liu}}, \bibinfo {author} {\bibfnamefont {P.}~\bibnamefont {Liu}}, \bibinfo {author} {\bibfnamefont {Y.}~\bibnamefont {Maeng}}, \bibinfo {author} {\bibfnamefont {K.}~\bibnamefont {Majmudar}}, \bibinfo {author} {\bibfnamefont {A.}~\bibnamefont {Malyshev}}, \bibinfo {author} {\bibfnamefont {J.}~\bibnamefont {Manela}}, \bibinfo {author} {\bibfnamefont {J.}~\bibnamefont {Marecek}}, \bibinfo {author} {\bibfnamefont {M.}~\bibnamefont {Marques}}, \bibinfo {author} {\bibfnamefont {D.}~\bibnamefont
  {Maslov}}, \bibinfo {author} {\bibfnamefont {D.}~\bibnamefont {Mathews}}, \bibinfo {author} {\bibfnamefont {A.}~\bibnamefont {Matsuo}}, \bibinfo {author} {\bibfnamefont {D.~T.}\ \bibnamefont {McClure}}, \bibinfo {author} {\bibfnamefont {C.}~\bibnamefont {McGarry}}, \bibinfo {author} {\bibfnamefont {D.}~\bibnamefont {McKay}}, \bibinfo {author} {\bibfnamefont {D.}~\bibnamefont {McPherson}}, \bibinfo {author} {\bibfnamefont {S.}~\bibnamefont {Meesala}}, \bibinfo {author} {\bibfnamefont {T.}~\bibnamefont {Metcalfe}}, \bibinfo {author} {\bibfnamefont {M.}~\bibnamefont {Mevissen}}, \bibinfo {author} {\bibfnamefont {A.}~\bibnamefont {Meyer}}, \bibinfo {author} {\bibfnamefont {A.}~\bibnamefont {Mezzacapo}}, \bibinfo {author} {\bibfnamefont {R.}~\bibnamefont {Midha}}, \bibinfo {author} {\bibfnamefont {Z.}~\bibnamefont {Minev}}, \bibinfo {author} {\bibfnamefont {A.}~\bibnamefont {Mitchell}}, \bibinfo {author} {\bibfnamefont {N.}~\bibnamefont {Moll}}, \bibinfo {author} {\bibfnamefont {J.}~\bibnamefont {Montanez}},
  \bibinfo {author} {\bibfnamefont {G.}~\bibnamefont {Monteiro}}, \bibinfo {author} {\bibfnamefont {M.~D.}\ \bibnamefont {Mooring}}, \bibinfo {author} {\bibfnamefont {R.}~\bibnamefont {Morales}}, \bibinfo {author} {\bibfnamefont {N.}~\bibnamefont {Moran}}, \bibinfo {author} {\bibfnamefont {M.}~\bibnamefont {Motta}}, \bibinfo {author} {\bibnamefont {MrF}}, \bibinfo {author} {\bibfnamefont {P.}~\bibnamefont {Murali}}, \bibinfo {author} {\bibfnamefont {J.}~\bibnamefont {M{\"u}ggenburg}}, \bibinfo {author} {\bibfnamefont {D.}~\bibnamefont {Nadlinger}}, \bibinfo {author} {\bibfnamefont {K.}~\bibnamefont {Nakanishi}}, \bibinfo {author} {\bibfnamefont {G.}~\bibnamefont {Nannicini}}, \bibinfo {author} {\bibfnamefont {P.}~\bibnamefont {Nation}}, \bibinfo {author} {\bibfnamefont {E.}~\bibnamefont {Navarro}}, \bibinfo {author} {\bibfnamefont {Y.}~\bibnamefont {Naveh}}, \bibinfo {author} {\bibfnamefont {S.~W.}\ \bibnamefont {Neagle}}, \bibinfo {author} {\bibfnamefont {P.}~\bibnamefont {Neuweiler}}, \bibinfo {author}
  {\bibfnamefont {J.}~\bibnamefont {Nicander}}, \bibinfo {author} {\bibfnamefont {P.}~\bibnamefont {Niroula}}, \bibinfo {author} {\bibfnamefont {H.}~\bibnamefont {Norlen}}, \bibinfo {author} {\bibnamefont {NuoWenLei}}, \bibinfo {author} {\bibfnamefont {L.~J.}\ \bibnamefont {O'Riordan}}, \bibinfo {author} {\bibfnamefont {O.}~\bibnamefont {Ogunbayo}}, \bibinfo {author} {\bibfnamefont {P.}~\bibnamefont {Ollitrault}}, \bibinfo {author} {\bibfnamefont {R.}~\bibnamefont {Otaolea}}, \bibinfo {author} {\bibfnamefont {S.}~\bibnamefont {Oud}}, \bibinfo {author} {\bibfnamefont {D.}~\bibnamefont {Padilha}}, \bibinfo {author} {\bibfnamefont {H.}~\bibnamefont {Paik}}, \bibinfo {author} {\bibfnamefont {S.}~\bibnamefont {Pal}}, \bibinfo {author} {\bibfnamefont {Y.}~\bibnamefont {Pang}}, \bibinfo {author} {\bibfnamefont {V.~R.}\ \bibnamefont {Pascuzzi}}, \bibinfo {author} {\bibfnamefont {S.}~\bibnamefont {Perriello}}, \bibinfo {author} {\bibfnamefont {A.}~\bibnamefont {Phan}}, \bibinfo {author} {\bibfnamefont
  {F.}~\bibnamefont {Piro}}, \bibinfo {author} {\bibfnamefont {M.}~\bibnamefont {Pistoia}}, \bibinfo {author} {\bibfnamefont {C.}~\bibnamefont {Piveteau}}, \bibinfo {author} {\bibfnamefont {P.}~\bibnamefont {Pocreau}}, \bibinfo {author} {\bibfnamefont {A.}~\bibnamefont {Pozas-Kerstjens}}, \bibinfo {author} {\bibfnamefont {M.}~\bibnamefont {Prokop}}, \bibinfo {author} {\bibfnamefont {V.}~\bibnamefont {Prutyanov}}, \bibinfo {author} {\bibfnamefont {D.}~\bibnamefont {Puzzuoli}}, \bibinfo {author} {\bibfnamefont {J.}~\bibnamefont {P{\'e}rez}}, \bibinfo {author} {\bibnamefont {Quintiii}}, \bibinfo {author} {\bibfnamefont {R.~I.}\ \bibnamefont {Rahman}}, \bibinfo {author} {\bibfnamefont {A.}~\bibnamefont {Raja}}, \bibinfo {author} {\bibfnamefont {N.}~\bibnamefont {Ramagiri}}, \bibinfo {author} {\bibfnamefont {A.}~\bibnamefont {Rao}}, \bibinfo {author} {\bibfnamefont {R.}~\bibnamefont {Raymond}}, \bibinfo {author} {\bibfnamefont {R.~M.-C.}\ \bibnamefont {Redondo}}, \bibinfo {author} {\bibfnamefont {M.}~\bibnamefont
  {Reuter}}, \bibinfo {author} {\bibfnamefont {J.}~\bibnamefont {Rice}}, \bibinfo {author} {\bibfnamefont {M.}~\bibnamefont {Riedemann}}, \bibinfo {author} {\bibfnamefont {M.~L.}\ \bibnamefont {Rocca}}, \bibinfo {author} {\bibfnamefont {D.~M.}\ \bibnamefont {Rodr{\'\i}guez}}, \bibinfo {author} {\bibnamefont {RohithKarur}}, \bibinfo {author} {\bibfnamefont {M.}~\bibnamefont {Rossmannek}}, \bibinfo {author} {\bibfnamefont {M.}~\bibnamefont {Ryu}}, \bibinfo {author} {\bibfnamefont {T.}~\bibnamefont {SAPV}}, \bibinfo {author} {\bibnamefont {SamFerracin}}, \bibinfo {author} {\bibfnamefont {M.}~\bibnamefont {Sandberg}}, \bibinfo {author} {\bibfnamefont {H.}~\bibnamefont {Sandesara}}, \bibinfo {author} {\bibfnamefont {R.}~\bibnamefont {Sapra}}, \bibinfo {author} {\bibfnamefont {H.}~\bibnamefont {Sargsyan}}, \bibinfo {author} {\bibfnamefont {A.}~\bibnamefont {Sarkar}}, \bibinfo {author} {\bibfnamefont {N.}~\bibnamefont {Sathaye}}, \bibinfo {author} {\bibfnamefont {B.}~\bibnamefont {Schmitt}}, \bibinfo {author}
  {\bibfnamefont {C.}~\bibnamefont {Schnabel}}, \bibinfo {author} {\bibfnamefont {Z.}~\bibnamefont {Schoenfeld}}, \bibinfo {author} {\bibfnamefont {T.~L.}\ \bibnamefont {Scholten}}, \bibinfo {author} {\bibfnamefont {E.}~\bibnamefont {Schoute}}, \bibinfo {author} {\bibfnamefont {J.}~\bibnamefont {Schwarm}}, \bibinfo {author} {\bibfnamefont {I.~F.}\ \bibnamefont {Sertage}}, \bibinfo {author} {\bibfnamefont {K.}~\bibnamefont {Setia}}, \bibinfo {author} {\bibfnamefont {N.}~\bibnamefont {Shammah}}, \bibinfo {author} {\bibfnamefont {Y.}~\bibnamefont {Shi}}, \bibinfo {author} {\bibfnamefont {A.}~\bibnamefont {Silva}}, \bibinfo {author} {\bibfnamefont {A.}~\bibnamefont {Simonetto}}, \bibinfo {author} {\bibfnamefont {N.}~\bibnamefont {Singstock}}, \bibinfo {author} {\bibfnamefont {Y.}~\bibnamefont {Siraichi}}, \bibinfo {author} {\bibfnamefont {I.}~\bibnamefont {Sitdikov}}, \bibinfo {author} {\bibfnamefont {S.}~\bibnamefont {Sivarajah}}, \bibinfo {author} {\bibfnamefont {M.~B.}\ \bibnamefont {Sletfjerding}}, \bibinfo
  {author} {\bibfnamefont {J.~A.}\ \bibnamefont {Smolin}}, \bibinfo {author} {\bibfnamefont {M.}~\bibnamefont {Soeken}}, \bibinfo {author} {\bibfnamefont {I.~O.}\ \bibnamefont {Sokolov}}, \bibinfo {author} {\bibfnamefont {I.}~\bibnamefont {Sokolov}}, \bibinfo {author} {\bibnamefont {SooluThomas}}, \bibinfo {author} {\bibnamefont {Starfish}}, \bibinfo {author} {\bibfnamefont {D.}~\bibnamefont {Steenken}}, \bibinfo {author} {\bibfnamefont {M.}~\bibnamefont {Stypulkoski}}, \bibinfo {author} {\bibfnamefont {S.}~\bibnamefont {Sun}}, \bibinfo {author} {\bibfnamefont {K.~J.}\ \bibnamefont {Sung}}, \bibinfo {author} {\bibfnamefont {H.}~\bibnamefont {Takahashi}}, \bibinfo {author} {\bibfnamefont {T.}~\bibnamefont {Takawale}}, \bibinfo {author} {\bibfnamefont {I.}~\bibnamefont {Tavernelli}}, \bibinfo {author} {\bibfnamefont {C.}~\bibnamefont {Taylor}}, \bibinfo {author} {\bibfnamefont {P.}~\bibnamefont {Taylour}}, \bibinfo {author} {\bibfnamefont {S.}~\bibnamefont {Thomas}}, \bibinfo {author} {\bibfnamefont
  {M.}~\bibnamefont {Tillet}}, \bibinfo {author} {\bibfnamefont {M.}~\bibnamefont {Tod}}, \bibinfo {author} {\bibfnamefont {M.}~\bibnamefont {Tomasik}}, \bibinfo {author} {\bibfnamefont {E.}~\bibnamefont {de~la Torre}}, \bibinfo {author} {\bibfnamefont {K.}~\bibnamefont {Trabing}}, \bibinfo {author} {\bibfnamefont {M.}~\bibnamefont {Treinish}}, \bibinfo {author} {\bibnamefont {TrishaPe}}, \bibinfo {author} {\bibfnamefont {D.}~\bibnamefont {Tulsi}}, \bibinfo {author} {\bibfnamefont {W.}~\bibnamefont {Turner}}, \bibinfo {author} {\bibfnamefont {Y.}~\bibnamefont {Vaknin}}, \bibinfo {author} {\bibfnamefont {C.~R.}\ \bibnamefont {Valcarce}}, \bibinfo {author} {\bibfnamefont {F.}~\bibnamefont {Varchon}}, \bibinfo {author} {\bibfnamefont {A.~C.}\ \bibnamefont {Vazquez}}, \bibinfo {author} {\bibfnamefont {V.}~\bibnamefont {Villar}}, \bibinfo {author} {\bibfnamefont {D.}~\bibnamefont {Vogt-Lee}}, \bibinfo {author} {\bibfnamefont {C.}~\bibnamefont {Vuillot}}, \bibinfo {author} {\bibfnamefont {J.}~\bibnamefont
  {Weaver}}, \bibinfo {author} {\bibfnamefont {J.}~\bibnamefont {Weidenfeller}}, \bibinfo {author} {\bibfnamefont {R.}~\bibnamefont {Wieczorek}}, \bibinfo {author} {\bibfnamefont {J.~A.}\ \bibnamefont {Wildstrom}}, \bibinfo {author} {\bibfnamefont {E.}~\bibnamefont {Winston}}, \bibinfo {author} {\bibfnamefont {J.~J.}\ \bibnamefont {Woehr}}, \bibinfo {author} {\bibfnamefont {S.}~\bibnamefont {Woerner}}, \bibinfo {author} {\bibfnamefont {R.}~\bibnamefont {Woo}}, \bibinfo {author} {\bibfnamefont {C.~J.}\ \bibnamefont {Wood}}, \bibinfo {author} {\bibfnamefont {R.}~\bibnamefont {Wood}}, \bibinfo {author} {\bibfnamefont {S.}~\bibnamefont {Wood}}, \bibinfo {author} {\bibfnamefont {S.}~\bibnamefont {Wood}}, \bibinfo {author} {\bibfnamefont {J.}~\bibnamefont {Wootton}}, \bibinfo {author} {\bibfnamefont {D.}~\bibnamefont {Yeralin}}, \bibinfo {author} {\bibfnamefont {D.}~\bibnamefont {Yonge-Mallo}}, \bibinfo {author} {\bibfnamefont {R.}~\bibnamefont {Young}}, \bibinfo {author} {\bibfnamefont {J.}~\bibnamefont {Yu}},
  \bibinfo {author} {\bibfnamefont {C.}~\bibnamefont {Zachow}}, \bibinfo {author} {\bibfnamefont {L.}~\bibnamefont {Zdanski}}, \bibinfo {author} {\bibfnamefont {H.}~\bibnamefont {Zhang}}, \bibinfo {author} {\bibfnamefont {C.}~\bibnamefont {Zoufal}}, \bibinfo {author} {\bibnamefont {Zoufalc}}, \bibinfo {author} {\bibnamefont {a~kapila}}, \bibinfo {author} {\bibnamefont {a~matsuo}}, \bibinfo {author} {\bibnamefont {bcamorrison}}, \bibinfo {author} {\bibnamefont {brandhsn}}, \bibinfo {author} {\bibnamefont {nick bronn}}, \bibinfo {author} {\bibnamefont {brosand}}, \bibinfo {author} {\bibnamefont {chlorophyll zz}}, \bibinfo {author} {\bibnamefont {csseifms}}, \bibinfo {author} {\bibnamefont {dekel.meirom}}, \bibinfo {author} {\bibnamefont {dekelmeirom}}, \bibinfo {author} {\bibnamefont {dekool}}, \bibinfo {author} {\bibnamefont {dime10}}, \bibinfo {author} {\bibnamefont {drholmie}}, \bibinfo {author} {\bibnamefont {dtrenev}}, \bibinfo {author} {\bibnamefont {ehchen}}, \bibinfo {author} {\bibnamefont
  {elfrocampeador}}, \bibinfo {author} {\bibnamefont {faisaldebouni}}, \bibinfo {author} {\bibnamefont {fanizzamarco}}, \bibinfo {author} {\bibnamefont {gabrieleagl}}, \bibinfo {author} {\bibnamefont {gadial}}, \bibinfo {author} {\bibnamefont {galeinston}}, \bibinfo {author} {\bibnamefont {georgios ts}}, \bibinfo {author} {\bibnamefont {gruu}}, \bibinfo {author} {\bibnamefont {hhorii}}, \bibinfo {author} {\bibnamefont {hykavitha}}, \bibinfo {author} {\bibnamefont {jagunther}}, \bibinfo {author} {\bibnamefont {jliu45}}, \bibinfo {author} {\bibnamefont {jscott2}}, \bibinfo {author} {\bibnamefont {kanejess}}, \bibinfo {author} {\bibnamefont {klinvill}}, \bibinfo {author} {\bibnamefont {krutik2966}}, \bibinfo {author} {\bibnamefont {kurarrr}}, \bibinfo {author} {\bibnamefont {lerongil}}, \bibinfo {author} {\bibnamefont {ma5x}}, \bibinfo {author} {\bibnamefont {merav aharoni}}, \bibinfo {author} {\bibnamefont {michelle4654}}, \bibinfo {author} {\bibnamefont {ordmoj}}, \bibinfo {author} {\bibnamefont {sagar
  pahwa}}, \bibinfo {author} {\bibnamefont {rmoyard}}, \bibinfo {author} {\bibnamefont {saswati qiskit}}, \bibinfo {author} {\bibnamefont {scottkelso}}, \bibinfo {author} {\bibnamefont {sethmerkel}}, \bibinfo {author} {\bibnamefont {shaashwat}}, \bibinfo {author} {\bibnamefont {sternparky}}, \bibinfo {author} {\bibnamefont {strickroman}}, \bibinfo {author} {\bibnamefont {sumitpuri}}, \bibinfo {author} {\bibnamefont {tigerjack}}, \bibinfo {author} {\bibnamefont {toural}}, \bibinfo {author} {\bibnamefont {tsura crisaldo}}, \bibinfo {author} {\bibnamefont {vvilpas}}, \bibinfo {author} {\bibnamefont {welien}}, \bibinfo {author} {\bibnamefont {willhbang}}, \bibinfo {author} {\bibnamefont {yang.luh}}, \bibinfo {author} {\bibnamefont {yotamvakninibm}},\ and\ \bibinfo {author} {\bibfnamefont {M.}~\bibnamefont {{\v{C}}epulkovskis}},\ }\href {https://doi.org/10.5281/zenodo.2562110} {\bibinfo {title} {Qiskit: An open-source framework for quantum computing}} (\bibinfo {year} {2019})\BibitemShut {NoStop}%
\bibitem [{IBM(2023)}]{IBMQExp}%
  \BibitemOpen
  \href@noop {} {\bibinfo {title} {Ibm quantum experience [online]}},\ \bibinfo {howpublished} {Available at \url{https://quantum-computing.ibm.com/}} (\bibinfo {year} {2023})\BibitemShut {NoStop}%
\bibitem [{\citenamefont {Carmichael}(1999)}]{carmichaeltxb}%
  \BibitemOpen
  \bibfield  {author} {\bibinfo {author} {\bibfnamefont {H.~J.}\ \bibnamefont {Carmichael}},\ }\href@noop {} {\emph {\bibinfo {title} {Statistical methods in quantum optics 1: master equations and Fokker-Planck equations}}},\ Vol.~\bibinfo {volume} {1}\ (\bibinfo  {publisher} {Springer Science \& Business Media},\ \bibinfo {year} {1999})\BibitemShut {NoStop}%
\bibitem [{\citenamefont {Agarwal}(2012)}]{Agarwalltxb}%
  \BibitemOpen
  \bibfield  {author} {\bibinfo {author} {\bibfnamefont {G.~S.}\ \bibnamefont {Agarwal}},\ }\href@noop {} {\emph {\bibinfo {title} {Quantum optics}}}\ (\bibinfo  {publisher} {Cambridge University Press},\ \bibinfo {year} {2012})\BibitemShut {NoStop}%
\bibitem [{\citenamefont {Breuer}\ \emph {et~al.}(2002)\citenamefont {Breuer}, \citenamefont {Petruccione} \emph {et~al.}}]{opendynamicstext}%
  \BibitemOpen
  \bibfield  {author} {\bibinfo {author} {\bibfnamefont {H.-P.}\ \bibnamefont {Breuer}}, \bibinfo {author} {\bibfnamefont {F.}~\bibnamefont {Petruccione}}, \emph {et~al.},\ }\href@noop {} {\emph {\bibinfo {title} {The theory of open quantum systems}}}\ (\bibinfo  {publisher} {Oxford University Press on Demand},\ \bibinfo {year} {2002})\BibitemShut {NoStop}%
\bibitem [{\citenamefont {Kraus}\ \emph {et~al.}(1983)\citenamefont {Kraus}, \citenamefont {B{\"o}hm}, \citenamefont {Dollard},\ and\ \citenamefont {Wootters}}]{KrausRepresentationref}%
  \BibitemOpen
  \bibfield  {author} {\bibinfo {author} {\bibfnamefont {K.}~\bibnamefont {Kraus}}, \bibinfo {author} {\bibfnamefont {A.}~\bibnamefont {B{\"o}hm}}, \bibinfo {author} {\bibfnamefont {J.~D.}\ \bibnamefont {Dollard}},\ and\ \bibinfo {author} {\bibfnamefont {W.}~\bibnamefont {Wootters}},\ }\href@noop {} {\bibfield  {journal} {\bibinfo  {journal} {Lecture notes in physics}\ }\textbf {\bibinfo {volume} {190}} (\bibinfo {year} {1983})}\BibitemShut {NoStop}%
\bibitem [{\citenamefont {Grover}(1996)}]{GroverQA}%
  \BibitemOpen
  \bibfield  {author} {\bibinfo {author} {\bibfnamefont {L.~K.}\ \bibnamefont {Grover}},\ }\href@noop {} {\bibinfo {title} {Proceedings of the 28th annual acm symposium on the theory of computing}} (\bibinfo {year} {1996})\BibitemShut {NoStop}%
\bibitem [{\citenamefont {Grover}(1997)}]{Grover1997}%
  \BibitemOpen
  \bibfield  {author} {\bibinfo {author} {\bibfnamefont {L.~K.}\ \bibnamefont {Grover}},\ }\href {https://doi.org/10.1103/PhysRevLett.79.325} {\bibfield  {journal} {\bibinfo  {journal} {Phys. Rev. Lett.}\ }\textbf {\bibinfo {volume} {79}},\ \bibinfo {pages} {325} (\bibinfo {year} {1997})}\BibitemShut {NoStop}%
\bibitem [{\citenamefont {Montanaro}(2016)}]{montanaro2016quantum}%
  \BibitemOpen
  \bibfield  {author} {\bibinfo {author} {\bibfnamefont {A.}~\bibnamefont {Montanaro}},\ }\href@noop {} {\bibfield  {journal} {\bibinfo  {journal} {npj Quantum Information}\ }\textbf {\bibinfo {volume} {2}},\ \bibinfo {pages} {1} (\bibinfo {year} {2016})}\BibitemShut {NoStop}%
\bibitem [{\citenamefont {Jordan}(2011)}]{QAZoo}%
  \BibitemOpen
  \bibfield  {author} {\bibinfo {author} {\bibfnamefont {S.}~\bibnamefont {Jordan}},\ }\href@noop {} {\bibfield  {journal} {\bibinfo  {journal} {Retrieved June}\ }\textbf {\bibinfo {volume} {27}},\ \bibinfo {pages} {2013} (\bibinfo {year} {2011})}\BibitemShut {NoStop}%
\bibitem [{\citenamefont {Minato}\ \emph {et~al.}(2021)\citenamefont {Minato}, \citenamefont {Higa},\ and\ \citenamefont {Nagai}}]{2021ibm}%
  \BibitemOpen
  \bibfield  {author} {\bibinfo {author} {\bibfnamefont {Y.}~\bibnamefont {Minato}}, \bibinfo {author} {\bibfnamefont {K.}~\bibnamefont {Higa}},\ and\ \bibinfo {author} {\bibfnamefont {R.}~\bibnamefont {Nagai}},\ }\href {https://books.google.co.jp/books?id=mOdKzgEACAAJ} {\emph {\bibinfo {title} {IBM Quantum de manabu ryoshi konpyuuta (in Japanese)}}}\ (\bibinfo  {publisher} {Shuwa System},\ \bibinfo {year} {2021})\BibitemShut {NoStop}%
\bibitem [{\citenamefont {Sarovar}\ \emph {et~al.}(2020)\citenamefont {Sarovar}, \citenamefont {Proctor}, \citenamefont {Rudinger}, \citenamefont {Young}, \citenamefont {Nielsen},\ and\ \citenamefont {Blume-Kohout}}]{sarovar2020detecting}%
  \BibitemOpen
  \bibfield  {author} {\bibinfo {author} {\bibfnamefont {M.}~\bibnamefont {Sarovar}}, \bibinfo {author} {\bibfnamefont {T.}~\bibnamefont {Proctor}}, \bibinfo {author} {\bibfnamefont {K.}~\bibnamefont {Rudinger}}, \bibinfo {author} {\bibfnamefont {K.}~\bibnamefont {Young}}, \bibinfo {author} {\bibfnamefont {E.}~\bibnamefont {Nielsen}},\ and\ \bibinfo {author} {\bibfnamefont {R.}~\bibnamefont {Blume-Kohout}},\ }\href {https://doi.org/https://doi.org/10.22331/q-2020-09-11-321} {\bibfield  {journal} {\bibinfo  {journal} {Quantum}\ }\textbf {\bibinfo {volume} {4}},\ \bibinfo {pages} {321} (\bibinfo {year} {2020})}\BibitemShut {NoStop}%
\bibitem [{\citenamefont {Wright}\ \emph {et~al.}(2019)\citenamefont {Wright}, \citenamefont {Beck}, \citenamefont {Debnath}, \citenamefont {Amini}, \citenamefont {Nam}, \citenamefont {Grzesiak}, \citenamefont {Chen}, \citenamefont {Pisenti}, \citenamefont {Chmielewski}, \citenamefont {Collins} \emph {et~al.}}]{wright2019benchmarking}%
  \BibitemOpen
  \bibfield  {author} {\bibinfo {author} {\bibfnamefont {K.}~\bibnamefont {Wright}}, \bibinfo {author} {\bibfnamefont {K.~M.}\ \bibnamefont {Beck}}, \bibinfo {author} {\bibfnamefont {S.}~\bibnamefont {Debnath}}, \bibinfo {author} {\bibfnamefont {J.}~\bibnamefont {Amini}}, \bibinfo {author} {\bibfnamefont {Y.}~\bibnamefont {Nam}}, \bibinfo {author} {\bibfnamefont {N.}~\bibnamefont {Grzesiak}}, \bibinfo {author} {\bibfnamefont {J.-S.}\ \bibnamefont {Chen}}, \bibinfo {author} {\bibfnamefont {N.}~\bibnamefont {Pisenti}}, \bibinfo {author} {\bibfnamefont {M.}~\bibnamefont {Chmielewski}}, \bibinfo {author} {\bibfnamefont {C.}~\bibnamefont {Collins}}, \emph {et~al.},\ }\href@noop {} {\bibfield  {journal} {\bibinfo  {journal} {Nature communications}\ }\textbf {\bibinfo {volume} {10}},\ \bibinfo {pages} {5464} (\bibinfo {year} {2019})}\BibitemShut {NoStop}%
\bibitem [{\citenamefont {Developers}(2022)}]{cirq_developers_2022_6599601}%
  \BibitemOpen
  \bibfield  {author} {\bibinfo {author} {\bibfnamefont {C.}~\bibnamefont {Developers}},\ }\href {https://doi.org/10.5281/zenodo.6599601} {\bibinfo {title} {Cirq}} (\bibinfo {year} {2022}),\ \bibinfo {note} {{See full list of authors on Github: https://github .com/quantumlib/Cirq/graphs/contributors}}\BibitemShut {NoStop}%
\bibitem [{\citenamefont {Dicke}(1954)}]{superradiance1}%
  \BibitemOpen
  \bibfield  {author} {\bibinfo {author} {\bibfnamefont {R.~H.}\ \bibnamefont {Dicke}},\ }\href@noop {} {\bibfield  {journal} {\bibinfo  {journal} {Physical review}\ }\textbf {\bibinfo {volume} {93}},\ \bibinfo {pages} {99} (\bibinfo {year} {1954})}\BibitemShut {NoStop}%
\bibitem [{\citenamefont {Gross}\ and\ \citenamefont {Haroche}(1982)}]{GH82}%
  \BibitemOpen
  \bibfield  {author} {\bibinfo {author} {\bibfnamefont {M.}~\bibnamefont {Gross}}\ and\ \bibinfo {author} {\bibfnamefont {S.}~\bibnamefont {Haroche}},\ }\href@noop {} {\bibfield  {journal} {\bibinfo  {journal} {Physics reports}\ }\textbf {\bibinfo {volume} {93}},\ \bibinfo {pages} {301} (\bibinfo {year} {1982})}\BibitemShut {NoStop}%
\bibitem [{\citenamefont {Duan}\ and\ \citenamefont {Guo}(1998)}]{DuancollectivedecohePRA1998}%
  \BibitemOpen
  \bibfield  {author} {\bibinfo {author} {\bibfnamefont {L.-M.}\ \bibnamefont {Duan}}\ and\ \bibinfo {author} {\bibfnamefont {G.-C.}\ \bibnamefont {Guo}},\ }\href {https://doi.org/10.1103/PhysRevA.58.3491} {\bibfield  {journal} {\bibinfo  {journal} {Phys. Rev. A}\ }\textbf {\bibinfo {volume} {58}},\ \bibinfo {pages} {3491} (\bibinfo {year} {1998})}\BibitemShut {NoStop}%
\bibitem [{\citenamefont {Fortunato}\ \emph {et~al.}(2002)\citenamefont {Fortunato}, \citenamefont {Viola}, \citenamefont {Hodges}, \citenamefont {Teklemariam},\ and\ \citenamefont {Cory}}]{ViolagroupNJP2002}%
  \BibitemOpen
  \bibfield  {author} {\bibinfo {author} {\bibfnamefont {E.~M.}\ \bibnamefont {Fortunato}}, \bibinfo {author} {\bibfnamefont {L.}~\bibnamefont {Viola}}, \bibinfo {author} {\bibfnamefont {J.}~\bibnamefont {Hodges}}, \bibinfo {author} {\bibfnamefont {G.}~\bibnamefont {Teklemariam}},\ and\ \bibinfo {author} {\bibfnamefont {D.~G.}\ \bibnamefont {Cory}},\ }\href@noop {} {\bibfield  {journal} {\bibinfo  {journal} {New Journal of Physics}\ }\textbf {\bibinfo {volume} {4}},\ \bibinfo {pages} {5} (\bibinfo {year} {2002})}\BibitemShut {NoStop}%
\bibitem [{\citenamefont {Macchiavello}\ and\ \citenamefont {Palma}(2002)}]{PhysRevA.65.050301}%
  \BibitemOpen
  \bibfield  {author} {\bibinfo {author} {\bibfnamefont {C.}~\bibnamefont {Macchiavello}}\ and\ \bibinfo {author} {\bibfnamefont {G.~M.}\ \bibnamefont {Palma}},\ }\href {https://doi.org/10.1103/PhysRevA.65.050301} {\bibfield  {journal} {\bibinfo  {journal} {Phys. Rev. A}\ }\textbf {\bibinfo {volume} {65}},\ \bibinfo {pages} {050301} (\bibinfo {year} {2002})}\BibitemShut {NoStop}%
\bibitem [{\citenamefont {Yeo}\ and\ \citenamefont {Skeen}(2003)}]{TCADPRA2003}%
  \BibitemOpen
  \bibfield  {author} {\bibinfo {author} {\bibfnamefont {Y.}~\bibnamefont {Yeo}}\ and\ \bibinfo {author} {\bibfnamefont {A.}~\bibnamefont {Skeen}},\ }\href {https://doi.org/10.1103/PhysRevA.67.064301} {\bibfield  {journal} {\bibinfo  {journal} {Phys. Rev. A}\ }\textbf {\bibinfo {volume} {67}},\ \bibinfo {pages} {064301} (\bibinfo {year} {2003})}\BibitemShut {NoStop}%
\bibitem [{\citenamefont {Terhal}\ and\ \citenamefont {Burkard}(2005)}]{Cnoisepra2005}%
  \BibitemOpen
  \bibfield  {author} {\bibinfo {author} {\bibfnamefont {B.~M.}\ \bibnamefont {Terhal}}\ and\ \bibinfo {author} {\bibfnamefont {G.}~\bibnamefont {Burkard}},\ }\href {https://doi.org/10.1103/PhysRevA.71.012336} {\bibfield  {journal} {\bibinfo  {journal} {Phys. Rev. A}\ }\textbf {\bibinfo {volume} {71}},\ \bibinfo {pages} {012336} (\bibinfo {year} {2005})}\BibitemShut {NoStop}%
\bibitem [{\citenamefont {Novais}\ and\ \citenamefont {Baranger}(2006)}]{NoviasprlCnoise}%
  \BibitemOpen
  \bibfield  {author} {\bibinfo {author} {\bibfnamefont {E.}~\bibnamefont {Novais}}\ and\ \bibinfo {author} {\bibfnamefont {H.~U.}\ \bibnamefont {Baranger}},\ }\href {https://doi.org/10.1103/PhysRevLett.97.040501} {\bibfield  {journal} {\bibinfo  {journal} {Phys. Rev. Lett.}\ }\textbf {\bibinfo {volume} {97}},\ \bibinfo {pages} {040501} (\bibinfo {year} {2006})}\BibitemShut {NoStop}%
\bibitem [{\citenamefont {Aharonov}\ \emph {et~al.}(2006)\citenamefont {Aharonov}, \citenamefont {Kitaev},\ and\ \citenamefont {Preskill}}]{KitaevgroupCnoiseprl2006}%
  \BibitemOpen
  \bibfield  {author} {\bibinfo {author} {\bibfnamefont {D.}~\bibnamefont {Aharonov}}, \bibinfo {author} {\bibfnamefont {A.}~\bibnamefont {Kitaev}},\ and\ \bibinfo {author} {\bibfnamefont {J.}~\bibnamefont {Preskill}},\ }\href {https://doi.org/10.1103/PhysRevLett.96.050504} {\bibfield  {journal} {\bibinfo  {journal} {Phys. Rev. Lett.}\ }\textbf {\bibinfo {volume} {96}},\ \bibinfo {pages} {050504} (\bibinfo {year} {2006})}\BibitemShut {NoStop}%
\bibitem [{\citenamefont {Ban}\ \emph {et~al.}(2005)\citenamefont {Ban}, \citenamefont {Kitajima},\ and\ \citenamefont {Shibata}}]{ban2005decoherence}%
  \BibitemOpen
  \bibfield  {author} {\bibinfo {author} {\bibfnamefont {M.}~\bibnamefont {Ban}}, \bibinfo {author} {\bibfnamefont {S.}~\bibnamefont {Kitajima}},\ and\ \bibinfo {author} {\bibfnamefont {F.}~\bibnamefont {Shibata}},\ }\href@noop {} {\bibfield  {journal} {\bibinfo  {journal} {Journal of Physics A: Mathematical and General}\ }\textbf {\bibinfo {volume} {38}},\ \bibinfo {pages} {7161} (\bibinfo {year} {2005})}\BibitemShut {NoStop}%
\bibitem [{\citenamefont {Yu}\ and\ \citenamefont {Eberly}(2010)}]{yu2010entanglement}%
  \BibitemOpen
  \bibfield  {author} {\bibinfo {author} {\bibfnamefont {T.}~\bibnamefont {Yu}}\ and\ \bibinfo {author} {\bibfnamefont {J.}~\bibnamefont {Eberly}},\ }\href@noop {} {\bibfield  {journal} {\bibinfo  {journal} {Optics Communications}\ }\textbf {\bibinfo {volume} {283}},\ \bibinfo {pages} {676} (\bibinfo {year} {2010})}\BibitemShut {NoStop}%
\end{thebibliography}%


\begin{thebibliography}{4}%
\makeatletter
\providecommand \@ifxundefined [1]{%
 \@ifx{#1\undefined}
}%
\providecommand \@ifnum [1]{%
 \ifnum #1\expandafter \@firstoftwo
 \else \expandafter \@secondoftwo
 \fi
}%
\providecommand \@ifx [1]{%
 \ifx #1\expandafter \@firstoftwo
 \else \expandafter \@secondoftwo
 \fi
}%
\providecommand \natexlab [1]{#1}%
\providecommand \enquote  [1]{``#1''}%
\providecommand \bibnamefont  [1]{#1}%
\providecommand \bibfnamefont [1]{#1}%
\providecommand \citenamefont [1]{#1}%
\providecommand \href@noop [0]{\@secondoftwo}%
\providecommand \href [0]{\begingroup \@sanitize@url \@href}%
\providecommand \@href[1]{\@@startlink{#1}\@@href}%
\providecommand \@@href[1]{\endgroup#1\@@endlink}%
\providecommand \@sanitize@url [0]{\catcode `\\12\catcode `\$12\catcode `\&12\catcode `\#12\catcode `\^12\catcode `\_12\catcode `\%12\relax}%
\providecommand \@@startlink[1]{}%
\providecommand \@@endlink[0]{}%
\providecommand \url  [0]{\begingroup\@sanitize@url \@url }%
\providecommand \@url [1]{\endgroup\@href {#1}{\urlprefix }}%
\providecommand \urlprefix  [0]{URL }%
\providecommand \Eprint [0]{\href }%
\providecommand \doibase [0]{https://doi.org/}%
\providecommand \selectlanguage [0]{\@gobble}%
\providecommand \bibinfo  [0]{\@secondoftwo}%
\providecommand \bibfield  [0]{\@secondoftwo}%
\providecommand \translation [1]{[#1]}%
\providecommand \BibitemOpen [0]{}%
\providecommand \bibitemStop [0]{}%
\providecommand \bibitemNoStop [0]{.\EOS\space}%
\providecommand \EOS [0]{\spacefactor3000\relax}%
\providecommand \BibitemShut  [1]{\csname bibitem#1\endcsname}%
\let\auto@bib@innerbib\@empty
\bibitem [{IBM(2023)}]{IBMQExp}%
  \BibitemOpen
  \href@noop {} {\bibinfo {title} {Ibm quantum experience [online]}},\ \bibinfo {howpublished} {Available at \url{https://quantum-computing.ibm.com/}} (\bibinfo {year} {2023})\BibitemShut {NoStop}%
\bibitem [{\citenamefont {Rist{\`e}}\ \emph {et~al.}(2013)\citenamefont {Rist{\`e}}, \citenamefont {Bultink}, \citenamefont {Tiggelman}, \citenamefont {Schouten}, \citenamefont {Lehnert},\ and\ \citenamefont {DiCarlo}}]{riste2013millisecond}%
  \BibitemOpen
  \bibfield  {author} {\bibinfo {author} {\bibfnamefont {D.}~\bibnamefont {Rist{\`e}}}, \bibinfo {author} {\bibfnamefont {C.}~\bibnamefont {Bultink}}, \bibinfo {author} {\bibfnamefont {M.~J.}\ \bibnamefont {Tiggelman}}, \bibinfo {author} {\bibfnamefont {R.~N.}\ \bibnamefont {Schouten}}, \bibinfo {author} {\bibfnamefont {K.~W.}\ \bibnamefont {Lehnert}},\ and\ \bibinfo {author} {\bibfnamefont {L.}~\bibnamefont {DiCarlo}},\ }\href@noop {} {\bibfield  {journal} {\bibinfo  {journal} {Nature communications}\ }\textbf {\bibinfo {volume} {4}},\ \bibinfo {pages} {1913} (\bibinfo {year} {2013})}\BibitemShut {NoStop}%
\bibitem [{\citenamefont {P{\'e}rez}\ \emph {et~al.}(2022)\citenamefont {P{\'e}rez}, \citenamefont {Bonitati}, \citenamefont {Lee}, \citenamefont {Quaglioni},\ and\ \citenamefont {Wendt}}]{perez2022quantum}%
  \BibitemOpen
  \bibfield  {author} {\bibinfo {author} {\bibfnamefont {E.~A.~C.}\ \bibnamefont {P{\'e}rez}}, \bibinfo {author} {\bibfnamefont {J.}~\bibnamefont {Bonitati}}, \bibinfo {author} {\bibfnamefont {D.}~\bibnamefont {Lee}}, \bibinfo {author} {\bibfnamefont {S.}~\bibnamefont {Quaglioni}},\ and\ \bibinfo {author} {\bibfnamefont {K.~A.}\ \bibnamefont {Wendt}},\ }\href@noop {} {\bibfield  {journal} {\bibinfo  {journal} {Physical Review A}\ }\textbf {\bibinfo {volume} {105}},\ \bibinfo {pages} {032403} (\bibinfo {year} {2022})}\BibitemShut {NoStop}%
\bibitem [{\citenamefont {Volya}\ and\ \citenamefont {Mishra}(2023)}]{volya2023state}%
  \BibitemOpen
  \bibfield  {author} {\bibinfo {author} {\bibfnamefont {D.}~\bibnamefont {Volya}}\ and\ \bibinfo {author} {\bibfnamefont {P.}~\bibnamefont {Mishra}},\ }\href@noop {} {\bibfield  {journal} {\bibinfo  {journal} {arXiv preprint arXiv:2302.13518}\ } (\bibinfo {year} {2023})}\BibitemShut {NoStop}%
\end{thebibliography}%

\end{document}


\title{Supplementary Material: Quantum Error Mitigation via Quantum-Noise-Effect Circuit Groups } 
\author{Yusuke~Hama$^\ast$}  
\affiliation{Quemix Inc., 2-11-2 Nihombashi, Chuo-ku, Tokyo 103-0027, Japan}  
\author{Hirofumi~Nishi}  
\affiliation{
Laboratory for Materials and Structures, Institute of Innovative Research, Tokyo Institute of Technology, Yokohama 226-8503, Japan}
\affiliation{Quemix Inc., 2-11-2 Nihombashi, Chuo-ku, Tokyo 103-0027, Japan} 
\maketitle

This supplementary material presents our extended  QEM formalisms, the optimized variational parameters of QAOA, and the quantum circuit for the operation $O^\text{QAA}$ given in Eq. (14) in the main text.

\section{Extended QEM Formalisms}\label{appendix1}   
In this section, we discuss extensions of our QEM formalism to various types of quantum noise effects;  generalized amplitude damping (GAD), phase damping (PD), and their composite channels.   
Further, we discuss extensions to cases of higher-order quantum noise effects.
\subsection{QEM for Other Quantum Noises}\label{extdQEMSsub1}
By taking exactly the same approach for deriving the QEM scheme for the AD effects we can straightforwardly construct our QEM schemes for other quantum noises including 
GAD, PD, their composite channels, and stochastic Pauli noises. 
First let us show the case of GAD effects.   The quantum master equation for describing the GAD process is given by
  \begin{widetext} \begin{align}
\frac { \partial  \rho (t)}{\partial t }  =  \gamma\mathcal{L} _{ \text{GAD}} [\rho (t)] =  \gamma (\bar{n} + 1) \sum_{j=0}^{N_q-1} \left[  \tilde{\sigma}^-_j     \rho (t)   \tilde{\sigma}^+_j 
 - \frac{1}{2} \big{\{}   \tilde{\sigma}^+_j  \tilde{\sigma}^-_j ,   \rho (t)   \big{\}} \right] 
  +  \gamma \bar{n}  \sum_{j=0}^{N_q-1} \left[  \tilde{\sigma}^+_j     \rho (t)   \tilde{\sigma}^-_j 
 - \frac{1}{2} \big{\{}   \tilde{\sigma}^-_j  \tilde{\sigma}^+_j ,   \rho (t)   \big{\}} \right] ,
 \label{GADQME}
\end{align}  \end{widetext}
where $\mathcal{L} _{ \text{GAD}}$ is the Lindblad superoperator of the GAD process and $ \bar{n}  =   (e^{ \beta \epsilon} - 1)^{-1}  $  is the Bose-Einstein distribution function for the energy $\epsilon$. The quantity $\beta$ is the inverse temperature.   
Note that by taking the zero-temperature limit $(\beta \to \infty)$ in Eq. \eqref{GADQME}, we obtain the quantum master equation (1) in the main text.
As we did in the case of the AD effect, let us estimate the GAD effect in the first order in $\tau$ and write it by $\Delta^{\text{GAD}}_1 \rho_{d\cdots1}$ like $\Delta^{\text{AD}}_1 \rho_{d\cdots1}$ in Eq. (2) in the main text.
From the quantum master equation  \eqref{GADQME}, the GAD effect $\Delta^{\text{GAD}}_1 \rho_{d\cdots1}$ is evaluated as
\begin{align}
\Delta^{\text{GAD}}_1 \rho_{d\cdots1} & = \sum_{k=1}^{d} \Delta^{\text{GAD}}_{1,k} \rho_{d\cdots1}, \notag\\
 \Delta^{\text{GAD}}_{1,k} \rho_{d\cdots1} & =  \left(\prod_{l=k+1}^d U_l  \right) \cdot \tilde{\rho}^{\text{GAD}}_{k\cdots1}      \cdot \left(\prod_{l=k+1}^d U_l \right)^\dagger , 
 \label{GADdeviationDM1}
\end{align}
where
\begin{align}
\tilde{\rho}^{\text{GAD}}_{k\cdots1} &  = \mathcal{L} _{ \text{GAD}} [ \rho_{k\cdots1} ]  \notag\\
& = \sum_{j=0}^{N_q-1}  (\bar{n} + 1) \Big{[}  \tilde{\sigma}^-_j    \rho_{k\cdots1}   \tilde{\sigma}^+_j 
 - \frac{1}{2} \big{\{} P^1_{j}  ,  \rho_{k\cdots1}   \big{\}} \Big{]} \notag\\
 & + \bar{n} \Big{[}   \tilde{\sigma}^+_j    \rho_{k\cdots1}   \tilde{\sigma}^-_j 
 - \frac{1}{2} \big{\{} P^0_{j}  ,  \rho_{k\cdots1}   \big{\}}  \Big{]} \notag\\
& =  (\bar{n} + 1) \tilde{\rho}^{\text{GAD,EM}}_{k\cdots1} + \bar{n}  \tilde{\rho}^{\text{GAD,AB}}_{k\cdots1} .
   \label{GADdeviationDM2}
\end{align}
 Physically, the quantity $\tilde{\rho}^{\text{GAD,EM}}_{k\cdots1} $ in the above equation describes emission process of  the register bits whereas $\tilde{\rho}^{\text{GAD,AB}}_{k\cdots1} $ represents absorption process. Like Eq. (9) in the main text, we rewrite these two quantities in terms of the identity operator, $Z$ gate, and the non-unitary operators $\big{\{} P^0, P^1, \tilde{\sigma}^\pm       \big{\}}$.
As a result, we have
\begin{align}
\tilde{\rho}^{\text{GAD,EM}}_{k\cdots1} & = \frac{-\rho_{k\cdots1} + Z_j  \rho_{k\cdots1}  Z_j + \tilde{\sigma}^-_j  \rho_{k\cdots1}  \tilde{\sigma}^+_j  - P^1_j  \rho_{k\cdots1}  P^1_j}{4}, \notag\\
\tilde{\rho}^{\text{GAD,AB}}_{k\cdots1}  & = \frac{-\rho_{k\cdots1} + Z_j  \rho_{k\cdots1}  Z_j + \tilde{\sigma}^+_j  \rho_{k\cdots1}  \tilde{\sigma}^-_j  - P^0_j  \rho_{k\cdots1}  P^0_j}{4}.
 \label{GADdeviationDM3}
\end{align}
Note that in the zero-temperature limit $(\beta \to \infty)$ we have $ \bar{n}  \to 0$ and $ \tilde{\rho}^{\text{GAD,EM}}_{k\cdots1} \to  \tilde{\rho}^{\text{AD}}_{k\cdots1}$.
By using Eq. (6) in the main text with setting $\vartheta=\pi$ , the operation of $\tilde{\sigma}^+_j $ can be generated by the AD-effect circuit B with post-selecting the measurement outcome of the ancilla bit to be $|1\rangle$  
while the operation of $P^0_j$ is created by the AD-effect circuit A with post-selecting the output state $|0\rangle_{Q_\text{a}}$.
Thus, as in the case of the AD effect, we can perform QEM for GAD by four types of quantum circuits mentioned previously, the original quantum circuit given by $U^{ \text{QC}}$, 
the quantum circuits with the additional $Z$-gate operations, and the AD-effect circuits A and B. 
Therefore, the quantum-noise-effect circuit group for GAD effect is equivalent to that for AD effect and the maximum number of the quantum circuits which we need to perform QEM for GAD effect is $3dN_q+1$. 
The QEM formula for the GAD effect is given by
\begin{align} 
 \langle \hat{O}  \rangle^{\text{QEM}}_{\rho  _{d\cdots1}} 
 =   \langle \hat{O}  \rangle_{\rho _{d\cdots1}} + \tau  \Big{(}   \langle \hat{O}  \rangle _{ \delta^{\text{GAD}}_1 ( \rho  _{d\cdots1} )} - \langle \hat{O}  \rangle_{ \Delta^{\text{GAD}}_1 \rho_{d\cdots1} }   \Big{)}, 
\label{QEMformula2} 
\end{align} 
where the symbol $ \delta^{\text{GAD}}_1$ describes the GAD effect on a real device in the first order in $\tau$ and
\begin{align} 
 \langle \hat{O}  \rangle_{ \Delta^{\text{GAD}}_1 \rho_{d\cdots1} }  = (\bar{n} + 1)   \langle \hat{O}  \rangle_{ \Delta^{\text{GAD,EM}}_1 \rho_{d\cdots1} } +
 \bar{n}    \langle \hat{O}  \rangle_{ \Delta^{\text{GAD,AB}}_1 \rho_{d\cdots1} } .
\label{QEMformula3}
\end{align} 
Our QEM scheme can also be extended to the case when both GAD and phase damping (PD) occur.
The quantum master equation which describes such a process is given by
 \begin{align}
\frac { \partial  \rho (t)}{\partial t }  & =  \gamma\mathcal{L} _{ \text{GAD}} [\rho (t)] +  \gamma _{\text{PD}} \mathcal{L} _{ \text{PD}} [\rho (t)]  \notag\\
& =  \gamma (\bar{n} + 1) \sum_{j=0}^{N_q-1} \left[  \tilde{\sigma}^-_j     \rho (t)   \tilde{\sigma}^+_j 
 - \frac{1}{2} \big{\{}   \tilde{\sigma}^+_j  \tilde{\sigma}^-_j ,   \rho (t)   \big{\}} \right] \notag\\
 & +  \gamma \bar{n}  \sum_{j=0}^{N_q-1} \left[  \tilde{\sigma}^+_j     \rho (t)   \tilde{\sigma}^-_j 
 - \frac{1}{2} \big{\{}   \tilde{\sigma}^-_j  \tilde{\sigma}^+_j ,   \rho (t)   \big{\}} \right] \notag\\
 & +  \gamma _{\text{PD}} \sum_{j=0}^{N_q-1} \left[  Z_j     \rho (t)  Z_j 
 -  \rho (t)   \right], 
 \label{GADPDQME}
\end{align} 
where  $\mathcal{L} _{ \text{PD}}$ is the Lindblad superoperator of PD process and $\gamma _{\text{PD}}$ is the strength of PD. We introduce the dimensionless quantity defined by $\tau _{\text{PD}} = \gamma  _{\text{PD}}\cdot \Delta t$ and assume to be in the same order as $\tau$. 
Like Eq. (2) in the main text, we represent the solution of quantum master equation \eqref{GADPDQME} as the perturbative series of $\tau $ and $\tau _{\text{PD}} $ given by
\begin{align}
 \rho (T) &= \rho_{d\cdots1} + \tau \Delta^{\text{GAD}}_1 \rho_{d\cdots1}  + \tau _{\text{PD}} \Delta^{\text{PD}}_1  \rho_{d\cdots1} \notag\\
 & + \mathcal{O}(\tau^2, \tau _{\text{PD}}^2,  \tau \tau _{\text{PD}}),
 \label{GADPDDM}
\end{align}
 where   $\Delta^{\text{PD}}_1  \rho_{d\cdots1} $ is 
 \begin{align}
\Delta^{\text{PD}}_1 \rho_{d\cdots1} & = \sum_{k=1}^{d} \Delta^{\text{PD}}_{1,k} \rho_{d\cdots1}, \notag\\
 \Delta^{\text{PD}}_{1,k} \rho_{d\cdots1} & =  \left(\prod_{l=k+1}^d U_l  \right)  \tilde{\rho}^{\text{PD}}_{k\cdots1}     \left(\prod_{l=k+1}^d U_l \right)^\dagger , \notag\\
 \tilde{\rho}^{\text{PD}}_{k\cdots1} & =    \mathcal{L} _{ \text{PD}} [ \rho_{k\cdots1} ]    =\sum_{j=0}^{N_q-1}  Z_j  \rho_{k\cdots1}  Z_j -\rho_{k\cdots1}  .
 \label{PDdeviationDM1}
\end{align} 
As we see in the above equation, to evaluate $\Delta^{\text{PD}}_1 \rho_{d\cdots1}$ the only additional operation which we need is the $Z$-gate operation.  
Therefore, the quantum-noise-effect circuit group for the composite channel of the GAD and PD effects is equivalent to that of GAD (AD) effect. 
 Let us denote the composition of the GAD and PD effects which occur on a real device by the symbol $\delta^{\text{GAD$\cdot$\text{PD}}}$.     
 In the first-order approximation with respect to $\tau$ and $ \tau _{\text{PD}} $ we can express $\delta^{\text{GAD$\cdot$\text{PD}}} ( \rho  _{d\cdots1} )$  as    
$\delta^{\text{GAD$\cdot$\text{PD}}} ( \rho  _{d\cdots1} ) = 
\tau  \delta^{\text{GAD}}_1 ( \rho  _{d\cdots1} ) +  \tau _{\text{PD}}  \delta^{\text{PD}}_1 ( \rho  _{d\cdots1} )$.
Our QEM formula is given by
  \begin{widetext}  \begin{align} 
 \langle \hat{O}  \rangle^{\text{QEM}}_{\rho  _{d\cdots1}} & \equiv 
\langle \hat{O}  \rangle_{\rho^{\text{real}}  _{d\cdots1}} - \tau   \langle \hat{O}  \rangle_{ (\Delta^{\text{GAD}}_1 \rho_{d\cdots1})^{\text{real}} }
- \tau _{\text{PD}}   \langle \hat{O}  \rangle_{ (\Delta^{\text{PD}}_1 \rho_{d\cdots1})^{\text{real}} }
 \notag\\
& =   \langle \hat{O}  \rangle_{\rho _{d\cdots1}} 
+ \tau  \Big{(}   \langle \hat{O}  \rangle _{ \delta^{\text{GAD}}_1 ( \rho  _{d\cdots1} )} - \langle \hat{O}  \rangle_{ \Delta^{\text{GAD}}_1 \rho_{d\cdots1} }   \Big{)} 
+ \tau _{\text{PD}}  \Big{(}   \langle \hat{O}  \rangle _{ \delta^{\text{PD}}_1 ( \rho  _{d\cdots1} )} - \langle \hat{O}  \rangle_{ \Delta^{\text{PD}}_1 \rho_{d\cdots1} }   \Big{)} +  \mathcal{O}(\tau^2, \tau _{\text{PD}}^2,  \tau \tau _{\text{PD}}). \label{QEMformula4}
\end{align} \end{widetext}
    We end this section by noting that our QEM scheme can be extended to the mitigation of stochastic Pauli noise effects (bit flip, phase flip, bit-phase flip, depolarizing channel):
     Here we do not show detailed analysis for them.
  Like the PD effect, the stochastic Pauli noise effects are represented by the identity operator and $X,Y,$ and $Z$ gates. 
  For these cases, to conduct QEM we just need to prepare quantum circuits composed with additional $X,Y,$ and $Z$ gate operations like $ \sum_{j=0}^{N_q-1}  Z_j  \rho_{k\cdots1}  Z_j$ in Eq. (9) in the main text
  and the ancilla bits are not needed. 
\subsection{QEM for Higher-Order AD Effects}\label{extdQEMSsub2}
When a quantum state at a time $t_0$, which we denote by $\rho(t_0)$, is subject to the AD effect during the time interval $\Delta t$, 
according to the quantum master equation (1) in the main text the quantum state at the time $t = t_0 + \Delta t$ is given by
$\rho(t_0 + \Delta t) = \rho(t_0 ) +  \mathcal{L} _{ \text{AD}} [ \rho(t_0 ) ] \tau  +  \mathcal{L}^2 _{ \text{AD}} [ \rho(t_0 ) ]  \frac{\tau^2}{2!} + \mathcal{O}(\tau^3)$. 
Here  $ \mathcal{L}^2  _{ \text{AD}}$ describes the twice operation of  $ \mathcal{L}  _{ \text{AD}}$ and  it is equivalent to the second-order time derivate obeying the quantum master equation (1) in the main text.
In other words, we have solved the quantum master equation (1) in the main text up to the second order in $ \Delta t$.
Let us analyze the quantum state generated by the unitary operation $U^{ \text{QC}}= \prod_{k=1}^d  U_{ k} =U_{ d} \cdot U_{ d-1} \cdots U_{ 2} \cdot U_{ 1}$ under the influence of the AD effect and write it by $ \rho (T)$.
As similar to the second line of the right-hand side of Eq.  (2) in the main text, up to $\mathcal{O}(\tau^2)$ the density matrix $ \rho (T)$ is calculated as
\begin{align}
 \rho (T) 
  = \rho_{d\cdots1} + \tau \Delta^{\text{AD}}_1 \rho_{d\cdots1}  +  \frac{\tau^2}{2!}  \Delta^{\text{AD}}_2 \rho_{d\cdots1}  + \mathcal{O}(\tau^3),
 \label{ADDMsecondorder}
\end{align}
where $\Delta^{\text{AD}}_2 \rho_{d\cdots1} $ denotes the theoretically-evaluated second-order AD effect associated with the unitary operation $U^{ \text{QC}}$ and is given by 
\begin{widetext}  \begin{align} 
 \Delta_2 \rho  _{d\cdots1} & = 
 \sum_{k=1}^{d} \sum_{j_1, j_2=0}^{N_q-1}  \left( \prod_{l=k+1}^d U_l \right) \mathcal{L}  ^{\text{AD}}_{j_1} \left[    \mathcal{L}  ^{\text{AD}}_{j_2} [\rho  _{k\cdots1} ]  \right]  \left( \prod_{l=k+1}^d U_l \right)^\dagger \notag\\
&+2 
\sum_{k_1=2}^{d-1}  \sum_{k_2=1}^{k_1-1}
 \sum_{j_1, j_2=0}^{N_q-1}  
  \left( \prod_{l_1=k_1+1}^{d} U_{l_1} \right)
\mathcal{L}  ^{\text{AD}}_{j_1} \left[   \left( \prod_{l_2=k_2+1}^{k_1} U_{l_2} \right)
\mathcal{L}  ^{\text{AD}}_{j_2} [ \rho  _{k_2\cdots1} ]
 \left( \prod_{l_2=k_2+1}^{k_1} U_{l_2} \right)^\dagger
\right]   \left( \prod_{l_1=k_1+1}^{d} U_{l_1} \right)^\dagger. \label{appendixsecondorderADeffect1}
  \end{align} \end{widetext}
The term in the first line of the right-hand side corresponds to the second-order derivative.
By using Eq.  (9) in the main text, $ \Delta_2 \rho  _{d\cdots1} $ in Eq.  \eqref{appendixsecondorderADeffect1} is  represented by the operators $\{ \boldsymbol{1}_{2\times2}, Z, \tilde{\sigma}^\pm, P^1\}$ as
\begin{widetext}  \begin{align} 
 \Delta_2 \rho  _{d\cdots1} & = \sum_{k=1}^{d} \sum_{j_1,j_2}^{N_q-1}  \sum_{p_1, p_2} c_{p_1} c_{p_2}  \left( \prod_{l=k+1}^d U_l  \right) s_{Q_{j_1},p_1} s_{Q_{j_2},p_2}  \rho_{k\cdots1} s_{Q_{j_2},p_2}^\dagger   s_{Q_{j_1},p_1} ^\dagger
  \left( \prod_{l=k+1}^d U_l  \right)^\dagger  \notag\\
&+ 2
 \sum_{k_1=2}^{d}  \sum_{k_2=1}^{k_1-1}
 \sum_{j_1,j_2}^{N_q-1}  \sum_{p_1, p_2} c_{p_1} c_{p_2} 
  \left( \prod_{l_1=k_1+1}^{d} U_{l_1} \right) s_{Q_{j_1},p_1}  \left( \prod_{l_2=k_2+1}^{k_1} U_{l_2} \right) s_{Q_{j_2},p_2} 
\rho_{k_2\cdots1}   \notag\\
  & \times   s_{Q_{j_2},p_2} ^\dagger   \left( \prod_{l_2=k_2+1}^{k_1} U_{l_2} \right)^\dagger s_{Q_{j_1},p_1} ^\dagger  \left( \prod_{l_1=k_1+1}^{d} U_{l_1} \right)^\dagger,
   \label{appendixsecondorderADeffect2}
  \end{align} \end{widetext} 
where the subscripts $p_1$ and $ p_2$ are used for labeling the four operators $\{ \boldsymbol{1}_{2\times2}, Z,  \tilde{\sigma}^-, P^1\}$, and correspondingly,
$s_{Q_{j_a},p_a}$ $(a=1,2)$ is the operator acting on the qubit $Q_{j_a}$ and $s_{Q_{j_a},p_a} \in \{ \boldsymbol{1}_{2\times2, j_a}, Z_{j_a}, \tilde{\sigma}^-_{j_a}, P^1_{j_a} \}$.
For instance,  $s_{Q_{j_a}, Z}$ is the $Z$-gate operation acting on the qubit $Q_{j_a}.$
The coefficient $ c_{p_a} $ $(a=1,2)$ is assigned when the operator $s_{Q_{j_a},p_a}$ is acted on the qubit $Q_{j_a}.$
The explicit form of the coefficients $ c_{p_a} $ are $  \{ c_{\boldsymbol{1}_{2\times2}}, c_{Z}, c_{\tilde{\sigma}^-}, c_{P^1}\} = \{ -\frac{1}{4}, \frac{1}{4}, 1,-1\}$.
By representing the second-order effect  $\Delta_2 \rho  _{d\cdots1}$ in the way given in Eq. \eqref{appendixsecondorderADeffect2},
 we can programmably evaluate $\Delta_2 \rho  _{d\cdots1}$ by the unitary operators $U_l$ $(l=1,\ldots,d)$ which compose the quantum algorithm under consideration,  the additional operations $s_{Q_{j_a},p_a} \in \{ \boldsymbol{1}_{2\times2}, Z, \tilde{\sigma}^-, P^1\}$, and the coefficients $ c_{p_a} \in  \{ -\frac{1}{4}, \frac{1}{4}, 1,-1\}$. In other words, we can create the AD-effect (quantum-noise-effect) circuit group for evaluating $\Delta_2 \rho  _{d\cdots1}$.
Note that the quantum circuits for evaluating the terms in the first and second lines of Eq.  \eqref{appendixsecondorderADeffect2} is equivalent to the AD-effect-circuit group for evaluating $\Delta_1 \rho  _{d\cdots1}$
for $j_1=j_2.$  
By using Eq. (8) in the main text, we can derive the formula of QEM for the second-order AD effect and is given by
 \begin{widetext}   \begin{align}  
 \langle \hat{O}  \rangle^{\text{QEM}}_{\rho  _{d\cdots1}} & \equiv 
\langle \hat{O}  \rangle_{\rho^{\text{real}}  _{d\cdots1}} - \tau  \langle \hat{O}  \rangle_{ (\Delta^{\text{AD}}_1 \rho_{d\cdots1})^{\text{real}} } 
-   \frac{ \tau^2}{2!}   \langle \hat{O}  \rangle_{ (\Delta^{\text{AD}}_1 \rho_{d\cdots1})^{\text{real}} }  +\tau^2    \langle \hat{O}  \rangle_{ \Delta^{\text{AD}}_1 \big{(}(\Delta^{\text{AD}}_1 \rho_{d\cdots1})\big{)}^{\text{real}} }       \notag\\
& =   \langle \hat{O}  \rangle_{\rho _{d\cdots1}} 
+ \tau  \Big{(}   \langle \hat{O}  \rangle _{ \delta^{\text{AD}}_1 ( \rho  _{d\cdots1} )} - \langle \hat{O}  \rangle_{ \Delta^{\text{AD}}_1 \rho_{d\cdots1} }   \Big{)} \notag\\ 
&   +  \frac{ \tau^2}{2!}               \Big{(}   \langle \hat{O}  \rangle _{ \delta^{\text{AD}}_2 ( \rho  _{d\cdots1} )} - \langle \hat{O}  \rangle_{ \Delta^{\text{AD}}_2 \rho_{d\cdots1} }   \Big{)}
- \tau^2  \Big{(}   
\langle \hat{O}  \rangle_{ \delta^{\text{AD}}_1 \big{(}(\Delta^{\text{AD}}_1 \rho_{d\cdots1})\big{)} } - \langle \hat{O}  \rangle_{ \Delta^{\text{AD}}_1 \big{(}(\Delta^{\text{AD}}_1 \rho_{d\cdots1})\big{)} }
  \Big{)} + \mathcal{O}(\tau^3). \label{appendixQEMformula1}
\end{align}   \end{widetext} 
The first term in the second line of the right-hand side of the above equation is the ideal expectation value while the second term describes the QEM for the first-order AD effect; compare with Eq. (8) in the main text.
On the other hand, the terms in the third line of the right-hand side describes the conduction of the QEM for the second-order AD effect.  
The first term describes the QEM for the intrinsic second-order AD effect while the second term describes the mitigation for the first-order AD effect $\delta^{\text{AD}}_1 \big{(}(\Delta^{\text{AD}}_1 \rho_{d\cdots1})\big{)}$
and such an error has been neglected in the first-order perturbation theory. 
To calculate $\Delta^{\text{AD}}_1 \big{(}(\Delta^{\text{AD}}_1 \rho_{d\cdots1})\big{)},$ first we need to rewrite the terms in $\Delta^{\text{AD}}_1 \rho_{d\cdots1}$ which are expressed by $U_k$ and the single-time additional operations of $\tilde{\sigma}^-$ and $P^1$ as 
the terms given by the unitary operations $U_k\otimes \boldsymbol{1}_{Q_\text{a}}$ and the gate operations composing the AD circuits A and B, respectively: see also Fig. 3 in the main text. 
Then we calculate $\Delta^{\text{AD}}_1 \big{(}(\Delta^{\text{AD}}_1 \rho_{d\cdots1})\big{)}$ which is represented in the way that the operations $\{Z,\tilde{\sigma}^-,P^1\}$ are added for the series of unitary operations given by $U_k\otimes \boldsymbol{1}_{Q_\text{a}}$,
$Z$ gate, and the quantum gates comprising the AD circuits A and B.   
Consequently, we have derived the QEM for the second-order AD effect and is represented by Eq. \eqref{appendixQEMformula1}.
 
The above argument can be extended into the formulations of QEM schemes for high-order AD effects.  
By using Eq.  (9) in the main text, we can evaluate the higher-order AD effects $\Delta^{\text{AD}}_p \rho_{d\cdots1}$ $(p \geq3)$
and represented them in programmable ways in terms of the operators $\{ \boldsymbol{1}_{2\times2}, Z, \tilde{\sigma}^-, P^1\}$ like Eq. \eqref{appendixsecondorderADeffect2}
and create $p$-th-order quantum-noise circuit groups. Then we can derive the QEM formula for the higher-order AD effects $\delta^{\text{AD}}_p \rho_{d\cdots1}$.
In order to do this, we need to evaluate not only $\Delta^{\text{AD}}_p \rho_{d\cdots1}$ but also $ \Delta^{\text{AD}}_k \big{(}\Delta^{\text{AD}}_{p-k} \rho_{d\cdots1}\big{)}$ such that $1 \leq k < p$,
which is coming from the conduction of QEM for $p-k$-th order AD effects $\delta^{\text{AD}}_k \rho_{d\cdots1} \big{(}\Delta^{\text{AD}}_{p-k} \rho_{d\cdots1}\big{)}$.
Furthermore, by applying the argument given above we can construct QEM formulas for higher-order effects of GAD, PD, their mixture, and the stochastic Pauli noises.

For evaluating the second-order effect $\Delta^{\text{AD}}_2 \rho_{d\cdots1}$, we perform the additional operators $\{Z,\tilde{\sigma}^-,P^1\}$ twice as shown in Eq. (\ref{appendixsecondorderADeffect2})
and the number of additional quantum circuits to create a second-order AD-effect quantum circuit group is $\mathcal{O}(\{dN_q\}^2)$ and maximally we need two ancilla bits. 
Similarly, in the case of $p$-th order AD-effect circuit groups, 
we need to perform $p$ additional operations of $\{Z,\tilde{\sigma}^\pm,P^0,P^1\}$, and correspondingly, 
we obtain the additional quantum circuits whose number is $\mathcal{O}(\{dN_q\}^p)$ and we need maximally $p$ ancilla bits.

\section{Quantum Circuit for $O_{\text{QAA}}$}\label{appendix2}  
\begin{figure*}[!htb] 
\centering
\includegraphics[width=0.7 \textwidth]{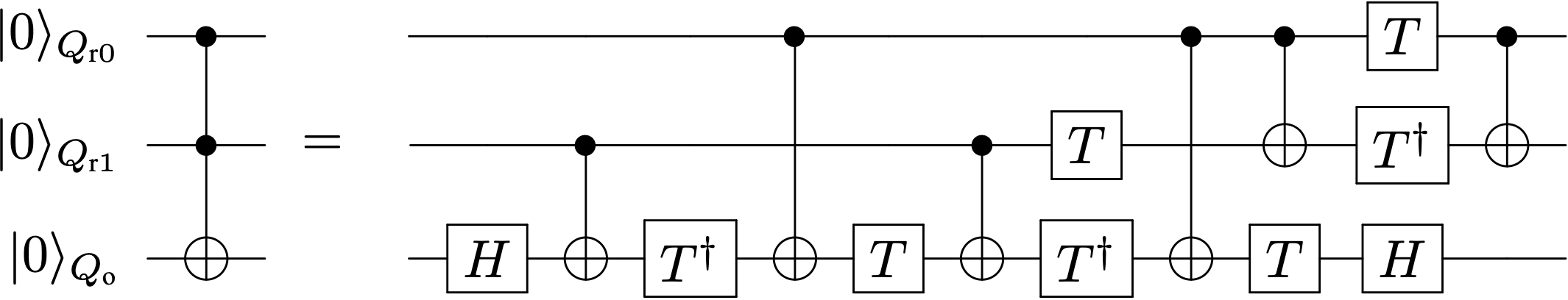}
\caption{ Quantum circuit for $ O^{ \text{QAA}}$ in Eq. (14). }
\label{toffoliQcirc}
\end{figure*}

In Fig. \ref{toffoliQcirc}, we show the quantum circuit for $O_{\text{QAA}}$ [33]
\section{Optimized Variational Parameters of QAOA}\label{appendix3}  
We list the optimized numerical values of the variational parameters of  QAOA which are given by
  \begin{align}
\vartheta^{\text{QAOA}}_1  &= 2.023075,\notag\\
\vartheta^{\text{QAOA}}_2  &= 2.130055,\notag\\
\varphi^{\text{QAOA}}_1    &= 1.011537,\notag\\
\varphi^{\text{QAOA}}_2    &= 1.118518.   \label{QAOAvariationalpara}
\end{align}

\section{Measurement Data of $T_1$ and $T_2$ Times}\label{SMT1T2data}
\begin{figure*}[!t]  
\centering
\includegraphics[width=0.9 \textwidth]{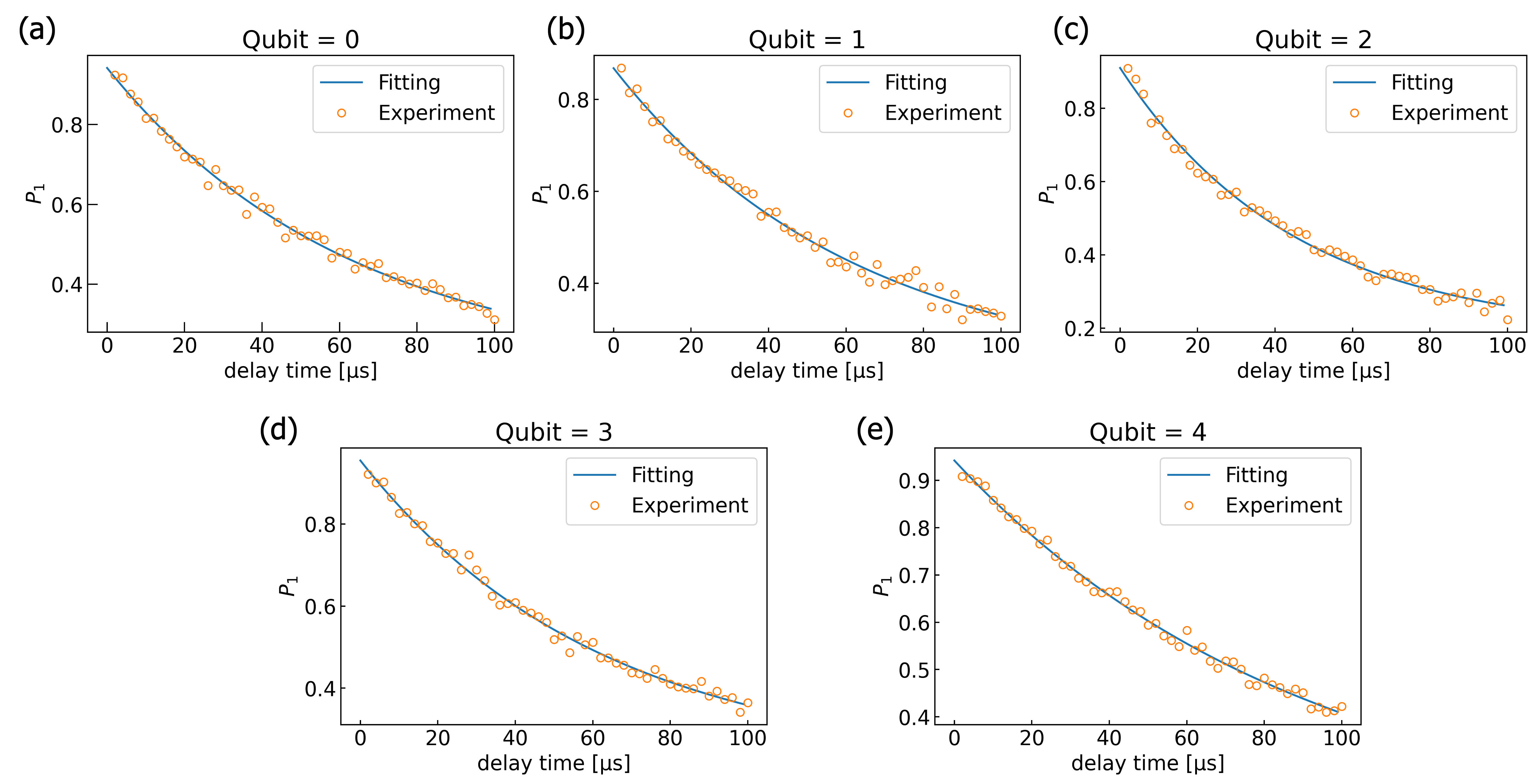}
\caption{ Experimental data of the expectation values $\langle P^1 \rangle(t_a^{\text{tot},T_1})$ for the qubits $Q_j$ ($j=0,\ldots,4$) in ibmq$\_$belem. 
The data was taken at 12:33, 09/11/2023.}
\label{T1datapointsibmqbelem}
\end{figure*}
\begin{figure*}[!t]  
\centering
\includegraphics[width=0.9 \textwidth]{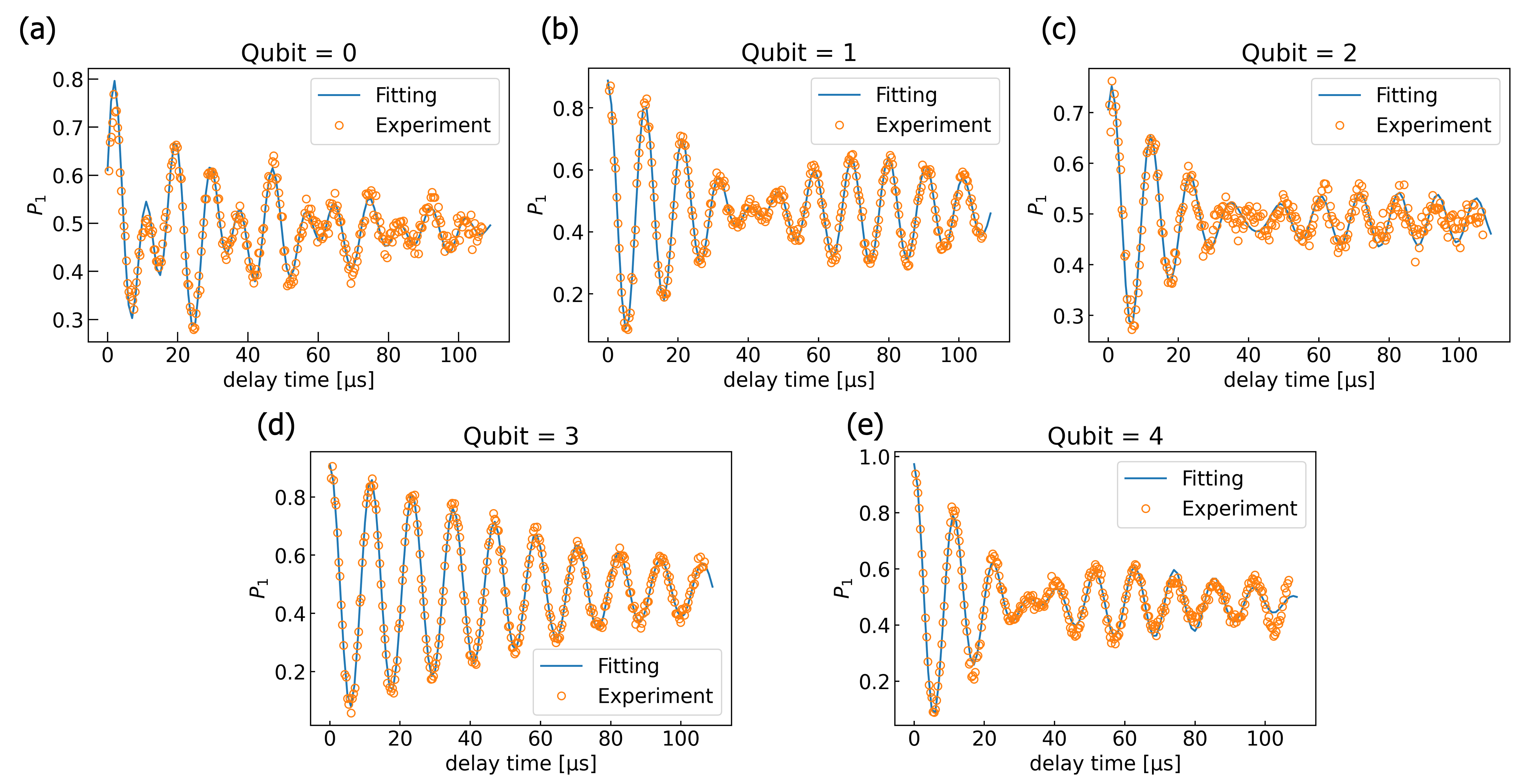}
\caption{Experimental data of the expectation values $\langle P^1 \rangle(t_a^{\text{tot},T_2})$ for the qubits $Q_j$ ($j=0,\ldots,4$) in ibmq$\_$belem. The data was taken at 12:33, 09/11/2023.}
\label{T2datapointsibmqbelem}
\end{figure*}

In this section, we explain how to extract the experimental values of $T_1$ and $T_2$ times with showing our data plots, 
Fig. \ref{T1datapointsibmqbelem} ($T_1$ data points) and Fig. \ref{T2datapointsibmqbelem} ($T_2$ data points). The experiment is done by ibmq$\_$belem (five-qubit machine).
We note that for both cases the relaxation processes given by the relaxation times $t^\text{relax}_a$  were generated by using a function (command) called ``delay" \cite{IBMQExp}. 

Let us explain from the measurements of the $T_1$ times. In Fig. \ref{T1datapointsibmqbelem}, we plot the experimental data of the expectation values $\langle P^1 \rangle(t_a^{\text{tot},T_1})$ for the qubits $Q_j$ ($j=0,\ldots,4$). Here we have taken $t_a^\text{relax} = 2 \times a$ ($\mu$s) with $a=1,\ldots, 50$  and $N_\text{QC}=2^{10}.$ We fit these data points by an exponential function $ \langle P^1 \rangle(t_a^{\text{tot},T_1}) = \exp\left( -\frac{t_a^{\text{tot},T_1}}{T_1}\right)$ and extract the values of $T_1$. In Table  \ref{T1andT2datatableibmqbelem} we list these experimental values. 
In addition,  we show the data which are open to the public in the parentheses \cite{IBMQExp}. As a result, our experimental data agrees well with them.
\begin{table}[htbp]
\centering 
  \begin{tabular}{cc||c|c|}  \hline
    Qubit & & $T_1$ ($\mu$s) & $T_2$ ($\mu$s) \\ \hline
    $Q_0$ & & 62.93 (109.91) & 50.00 (178.67)  \\ \hline
    $Q_1$ & & 62.31 (81.12) & 89.38 (95.21)  \\ \hline
    $Q_2$ & & 45.03 (79.47) & 50.00 (60.13) \\ \hline
    $Q_3$ & & 61.95 (98.60) & 78.29 (148.77)  \\ \hline
    $Q_4$ & &  90.69 (93.07) & 50.00 (146.12)  \\ \hline
     \end{tabular}
    \caption{Experimental values of $T_1$ and $T_2$ times for ibmq$\_$belem.  The values in the parentheses are the data which are open to the public \cite{IBMQExp}. 
 Our measurement  data was taken at 12:33, 09/11/2023. The data open to the public were taken at 05:08, 09/11/2023.} 
   \label{T1andT2datatableibmqbelem}
\end{table}

Next, let us discuss  the measurement data of the $T_2$ times presented in Fig. \ref{T2datapointsibmqbelem}. 
Here we plot the experimental data points of  $\langle P^1 \rangle(t_a^{\text{tot},T_2})$ for the qubits $Q_j$ and we have taken $t_a^\text{relax} = \frac{16}{45}a$ ($\mu$s) with $a=1,\ldots, 300$ and $N_\text{QC}=2^{10}$. 
In this case, instead of a simple exponential decay curve, we fit the data points by a function
\begin{align}
f_{T_2}(t_a^{\text{tot},T_2}) = e^{-\frac{t_a^{\text{tot},T_2}}{T_2}}\left[
a_1\cos(2\pi f_1 t_a^{\text{tot},T_2} +\phi_1) + a_2\cos(2\pi f_2 t_a^{\text{tot},T_2} +\phi_2)
\right]+b,
\label{T2fittingfunction} 
\end{align}
where $a_{1,2},\phi_{1,2}$, and $b$ are real numbers. 
The reasons we fit  $\langle  P^1 \rangle(t_a^{\text{tot},T_2})$ with $f_{T_2}(t_a^{\text{tot},T_2}) $ in Eq. \eqref{T2fittingfunction} are the following. 
First, a detuning, which is a difference between the qubit frequency and a driving frequency (frequency of a pulse which generates a gate operation),  is not actually zero, 
and instead of an exponential decay the plots of an expectation value of $ P^1$ must be fitted by a damped oscillation curve \cite{IBMQExp}. 
Here we fit the the oscillation part by a cosine curve. When the detuning, however, is small it is not effective to fit the plots by the damped oscillation function since it is hard to distinguish whether the decreasing behaviors of the plots are coming from the exponential decay (decoherence) or from the oscillation. We avoid such an issue by enhancing the magnitude of the detuning explicitly. Thus, by writing the unitary operation for the measurement of  the $T_2$ times by $U^{\text{meas},T_2}$,
the actual series of the gate operations which have been exploited for the $T_2$  measurements are not the two Hadamard gates but $U^{\text{meas},T_2} = H\cdot \boldsymbol{I} \cdot R_z(\theta^\text{det}) \cdot H$ with $\theta^\text{det}=\frac{16a}{225}\pi$ ($a=1,\ldots, 300$).
The rotation operation $R_z(\theta^\text{det})$ plays the role of the enhancement of the detuning whereas $\boldsymbol{I}$ represents the relaxation part and is generated by the delay command as mentioned above.  
Second, we have examined that the fitting by a damped oscillation function with an oscillation part given by a single type of amplitude, frequency, and phase is not a good fitting, and instead,  using  the function $f_{T_2}(t_a^{\text{tot},T_2})$  in Eq. \eqref{T2fittingfunction}, which is given by two types of amplitudes, frequencies, and phases, shows much better fitting. We consider the physical reasoning for this as follows. In \cite{riste2013millisecond}, two different two-level systems having opposite charge parities induced by quasiparticle tunneling have been observed in a transmon qubit system,  and in such a case there exist two different frequencies of the two-level systems. 
Since the ibmq$\_$belem is a transmon qubit system \cite{perez2022quantum, volya2023state} we consider that the similar phenomenon has been observed in our experiment and we fitted the data points with the function $f_{T_2}(t_a^{\text{tot},T_2}) $ in Eq. \eqref{T2fittingfunction}.
  As a result, we obtain the values of  $T_{2,j}$. 
  In Table \ref{T1andT2datatableibmqbelem}, we display the experimental data of $T_2$ as well as the data open to the public \cite{IBMQExp} written in the parenthesis.
The $T_2$ times for $Q_1$ and $Q_2$ show good agreement whereas  those for the other qubits have moderate differences although the orders are the same.

\section{Quantum Device Properties}\label{QDprop}
\begin{figure}[t] 
\centering
\includegraphics[width=0.75 \textwidth]{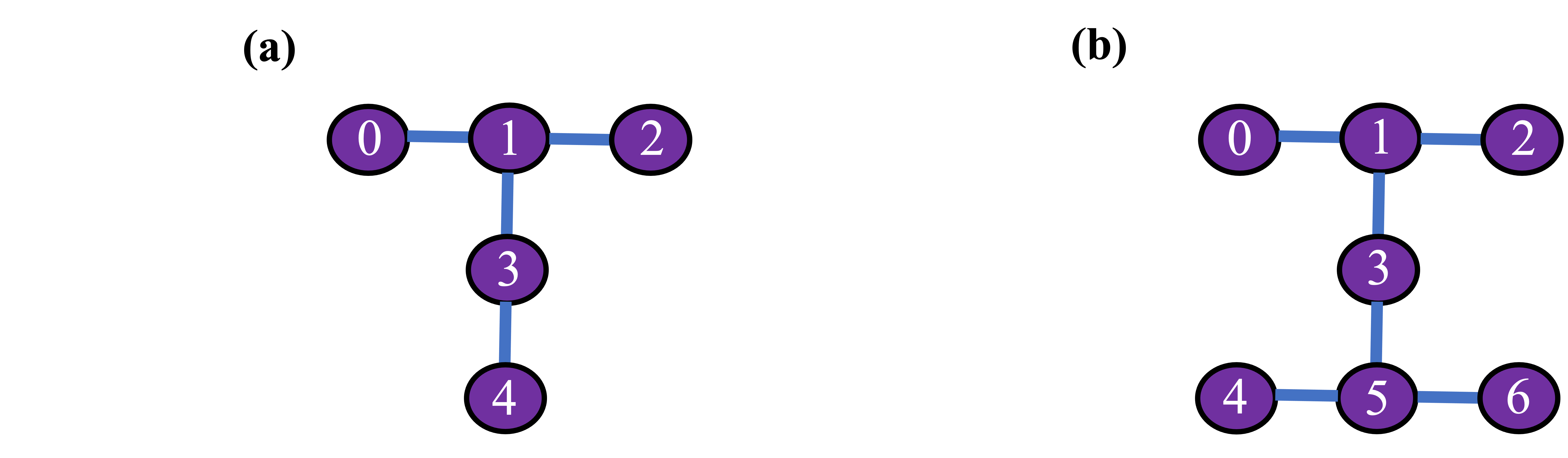}
\caption{Illustration of the spatial configuration for qubits  for  (a) ibmq$\_$belem and (b) ibm$\_$perth.} 
\label{IBM_5and7Qmachines}
\end{figure} 
In this section, we present the data of the physical properties of  ibmq$\_$belem and ibm$\_$perth (seven-qubit machine) which is open to the public  \cite{IBMQExp}:
In Fig. \ref{IBM_5and7Qmachines}, we display illustrations of spatial configurations for qubits in these machines.  
First, we list the single-qubit  properties of  ibmq$\_$belem, qubit frequencies  and gate operation times in Table \ref{singleibmqbelem}.
Here $t_{id},t_{Rz},t_{\sqrt{X}},$ and $t_X$ denote the gate times of identity gate, virtual $Z$ gate, $\sqrt{X}$ gate, and $X$ gate, respectively.  
In addition, we present the two-qubit gate properties, CNOT operation times denoted by $t_{CX}$.
The symbol  $\big{[}Q_i,Q_j\big{]}$ ($i,j=0,1,2,3,4$ with  $i\neq j$) represent nearest-neighboring qubits $Q_i$ and $Q_j$.  
The gate operation times in Table \ref{singleibmqbelem} and Table \ref{twoibmqbelem} are used for the implementation of our QEM for the quantum algorithm $U^{ \text{QC}}_{ \text{imp},2}$: see the caption of Fig. 22 in the main text.
\begin{table}[htbp]
  \begin{minipage}[c]{0.5\hsize}
    \centering
   \begin{tabular}{cc||c|c|c|c|c|}  \hline
    Qubit && $\omega_{01}$ (GHz) & $t_{id}$ (ns) &$t_{Rz}$ (ns)  & $t_{\sqrt{X}}$ (ns) & $t_X$ (ns)  \\ \hline
   $Q_0$   && 5.09 & 35.56 & 35.56 & 35.56 & 35.56  \\ \hline
  $Q_1$  && 5.25 & 35.56 & 35.56 & 35.56 & 35.56 \\ \hline
  $Q_2$   && 5.36  & 35.56 & 35.56 & 35.56 & 35.56 \\ \hline
  $Q_3$  && 5.17  & 35.56 & 35.56 & 35.56 & 35.56 \\ \hline
  $Q_4$  && 5.26 & 35.56 & 35.56 & 35.56 & 35.56  \\ \hline
     \end{tabular}
    \caption{List of the single-qubit properties of ibmq$\_$belem which are open to the public \cite{IBMQExp}.    The above data was taken at 05:08, 09/11/2023.} 
    \label{singleibmqbelem}
  \end{minipage}
  \begin{minipage}[c]{0.5\hsize}
    \centering
    \begin{tabular}{cc||c|}  \hline
    Qubits &&   $t_{CX}$ (ns)   \\ \hline
   $\big{[}Q_0,Q_1\big{]}$ ($\big{[}Q_1,Q_0\big{]}$)  && 810.67 (775.11)   \\ \hline
  $\big{[}Q_1,Q_2\big{]}$ ($\big{[}Q_2,Q_1\big{]}$) && 419.56 (384.00) \\ \hline
  $\big{[}Q_1,Q_3\big{]}$  ($\big{[}Q_3,Q_1\big{]}$)   && 440.89 (405.33)  \\ \hline
  $\big{[}Q_3,Q_4\big{]}$  ($\big{[}Q_4,Q_3\big{]}$)   && 526.22 (490.67)  \\ \hline
     \end{tabular}
    \caption{List of the two-qubit properties of ibmq$\_$belem which are open to the public \cite{IBMQExp}.    The above data was taken at 05:08,09/11/2023.}
    \label{twoibmqbelem}
  \end{minipage}
\end{table}
Similarly, in Table  \ref{singleibmperth09212023}-\ref{twoibmperth09262023} we list the physical properties of ibm$\_$perth: 
Table \ref{singleibmperth09212023} and Table \ref{singleibmperth09262023} present the data of the single-qubit properties while 
Table \ref{twoibmperth09212023}  and Table \ref{twoibmperth09262023} show those of the two-qubit properties. 
We have used the data of the gate operation times taken in 09/26/2023 for the implementation of our QEM for the quantum algorithm $U^{ \text{QC}}_{ \text{pre}1}$ (Table \ref{singleibmperth09262023} and Table \ref{twoibmperth09262023})
while we have used the ones taken in 09/21/2023 for  $U^\text{QC}_\text{imp,2}$ (Table \ref{singleibmperth09212023} and Table \ref{twoibmperth09212023}): see also the captions of Fig. 20 and Fig. 21 in Sec. VB in the main text. 
\begin{table}[htbp]
  \begin{minipage}[c]{0.5\hsize}
    \centering
   \begin{tabular}{cc||c|c|c|c|c|}  \hline
    Qubit && $\omega_{01}$ (GHz) & $t_{id}$ (ns) &$t_{Rz}$ (ns)  & $t_{\sqrt{X}}$ (ns) & $t_X$ (ns)  \\ \hline
   $Q_0$   && 5.16 & 35.56 & 35.56 & 35.56 & 35.56  \\ \hline
  $Q_1$  && 5.03 & 35.56 & 35.56 & 35.56 & 35.56 \\ \hline
  $Q_2$   && 4.86  & 35.56 & 35.56 & 35.56 & 35.56 \\ \hline
  $Q_3$  && 5.13  & 35.56 & 35.56 & 35.56 & 35.56 \\ \hline
  $Q_4$  && 5.16 & 35.56 & 35.56 & 35.56 & 35.56  \\ \hline
  $Q_5$  && 4.98  & 35.56 & 35.56 & 35.56 & 35.56 \\ \hline
  $Q_6$  && 5.16 & 35.56 & 35.56 & 35.56 & 35.56  \\ \hline
     \end{tabular}
    \caption{List of the single-qubit properties of ibm$\_$perth which are open to the public \cite{IBMQExp}.    The above data was taken at 07:32,  09/21/2023.} 
    \label{singleibmperth09212023}
  \end{minipage}
  \begin{minipage}[c]{0.5\hsize}
    \centering
   \begin{tabular}{cc||c|c|c|c|c|}  \hline
    Qubit && $\omega_{01}$ (GHz) & $t_{id}$ (ns) &$t_{Rz}$ (ns)  & $t_{\sqrt{X}}$ (ns) & $t_X$ (ns)  \\ \hline
   $Q_0$   && 5.16 & 35.56 & 35.56 & 35.56 & 35.56  \\ \hline
  $Q_1$  && 5.03 & 35.56 & 35.56 & 35.56 & 35.56 \\ \hline
  $Q_2$   && 4.86  & 35.56 & 35.56 & 35.56 & 35.56 \\ \hline
  $Q_3$  && 5.13  & 35.56 & 35.56 & 35.56 & 35.56 \\ \hline
  $Q_4$  && 5.16 & 35.56 & 35.56 & 35.56 & 35.56  \\ \hline
  $Q_5$  && 4.98  & 35.56 & 35.56 & 35.56 & 35.56 \\ \hline
  $Q_6$  && 5.16 & 35.56 & 35.56 & 35.56 & 35.56  \\ \hline
     \end{tabular}
    \caption{List of the single-qubit properties of ibm$\_$perth which are open to the public \cite{IBMQExp}.    The above data was taken at 04:30,  09/26/2023.} 
    \label{singleibmperth09262023}
  \end{minipage}
  \begin{minipage}[c]{0.5\hsize}
    \centering
    \begin{tabular}{cc||c|}  \hline
    Qubits &&   $t_{CX}$ (ns)   \\ \hline
   $\big{[}Q_0,Q_1\big{]}$ ($\big{[}Q_1,Q_0\big{]}$)  && 391.11 (426.67)   \\ \hline
  $\big{[}Q_1,Q_2\big{]}$ ($\big{[}Q_2,Q_1\big{]}$) && 640.00 (604.44) \\ \hline
  $\big{[}Q_1,Q_3\big{]}$  ($\big{[}Q_3,Q_1\big{]}$)   && 369.78 (334.22)  \\ \hline
  $\big{[}Q_3,Q_5\big{]}$  ($\big{[}Q_5,Q_3\big{]}$)   && 284.44 (320.00)  \\ \hline
  $\big{[}Q_4,Q_5\big{]}$  ($\big{[}Q_5,Q_4\big{]}$)   && 590.22 (625.78)  \\ \hline
   $\big{[}Q_5,Q_6\big{]}$  ($\big{[}Q_6,Q_5\big{]}$)   && 640.00 (604.44)  \\ \hline
     \end{tabular}
    \caption{List of the two-qubit properties of ibm$\_$perth which are open to the public \cite{IBMQExp}.    The above data was taken at 07:32,  09/21/2023.}
    \label{twoibmperth09212023}
  \end{minipage}
  \begin{minipage}[c]{0.5\hsize}
    \centering
    \begin{tabular}{cc||c|}  \hline
    Qubits &&   $t_{CX}$ (ns)   \\ \hline
   $\big{[}Q_0,Q_1\big{]}$ ($\big{[}Q_1,Q_0\big{]}$)  && 391.11 (426.67)   \\ \hline
  $\big{[}Q_1,Q_2\big{]}$ ($\big{[}Q_2,Q_1\big{]}$) && 640.00 (604.44) \\ \hline
  $\big{[}Q_1,Q_3\big{]}$  ($\big{[}Q_3,Q_1\big{]}$)   && 369.78 (334.22)  \\ \hline
  $\big{[}Q_3,Q_5\big{]}$  ($\big{[}Q_5,Q_3\big{]}$)   && 284.44 (320.00)  \\ \hline
  $\big{[}Q_4,Q_5\big{]}$  ($\big{[}Q_5,Q_4\big{]}$)   && 590.22 (625.78)  \\ \hline
   $\big{[}Q_5,Q_6\big{]}$  ($\big{[}Q_6,Q_5\big{]}$)   && 640.00 (604.44)  \\ \hline
     \end{tabular}
    \caption{List of the two-qubit properties of ibm$\_$perth which are open to the public \cite{IBMQExp}.    The above data was taken at 04:30,  09/26/2023.}
    \label{twoibmperth09262023}
  \end{minipage}
\end{table}
%

\section{References}
\bibliographystyle{apsrev4-2} 
\bibliography{cleanQEMlistref.bib}